\documentclass{aa}  
\usepackage{natbib,graphicx}
\usepackage{txfonts}
\usepackage{bm}
\usepackage{lmodern}

\usepackage{color}
\usepackage{txfonts}
\usepackage{appendix}
\usepackage{textcomp}
 \usepackage{array}  		
\usepackage{subfig}
\usepackage{amsmath}
\usepackage{amssymb}
\usepackage{hyperref}
\usepackage[final]{pdfpages}
\usepackage{media9}

\hypersetup{
  colorlinks=true,
  urlcolor=blue,
  citecolor= blue,
  linkcolor = blue
}

\begin{document}
\label{firstpage}

\title{Dark matter-baryon scaling relations from Einasto halo fits \\ to SPARC galaxy rotation curves}
\setcounter{footnote}{4}
\author{
Amir Ghari \inst{1,2} \and 
Benoit Famaey \inst{1} \and 
Chervin Laporte \inst{3}\thanks{CITA National Fellow} \and
Hosein Haghi \inst{2,4}
}

\institute{$^1$Universit\'e de Strasbourg, CNRS, UMR 7550, Observatoire astronomique de Strasbourg, Strasbourg, France
\\$^2$Department of Physics, Institute for Advanced Studies in Basic Sciences, 11365-9161, Zanjan, Iran
\\$^3$Departement of Physics and Astronomy, University of Victoria, Victoria, BC, V8P 1A1, Canada
\\$^4$Helmholtz-Institut für Strahlen- und Kernphysik (HISKP), Universität Bonn, Bonn, Germany
}

\date{Received xxx; accepted yyy}

\abstract
{Dark matter-baryon scaling relations in galaxies are important in order to constrain galaxy formation models. Here, we provide a modern quantitative assessment of those relations, by modelling the rotation curves of galaxies from the Spitzer Photometry and Accurate Rotation Curves (SPARC) database with the Einasto dark halo model. We focus in particular on the comparison between the original SPARC parameters, with constant mass-to-light ratios for bulges and disks, and the parameters for which galaxies follow the tightest radial acceleration relation. We show that fits are improved in the second case, and that the pure halo scaling relations also become tighter. We report that the density at the radius where the slope is $-2$ is strongly anticorrelated to this radius, and to the Einasto index. The latter is close to unity for a large number of galaxies, indicative of large cores. In terms of dark matter-baryon scalings, we focus on relations between the core properties and the extent of the baryonic component, which are relevant to the cusp-core transformation process. We report a positive correlation between the core size of halos with small Einasto index and the stellar disk scale-length, as well as between the averaged dark matter density within 2~kpc and the baryon-induced rotational velocity at that radius. This finding is related to the consequence of the radial acceleration relation on the diversity of rotation curve shapes, quantified by the rotational velocity at 2~kpc. While a tight radial acceleration relation slightly decreases the observed diversity compared to the original SPARC parameters, the diversity of baryon-induced accelerations at 2~kpc is sufficient to induce a large diversity, incompatible with current hydrodynamical simulations of galaxy formation, while maintaining a tight radial acceleration relation.}


\maketitle


\section{Introduction}

The nature of the dark sector of the Universe arguably represents one of the deepest mystery of modern physics. Despite earlier hints \citep[e.g.,][]{Zwicky33}, it is the flattening of rotation curves (RCs) of disk galaxies which provided the first clean observational evidence for mass discrepancies in galactic systems \citep{Bosma1978, Rubin1978}. Many different observational probes -- especially on cosmological scales \citep[e.g.,][]{Planck2016} -- then led to the development of the current standard cosmological framework ($\Lambda$CDM), in which one needs to add non-baryonic dark matter (DM) particles to the baryonic content of the Universe, which can thus in principle explain the RC shapes. 

However, while arguably very successful on large scales, the current $\Lambda$CDM picture is nevertheless, at face value, facing some challenges \citep[e.g.,][]{Bullock2017}, notably to explain the shapes of RCs in some galaxies.

From a theoretical perspective, the standard paradigm proposes that galaxies form and evolve in virialised haloes \citep[e.g.,][]{White78} of cold DM (CDM). Quantum fluctuations from the vacuum state create the seed perturbations in the early universe which are stretched by inflation, and because DM forms a collisionless and pressureless fluid, it is then able to collapse under gravity, providing the first potential wells in which the baryons later cool into. The evolution of the DM fluid is modelled by numerical simulations \citep{Davis1985}. DM-only (DMO) cosmological simulations treat 100 percent of the matter in the Universe as collisionless DM. CDM haloes simulated with DMO numerical models have density profiles that are well described to first order by the NFW profile \citep{nav1996} with a central cusp (${\rm d \, ln} \rho/{\rm d \, ln} r = -1$ in the center, where $\rho$ is the DM density), while observations rather point to large constant density cores of DM in the central parts of a large number (but not all) of rotationally-supported galaxies~\citep{deblok2001,Gentilecore2004}. This problem of DMO simulations becomes even more stringent when realizing that galaxies residing in halos of similar maximum circular velocity display a wide range of RC shapes in the central parts \citep{Oman2015}. This diversity of rotation curve shapes is to be contrasted with the uniformity found in the relation between the baryonic gravitational acceleration and the total gravitational acceleration \citep{RAR1, RAR2, RAR3, Desmond1, Desmond2}. 

Despite the NFW profile being widely used, one has to note that \citet{Navarro2004} and \citet{Maccio2008} have proposed another model that fits the density profiles of halos in DMO simulations better in the inner regions, a profile which had been previously introduced for the distribution of stellar light and mass in galaxies by \citet{Einasto1965}, and which does not harbour a central cusp but a slope ${\rm d \,ln} \rho/{\rm d \, ln} r \propto -r^{1/n}$ in the center, meaning that the slope is indeed zero very close to the nucleus. For halos of the order and below the typical mass of the Milky Way halo in DMO simulations, the Einasto index is $n \sim 6$. Nevertheless, as first shown by \citet{Chemin2011}, fitting such profiles to observed rotation curves often leads to values of $n$ in disagreement with DMO simulations, implying much larger cores in observations than in simulations, as expected from the previously known core-cusp issue. Here we reproduce such a study by using the latest and most up-to-date database on galaxy rotation curves, the Spitzer Photometry and Accurate Rotation Curves (SPARC) database compiled by \citet{Lelli2016}. 

Upon the completion of this manuscript, \citet{Lieinasto2018} presented a study on scaling relations between DM halo parameters and disc galaxy luminosities from the SPARC database. The authors chose two DM density profile parametrisations for their exploration: the Einasto profile, like us, and general ($\alpha,\beta, \gamma$) models \citep{Zhao96} with parameters motivated by empirical relations as functions of stellar-to-halo mass described in \cite{DiCintio2014} who analysed results from hydrodynamical cosmological simulations of galaxies drawn from the MAGICC project \citep{Brook2012,Stinson2013}. Using these models they fit their rotations curves through a Markov Chain Monte Carlo method, taking lognormal priors on near-infrared stellar mass-to-light ratios, Gaussian priors on the inclinations and distances as well as $\Lambda$CDM priors for the DM halos. The authors in particular showed that the density at the radius where the slope of the Einasto profiles becomes $-2$ displays no correlation with luminosity.  

Our work, while sharing many similarities in data and methodology with \citet{Lieinasto2018}, takes a different turn instead in examining the correlations of inferred DM halo parameters with the extent of the baryonic components. Such scaling relations are highly relevant to constraining the cusp-core transformation process. We also choose to fix the baryonic galactic parameters in two extreme cases which we propose to compare: (i) the original SPARC parameters with constant mass-to-light ratios for bulges and disks \citep{Lelli2016,RAR1}, and (ii) the parameters for which galaxies follow a very tight radial acceleration relation \citep[RAR, ][]{RAR3}. In Sec.~2 we briefly present the SPARC sample, while we present the form of the Einasto profile and our fitting procedure in Sec.~3. We then present our results on scaling relations in Sec.~4 and conclude in Sec.~5. 

\section{Data}

The SPARC database \citep{Lelli2016} comprises $\sim 200$  extended  HI rotation curves compiled from the literature, for which near-infrared photometry at 3.6 microns is available from Spitzer \citet{Spitzer} and \citet{SchombertMcGaugh2014}. It forms a representative sample of disk galaxies in the local Universe in terms of various properties such as luminosity, size and surface brightness. Here, we exclude from the sample the galaxies that have been given a bad quality label ($Q=3$) in the original database, leaving us with a sample of 160 galaxies.

The availability of near-infrared photometry allows one to fix in principle the stellar mass-to-light ratio as it is known that this quantity does not vary much in that band. Indeed, \citet{SchombertMcGaugh2014} constructed stellar population synthesis models of disk galaxies to show that the stellar mass-to-light ratio does not depend strongly on age, metallicity, or color, for a large range of models with different star formation histories. As a first step, and following \citet{RAR1}, we adopt here a mass-to-light ratio $\Upsilon_d = 0.5 \, {\rm M}_\odot/{\rm L}_\odot$ for disks and $\Upsilon_b = 0.7 \, {\rm M}_\odot/{\rm L}_\odot$ for bulges. In a subsequent paper, \citet[][hereafter L18a]{RAR3} adjusted, with well-chosen priors, the mass-to-light ratio $\Upsilon_d$ for disks and $\Upsilon_b$ for bulges, distances and inclinations for galaxies in the SPARC sample, showing that the radial acceleration relation (RAR) between the gravitational acceleration generated by baryons and the total gravitational acceleration could have a scatter as small as 0.057 dex, compatible with observational uncertainties on measured velocities themselves. As a second step, we adopt these parameters, such that, in both cases, only the DM halo parameters are left as free parameters. 

\section{The Einasto profile fits}

The Einasto halo model has been proposed by  \citet{Navarro2004} as a density profile for CDM halos in DMO $\Lambda$CDM simulations, and had been previously introduced for the distribution of stellar light and mass in early-type galaxies \citep{Einasto1965,Einasto1968,Einasto1969}. Its density profile reads:

\begin{equation}
\rho_{\rm E}(r) = \rho_{-2}\,\exp\left[-2n\,\left(\left ({r\over
    r_{-2}}\right)^{1/n}-1\right)\right] \ ,
\label{eq:rhoeinasto}
\end{equation}
where, $r_{-2}$ is the radius at which the density profile has a slope of $-2$, and $\rho_{-2}$ is the density at that radius. The third parameter, $n$, is the Einasto index and sets the general shape of the density profile. As shown in, e.g., \citet{Chemin2011}, increasing the Einasto index at fixed $r_{-2}$ and $\rho_{-2}$ increases and steepens the density  profile in the central part of the halo, making it more cuspy, the central density being given by $\rho_0 = \rho_{-2}\, \exp(2n)$. Hence $n \sim 6$ DMO halos are indeed cuspy, even though they display a tiny core very close to the center. Conversely, at fixed $n \sim 1$, the profile is more cored, and the characteristic radius $r_{-2}$ is a good proxy to give an idea of the core-size.

\citet{Chemin2011} have used for the first time the Einasto model for mass decompositions of RCs on a sample of spiral galaxies from the THINGS survey. They found that the RCs are significantly better fit with the Einasto halo than with either the isothermal or NFW halo models. Here, in the same spirit, we use the Einasto DM halo profile for the decomposition of rotation curves of the SPARC database. This will allow us to display up-to-date scaling relations for DM and baryons, complementary to the recent study of \citet{Lieinasto2018}.

For finding the structural parameters of the fitted DM halo model, we decompose the observed rotation curves into four components: the stellar disk and bulge, the gaseous disk and the DM halo. Only the (spherical) DM halo parameters are free in our fits. The observed rotation curves can be written as:
\begin{equation}
V_{\rm rot}^2=\Upsilon_d V_{d}^2+\Upsilon_b V_{b}^2+V_{\rm gas}^2+V_{\rm halo}^2
\end{equation} 
where $V_d$ is the rotation velocity of the stellar disk, and $V_{b}$ of the stellar bulge, both for a stellar mass-to-light ratio of one. Finally, $V_{\rm halo}$ is the rotation velocity of the DM halo which we are fitting to the data with three free parameters. We fix all the other galaxy parameters (including distance and inclination) (i) to the original SPARC values \citep{Lelli2016} with fixed mass-to-light ratios $\Upsilon_d = 0.5 \, {\rm M}_\odot/{\rm L}_\odot$ and $\Upsilon_b = 0.7 \, {\rm M}_\odot/{\rm L}_\odot$, which we call the ``MLconst'' case hereafter, and (ii) to the L18a galaxy parameters. 

\subsection{DMO-inspired fits with virial mass as only parameter}
For illustrative purposes, at first we fixed the Einasto index to $n=6$ and use the $c-M_{200}$ relation of DMO simulations in \citet{DuttonMaccio2014},
\begin{equation}
{\rm log}(c)=0.997-0.13 \times {\rm log}(\frac{M_{200}}{10^{12}h^{-1}}).
\end{equation}
Then, the virial mass $M_{200}$ is the only free parameter, which we can use to fit the last observed point of the rotation curve. Note that there is in fact non-negligible scatter in this DMO relation, as well as small variations of $n$, which we do not account for here. These fits are mainly made here for illustratory purpose, showing what typical rotation curve shape is expected at a given virial mass in the absence of any feedback from the baryons. As expected, this leads to halos that are generally too cuspy. We show this curve for the L18a case in the Appendix (Fig. \ref{figa1}) as the blue dashed curve. It is clear that many of these fits are not satisfactory, especially for low-mass galaxies. 

\subsection{Three-parameters fits}

\begin{figure*}
\begin{center}
\includegraphics[width=165mm]{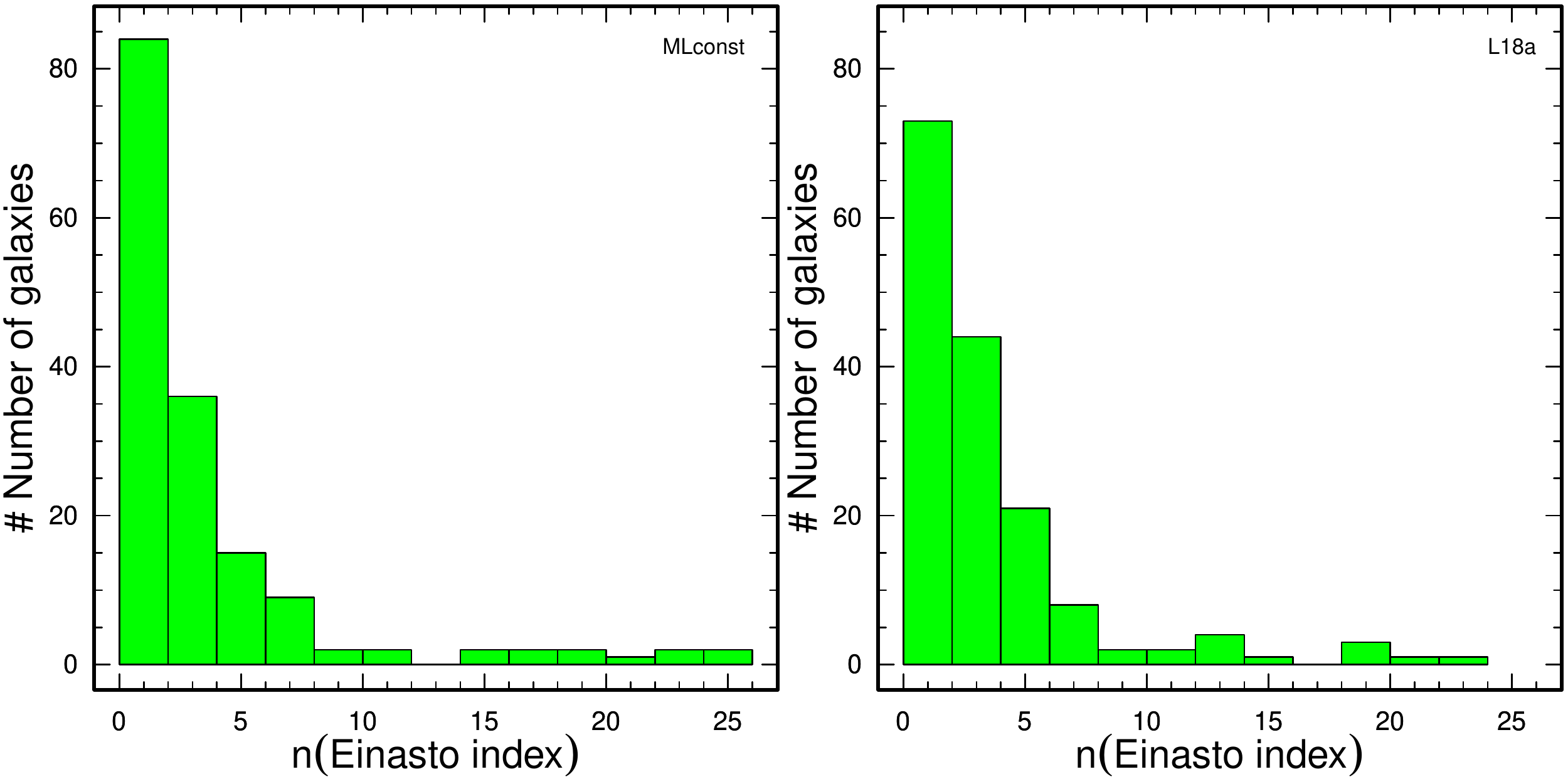}
  
\caption{Histogram of Einasto indices $n$ for the fits to the 160 galaxies with $Q<3$ in the SPARC sample. Left panel: original SPARC parameters with constant mass-to-light ratios for bulges and
disks \citep[``MLconst'',][]{Lelli2016, RAR1}. Right panel: parameters from \citet[][L18a]{RAR3}.}\label{hist_n}
\end{center}
\end{figure*}

\begin{figure*}
\begin{center}
\includegraphics[width=170mm]{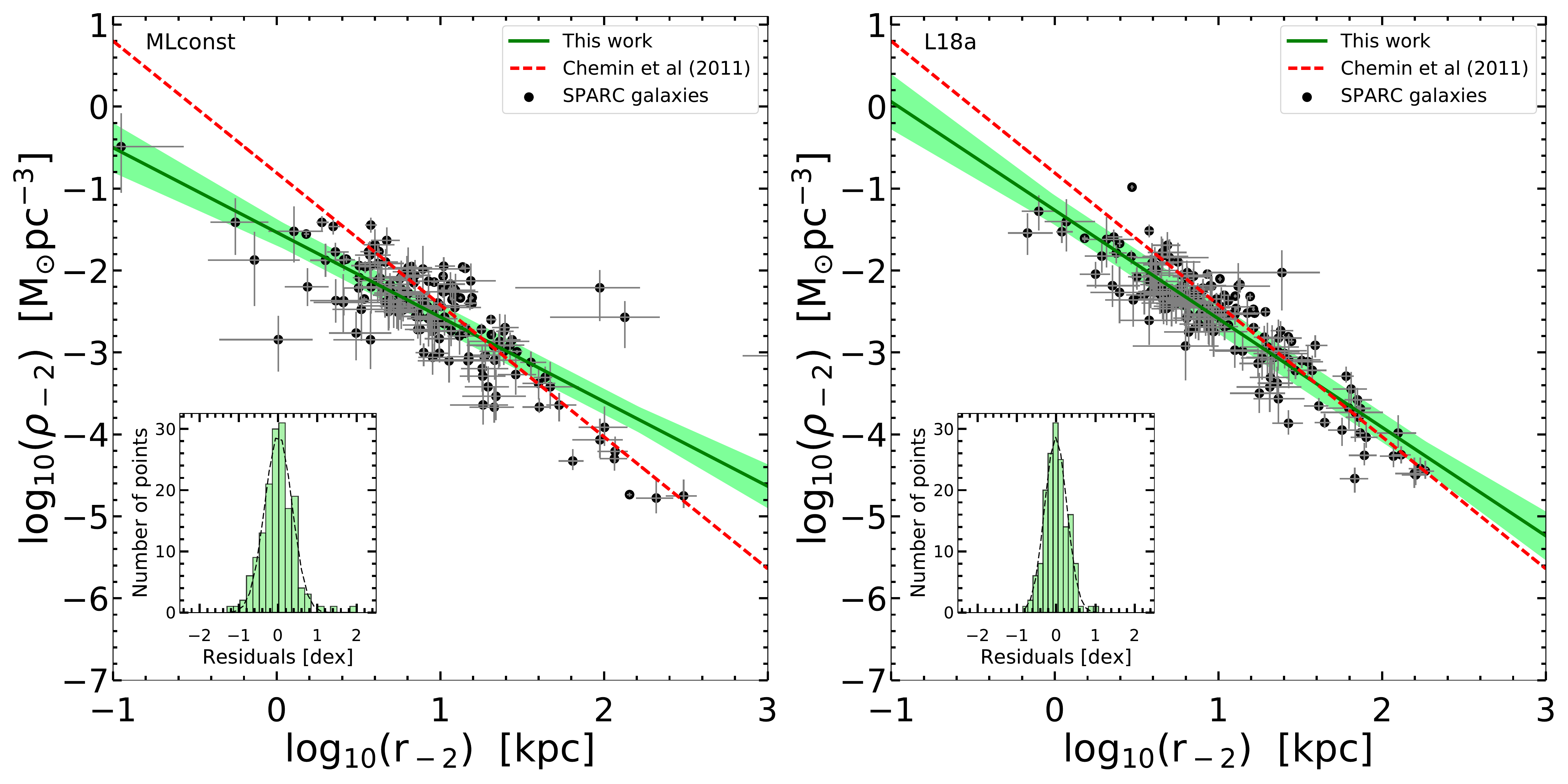}
  
\caption{The anti-correlation between the characteristic halo radius, $r_{-2}$, and the characteristic halo density, $\rho_{-2}$ for the 160 galaxies of the SPARC sample (left: ``MLconst''; right: L18a). In this plot the black dots with error bars are the best fit parameters derived from fitting the rotation curves, the red doted line is the line of \citet{Chemin2011}, and the green solid line is the best fit line to the data. The green shaded area represents the 1 sigma uncertainty on this best fit line (confidence interval, which does not represent the intrinsic scatter around the relation).}\label{scaling_relation}
\end{center}
\end{figure*}

We then use the Einasto DM halo profile from Eq.~\ref{eq:rhoeinasto} with all three parameters being free in the fit. We  use the affine Invariant Markov chain Monte Carlo Ensemble sampler, from the open source Python package, emcee \citep{Foreman-Mackey2013} to fit the observational velocity curve to the theoretical model by finding the peak of the likelihood function. Here, we do not use any $\Lambda$CDM inspired prior, and thus simply assume a flat linear prior for the three free parameters. The MCMC sampler then estimates the posterior probability distribution for all these parameters. 

The best fit values of the parameters $\rho_{-2}$, $r_{-2}$ and $n$ are chosen at the maximum likelihood for each galaxy. We can then compare the quality of fits in the ``MLconst'' and L18a cases, by comparing the global likelihoods. Interestingly, the ``MLconst'' case, which is most often used in the literature to compare simulations to data, gives less good fits than the L18a case, despite the L18a marginalization making absolutely no assumption on the dark matter halo parameters themselves. The global likelihood ratio is of the order of 3 -- and there are 15 galaxies with a likelihood ratio higher than 10 -- in favour of the L18a case, which is not highly significant, but still indicates a slight global preference for the L18a parameters. 

The results of the best fit parameters in the L18a case, together with their associated uncertainties, for all 160 galaxies are listed in Table~\ref{rotation_curve_result} in the Appendix. The rotation curves of these galaxies are shown in Fig.~\ref{figa1}. 

\section{Scaling relations}

\begin{figure*}
\begin{center}
\includegraphics[width=170mm]{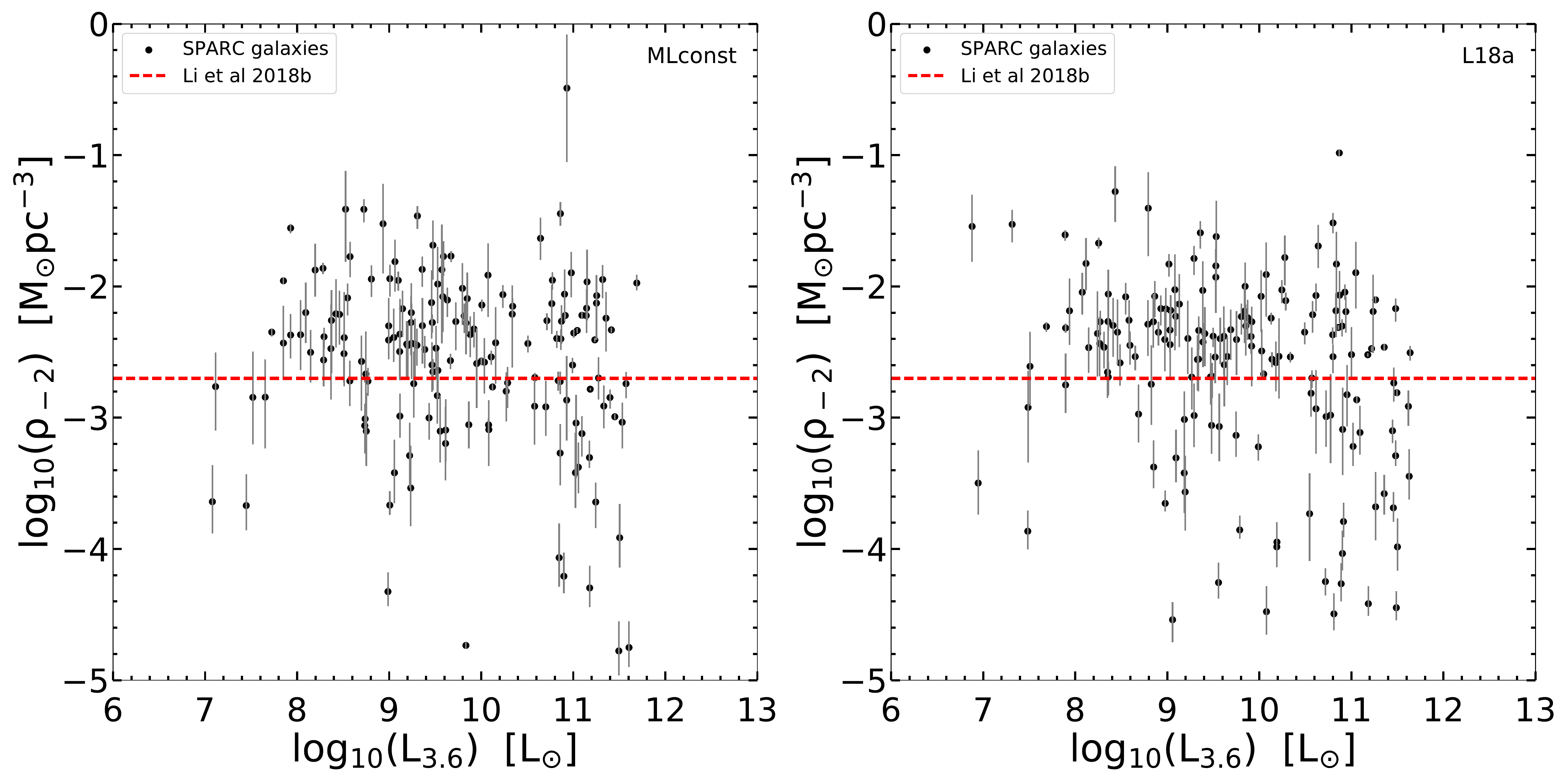}\hfill 
\includegraphics[width=170mm]{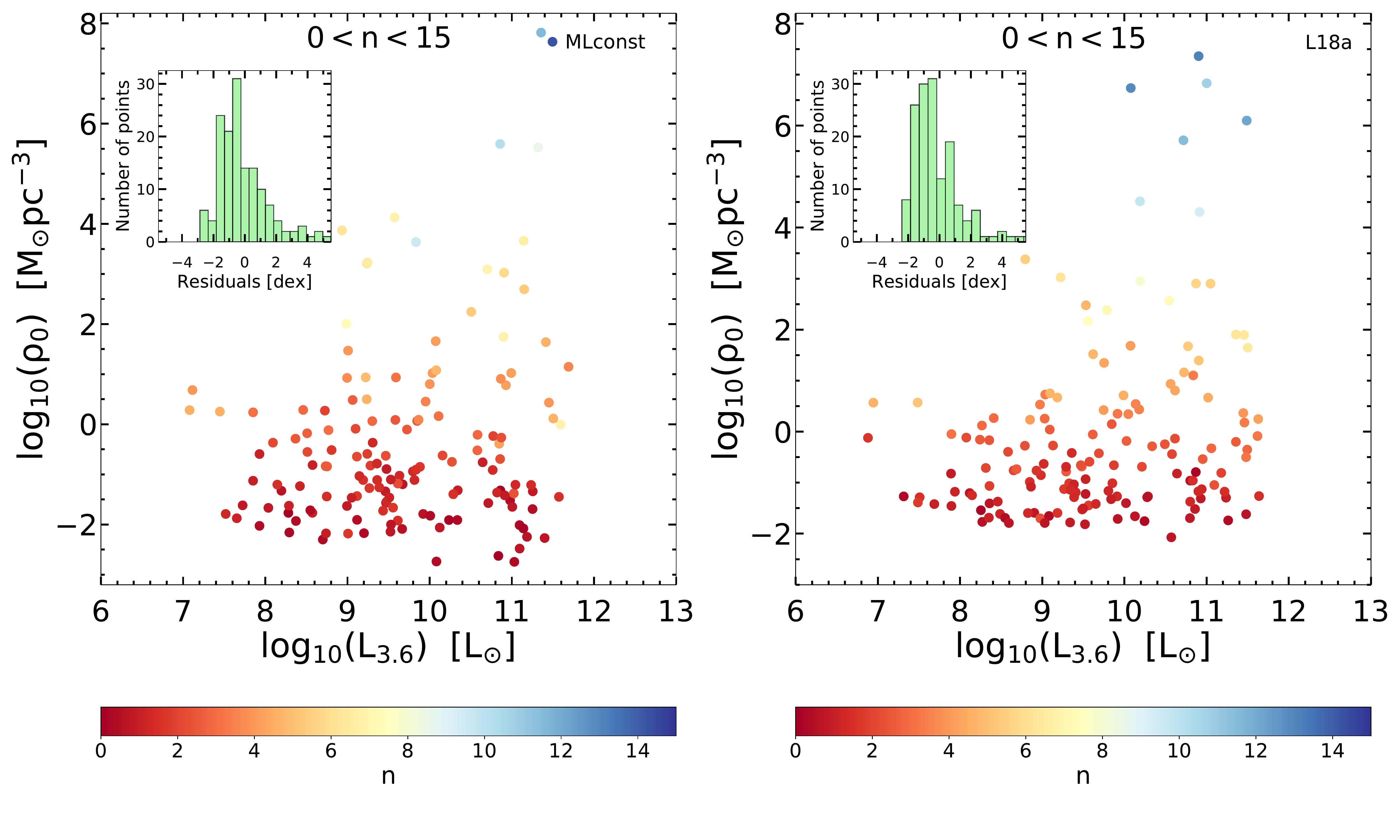}
 
\caption{Top panels: the characteristic density $\rho_{-2}$ at the radius of density slope $-2$ versus the 3.6 micron luminosity (left: ``MLconst''; right: L18a), compared to the density scale of \citet{Lieinasto2018}. Bottom panels: the central density $\rho_0$ versus the 3.6 micron luminosity (left: ``MLconst''; right: L18a), colour-coded by the value of the Einasto index $n$.} \label{DM_L}
\end{center}
\end{figure*}

\begin{figure*}
\begin{center}
\includegraphics[width=170mm]{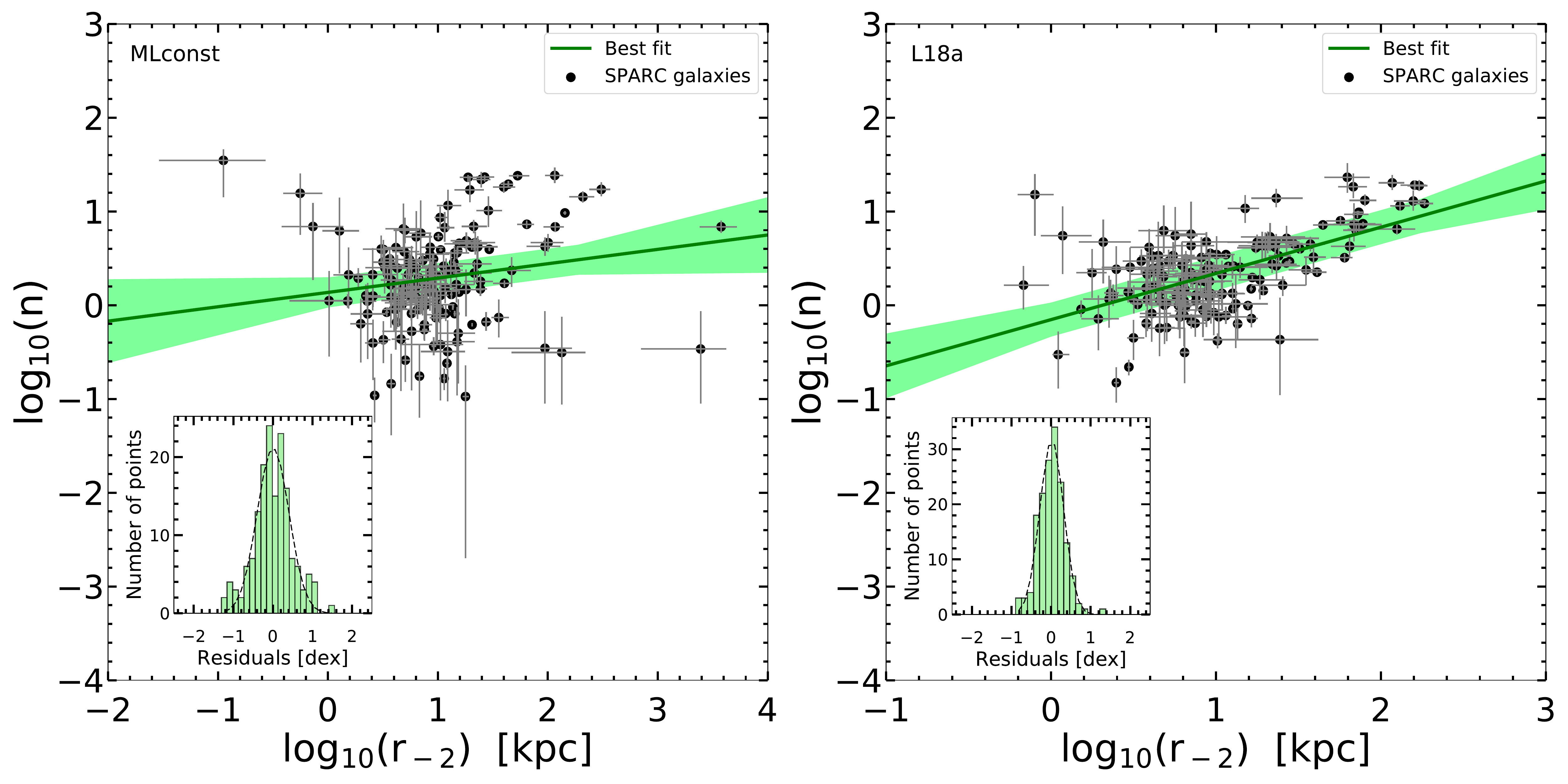}\hfill
\includegraphics[width=170mm]{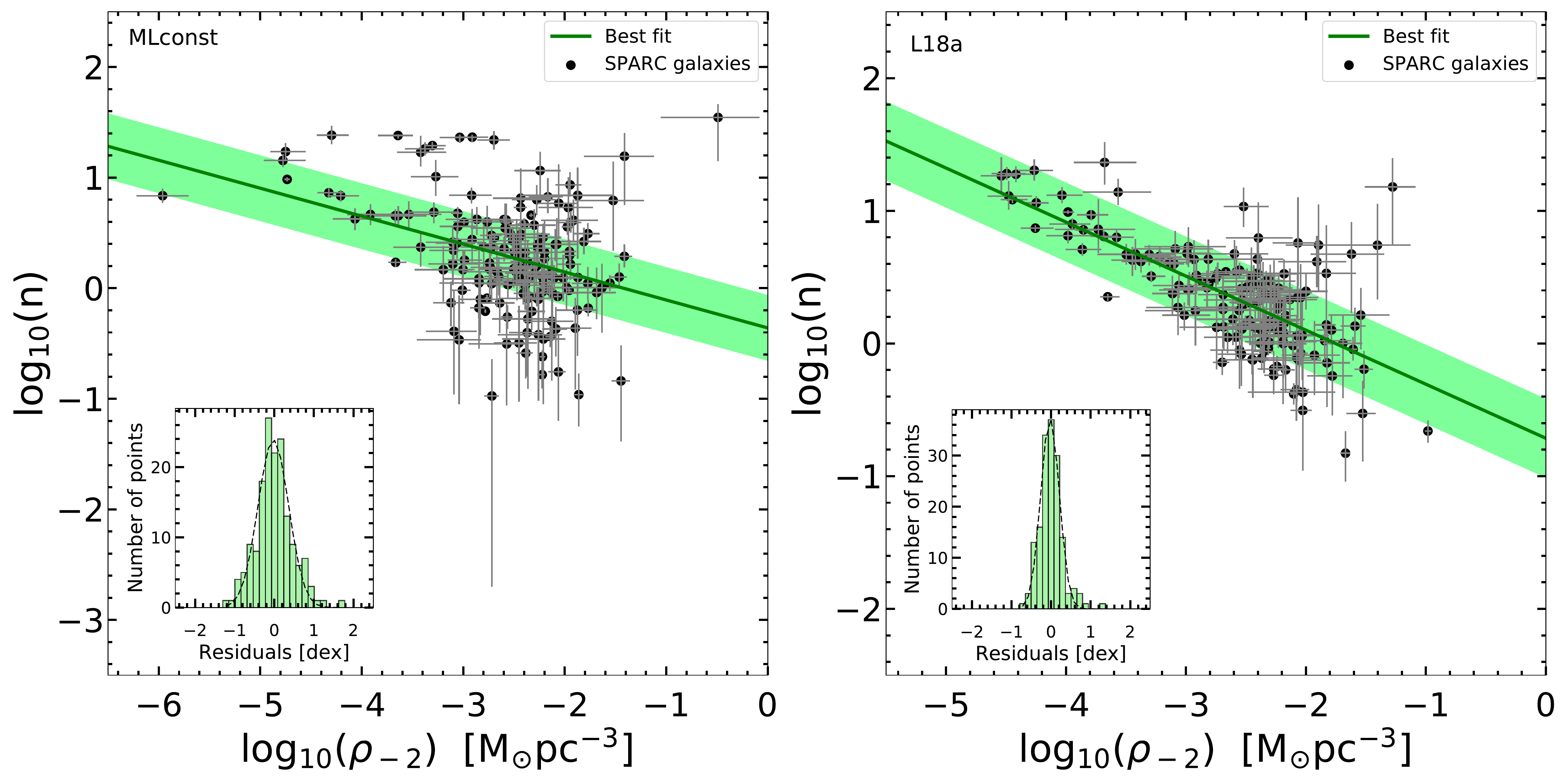}  
\caption{Correlation between Einasto Index $n$ and dark matter halo parameters $r_{-2}$ (top) and $\rho_{-2}$ (bottom), in the ``MLconst'' (left) and L18a (right) cases.}\label{Einasto_DM_corr}
\end{center}
\end{figure*}

\begin{figure*}
\begin{center}
\includegraphics[width=170mm]{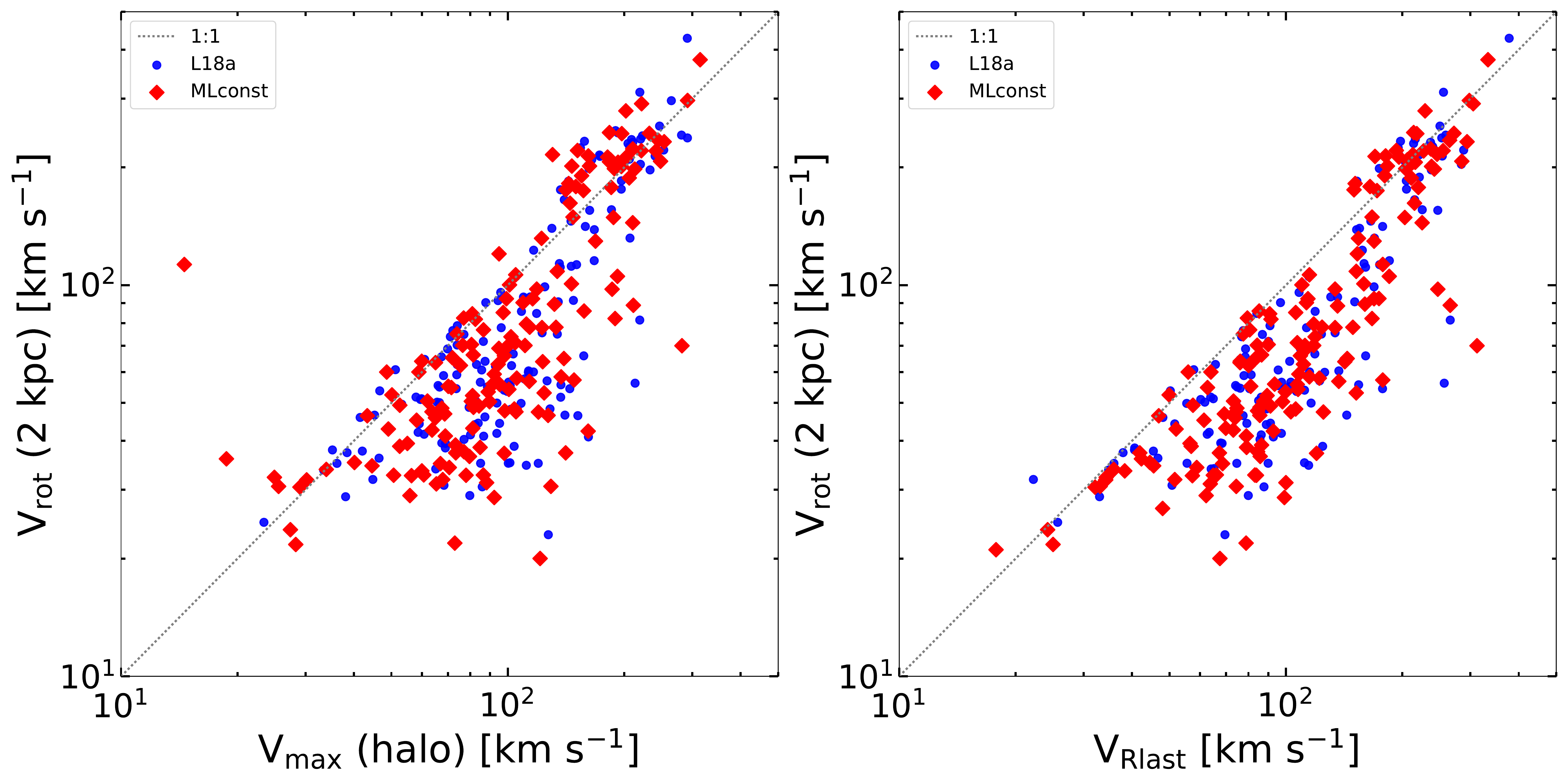}\hfill
\includegraphics[width=170mm]{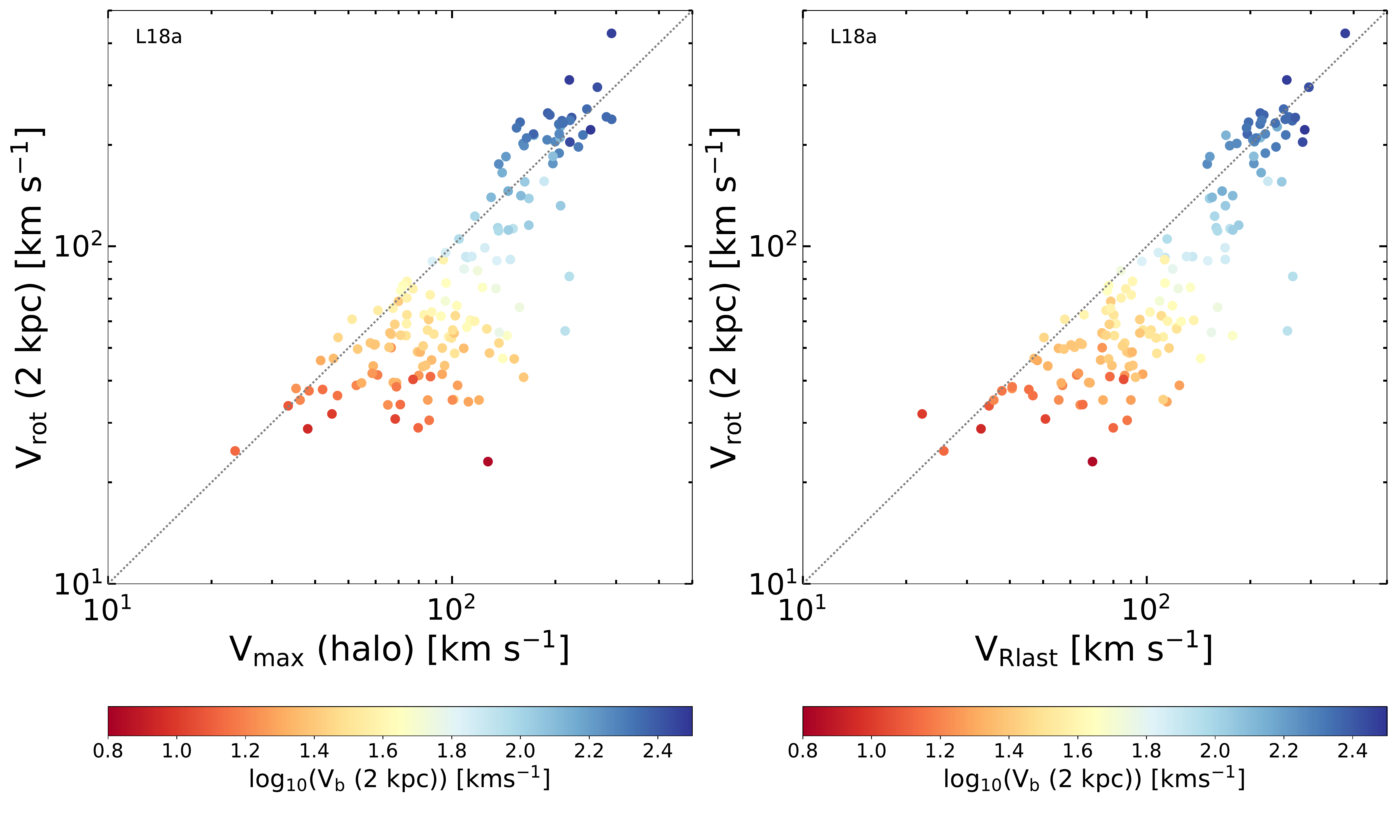}\
\caption{Top panels: the measured circular velocity at 2~kpc from the center $V_{\rm rot}(\rm 2 kpc)$ versus the maximum circular velocity of the DM halo $V_{\rm max}(\rm halo)$ (left) and the last observed point of the rotation curve $V_{\rm Rlast}$ (right). The ``MLconst'' values are denoted with red diamonds and the L18a values with blue circles. The grey dotted line is the 1:1 line. Bottom panels: The same plots for the L18a points color-coded by the log of the baryon-induced velocity at 2~kpc, ${\rm log}[V_b(2 \, {\rm kpc})]$.}\label{diversity_plot}
\end{center}
\end{figure*}

\begin{figure*}
\begin{center}
\includegraphics[width=170mm]{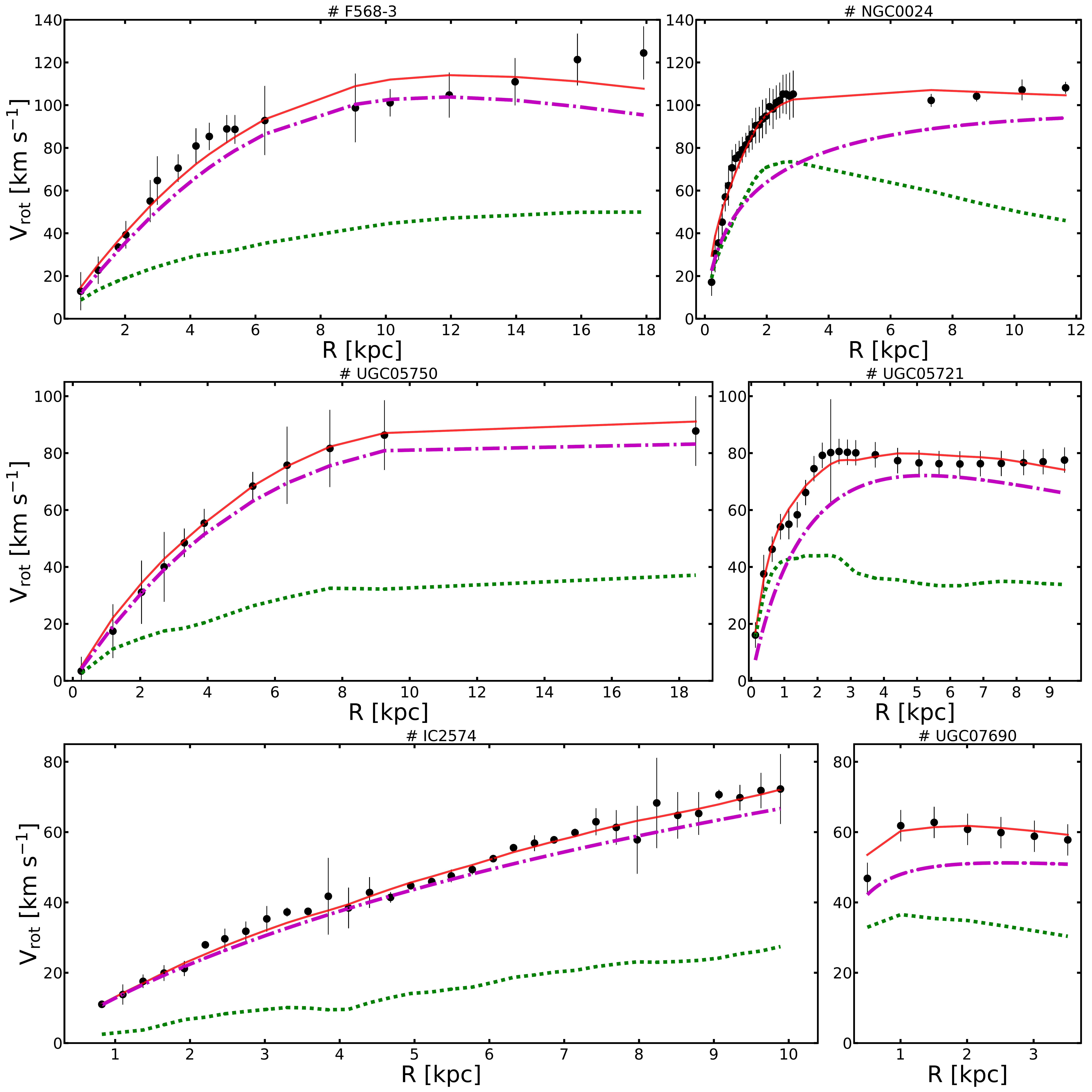}
  
\caption{The BTFR twins ``paradox''. Rotation curves of three pairs of galaxies with L18a parameters, each pair having very similar asymptotic circular velocities, but very different velocities at 2~kpc from the center. In these plots the black dots with error bars are the observed rotation curves of each galaxy. The green doted curves are the baryonic contribution to the rotation curves, the violet dot-dashed curves are the rotation curves of the DM halo and the red solid curves are the resulting best-fit rotation curves to the observed data.}\label{RC_selected_galaxies}
\end{center}
\end{figure*}

\begin{figure*}
\begin{center}
\includegraphics[width=170mm]{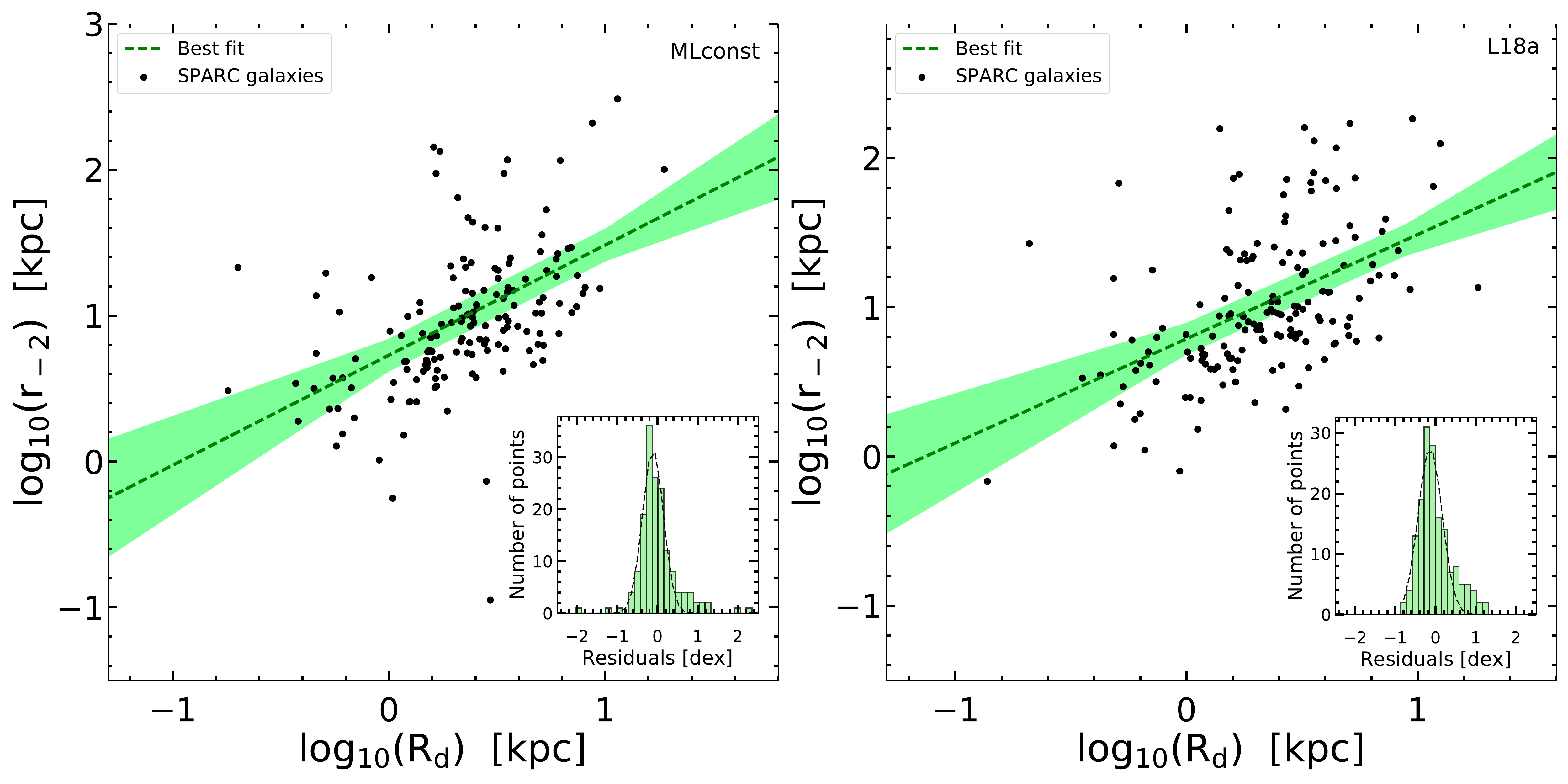}\hfill
\includegraphics[width=170mm] {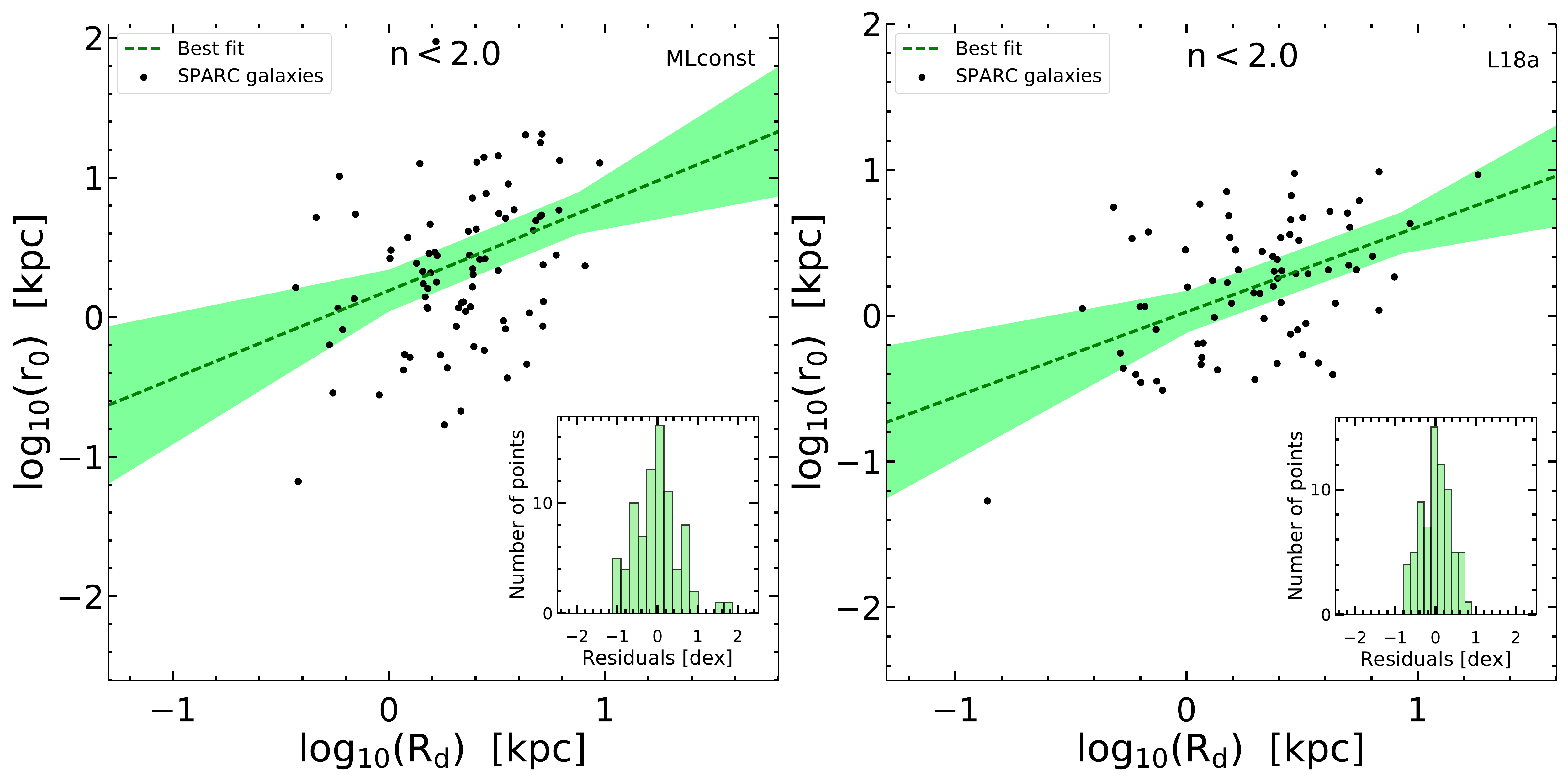}
\caption{Top panels: correlation between the stellar disk scale length $R_{\rm d}$ and the dark matter halo parameter $r_{-2}$ (left: ``MLconst''; right: L18a). Bottom panels: for cored profiles with $n<2$, correlation between the stellar disk scale-length and the core size $r_0$ defined as the radius at which the DM density becomes half the central one (left: ``MLconst''; right: L18a).}\label{DM_Rd}
\end{center}
\end{figure*}

\begin{figure*}
\begin{center}
\includegraphics[width=170mm]{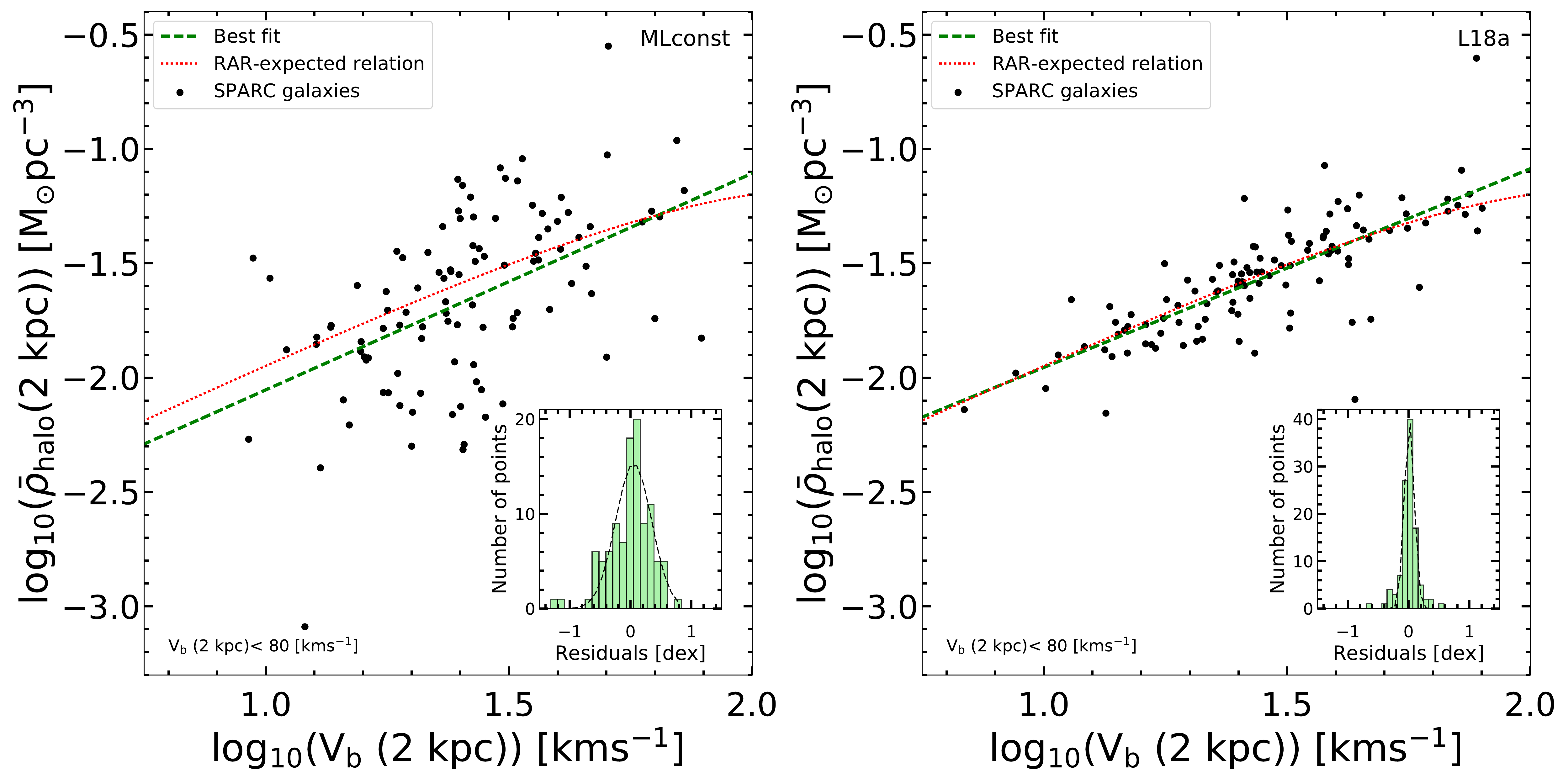}
\caption{The ``strong'' version of the old core-cusp problem. Correlation between the averaged DM density within 2~kpc $\bar{\rho}(2 \, {\rm kpc})$ and the baryon-induced velocity at 2~kpc, $V_b(2 \, {\rm kpc}$) (left: ``MLconst''; right: L18a). The red-dotted curve is the correlation induced by the RAR in Eq.~\ref{eq:RAR}. Note how this plot is closely related to the color gradient visible on Fig.~\ref{diversity_plot}, as the average DM density within 2~kpc simply reads as $\bar{\rho} = (V_{\rm rot}^2-V_{\rm b}^2)/GR^2$, where $R=2 \, {\rm kpc}$.}\label{averagedrho}
\end{center}
\end{figure*}

\begin{figure*}
\begin{center}
\includegraphics[width=170mm]{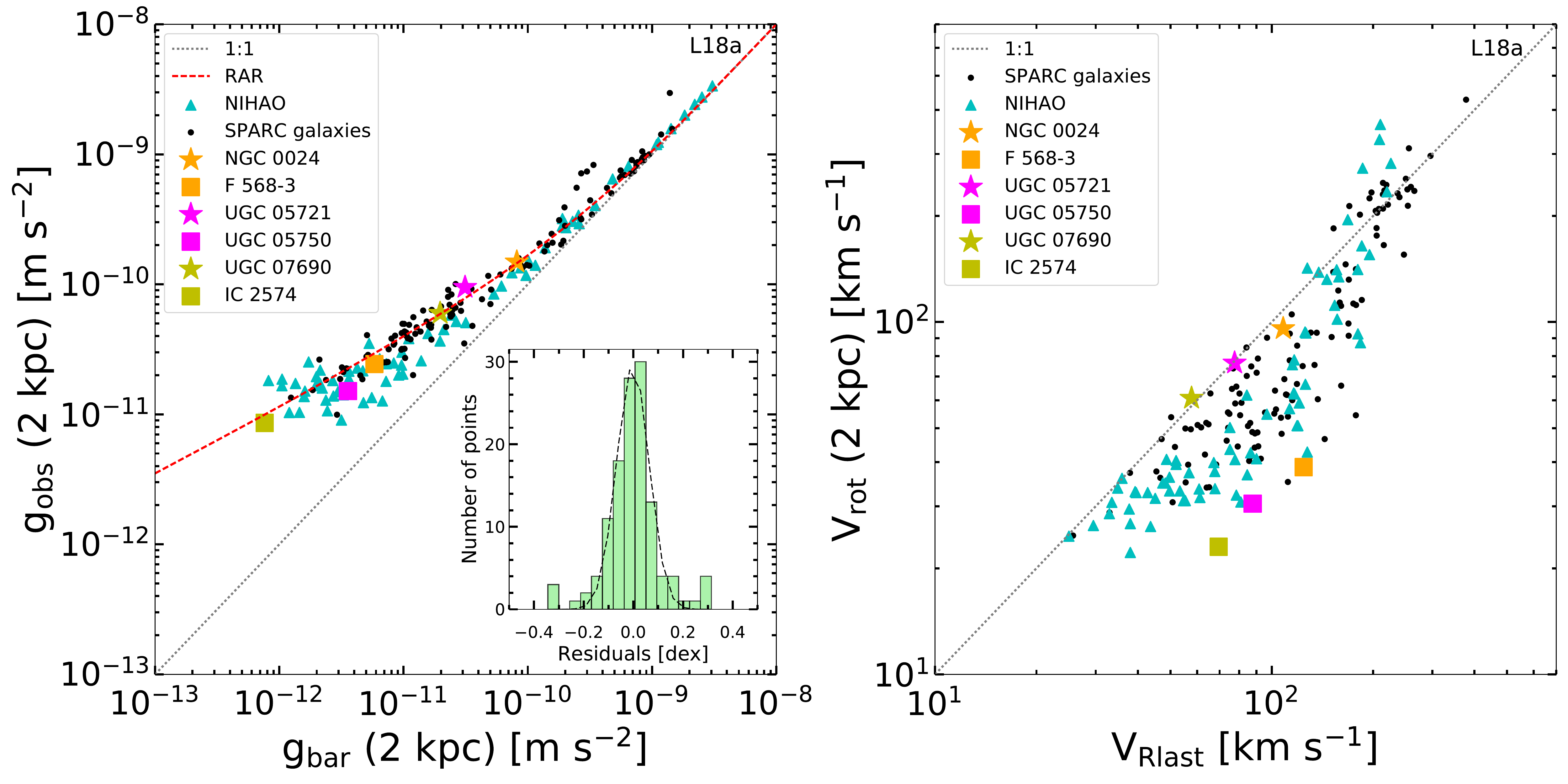}
  
\caption{Left: The radial acceleration relation (RAR) at 2 kpc (see text) with L18a parameters. The 6 representative galaxies plotted on Fig.~\ref{RC_selected_galaxies} are displayed as green, magenta and orange squares and stars on this plot. The grey dotted line is the 1:1 line. The red dashed curve is the RAR from Eq.~\ref{eq:RAR}. UGC~7690 and NGC~024 fall particularly well on the relation. Also overplotted as cyan triangles are the points of the RAR at 2~kpc from the NIHAO simulations of Dutton et al. (in prep.). Right: Diversity plot: the circular velocity at 2~kpc from the center $V_{\rm rot}(\rm 2 kpc)$ versus the circular velocity at the last observed point $V_{\rm Rlast}$ for the same galaxies. The cyan triangles are the predictions of the NIHAO simulations from figure 2 of \citet{SantosSantos2018}. UGC~7690 and NGC~024 appear to have too high velocities $V_{\rm rot}(\rm 2 kpc)$ and hence to high central DM concentrations ($n=5.5$ and $4.7$, respectively) compared to simulations with very efficient core formation.}\label{RAR_2kpc}
\end{center}
\end{figure*}

\begin{figure*}
\begin{center}
\includegraphics[width=175mm]{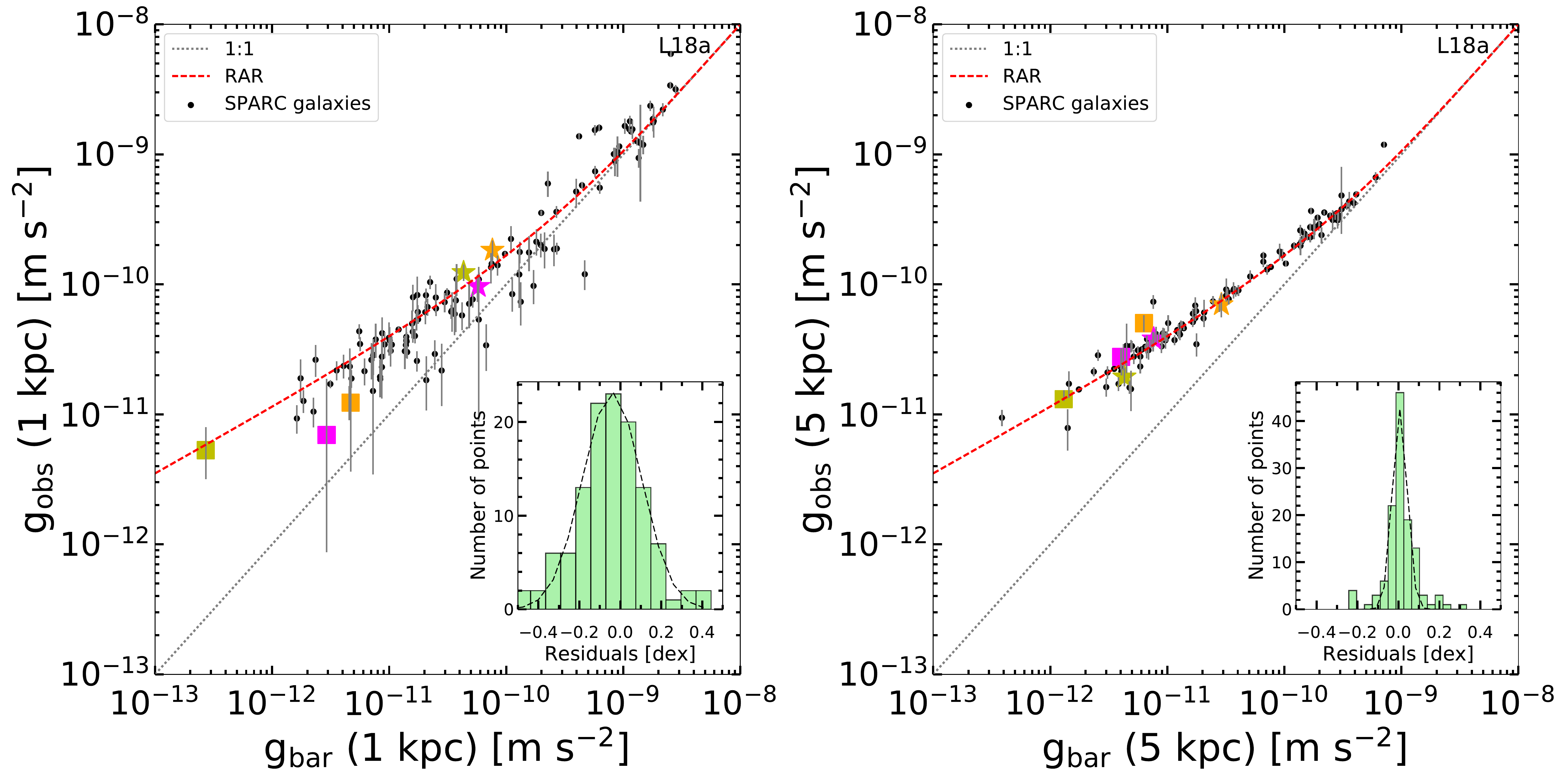}
\caption{Left: The RAR at 1~kpc with L18a parameters. Observational errors from the rotation curve measurements are also plotted. Note that those large errors seem to be associated with a downward scatter in $g_{\rm obs}$ w.r.t. the RAR. Right: The RAR at 5~kpc. The observational errors are small and the RAR is particularly tight at this radius. The 6 representative galaxies of Fig.~\ref{RC_selected_galaxies} are displayed as in Fig.~\ref{RAR_2kpc}.}\label{RAR_1_5kpc}
\end{center}
\end{figure*}

\begin{figure}
\begin{center}
\includegraphics[width=90mm]{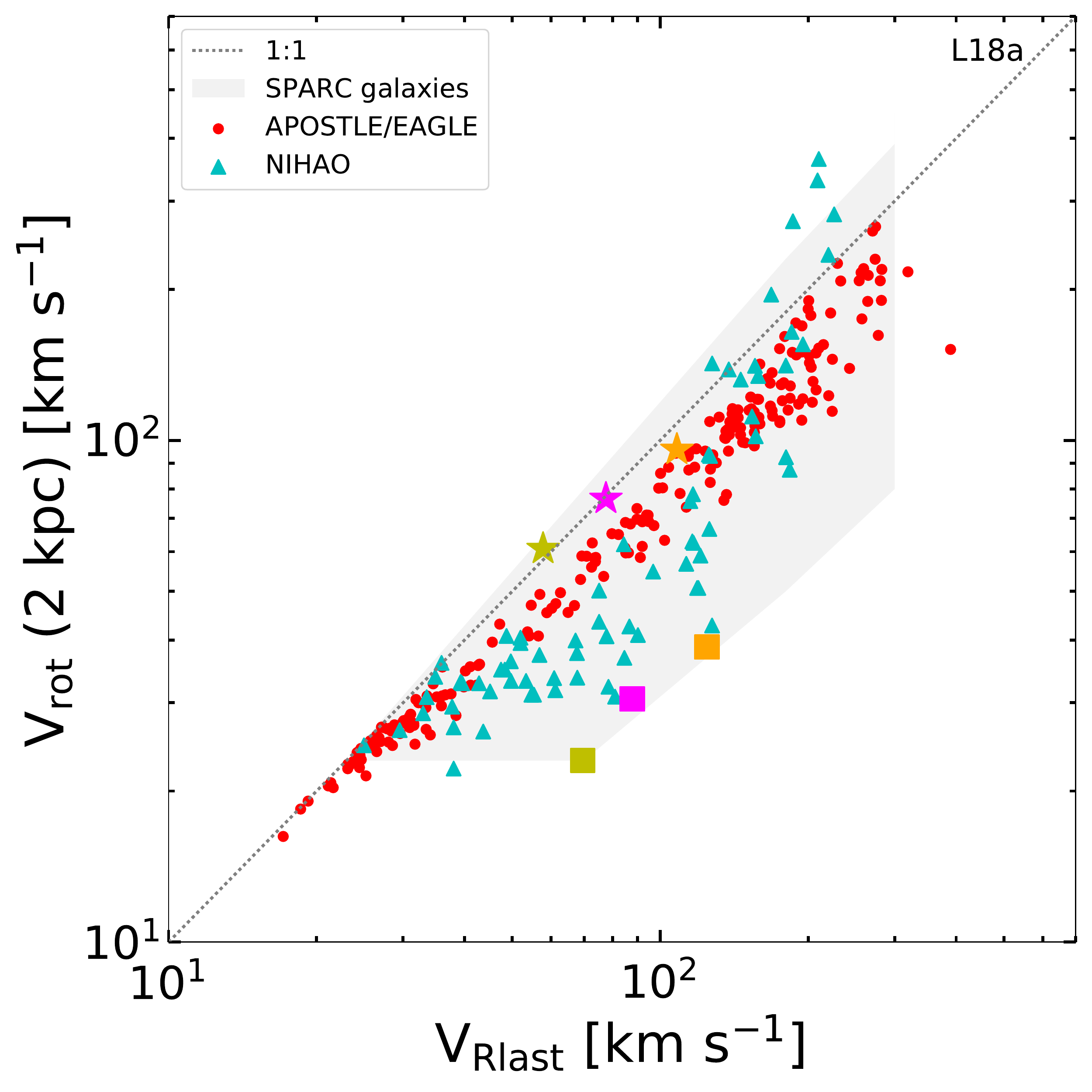}

  \caption{Diversity plot from simulations: the circular velocity at 2~kpc from the center $V_{\rm rot}(\rm 2 kpc)$ versus the circular velocity at the last observed point $V_{\rm Rlast}$ from the NIHAO (cyan triangles) and APOSTLE/EAGLE (red points) projects. The grey shaded area indicates the region spanned by observed galaxies in SPARC (i.e., the black points of Fig.~\ref{RAR_2kpc}). The six representative galaxies are plotted as in Fig.~\ref{RAR_2kpc}.}\label{nihao_apostle_diversity}
\end{center}
\end{figure}

We now explore the scaling relations between the free parameters of the DM halo itself, as well as DM-baryons scaling relations. 

\subsection{Dark matter halo scaling relations}

We start by plotting the histogram of Einasto indices on Fig.~\ref{hist_n}, for both cases. The first interesting thing to note here is that the majority of Einasto indices is comprised between 0 and 2, even though the number of very small $n$ is slighlty reduced in the L18a case. This is in line with the findings of \citet{Chemin2011} and indicates that the profiles tend to be more cored than in DMO simulations.

Second, like \citet{Chemin2011}, we find a tight and strong anti-correlation between the characteristic halo radius, $r_{\rm -2}$ and the characteristic halo density,  $\rho_{\rm -2}$ (Fig.~\ref{scaling_relation}). Interestingly, the slope of this relation becomes closer to the \citet{Chemin2011} one for the L18a parameters. The scatter of the relation also becomes smaller in the L18a case, decreasing from 0.35 dex in the ``MLconst'' case to 0.28 dex in the L18a case.

The best fit relation for the L18a parameters is 
\begin{equation}
{\rm log}(\rho_{-2})=(-1.32 \pm0.15) \times {\rm log}(r_{-2})-(1.27 \pm 0.18).
\end{equation}
This anti-correlation is reminiscent of the scaling relation found with Burkert halos between the central density and core radius \citep{Donato2009}, although, as shown also by \citet{Lieinasto2018}, it does not imply a constant surface density scale $\rho_{-2} \times r_{-2}$ independent of luminosity. Indeed, we find a linear relation between the value of ${\rm log}(\rho_{-2} \times r_{-2})$ and the log of the 3.6 micron luminosity, with a slope $0.13$  in the ``MLconst'' case, and a slope of $0.1$ in the L18a case. This relation is related to the absence of any significant correlation between the $\rho_{-2}$ density scale and the luminosity, already put forward by \citet{Lieinasto2018} (see Fig.~\ref{DM_L}). We nevertheless also note that, contrary to $\rho_{-2}$, the {\it central} volume density of DM haloes, $\rho_0$, does display some correlation with luminosity, due to the higher prevalence of high Einasto indices $n$ (and hence more cuspy halos) among galaxies of higher luminosities, as shown on the lower panels of Fig.~\ref{DM_L}.

The latter result brings us to the correlation of the Einasto index $n$ with $\rho_{-2}$ and $r_{-2}$, displayed on Fig.~\ref{Einasto_DM_corr}. The correlation between $r_{-2}$ and $n$ is rather weak in the ``MLconst'' case, with a slope consistent with zero and a scatter of 0.4 dex, but it becomes steeper with a scatter of 0.3 dex for the L18a parameters: 
\begin{equation}
{\rm log}(n)= (+0.5 \pm 0.15) \times {\rm log}(r_{-2})-(0.16 \pm 0.18).
\end{equation}
Similarly, the anticorrelation between $\rho_{-2}$ and $n$ has a small slope, with a large scatter of 0.4 dex in the ``MLconst'' case, but becomes steeper and with a smaller scatter of 0.23 dex for the L18a parameters:
\begin{equation}
{\rm log}(n)=(-0.4 \pm 0.11) \times {\rm log}(\rho_{-2})-(0.71 \pm 0.29). 
\end{equation}

\subsection{Dark matter-baryon scaling relations: diversity and uniformity}

We now concentrate on scaling relations between DM and baryons. The first well-known such scaling relation, not reproduced here, is of course the baryonic Tully-Fisher relation (BTFR) between the baryonic mass of disk galaxies and the asymptotic circular velocity, which has an intrinsic scatter smaller than 0.1 dex \citep{McGaugh2000, LelliBTF}, {\it a priori} posing a fine-tuning problem in the $\Lambda$CDM context \citep{Desmond2}. Semi-empirical models with cored DM haloes can reproduce the observed slope and normalization, but not this small scatter \citep{DiCintio}. But what is more, galaxies with the same asymptotic circular velocity -- hence twins of identical baryonic mass on the BTFR -- can display a very broad range of rotation curve shapes, which we could call the ``{\it BTFR twins paradox}''. This unexpected diversity of shapes of rotation curves has been nicely put forward in, e.g., \citet[][]{Oman2015}, where the circular velocity at 2~kpc from the center was shown to be extremely diverse for a given maximum circular velocity of the halo. We reproduce this diversity plot in Fig.~\ref{diversity_plot}, for both sets of galaxy parameters. It should be noted that the ``MLconst'' parameters are often used in the literature to compare directly the data to simulations. However, as we showed hereabove, the L18a parameters yield better fits to the SPARC galaxy rotation curves. It is therefore interesting to compare the diversity of rotation curve shapes in the two cases: whilst the global picture remains the same, it is however interesting to note that some of the most extreme low values of $V_{\rm rot}(2 \, {\rm kpc})$ in the ``MLconst'' case become more in line with the general population when switching to the L18a parameters. Hence some of the most extreme cored DM profile cases which are challenging to reproduce in simulations can be easier to reproduce with the L18a parameters.

Nevertheless, one important information missing from the usually presented diversity plot is the close relation between the value of the observed velocity at 2~kpc and the {\it baryon}-induced velocity at this radius, produced in the absence of DM. This is illustrated on the bottom row of Fig.~\ref{diversity_plot}, where the points are color-coded by ${\rm log}[V_b(2 \, {\rm kpc})]$. One can clearly see how smaller baryonic velocities at 2~kpc are associated with smaller total velocities $V_{\rm rot}(2 \, {\rm kpc})$ at this radius. Of course, while this figure clearly shows that the baryons are an important part of the story, the fact remains that the central mass deficit is enormous relative to a cuspy DM halo, sometimes higher than the total baryonic mass of the galaxy itself \citep{Oman2015}.

In order to display a few concrete examples of this relation between the baryon-induced velocities and the diversity of measured velocities at 2~kpc, we show for the L18a parameters (and on the same $R$-axis scale) on Fig.~\ref{RC_selected_galaxies} the rotation curves of three galaxy pairs with similar asymptotic velocities, hence ``BTFR twins'', but very different measured velocities at 2~kpc from the center. In the case of IC~2574, note that we added a prior on the size of the core not being larger than 5 times the last observed point of the rotation curve. We alternatively added to the SPARC data one outer data point, coming from the THINGS survey \citep{Oh2008} at large radius, recalibrated for the exact same distance and inclination as quoted in L18a. Those two approaches gave almost exactly the same fit.

What is immediately striking on Fig.~\ref{RC_selected_galaxies} is that the more compact galaxies in terms of baryons are also more compact in terms of their DM distribution. This is of course expected from adiabatic contraction of DM halos, but the problem is that this must happen after an efficient core formation mechanism, and that going back to very steep DM distributions for the highest baryonic surface densities, through adiabatic contraction, can then be complicated, as we discuss below in the context of hydrodynamical simulations of galaxy formation. Here, the gravitational acceleration generated by baryons at 2~kpc is much smaller in F568--3, UGC~5750 and IC~2574 than in NGC~024, UGC~5721 and UGC~7690, and so is the measured circular velocity, and hence the total gravitational acceleration at 2~kpc. This is highly suggestive of the fact that the diversity of rotation curve shapes is driven by the diversity of baryonic distributions at a given maximum velocity scale, together with a tight baryon-DM scaling relation. This relation is the RAR. What this means is that, while a tight RAR as obeyed in the L18a case slightly decreases the diversity compared to the original ``MLconst'' case, the diversity of {\it baryon}-induced accelerations at 2 kpc of the SPARC galaxies is sufficient to induce a large diversity of rotation curve shapes with a tight RAR.

By fitting the following function proposed by \citet{McGaugh2008}, \citet{FamaeyMcGaugh2012} and \citet{McGaughSchombert2014}
\begin{equation}\label{eq:RAR}
 g_{\rm obs} = \mathcal{F}(g_{\rm bar}) = \frac{g_{\rm bar}}{1 - e^{-\sqrt{g_{\rm bar}/g_\dag} } },
\end{equation}
with the only free parameter $g_\dag$,  \citealt{RAR1} and \citet{RAR2} found a good fit to the RAR of the SPARC sample of galaxies with $g_\dag = (1.20 \pm 0.02) \times 10^{-10}$ m~s$^{-2}$. With the L18a parameters, the RAR scatter is as small as 0.057 dex. In the outer flat part of rotation curves, this relation is equivalent to the BTFR \citep[e.g.][]{LelliBTF}, but it is important to remember that the BTFR does {\it not} imply the RAR in the central parts of galaxies, where the rotation curves are rising, and where the RAR is more closely connected to the size of the DM cores. 

Indeed, the RAR implies that the rotation curve shape is actually correlated with the surface density of the baryons \citep[see Figure 15 of][]{FamaeyMcGaugh2012}, as for instance illustrated in \citet{Lelli2013, Lelli2016b}. Also, the rotation curve shapes of late-type spiral galaxies in the low-acceleration regime should all be similar when expressed in units of disk scale-length \citep{Swaters2009,McGaughlaw2014}, at least for perfect exponential disks. Yet in other words, one also expects that the size of the DM core will be directly correlated with the disk scale-length, a finding already put forward by \citet{Donato2004}. However, gaseous distributions are not exponential, so that any plotted correlation using the stellar disk scale-length as a proxy for the extent and ``fluffiness'' of a disk with a certain baryonic mass will be much less tight than the RAR itself.

Fig. \ref{DM_Rd} indeed shows the correlation between the DM halo parameter $r_{-2}$ and the stellar disk scale length $R_{\rm d}$, in the ``MLconst'' and L18a cases. The scatter is similar in both cases, of the order of 0.35 dex. The L18a relation is
\begin{equation}
{\rm log}(r_{-2})=(0.7 \pm 0.22) \times {\rm log}(R_{\rm d})+(0.79 \pm 0.1).
\end{equation}
If we restrict ourselves to cored profiles with $n<2$, and define the core size $r_0$ as the radius at which the density becomes half the central one $\rho_0$, we end up with a similar correlation, which we display on the lower panels of Fig.~\ref{DM_Rd}. In that case, the scatter is reduced from 0.5 dex in the ``MLconst'' case to 0.4 dex in the L18a case. The L18a correlation of the lower-right panel of Fig.~\ref{DM_Rd} for $n<2$ reads
\begin{equation}
{\rm log}(r_{0})=(0.59 \pm 0.3) \times {\rm log}(R_{\rm d})+(0.04 \pm 0.14).
\label{eq:r0Rd}
\end{equation}

Finally, a way to illustrate the influence of baryons on the diversity of central DM profiles in rotationally-supported galaxies is to look at the averaged DM density within 2~kpc, and compare it to the baryon-induced velocity at that radius. We indeed found on Fig.~\ref{averagedrho}, in both the ``MLconst'' and L18a cases, a very well-defined relation for galaxies with low-enough baryonic velocities at 2~kpc, $V_b(2 \, {\rm kpc})<80\,{\rm km}\,{\rm s}^{-1}$, meaning that the baryons are not heavily dominating the measured circular velocity at 2~kpc (which would imply a very bad constraint on the enclosed DM mass). The log-log slope of the relation is 0.93 in the ``MLconst" case, with a scatter of 0.3 dex, and 0.87 in the L18a case with a scatter of only 0.08 dex. In the L18a case, the relation for $V_b(2 \, {\rm kpc})<80\,{\rm km}\,{\rm s}^{-1}$ reads
\begin{equation}
{\rm log}(\bar{\rho}(2 \, {\rm kpc}))=(0.87 \pm 0.4) \times {\rm log}(V_b(2 \, {\rm kpc}))-(2.84 \pm 0.58).
\label{eq:averaged}
\end{equation}
Let us note that such a relation can actually directly be inferred from the RAR, a relation which we overplot as a dotted red curve on Fig.~\ref{averagedrho}. Let us note that this Figure is in fact closely related to the color gradient visible on Fig.~\ref{diversity_plot}, as the average DM density within 2~kpc also reads as $\bar{\rho} = (V_{\rm rot}^2-V_{\rm b}^2)/GR^2$, where $R=2 \, {\rm kpc}$.

Eq.~\ref{eq:r0Rd} and Eq.~\ref{eq:averaged} can be considered to represent the {\it strong} version of the old core-cusp problem. Not only do late-type low-mass galaxies often require DM cores, the averaged density and size of these cores are closely correlated to the baryonic distribution. Together with the diversity of baryonic surface density distributions at a given DM halo mass scale, this is at the root of the diversity of rotation curve shapes.

This diversity of rotation curve shapes is actually an important constraint to take into account in hydrodynamical simulations of galaxy formation, closely related to the ability of simulations at transforming initially cuspy DM profiles into cores \citep[e.g.,][]{Navarro96, Pontzen12}. As shown in \citet[][]{Oman2015}, cosmological hydrodynamical simulations from the EAGLE and APOSTLE projects are unable to produce large cores as required by the low values of $V_{\rm rot}(2 \, {\rm kpc})$ in a large number of galaxies. This issue was explored further by \citet{BenitezLlambay2018} who showed that the core formation crucially depends on the star formation gas density treshold. They showed that high tresholds are needed for baryons to collapse at the center at densities comparable to DM, before being ejected by feedback, in order for cores to form under the influence of a small number of distinctive gas blowouts. A simulation with high star formation gas density threshold resulted in galaxies with sizeable cores over a limited halo mass range, but still insufficient variety in mass profiles to explain the observed diversity of galaxy rotation curves. Other simulations do create cores more easily than those based on EAGLE. For instance, \citet{DiCintio2014} used a suite of simulated galaxies drawn from the MaGICC project to investigate the effects of baryonic feedback on the density profiles of DM haloes, producing cored profiles. \citet{Read2016, Read2018} also used high resolution simulations of isolated dwarf galaxies to study the physics of DM cusp-core transformation, which was argued to be linked to short potential fluctuations associated with bursty star formation over a long period \citep[in contradiction to][]{BenitezLlambay2018}. Most of these simulations also showed how the general shape of the RAR was naturally obtained when including baryonic feedback \citep[e.g.,][]{Ludlow, Keller}. Perhaps the most impressive simulations in that sense are the recent NIHAO simulations \citep[e.g.,][]{SantosSantos2018}, which are argued (Dutton et al. in prep.\footnote{see darkmatter2018.weebly.com/program.html}) to produce a tight RAR with a scatter of order 0.08 dex over the entire distance range. Nevertheless, we argue hereafter that the real challenge is to produce {\it both} a tight RAR and the observed diversity of rotation curve shapes. To illustrate the relation between a tight RAR and the diversity problem, we plot, on Fig.~\ref{RAR_2kpc}, 124 galaxies with L18a parameters (rejecting galaxies with uncertainty$>10\%$ in $V_{\rm rot}(2 \, {\rm kpc})$ and those with $g_{\rm obs}< g_{\rm bar}$ at 2~kpc), compared to the RAR, {\it at 2~kpc only}. This means that, contrary to the traditional RAR plot, each point on this plot represents a different galaxy. Eq.~\ref{eq:RAR} is overplotted together with those points. The scatter around the relation is $\sim 0.06$ dex. The six representative galaxies plotted on Fig.~\ref{RC_selected_galaxies} are displayed as squares and stars on  this plot. These selected galaxies can then illustrate the effect of the RAR on the diversity plot, which we reproduce for the same set of 124 galaxies on the right panel of Fig.~\ref{RAR_2kpc}. Note that, in this case, the scatter of $V_{\rm rot}$ around the median at a given $V_{\rm Rlast}$ is $\sim 0.2$ dex. This can then be compared directly to the points from Figure~2 of \citet{SantosSantos2018}. Interestingly, it appears that the NIHAO simulations are very efficient at producing cores reproducing the right $V_{\rm rot}(2 \, {\rm kpc})$ for galaxies sitting at the bottom of the RAR, with a low baryonic acceleration at 2~kpc. But the three high-baryonic acceleration galaxies of Fig.~\ref{RC_selected_galaxies}, NGC~024, UGC~5721 and UGC~7690 seem to all have too high observed $V_{\rm rot}(2 \, {\rm kpc})$ to be reproduced by the NIHAO simulations. On the left panel of Fig.~\ref{RAR_2kpc}, we also overplot the NIHAO points on the RAR at 2~kpc from Dutton et al. (in prep.), spanning the same baryonic acceleration range as our sample. Interestingly, the NIHAO simulations seem to slightly underpredict the observed RAR at 2~kpc in the regime of baryonic accelerations ranging from $\sim 10^{-11}$ to $10^{-10} \, {\rm m s}^{-2}$, corresponding to the regime of our three representative high baryonic acceleration galaxies. One interesting thing to note is that \citet{Oman2019} showed that `observing' and modelling the same simulated galaxies as in \citet{Oman2015} to measure their rotation curves, rather than simply measuring the rotation curve from the mass profile, introduced an additional downward scatter in the diversity plot. Interestingly, when considering the observed RAR for the same galaxies as in Fig.~\ref{RAR_2kpc} at different radii, one clearly sees that the scatter increases downwards when the quoted observational errors increase (Fig.~\ref{RAR_1_5kpc}). At 1~kpc, the observational errors become quite important (whilst non-circular motions might also become more important than in the outskirts), and many galaxies scatter down in the RAR. At 5~kpc, on the other hand, the quoted observational errors are smaller, and the tightness of the RAR becomes impressive. In terms of diversity, that could mean that results such as those from NIHAO are actually worse than they appear, since if the galaxies were realistically observed and modelled they could scatter down even more in terms of $V_{\rm rot} (2 \, {\rm kpc})$. Additionally, it would be important to check whether the relation between $V_{\rm max}$ and $V_{\rm Rlast}$ in NIHAO are in line with observational fits. Interestingly, as originally shown by \citet{Oman2015}, galaxies such as NGC~024, UGC~5721 and UGC~7690 can in principle be more closely reproduced in the diversity plot by simulations from the EAGLE and APOSTLE projects, which in turn cannot reproduce the smaller $V_{\rm rot}(2 \, {\rm kpc})$ galaxies. This is illustrated in Fig.~\ref{nihao_apostle_diversity} where the diversity plot of the NIHAO and APOSTLE/EAGLE simulations are compared. Even taking into account observational errors, it is unlikely that the APOSTLE/EAGLE points will scatter down enough. On the other hand, NIHAO points will probably scatter down too much once such errors are taken into account. This might indicate that the answer lies somewhere between the two types of simulations, combining cores slightly more modest than in NIHAO with some contribution from rotation curve errors (to add some scatter and downward shift). But in any case, reproducing {\it both} a tight RAR and the observed diversity of rotation curve shapes thus raises an interesting challenge for simulations of galaxy formation.


\section{Discussion and conclusion}

In this paper, we used the Einasto DM halo profile for the decomposition of rotation curves, in the same spirit as \citet{Chemin2011} and \citet{Lieinasto2018}. Our study is complementary to the latter, as we compared here the quality of the fits and the scaling relations obtained for different sets of galaxy parameters, and focused in particular on scaling relations between the characteristics of the DM core of low-mass galaxies and their baryonic distributions. 

The two sets of galaxy parameters used in this study were the original SPARC parameters with constant mass-to-light ratios for bulges and disks \citep{Lelli2016,RAR1}, and the parameters for which galaxies follow the tightest radial acceleration relation (RAR, L18a). We found that the fits were globally slightly better in the latter case, and that the dark halo parameter scaling relations also became a bit tighter. In particular, we found for both sets of parameters a tight and strong anti-correlation between the characteristic halo radius, $r_{\rm -2}$ and the characteristic halo density,  $\rho_{\rm -2}$. The latter is also anti-correlated with the Einasto index $n$, but in a tighter and steeper way for the L18a parameters than for the ``MLconst'' ones.

In terms of dark matter-baryons scaling relations, we focused on relations between the core properties and the extent of the baryonic component, which are highly relevant to the cusp-core transformation process. For galaxies with Einasto index $n<2$, we found a positive correlation between the core size, defined as the radius where the density reaches half of its central value, and the stellar disk scale-length (Eq.~\ref{eq:r0Rd}), albeit with some significant scatter. A tighter relation has been found between the averaged dark matter density within 2~kpc and the baryon-induced rotational velocity at that radius (Eq.~\ref{eq:averaged}), for galaxies not too strongly dominated by baryons in the center, namely where this velocity does not exceed $80\,{\rm km}\,{\rm s}^{-1}$.  These scaling relations could be referred to as the {\it strong} version of the old core-cusp problem. 

These scaling relations are directly related to the consequence of the RAR on the diversity of rotation curve shapes \citep{Oman2015}, quantified by the observed rotational velocity at 2~kpc, at a given maximum or asymptotic velocity scale. This diversity of rotation curve shapes could be referred to as the BTFR twins ``paradox'', since the {\it a priori} expectation, without efficient feedback, would have been a tight relation between the rotational velocity at 2~kpc and the maximum halo velocity for dark matter dominated galaxies. We confirmed this observed diversity of rotation curve shapes, but pointed out how it was directly related to the tightness of the radial acceleration relation together with the diversity of baryon distributions in galactic disks. Indeed, while we showed that a tight RAR with the L18a parameters slightly decreases the observed diversity compared to the ``MLconst'' case, we also showed that, in the L18a case, the diversity of baryon-induced accelerations at 2~kpc toghether with the RAR are sufficient to induce a large diversity of rotation curve shapes.

Reproducing all the scaling relations displayed in this paper together, and in particular the RAR and the diversity of rotation curve shapes, side by side, is still challenging for current hydrodynamical simulations of galaxy formation in a cosmological context, as we illustrated in Figs.~\ref{RAR_2kpc}-\ref{nihao_apostle_diversity}. This challenge could perhaps imply something fundamental on the nature of dark matter \citep{Kamada, FamaeyKhouryRen} or even gravity \citep{Milgrom83, FamaeyMcGaugh2012, Smolin, Verlinde}. In the standard context, it might seem to call for an apparently fine-tuned feedback from the baryons, where galaxies would self-regulate their star formation, implying an interplay between gas inflow, star formation, feedback, and the final DM distribution, which would have to conspire to make galaxies find an attractor acceleration relation with little scatter over the entire acceleration range, while maintaining a high diversity of rotation curve shapes.






\section*{Acknowledgements}


The authors thank the referee, Kyle Oman, for excellent suggestions, and for providing us with the APOSTLE/EAGLE $V_{\rm Rlast}$ and $V_{\rm rot}(2 \, {\rm kpc})$ simulated datapoints. We also thank Aaron Dutton for providing us with the NIHAO RAR at 2 kpc. We acknowledge useful discussions with Federico Lelli and Stacy McGaugh.




\bibliographystyle{aa} 

\bibliography{ref}



\appendix

\section{Rotation curve fits results}

\longtab[1]{
\begin{longtable}{lcccccc}
\caption{Maximum posterior DM halo parameters of individual rotation curve fits to 160 SPARC galaxies with Q < 3 and fixed mass-to-light ratios, inclinations and distances from L18a. The first column is the name of the galaxies. The columns 2, 3 and 4 are the the parameters of the Einasto halo profile, together with the 68\% confidence interval. Column 5 is the maximum circular velocity of the DM halo $V_{\rm max}(\rm halo)$ and column 6 is the measured rotational velocity at 2~kpc from the center $V_{\rm rot}(\rm 2 kpc)$.}
\label{rotation_curve_result}\\
\hline\hline
Galaxy& $\rho_{-2}$& $r_{-2}$ & n & $V_{\rm max}(\rm halo)$ & $V_{\rm rot}(\rm 2 kpc)$   \\
{} &
{10$^{-3}$ M$_{\odot}$ pc$^{-3}$} &
{kpc} &
{} &
{km/s} &
{km/s}\\
\hline
\endfirsthead
\caption{Continued.}\\
\hline
Galaxy& $\rho_{-2}$& $r_{-2}$ & n & $V_{\rm max}(\rm halo)$ & $V_{\rm rot}(\rm 2 kpc)$   \\
{} &
{10$^{-3}$ M$_{\odot}$ pc$^{-3}$} &
{kpc} &
{} &
{km/s} &
{km/s}\\
\hline
\endhead
\hline
\endfoot
\hline
\endlastfoot

IC 2574  & $   0.222_{-  0.030}^{+   0.057} $& $  41.010_{-  7.758}^{+  5.842} $&$ 2.248_{-  0.161}^{+  0.120} $&  127.285 &     23.021 &   \\
UGC 04483  & $  28.642_{- 13.217}^{+  21.184} $& $   0.681_{-  0.165}^{+  0.292} $&$ 1.637_{-  0.736}^{+  0.994} $&   23.412 &     24.762 &   \\
D 564-8  & $   2.459_{-  1.229}^{+   2.059} $& $   3.771_{-  1.252}^{+  2.329} $&$ 1.523_{-  0.555}^{+  0.776} $&   38.074 &     28.791 &   \\
F 565-V2  & $   1.798_{-  0.919}^{+   1.764} $& $   9.336_{-  3.021}^{+  5.686} $&$ 1.323_{-  0.685}^{+  1.012} $&   79.761 &     28.991 &   \\
UGC 05750  & $   2.793_{-  1.205}^{+   1.567} $& $   8.333_{-  1.567}^{+  2.726} $&$ 0.891_{-  0.435}^{+  0.647} $&   85.878 &     30.516 &   \\
NGC 3741  & $   0.137_{-  0.038}^{+   0.060} $& $  26.772_{-  5.510}^{+  6.057} $&$ 5.106_{-  0.532}^{+  0.593} $&   68.390 &     30.792 &   \\
UGC 07577  & $   1.198_{-  0.744}^{+   1.089} $& $   6.286_{-  3.295}^{+  3.755} $&$ 1.765_{-  0.801}^{+  1.734} $&   44.779 &     31.869 &   \\
KK 98-251  & $  29.707_{-  8.084}^{+   8.561} $& $   1.104_{-  0.086}^{+  0.186} $&$ 0.297_{-  0.168}^{+  0.226} $&   33.406 &     33.684 &   \\
UGC 00891  & $   2.616_{-  0.862}^{+   1.012} $& $   6.406_{-  1.117}^{+  1.834} $&$ 1.116_{-  0.347}^{+  0.431} $&   65.114 &     33.900 &   \\
NGC 3109  & $   2.224_{-  0.811}^{+   1.033} $& $   7.553_{-  1.778}^{+  3.093} $&$ 1.113_{-  0.269}^{+  0.320} $&   70.772 &     33.976 &   \\
UGC 05005  & $   0.856_{-  0.392}^{+   0.672} $& $  18.468_{-  4.781}^{+  7.796} $&$ 1.866_{-  0.712}^{+  0.933} $&  111.659 &     34.637 &   \\
UGC 07559  & $   6.520_{-  2.938}^{+   5.000} $& $   2.247_{-  0.682}^{+  1.060} $&$ 1.160_{-  0.584}^{+  0.899} $&   36.165 &     35.034 &   \\
UGC 06818  & $   3.604_{-  1.652}^{+   2.664} $& $  10.396_{-  4.449}^{+  6.260} $&$ 0.751_{-  0.363}^{+  0.498} $&  119.880 &     35.042 &   \\
F 571-V1  & $   2.037_{-  0.882}^{+   1.399} $& $   9.144_{-  2.080}^{+  3.389} $&$ 1.802_{-  0.881}^{+  1.389} $&   85.056 &     35.053 &   \\
UGC 08837  & $   3.560_{-  1.031}^{+   0.967} $& $   8.745_{-  2.426}^{+  2.216} $&$ 0.757_{-  0.174}^{+  0.263} $&  100.299 &     35.060 &   \\
UGC 05999  & $   2.897_{-  1.063}^{+   1.134} $& $   9.709_{-  1.223}^{+  2.519} $&$ 0.831_{-  0.354}^{+  0.563} $&  101.288 &     35.160 &   \\
DDO 154  & $   4.948_{-  0.447}^{+   0.405} $& $   3.350_{-  0.135}^{+  0.174} $&$ 1.019_{-  0.109}^{+  0.128} $&   46.467 &     36.123 &   \\
D 512-2  & $   5.405_{-  3.132}^{+   5.798} $& $   2.486_{-  0.839}^{+  1.806} $&$ 2.417_{-  1.252}^{+  1.828} $&   38.402 &     37.318 &   \\
UGC 05918  & $   4.378_{-  2.319}^{+   4.800} $& $   3.016_{-  0.987}^{+  1.653} $&$ 2.535_{-  1.115}^{+  1.582} $&   42.059 &     37.660 &   \\
UGC 07866  & $   9.026_{-  2.920}^{+   3.668} $& $   1.773_{-  0.353}^{+  0.400} $&$ 2.220_{-  1.026}^{+  1.744} $&   35.204 &     37.914 &   \\
UGCA 444  & $   0.318_{-  0.135}^{+   0.247} $& $  17.767_{-  6.037}^{+  7.771} $&$ 4.680_{-  0.758}^{+  0.971} $&   68.960 &     38.360 &   \\
UGC 01281  & $   8.332_{-  2.210}^{+   2.342} $& $   3.160_{-  0.272}^{+  0.523} $&$ 0.448_{-  0.187}^{+  0.248} $&   52.713 &     38.733 &   \\
F 568-3  & $   5.381_{-  1.437}^{+   1.622} $& $   7.479_{-  0.786}^{+  1.215} $&$ 0.642_{-  0.183}^{+  0.234} $&  103.863 &     38.734 &   \\
UGCA 442  & $   3.418_{-  1.219}^{+   1.452} $& $   4.616_{-  0.719}^{+  1.253} $&$ 1.403_{-  0.432}^{+  0.553} $&   54.597 &     39.362 &   \\
D 631-7  & $   5.383_{-  0.913}^{+   1.153} $& $   5.021_{-  0.503}^{+  0.603} $&$ 0.576_{-  0.162}^{+  0.168} $&   69.013 &     39.407 &   \\
DDO 161  & $   2.039_{-  0.561}^{+   0.618} $& $   7.235_{-  0.972}^{+  1.464} $&$ 1.872_{-  0.380}^{+  0.454} $&   67.509 &     39.498 &   \\
UGC 06667  & $   3.772_{-  1.353}^{+   1.454} $& $   6.233_{-  0.896}^{+  1.685} $&$ 1.311_{-  0.412}^{+  0.574} $&   77.061 &     40.338 &   \\
UGC 04278  & $   0.969_{-  0.334}^{+   0.621} $& $  25.370_{-  9.014}^{+  9.683} $&$ 1.634_{-  0.385}^{+  0.404} $&  161.626 &     40.932 &   \\
UGC 05829  & $   1.065_{-  0.416}^{+   0.725} $& $  12.583_{-  3.795}^{+  4.696} $&$ 2.576_{-  0.669}^{+  0.896} $&   86.622 &     41.118 &   \\
F 583-1  & $   4.474_{-  0.923}^{+   0.904} $& $   6.151_{-  0.529}^{+  0.750} $&$ 0.872_{-  0.191}^{+  0.249} $&   80.075 &     41.435 &   \\
UGC 07608  & $   4.481_{-  1.940}^{+   3.465} $& $   4.565_{-  1.333}^{+  2.064} $&$ 1.125_{-  0.529}^{+  0.766} $&   60.767 &     41.580 &   \\
F 563-1  & $   2.774_{-  1.161}^{+   1.481} $& $   8.638_{-  1.483}^{+  2.634} $&$ 1.777_{-  0.529}^{+  0.787} $&   93.678 &     41.799 &   \\
DDO 170  & $   3.438_{-  1.061}^{+   1.080} $& $   4.815_{-  0.610}^{+  1.024} $&$ 2.017_{-  0.515}^{+  0.727} $&   58.635 &     42.054 &   \\
NGC 0055  & $   4.178_{-  0.734}^{+   0.672} $& $   6.462_{-  0.491}^{+  0.730} $&$ 1.005_{-  0.170}^{+  0.220} $&   82.290 &     44.019 &   \\
ESO 444-G084  & $   1.776_{-  0.692}^{+   1.318} $& $   6.567_{-  1.925}^{+  2.447} $&$ 3.111_{-  0.676}^{+  0.677} $&   59.030 &     44.207 &   \\
UGC 07089  & $   1.039_{-  0.442}^{+   0.783} $& $  14.021_{-  4.510}^{+  6.637} $&$ 2.534_{-  0.590}^{+  0.743} $&   95.240 &     44.332 &   \\
NGC 0100  & $   2.924_{-  1.154}^{+   1.206} $& $   7.715_{-  1.325}^{+  2.715} $&$ 1.286_{-  0.370}^{+  0.527} $&   83.853 &     44.392 &   \\
UGC 09992  & $  52.805_{- 21.937}^{+  29.457} $& $   0.798_{-  0.171}^{+  0.240} $&$15.132_{-  9.657}^{+  9.907} $&   41.518 &     45.904 &   \\
F 583-4  & $   0.379_{-  0.192}^{+   0.405} $& $  20.575_{-  7.672}^{+ 11.494} $&$ 4.710_{-  1.135}^{+  1.428} $&   87.268 &     46.052 &   \\
UGC 02023  & $   9.438_{-  6.172}^{+   8.152} $& $  24.411_{- 15.996}^{+ 17.290} $&$ 0.429_{-  0.319}^{+  0.771} $&  151.760 &     46.397 &   \\
\clearpage
DDO 064  & $  14.964_{-  5.633}^{+   8.384} $& $   1.935_{-  0.410}^{+  0.649} $&$ 0.715_{-  0.384}^{+  0.557} $&   45.246 &     46.531 &   \\
F 571-8  & $   2.004_{-  0.323}^{+   0.295} $& $  16.404_{-  1.033}^{+  1.522} $&$ 0.721_{-  0.143}^{+  0.185} $&  140.528 &     46.540 &   \\
UGC 05986  & $   5.732_{-  0.612}^{+   0.583} $& $   7.076_{-  0.257}^{+  0.328} $&$ 0.671_{-  0.108}^{+  0.134} $&  101.836 &     48.169 &   \\
NGC 5585  & $   0.055_{-  0.014}^{+   0.023} $& $  77.844_{- 15.275}^{+ 14.845} $&$ 7.395_{-  0.449}^{+  0.400} $&  128.583 &     48.280 &   \\
UGC 07524  & $   4.619_{-  1.324}^{+   1.287} $& $   5.887_{-  0.671}^{+  1.161} $&$ 1.376_{-  0.316}^{+  0.432} $&   80.845 &     48.600 &   \\
UGC 06399  & $   4.367_{-  1.907}^{+   2.589} $& $   5.958_{-  1.257}^{+  2.387} $&$ 1.348_{-  0.570}^{+  0.760} $&   79.429 &     48.721 &   \\
UGC 08550  & $   5.043_{-  2.138}^{+   3.138} $& $   3.522_{-  0.800}^{+  1.260} $&$ 2.951_{-  0.802}^{+  0.978} $&   53.195 &     49.617 &   \\
NGC 4068  & $   8.744_{-  4.121}^{+   4.662} $& $   6.027_{-  2.587}^{+  2.685} $&$ 0.750_{-  0.308}^{+  0.540} $&  108.220 &     49.867 &   \\
F 568-1  & $   6.469_{-  2.767}^{+   3.656} $& $   5.920_{-  0.931}^{+  1.933} $&$ 0.996_{-  0.445}^{+  0.641} $&   93.739 &     49.959 &   \\
UGC 00731  & $   3.671_{-  1.604}^{+   2.217} $& $   5.168_{-  1.145}^{+  1.928} $&$ 2.942_{-  0.761}^{+  0.962} $&   66.580 &     50.096 &   \\
UGC 05414  & $   5.383_{-  1.979}^{+   3.403} $& $   4.404_{-  1.257}^{+  1.629} $&$ 1.478_{-  0.439}^{+  0.533} $&   65.647 &     50.227 &   \\
NGC 0300  & $   2.046_{-  0.786}^{+   1.057} $& $   8.694_{-  1.749}^{+  2.755} $&$ 2.339_{-  0.573}^{+  0.676} $&   82.471 &     50.683 &   \\
DDO 168  & $  21.386_{-  2.035}^{+   2.109} $& $   2.488_{-  0.080}^{+  0.088} $&$ 0.149_{-  0.058}^{+  0.069} $&   59.457 &     51.025 &   \\
UGC 07125  & $   5.525_{-  2.076}^{+   2.322} $& $   3.859_{-  0.624}^{+  1.104} $&$ 2.144_{-  0.624}^{+  0.879} $&   59.816 &     51.257 &   \\
UGC 07323  & $   0.873_{-  0.344}^{+   0.582} $& $  21.897_{-  7.112}^{+  8.929} $&$ 2.734_{-  0.535}^{+  0.676} $&  137.026 &     51.708 &   \\
UGC 07603  & $   8.408_{-  2.604}^{+   2.621} $& $   3.170_{-  0.372}^{+  0.667} $&$ 1.147_{-  0.332}^{+  0.474} $&   57.898 &     51.750 &   \\
UGC 00634  & $   2.062_{-  0.652}^{+   0.552} $& $  10.893_{-  1.127}^{+  2.293} $&$ 1.335_{-  0.444}^{+  0.716} $&   99.725 &     53.507 &   \\
UGC 05764  & $  24.718_{-  2.454}^{+   2.077} $& $   1.520_{-  0.055}^{+  0.070} $&$ 0.903_{-  0.161}^{+  0.218} $&   46.655 &     53.689 &   \\
ESO 116-G012  & $   4.156_{-  1.105}^{+   1.028} $& $   7.552_{-  0.750}^{+  1.317} $&$ 1.281_{-  0.287}^{+  0.402} $&   97.831 &     53.897 &   \\
ESO 079-G014  & $   4.281_{-  0.646}^{+   0.593} $& $  11.476_{-  0.658}^{+  0.956} $&$ 0.775_{-  0.145}^{+  0.188} $&  144.640 &     54.376 &   \\
UGC 05716  & $   2.918_{-  0.621}^{+   0.626} $& $   6.552_{-  0.617}^{+  0.882} $&$ 2.043_{-  0.348}^{+  0.431} $&   73.574 &     54.384 &   \\
NGC 2915  & $   6.776_{-  1.746}^{+   1.775} $& $   4.214_{-  0.410}^{+  0.584} $&$ 1.619_{-  0.447}^{+  0.656} $&   70.941 &     54.523 &   \\
UGC 04499  & $   6.743_{-  2.732}^{+   2.981} $& $   3.985_{-  0.655}^{+  1.294} $&$ 1.517_{-  0.509}^{+  0.750} $&   66.598 &     54.894 &   \\
F 574-1  & $   5.886_{-  1.590}^{+   1.508} $& $   5.764_{-  0.561}^{+  0.942} $&$ 1.230_{-  0.305}^{+  0.430} $&   88.576 &     54.968 &   \\
UGC 11820  & $   0.421_{-  0.131}^{+   0.251} $& $  22.838_{-  5.504}^{+  5.658} $&$ 4.156_{-  0.739}^{+  0.648} $&  101.448 &     55.359 &   \\
UGC 12632  & $   5.153_{-  2.193}^{+   2.719} $& $   4.389_{-  0.824}^{+  1.508} $&$ 2.316_{-  0.762}^{+  1.077} $&   66.033 &     55.400 &   \\
UGC 09037  & $   2.915_{-  0.738}^{+   0.708} $& $  12.604_{-  1.292}^{+  2.098} $&$ 1.339_{-  0.280}^{+  0.388} $&  137.220 &     55.637 &   \\
ESO 563-G021  & $   4.822_{-  0.395}^{+   0.398} $& $  15.615_{-  0.481}^{+  0.558} $&$ 0.990_{-  0.081}^{+  0.090} $&  213.375 &     56.165 &   \\
F 568-V1  & $   4.169_{-  2.058}^{+   2.880} $& $   6.219_{-  1.451}^{+  2.634} $&$ 2.677_{-  1.241}^{+  2.016} $&   84.913 &     56.484 &   \\
UGC 12732  & $   0.494_{-  0.172}^{+   0.317} $& $  20.767_{-  5.270}^{+  5.953} $&$ 4.671_{-  0.816}^{+  0.864} $&  100.502 &     56.508 &   \\
UGC 01230  & $   5.780_{-  1.870}^{+   2.134} $& $   8.057_{-  1.037}^{+  1.358} $&$ 1.819_{-  0.589}^{+  0.930} $&  126.330 &     56.937 &   \\
NGC 4010  & $   9.404_{-  1.956}^{+   1.780} $& $   6.450_{-  0.366}^{+  0.455} $&$ 0.313_{-  0.166}^{+  0.264} $&  110.456 &     57.582 &   \\
UGC 10310  & $   7.338_{-  3.890}^{+   5.073} $& $   3.819_{-  0.837}^{+  1.924} $&$ 2.144_{-  0.957}^{+  1.532} $&   68.214 &     58.734 &   \\
UGC 06923  & $   9.323_{-  4.507}^{+   6.262} $& $   3.838_{-  0.896}^{+  2.057} $&$ 1.147_{-  0.618}^{+  0.891} $&   73.806 &     59.011 &   \\
UGC 11557  & $   8.403_{-  3.906}^{+   4.851} $& $   6.400_{-  1.172}^{+  2.823} $&$ 0.776_{-  0.412}^{+  0.632} $&  113.019 &     59.881 &   \\
NGC 1003  & $   0.139_{-  0.020}^{+   0.040} $& $  44.511_{-  5.825}^{+  3.921} $&$ 7.182_{-  0.610}^{+  0.544} $&  116.503 &     60.004 &   \\
NGC 3917  & $   7.790_{-  1.248}^{+   1.181} $& $   6.530_{-  0.404}^{+  0.569} $&$ 0.966_{-  0.167}^{+  0.212} $&  113.179 &     60.384 &   \\
F 563-V2  & $  11.762_{-  5.497}^{+   7.509} $& $   4.080_{-  0.626}^{+  1.245} $&$ 0.815_{-  0.408}^{+  0.665} $&   85.596 &     60.701 &   \\
UGC 07690  & $  39.453_{- 22.461}^{+  34.945} $& $   1.176_{-  0.311}^{+  0.585} $&$ 5.507_{-  3.357}^{+  5.786} $&   51.233 &     60.807 &   \\
UGC 06917  & $   5.000_{-  2.033}^{+   2.276} $& $   6.466_{-  1.110}^{+  2.166} $&$ 1.485_{-  0.500}^{+  0.716} $&   92.922 &     62.113 &   \\
\clearpage
NGC 0247  & $   0.733_{-  0.230}^{+   0.383} $& $  17.451_{-  4.015}^{+  4.670} $&$ 4.096_{-  0.525}^{+  0.475} $&  102.272 &     62.293 &   \\
UGC 06446  & $   3.942_{-  1.686}^{+   2.794} $& $   5.499_{-  1.336}^{+  1.942} $&$ 3.380_{-  1.106}^{+  1.335} $&   73.976 &     62.666 &   \\
NGC 4214  & $   0.029_{-  0.009}^{+   0.010} $& $  67.969_{- 12.906}^{+ 14.253} $&$18.336_{-  4.497}^{+  7.093} $&   83.016 &     62.702 &   \\
UGC 07399  & $   3.999_{-  1.444}^{+   1.776} $& $   6.636_{-  1.181}^{+  1.894} $&$ 2.081_{-  0.536}^{+  0.679} $&   87.346 &     63.879 &   \\
UGC 07151  & $  16.330_{-  4.088}^{+   3.992} $& $   2.377_{-  0.239}^{+  0.377} $&$ 1.264_{-  0.288}^{+  0.398} $&   60.944 &     64.602 &   \\
UGC 07261  & $   4.012_{-  1.870}^{+   3.755} $& $   4.825_{-  1.446}^{+  1.954} $&$ 6.241_{-  2.666}^{+  5.349} $&   67.377 &     65.574 &   \\
UGC 00128  & $   0.600_{-  0.129}^{+   0.147} $& $  29.520_{-  3.228}^{+  4.067} $&$ 4.528_{-  0.572}^{+  0.625} $&  157.180 &     65.970 &   \\
UGC 06930  & $   3.510_{-  1.777}^{+   2.565} $& $   8.157_{-  1.922}^{+  3.655} $&$ 3.230_{-  1.329}^{+  2.077} $&  103.301 &     66.710 &   \\
UGC 06628  & $  23.971_{- 13.054}^{+  20.973} $& $   2.072_{-  0.600}^{+  0.971} $&$ 4.718_{-  3.020}^{+  3.506} $&   69.889 &     68.725 &   \\
UGC 06983  & $   4.681_{-  1.928}^{+   2.002} $& $   6.698_{-  1.064}^{+  2.094} $&$ 2.193_{-  0.738}^{+  1.158} $&   95.707 &     68.852 &   \\
UGC 08286  & $   5.913_{-  1.708}^{+   1.811} $& $   4.555_{-  0.584}^{+  0.902} $&$ 2.612_{-  0.489}^{+  0.637} $&   73.960 &     70.313 &   \\
UGC 02259  & $   0.272_{-  0.134}^{+   0.242} $& $  23.210_{-  6.857}^{+ 10.335} $&$13.833_{-  2.767}^{+  3.582} $&   86.464 &     71.815 &   \\
UGC 08490  & $   6.569_{-  1.931}^{+   2.020} $& $   4.092_{-  0.489}^{+  0.762} $&$ 3.350_{-  0.720}^{+  0.944} $&   71.032 &     73.839 &   \\
UGC 00191  & $   4.650_{-  1.504}^{+   1.933} $& $   5.308_{-  0.915}^{+  1.305} $&$ 2.979_{-  0.490}^{+  0.550} $&   77.015 &     74.831 &   \\
UGC 03580  & $   0.113_{-  0.041}^{+   0.047} $& $  56.846_{- 10.210}^{+ 16.508} $&$ 7.941_{-  0.648}^{+  0.820} $&  134.297 &     74.983 &   \\
NGC 3972  & $   2.931_{-  1.529}^{+   2.789} $& $  10.872_{-  3.673}^{+  6.563} $&$ 2.124_{-  0.724}^{+  0.855} $&  122.668 &     75.545 &   \\
UGC 05721  & $  14.790_{-  2.883}^{+   2.841} $& $   2.934_{-  0.205}^{+  0.267} $&$ 1.380_{-  0.319}^{+  0.450} $&   72.081 &     76.546 &   \\
NGC 4183  & $   3.938_{-  1.821}^{+   2.536} $& $   7.063_{-  1.544}^{+  2.688} $&$ 4.321_{-  1.300}^{+  1.741} $&   96.166 &     77.856 &   \\
UGC 04325  & $  25.617_{-  6.185}^{+   5.925} $& $   2.293_{-  0.190}^{+  0.288} $&$ 1.352_{-  0.338}^{+  0.513} $&   74.054 &     78.782 &   \\
UGC 11455  & $   3.126_{-  0.413}^{+   0.402} $& $  19.362_{-  1.024}^{+  1.331} $&$ 1.430_{-  0.163}^{+  0.195} $&  219.389 &     81.435 &   \\
NGC 2976  & $  14.343_{-  6.094}^{+   9.634} $& $   5.013_{-  1.988}^{+  3.692} $&$ 1.050_{-  0.312}^{+  0.369} $&  118.721 &     84.677 &   \\
NGC 4559  & $   3.221_{-  1.272}^{+   1.486} $& $   9.041_{-  1.590}^{+  2.790} $&$ 2.659_{-  0.781}^{+  1.020} $&  108.467 &     85.704 &   \\
NGC 7793  & $  10.009_{-  4.278}^{+   5.189} $& $   4.166_{-  0.791}^{+  1.487} $&$ 2.472_{-  0.668}^{+  0.863} $&   87.708 &     90.253 &   \\
NGC 3198  & $   2.904_{-  0.273}^{+   0.264} $& $  11.875_{-  0.464}^{+  0.559} $&$ 2.591_{-  0.199}^{+  0.231} $&  135.058 &     90.673 &   \\
F 579-V1  & $  12.317_{-  6.449}^{+   9.315} $& $   3.928_{-  0.914}^{+  1.733} $&$ 4.137_{-  1.491}^{+  2.348} $&   94.381 &     91.310 &   \\
NGC 3726  & $   1.169_{-  0.637}^{+   1.140} $& $  19.930_{-  6.252}^{+ 11.005} $&$ 4.304_{-  1.321}^{+  1.693} $&  147.807 &     91.435 &   \\
NGC 3769  & $   2.629_{-  1.044}^{+   1.182} $& $  10.083_{-  1.741}^{+  2.865} $&$ 3.471_{-  1.156}^{+  1.958} $&  110.951 &     92.687 &   \\
NGC 4085  & $  16.594_{-  6.395}^{+   7.896} $& $   4.552_{-  0.853}^{+  1.833} $&$ 0.569_{-  0.283}^{+  0.418} $&  109.782 &     93.255 &   \\
NGC 2403  & $   2.150_{-  0.158}^{+   0.155} $& $  11.489_{-  0.430}^{+  0.494} $&$ 3.466_{-  0.134}^{+  0.146} $&  114.309 &     93.323 &   \\
NGC 0024  & $   2.531_{-  1.311}^{+   2.309} $& $   8.745_{-  2.546}^{+  4.343} $&$ 4.738_{-  1.142}^{+  1.294} $&   95.871 &     95.824 &   \\
NGC 3877  & $  30.472_{-  5.087}^{+   5.744} $& $   3.772_{-  0.240}^{+  0.272} $&$ 0.640_{-  0.182}^{+  0.239} $&  124.633 &     99.039 &   \\
NGC 6503  & $   2.789_{-  0.365}^{+   0.359} $& $   9.236_{-  0.527}^{+  0.658} $&$ 3.565_{-  0.419}^{+  0.498} $&  104.811 &    105.152 &   \\
NGC 1090  & $   4.481_{-  0.852}^{+   0.849} $& $   9.712_{-  0.781}^{+  1.064} $&$ 2.420_{-  0.369}^{+  0.467} $&  136.640 &    111.164 &   \\
NGC 4217  & $   6.426_{-  1.990}^{+   1.956} $& $   9.091_{-  1.081}^{+  1.951} $&$ 1.224_{-  0.354}^{+  0.487} $&  145.920 &    111.773 &   \\
NGC 4088  & $   1.045_{-  0.595}^{+   1.101} $& $  21.268_{-  7.093}^{+ 13.420} $&$ 5.359_{-  1.606}^{+  2.181} $&  150.608 &    112.843 &   \\
NGC 4100  & $   8.527_{-  2.060}^{+   1.980} $& $   7.043_{-  0.649}^{+  0.927} $&$ 2.225_{-  0.668}^{+  1.047} $&  135.957 &    113.577 &   \\
NGC 4157  & $   0.813_{-  0.448}^{+   0.879} $& $  26.817_{-  8.967}^{+ 15.265} $&$ 5.163_{-  1.613}^{+  1.885} $&  167.288 &    115.527 &   \\
NGC 4051  & $  14.806_{-  8.174}^{+  11.304} $& $   4.474_{-  1.042}^{+  2.048} $&$ 3.376_{-  2.321}^{+  4.397} $&  116.645 &    122.867 &   \\
NGC 3949  & $   0.186_{-  0.105}^{+   0.191} $& $  68.550_{- 26.605}^{+ 32.620} $&$ 7.252_{-  2.458}^{+  4.963} $&  207.004 &    131.944 &   \\
NGC 6015  & $   0.104_{-  0.010}^{+   0.013} $& $  73.416_{-  4.950}^{+  4.529} $&$ 9.786_{-  0.336}^{+  0.359} $&  167.406 &    138.659 &   \\
\clearpage
NGC 6946  & $   6.090_{-  1.649}^{+   1.626} $& $   8.014_{-  0.966}^{+  1.559} $&$ 2.043_{-  0.366}^{+  0.485} $&  130.006 &    139.752 &   \\
NGC 2903  & $   3.443_{-  0.224}^{+   0.222} $& $  12.808_{-  0.392}^{+  0.432} $&$ 2.606_{-  0.217}^{+  0.246} $&  158.666 &    141.299 &   \\
NGC 3893  & $   6.532_{-  2.412}^{+   2.323} $& $   8.903_{-  1.187}^{+  2.245} $&$ 1.412_{-  0.512}^{+  0.850} $&  145.681 &    145.805 &   \\
IC 4202  & $ 103.974_{-  6.181}^{+   6.572} $& $   2.963_{-  0.067}^{+  0.068} $&$ 0.219_{-  0.041}^{+  0.044} $&  162.802 &    155.362 &   \\
UGC 12506  & $   3.022_{-  0.962}^{+   1.865} $& $  15.019_{-  3.228}^{+  3.225} $&$10.765_{-  3.240}^{+  4.173} $&  185.369 &    155.912 &   \\
NGC 3953  & $  12.727_{-  5.933}^{+   9.177} $& $   5.656_{-  1.350}^{+  2.090} $&$ 5.522_{-  3.088}^{+  5.609} $&  139.974 &    165.349 &   \\
NGC 4138  & $  20.322_{-  6.553}^{+   9.162} $& $   4.875_{-  0.960}^{+  0.969} $&$ 1.002_{-  0.616}^{+  0.662} $&  136.848 &    175.357 &   \\
UGC 06786  & $   0.795_{-  0.154}^{+   0.169} $& $  32.248_{-  3.069}^{+  3.875} $&$ 3.989_{-  0.450}^{+  0.509} $&  196.540 &    175.980 &   \\
NGC 2683  & $   8.579_{-  3.319}^{+   4.530} $& $   7.056_{-  1.493}^{+  1.769} $&$ 5.722_{-  3.014}^{+  6.955} $&  143.579 &    184.474 &   \\
NGC 2998  & $   0.054_{-  0.014}^{+   0.024} $& $ 117.195_{- 20.608}^{+ 21.541} $&$20.180_{-  3.286}^{+  4.199} $&  196.666 &    184.833 &   \\
UGC 03205  & $   0.162_{-  0.039}^{+   0.063} $& $  72.112_{- 12.249}^{+ 11.996} $&$ 9.323_{-  0.846}^{+  0.854} $&  204.997 &    188.861 &   \\
NGC 7331  & $   0.206_{-  0.045}^{+   0.066} $& $  73.686_{- 10.858}^{+ 11.111} $&$ 6.427_{-  0.597}^{+  0.643} $&  233.249 &    197.100 &   \\
NGC 0289  & $   1.534_{-  0.520}^{+   0.570} $& $  19.070_{-  2.655}^{+  4.292} $&$ 4.318_{-  1.255}^{+  1.746} $&  162.071 &    198.795 &   \\
UGC 08699  & $   1.019_{-  0.419}^{+   0.590} $& $  23.136_{-  5.227}^{+  8.173} $&$ 4.782_{-  0.854}^{+  0.969} $&  161.001 &    201.786 &   \\
NGC 5985  & $   1.839_{-  0.514}^{+   0.572} $& $  23.983_{-  3.167}^{+  4.589} $&$ 3.354_{-  0.566}^{+  0.713} $&  220.272 &    203.722 &   \\
NGC 3521  & $   0.604_{-  0.175}^{+   0.313} $& $  37.370_{-  8.344}^{+  8.398} $&$ 4.474_{-  0.745}^{+  0.882} $&  199.606 &    204.271 &   \\
NGC 5055  & $   3.016_{-  0.097}^{+   0.095} $& $  16.610_{-  0.206}^{+  0.223} $&$ 1.971_{-  0.071}^{+  0.076} $&  189.173 &    206.882 &   \\
NGC 0891  & $   7.908_{-  0.507}^{+   0.487} $& $  10.205_{-  0.263}^{+  0.274} $&$ 0.417_{-  0.070}^{+  0.085} $&  164.879 &    209.442 &   \\
NGC 5907  & $   0.032_{-  0.008}^{+   0.014} $& $ 160.201_{- 28.459}^{+ 26.972} $&$19.084_{-  2.048}^{+  2.311} $&  206.233 &    209.942 &   \\
NGC 4013  & $   0.092_{-  0.024}^{+   0.045} $& $  79.864_{- 15.559}^{+ 13.980} $&$13.121_{-  1.734}^{+  2.029} $&  173.176 &    213.446 &   \\
NGC 6674  & $   0.038_{-  0.007}^{+   0.014} $& $ 170.883_{- 25.775}^{+ 20.335} $&$18.839_{-  2.376}^{+  2.861} $&  240.395 &    213.831 &   \\
NGC 5033  & $   4.868_{-  0.206}^{+   0.207} $& $  12.658_{-  0.275}^{+  0.271} $&$ 0.916_{-  0.103}^{+  0.120} $&  172.632 &    215.180 &   \\
UGC 06614  & $   0.771_{-  0.249}^{+   0.464} $& $  35.184_{-  7.621}^{+  8.277} $&$ 2.379_{-  0.751}^{+  0.858} $&  205.130 &    215.427 &   \\
NGC 2841  & $   0.264_{-  0.081}^{+   0.104} $& $  70.651_{- 11.760}^{+ 15.846} $&$ 6.311_{-  0.785}^{+  0.874} $&  253.045 &    221.563 &   \\
UGC 03546  & $   4.989_{-  0.823}^{+   0.644} $& $  10.845_{-  0.556}^{+  0.926} $&$ 1.302_{-  0.314}^{+  0.463} $&  154.126 &    224.497 &   \\
NGC 3992  & $   0.209_{-  0.093}^{+   0.177} $& $  62.548_{- 17.175}^{+ 22.943} $&$23.132_{-  7.386}^{+  9.756} $&  206.545 &    226.200 &   \\
NGC 5371  & $   0.033_{-  0.011}^{+   0.019} $& $ 157.174_{- 37.190}^{+ 30.321} $&$12.911_{-  2.734}^{+  3.843} $&  204.544 &    230.266 &   \\
NGC 2955  & $   6.778_{-  1.404}^{+   1.326} $& $  13.523_{-  0.719}^{+  0.988} $&$ 0.633_{-  0.202}^{+  0.304} $&  210.523 &    231.898 &   \\
UGC 02916  & $   9.038_{-  1.138}^{+   1.304} $& $   8.555_{-  0.553}^{+  0.612} $&$ 0.844_{-  0.160}^{+  0.173} $&  157.854 &    233.305 &   \\
NGC 5005  & $   6.439_{-  3.332}^{+   5.837} $& $  13.183_{-  5.353}^{+  7.463} $&$ 1.032_{-  0.684}^{+  1.058} $&  208.888 &    235.651 &   \\
NGC 0801  & $   0.056_{-  0.012}^{+   0.015} $& $ 130.564_{- 16.852}^{+ 19.318} $&$11.468_{-  1.327}^{+  1.474} $&  220.493 &    236.328 &   \\
UGC 11914  & $   0.513_{-  0.084}^{+   0.160} $& $  60.221_{- 10.498}^{+  6.956} $&$ 3.211_{-  0.324}^{+  0.321} $&  291.419 &    237.935 &   \\
UGC 02953  & $   1.549_{-  0.089}^{+   0.088} $& $  26.697_{-  0.760}^{+  0.845} $&$ 2.829_{-  0.118}^{+  0.126} $&  222.908 &    240.853 &   \\
NGC 6195  & $   0.104_{-  0.035}^{+   0.068} $& $ 125.123_{- 32.578}^{+ 36.431} $&$ 6.483_{-  0.809}^{+  0.803} $&  281.342 &    241.926 &   \\
UGC 05253  & $   3.355_{-  0.160}^{+   0.154} $& $  16.351_{-  0.357}^{+  0.393} $&$ 1.495_{-  0.089}^{+  0.096} $&  192.593 &    245.137 &   \\
NGC 7814  & $   1.497_{-  0.641}^{+   1.029} $& $  23.241_{-  6.048}^{+  9.061} $&$ 2.633_{-  0.795}^{+  0.901} $&  189.943 &    248.367 &   \\
UGC 06787  & $   0.036_{-  0.007}^{+   0.012} $& $ 183.521_{- 27.329}^{+ 24.166} $&$12.142_{-  0.736}^{+  0.606} $&  246.840 &    255.110 &   \\
UGC 02885  & $   0.357_{-  0.119}^{+   0.220} $& $  64.595_{- 14.781}^{+ 16.304} $&$ 4.253_{-  1.020}^{+  1.099} $&  264.791 &    296.261 &   \\
UGC 09133  & $   1.369_{-  0.088}^{+   0.084} $& $  27.905_{-  0.795}^{+  0.919} $&$ 2.921_{-  0.181}^{+  0.195} $&  219.488 &    311.334 &   \\
UGC 02487  & $   1.220_{-  0.351}^{+   0.395} $& $  38.987_{-  4.915}^{+  7.301} $&$ 3.255_{-  0.873}^{+  1.113} $&  291.172 &    427.767 &   \\ 
\end{longtable}
}
\clearpage

\begin{figure*}
\centering
\includegraphics[width=.33\textwidth]{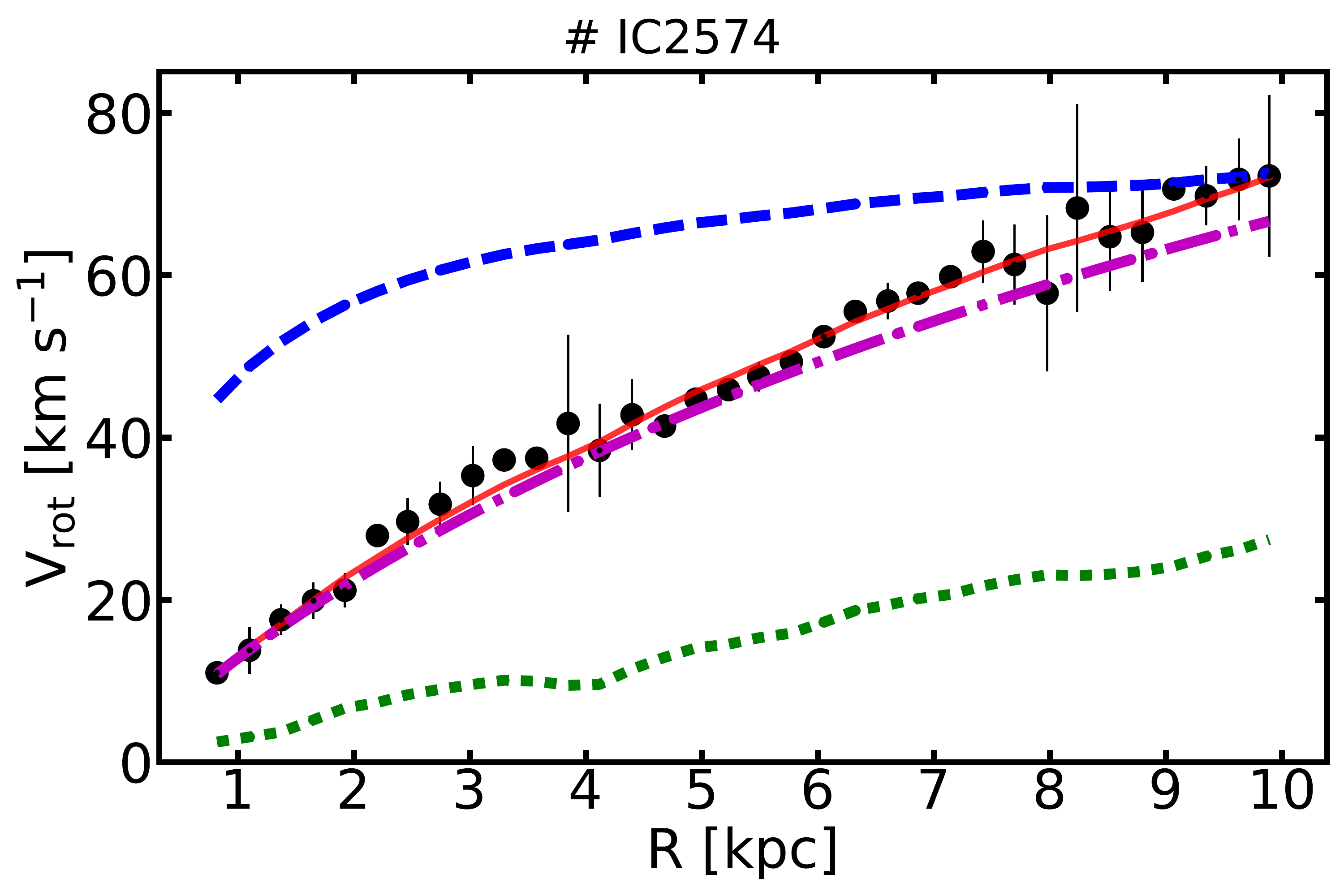}\hfill
\includegraphics[width=.33\textwidth]{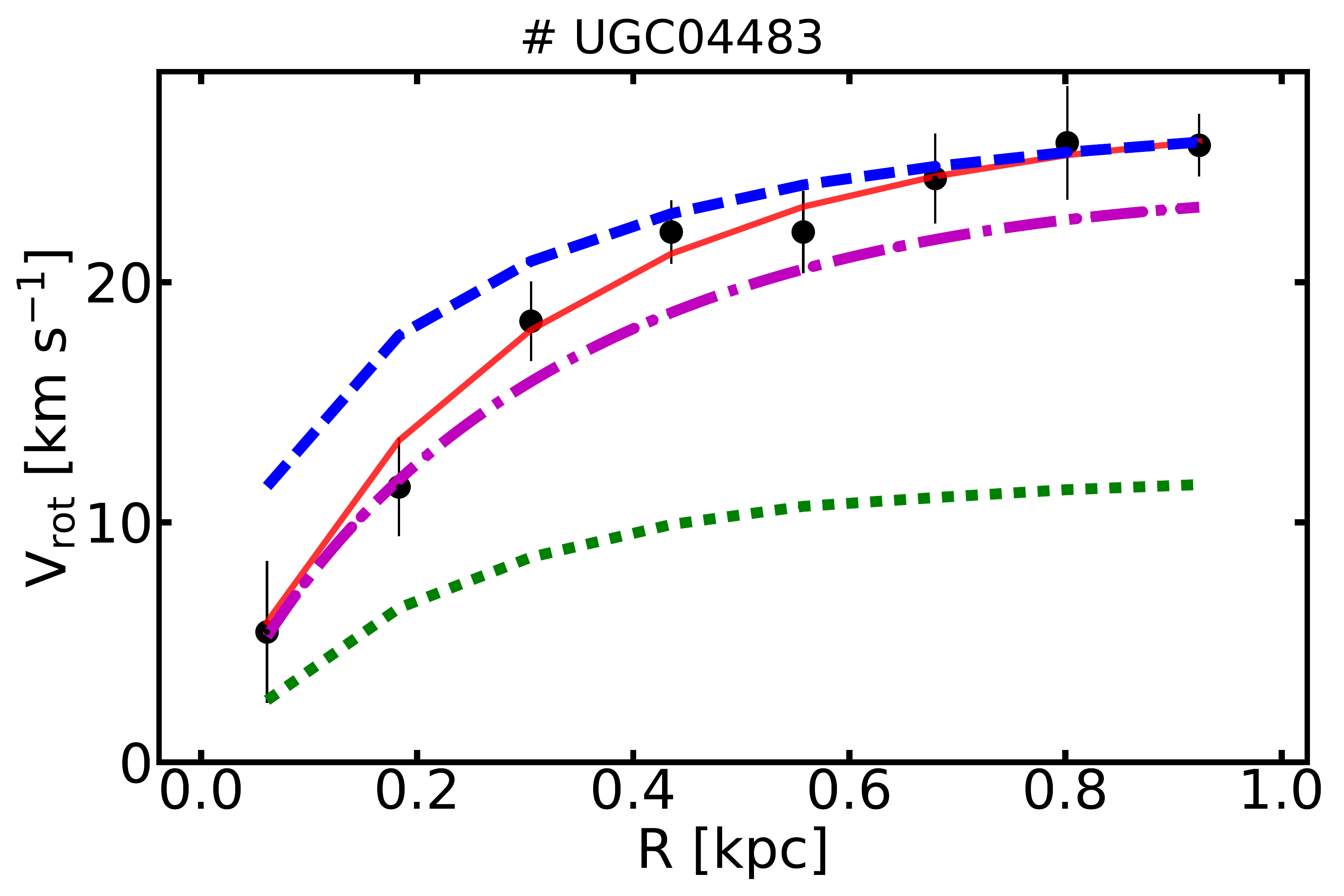}\hfill
\includegraphics[width=.33\textwidth]{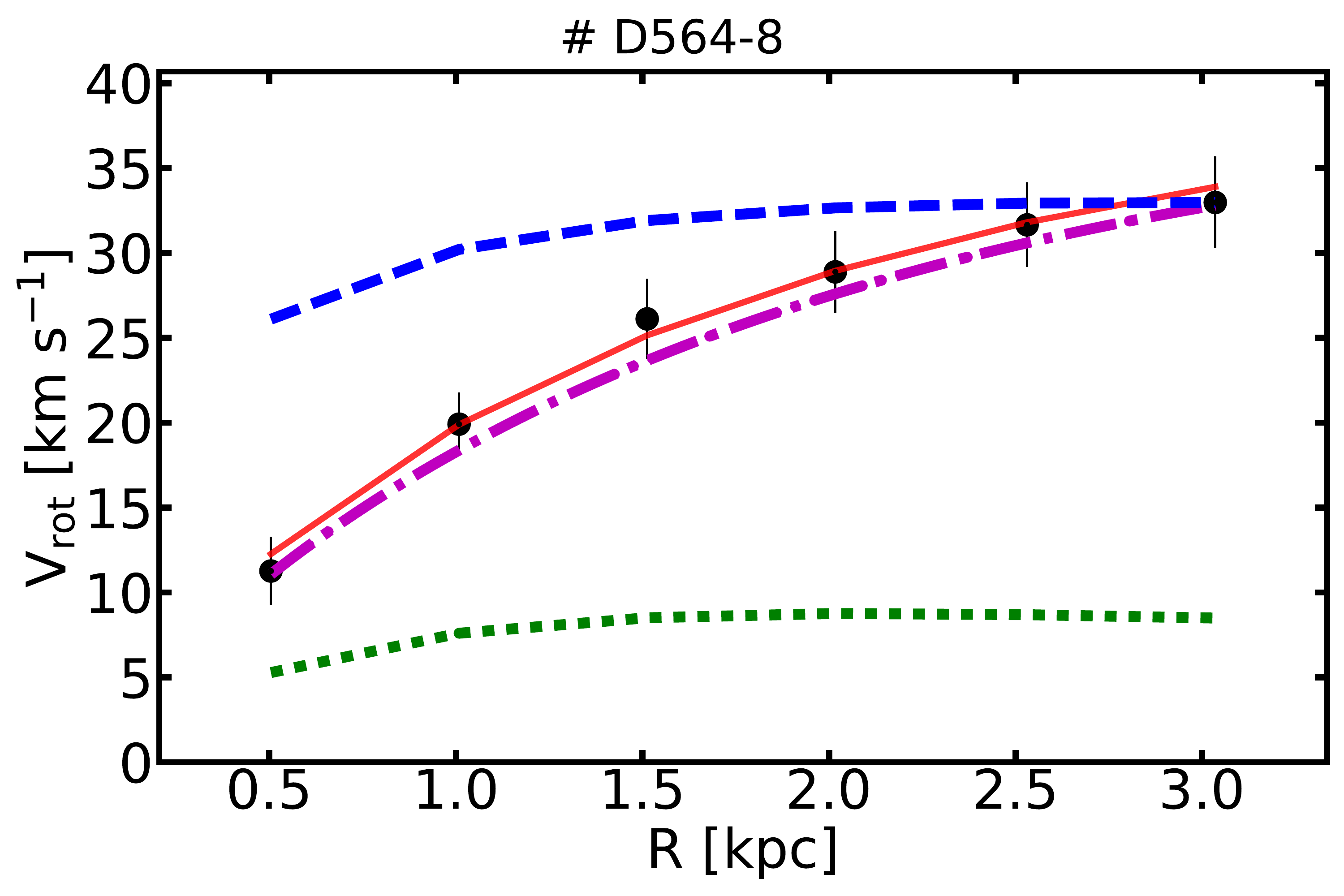}

\includegraphics[width=.33\textwidth]{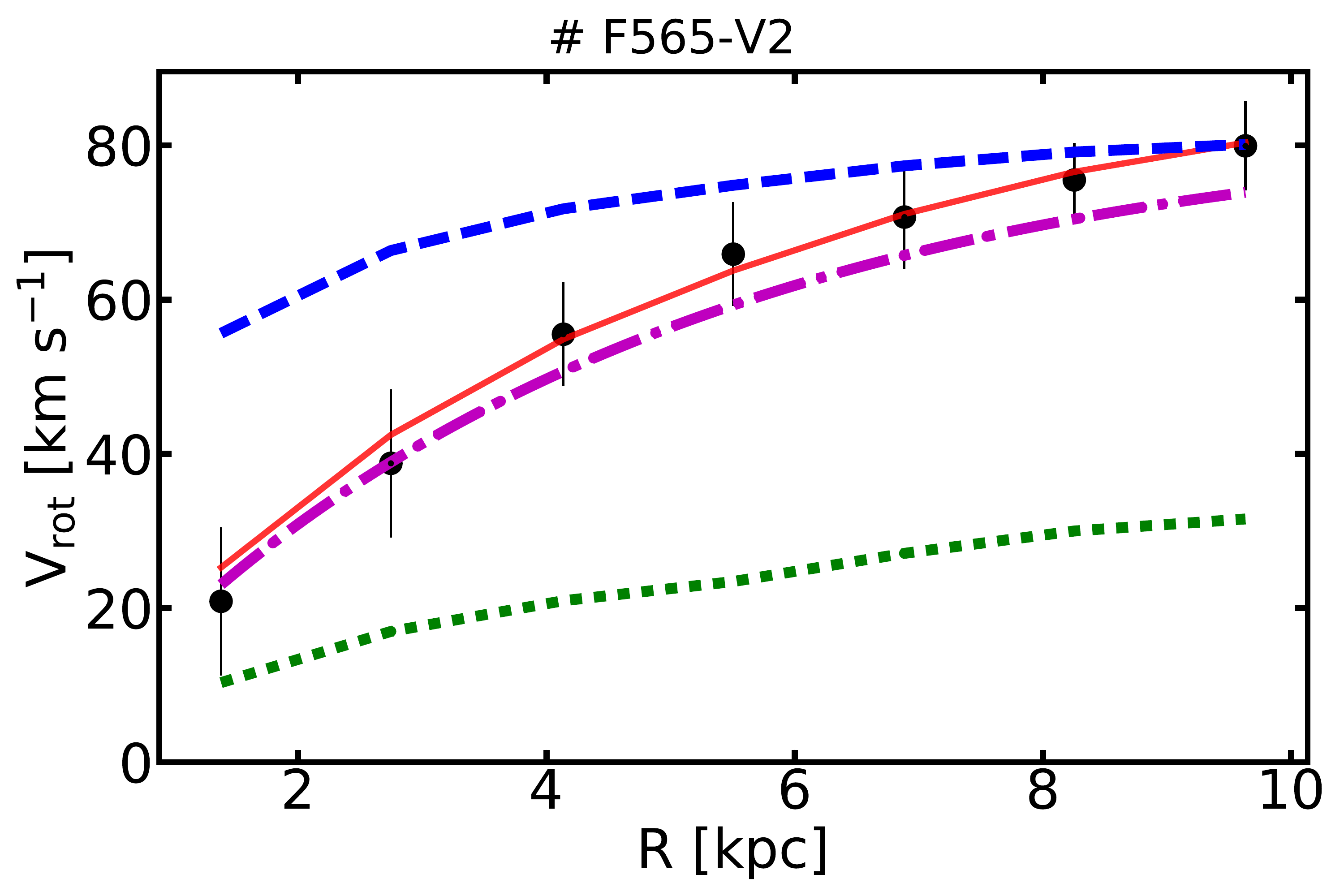}\hfill
\includegraphics[width=.33\textwidth]{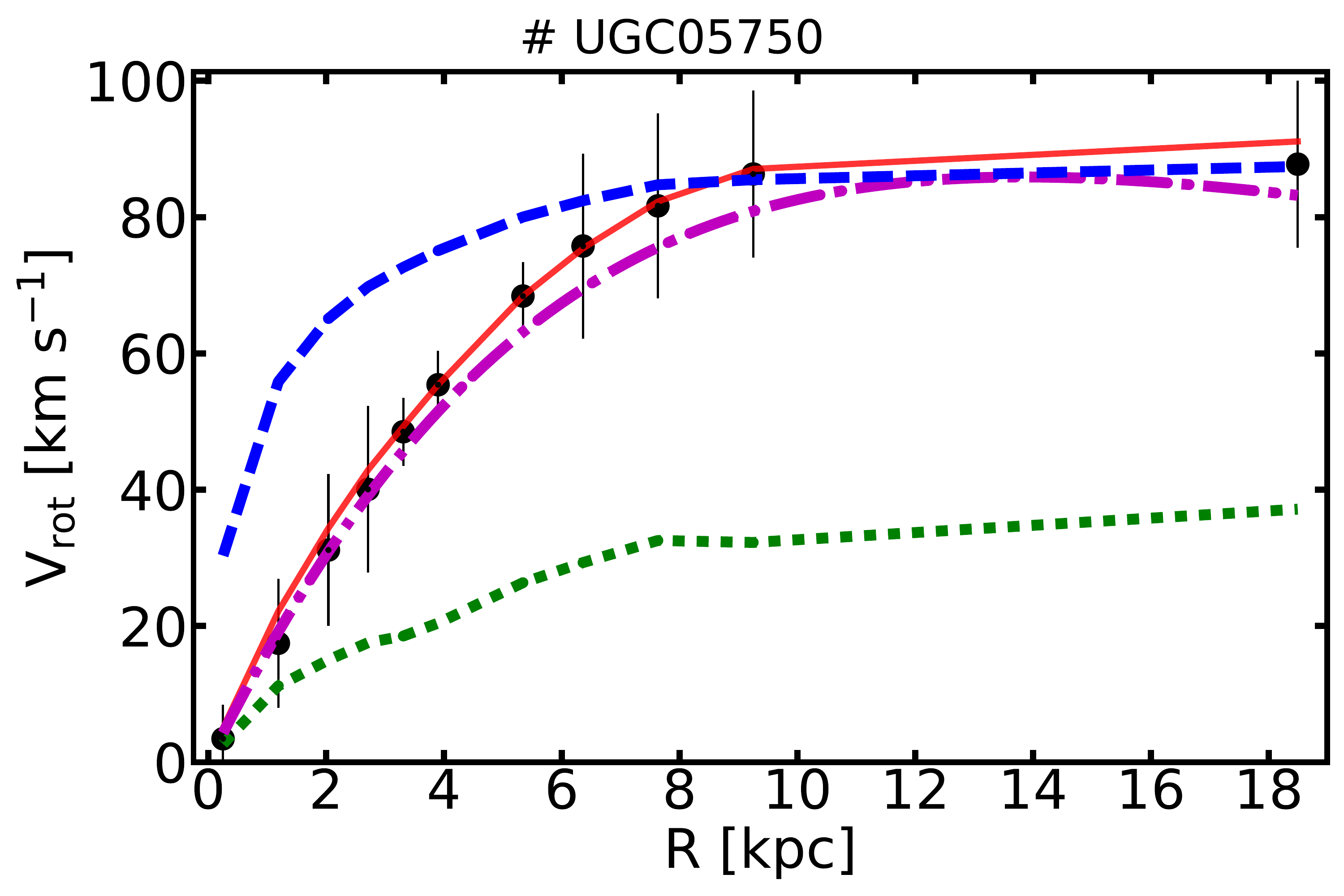}\hfill
\includegraphics[width=.33\textwidth]{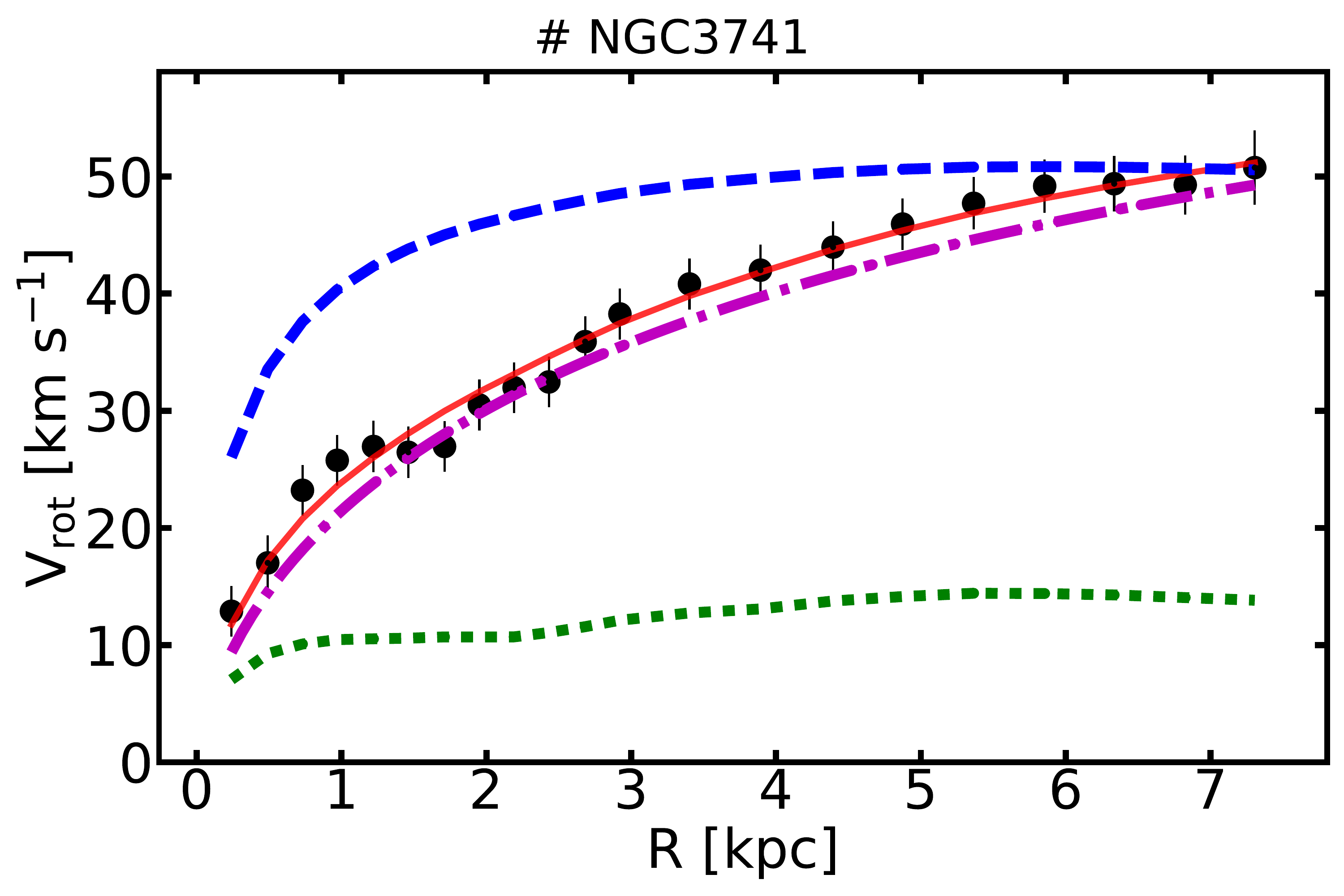}

\includegraphics[width=.33\textwidth]{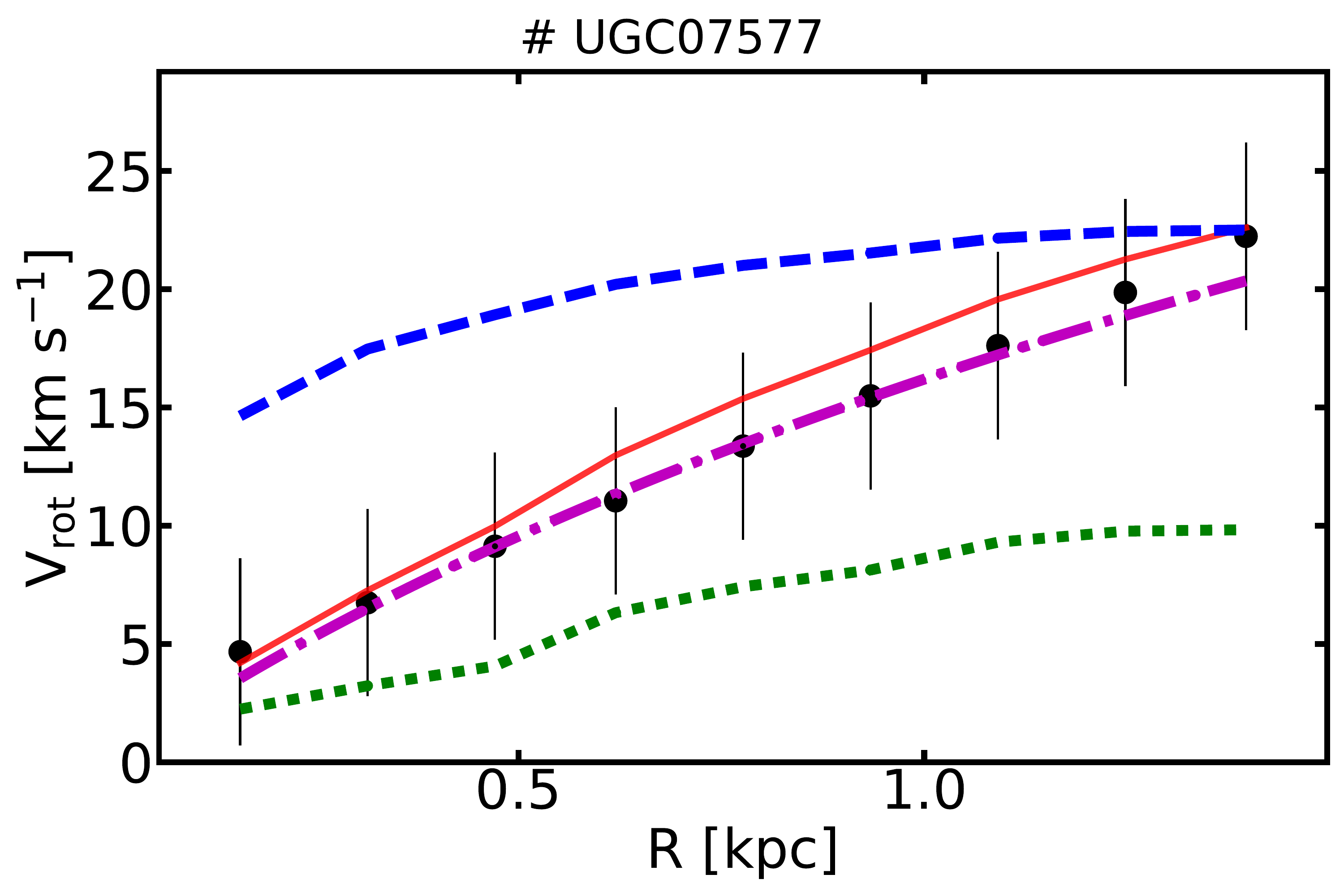}\hfill
\includegraphics[width=.33\textwidth]{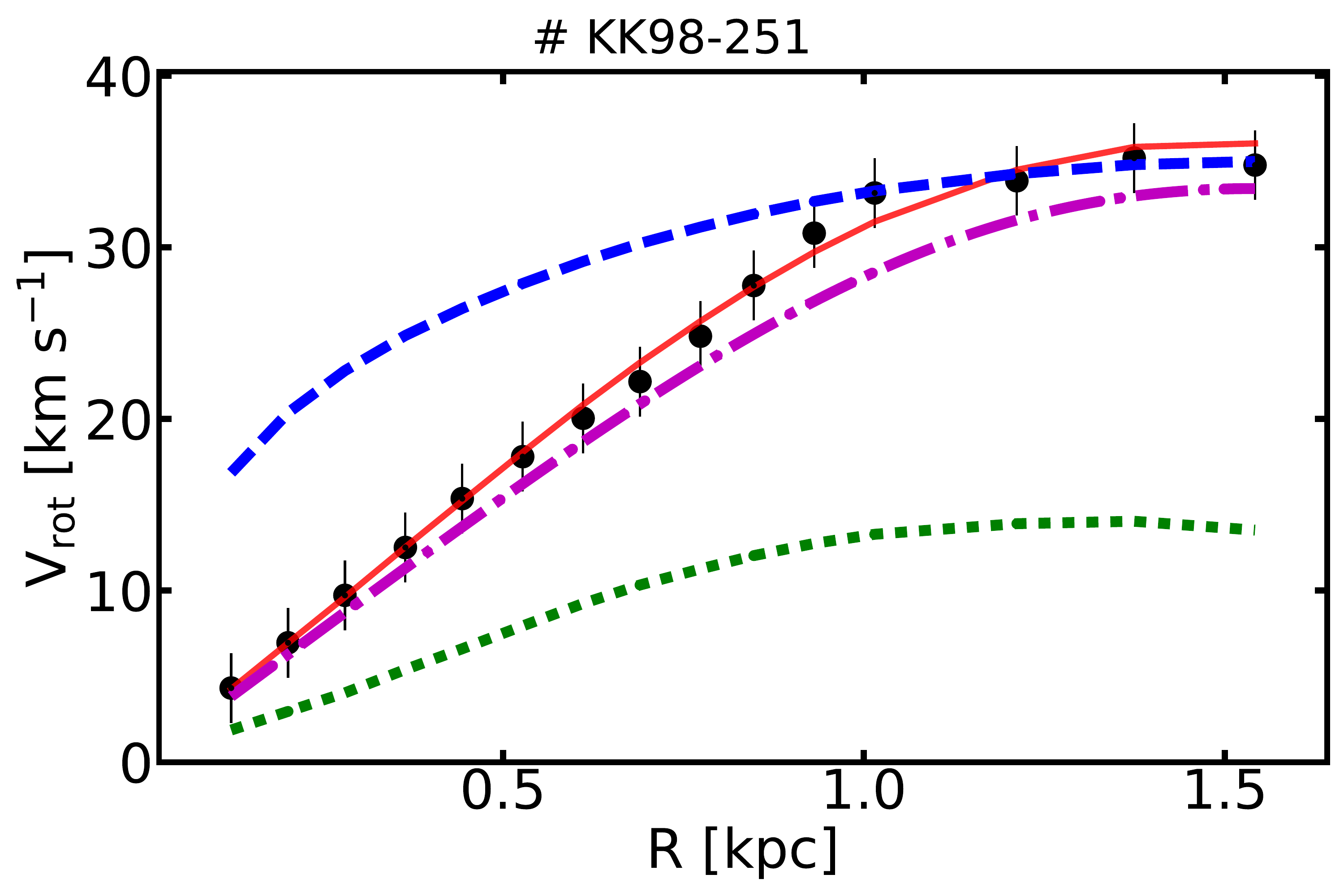}\hfill
\includegraphics[width=.33\textwidth]{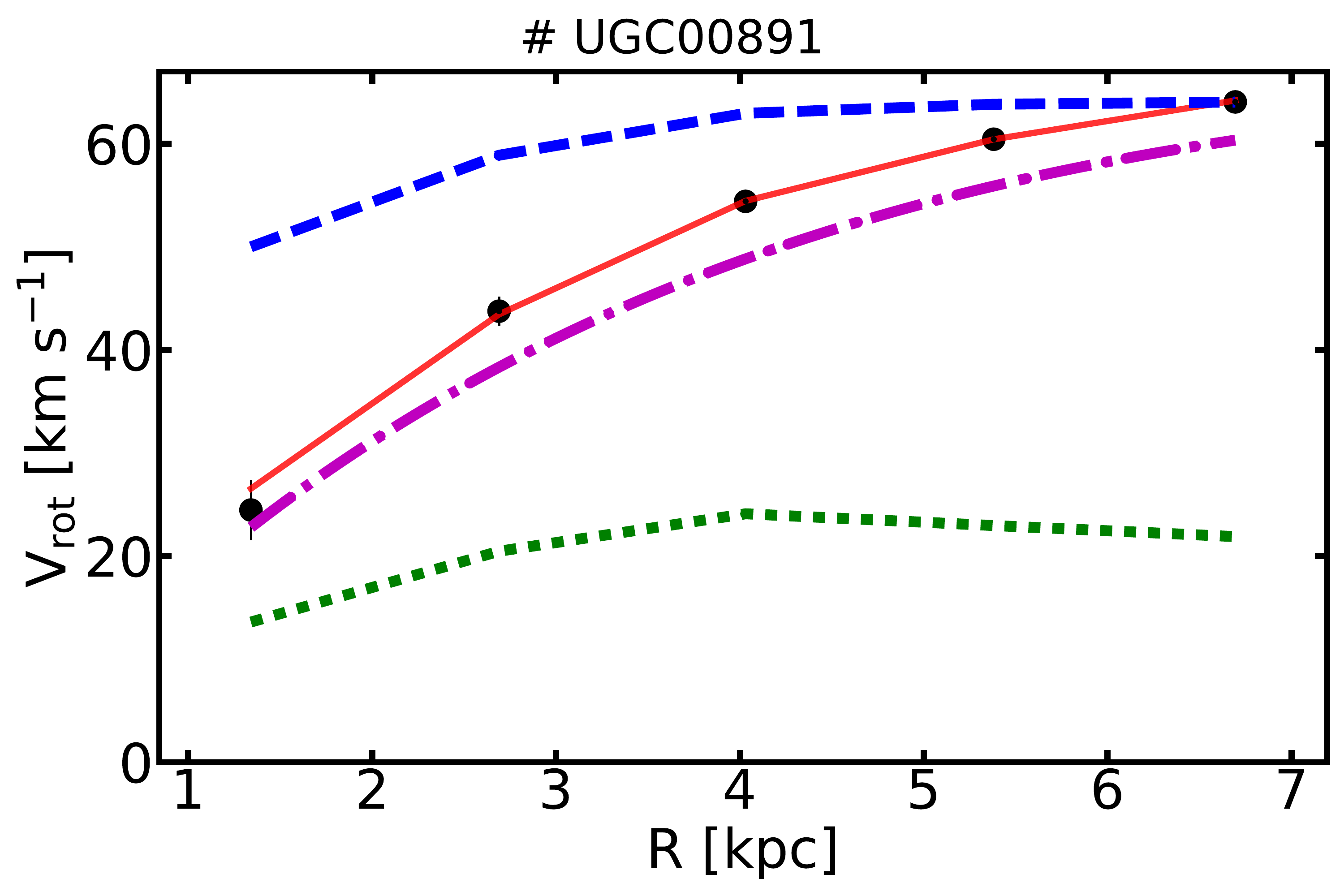}

\includegraphics[width=.33\textwidth]{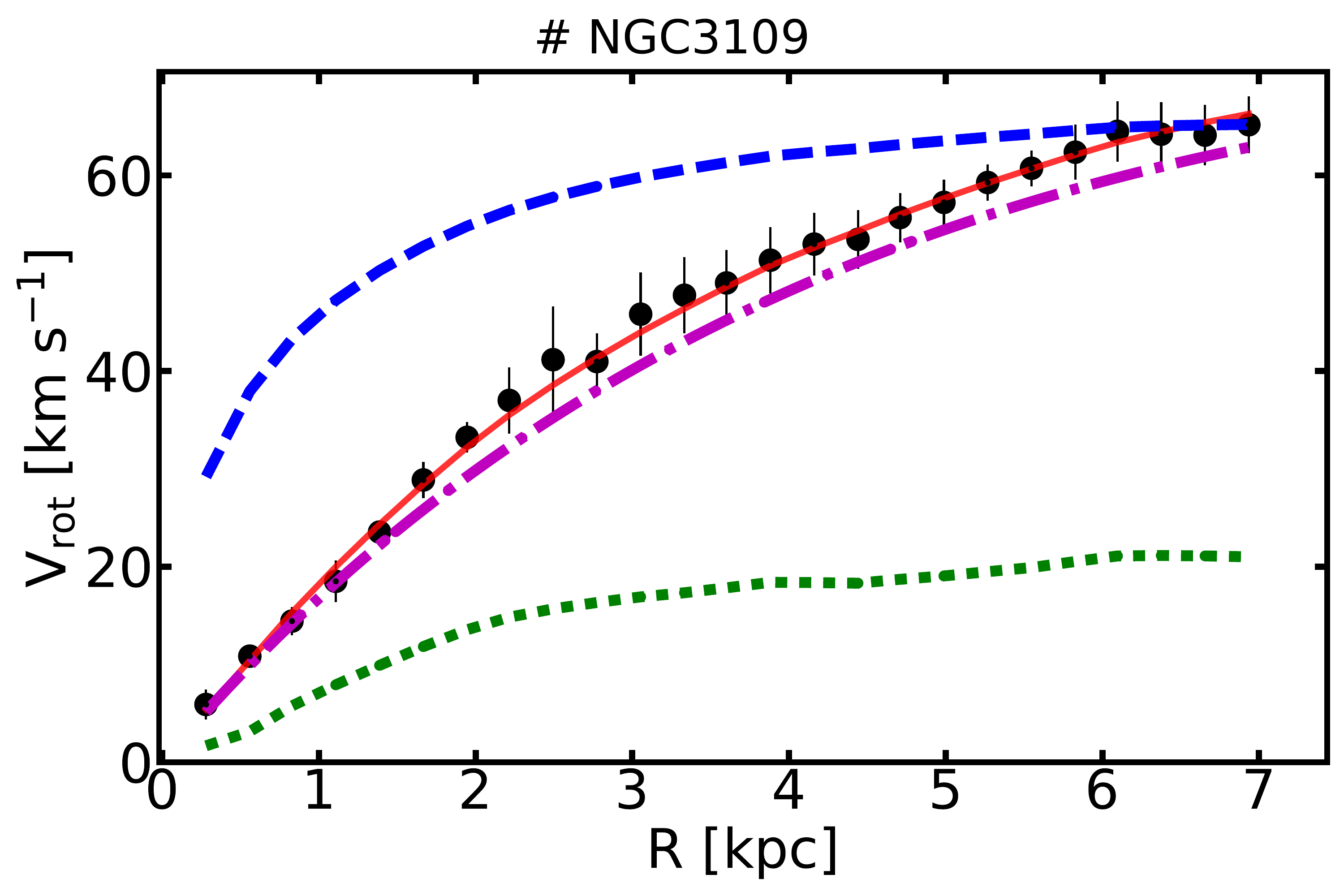}\hfill
\includegraphics[width=.33\textwidth]{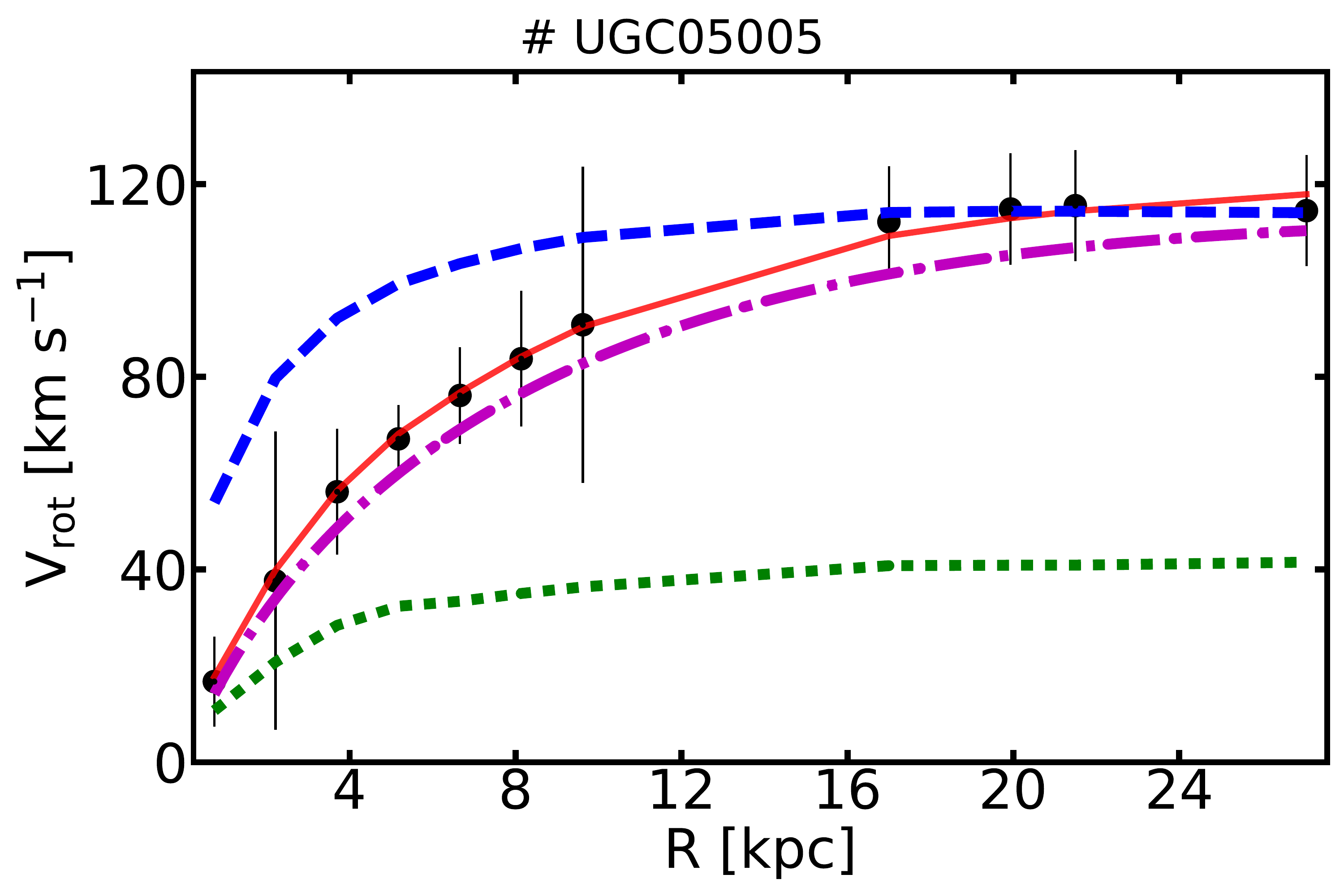}\hfill
\includegraphics[width=.33\textwidth]{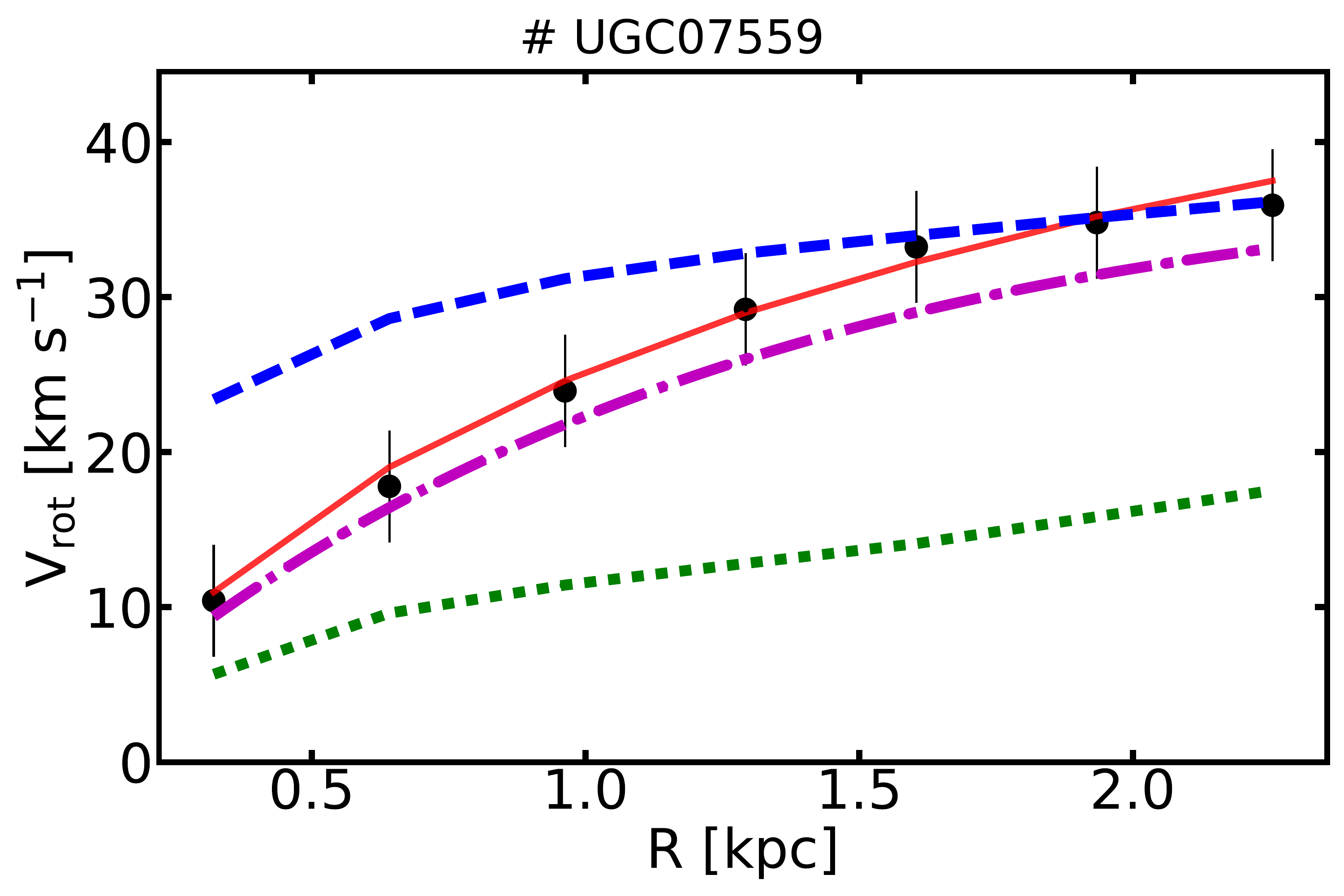}

\includegraphics[width=.33\textwidth]{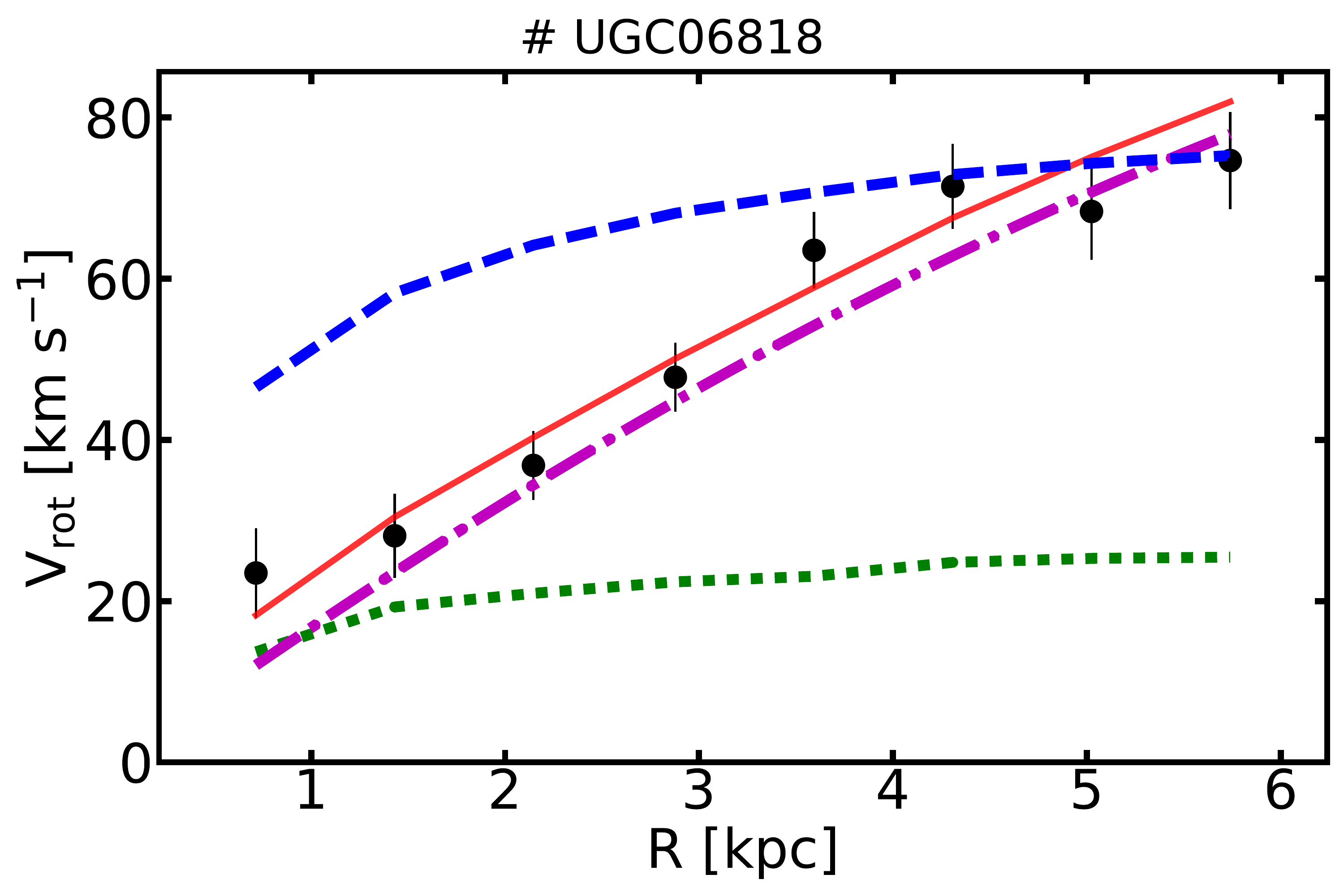}\hfill
\includegraphics[width=.33\textwidth]{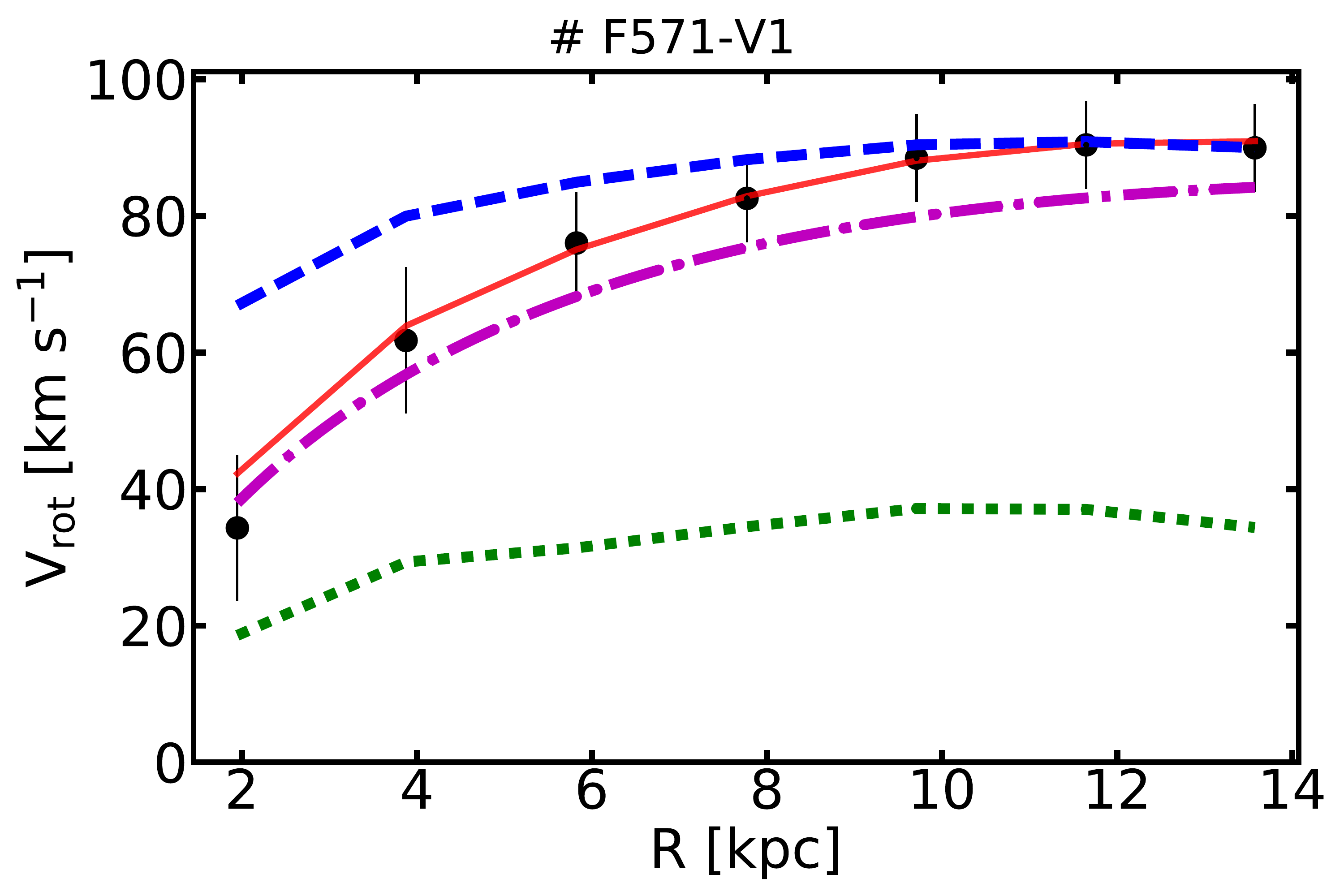}\hfill
\includegraphics[width=.33\textwidth]{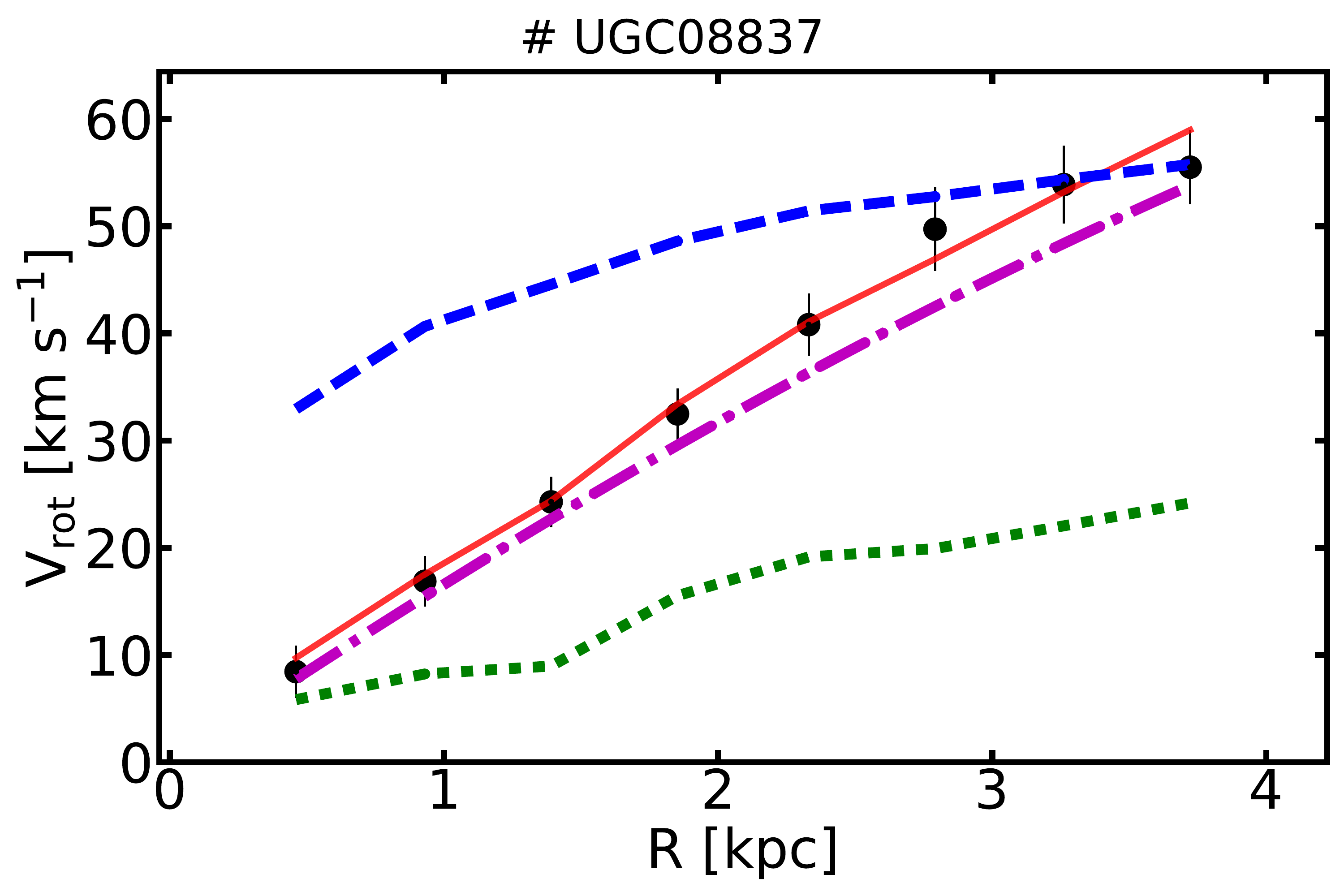}

\caption{Rotation curve fits for SPARC galaxies with L18a parameters. In each panel the black dots represent the measured rotation velocities and their observational uncertainties. The red curves are the best-fit curves to the data with our three-parameters fits. The green dashed curves represent the contribution of the baryonic matter to the rotation curves and the dark matter halo rotation velocities are plotted by violet dot-dashed curves. The blue dashed curves are there for visual comparison only, using $n = 6$ and the virial mass as the only free parameter to give an idea of the typical shape expected in the DMO case.}
\label{figa1}
\end{figure*}

\begin{figure*}
\centering
\ContinuedFloat

\includegraphics[width=.33\textwidth]{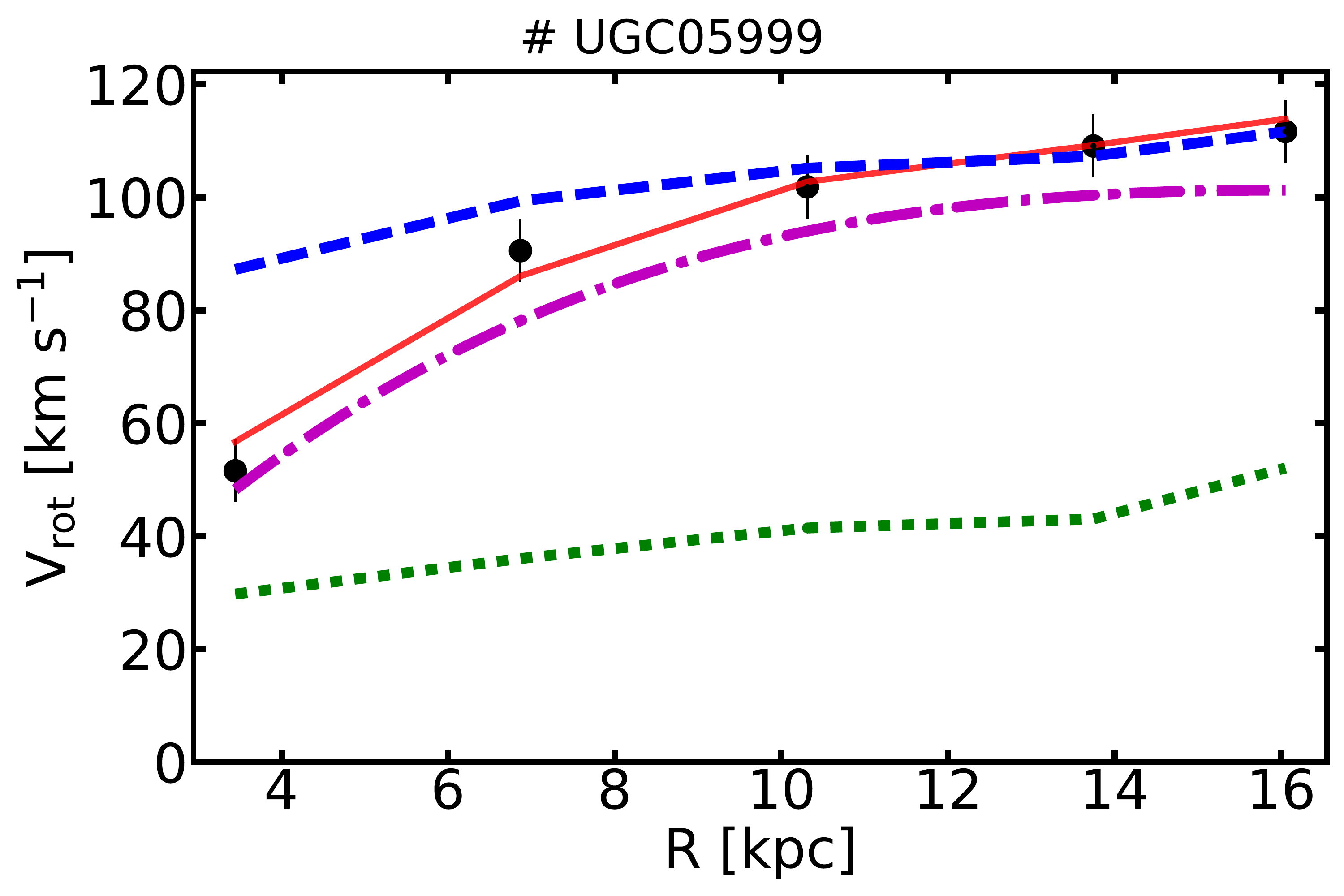}\hfill
\includegraphics[width=.33\textwidth]{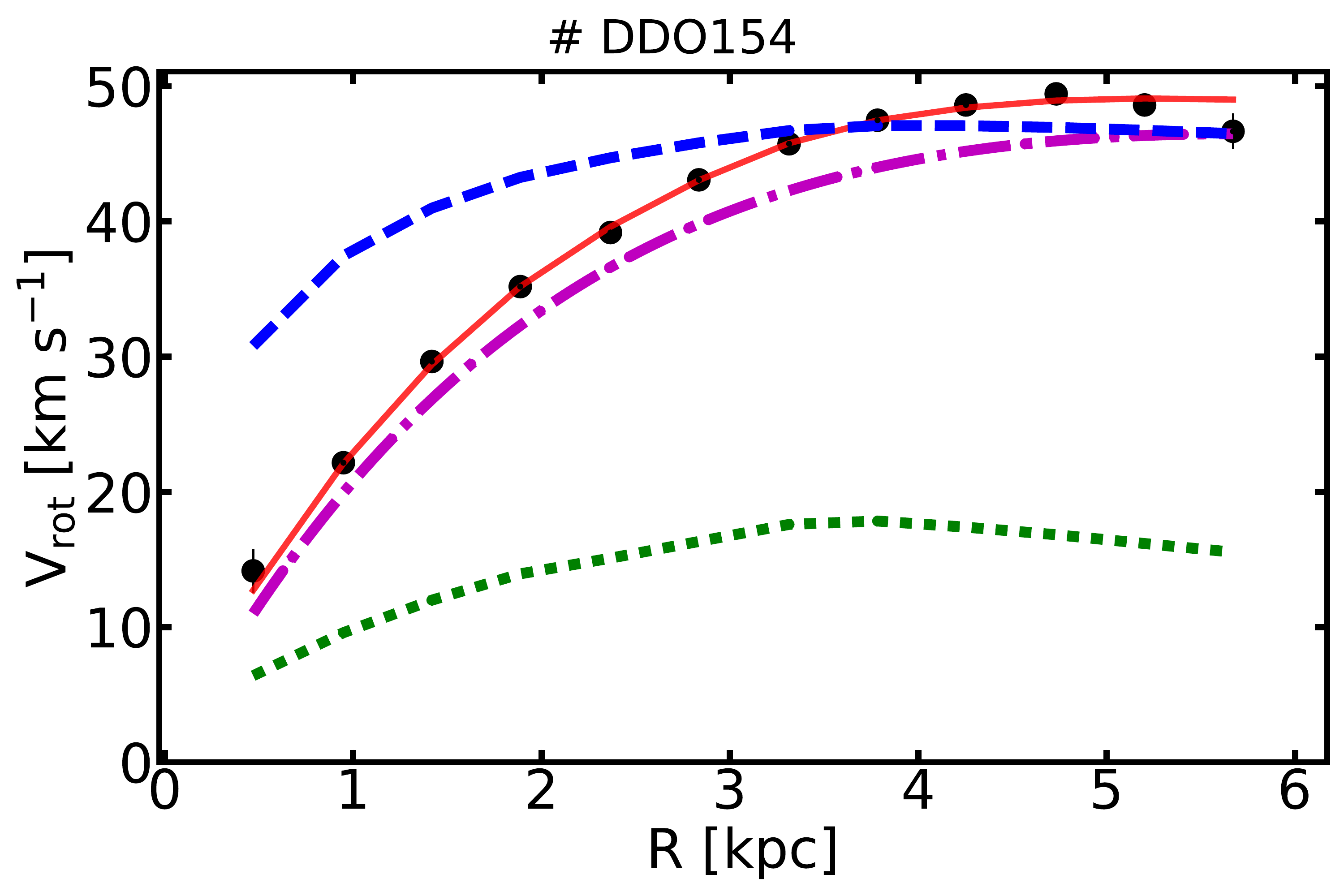}\hfill
\includegraphics[width=.33\textwidth]{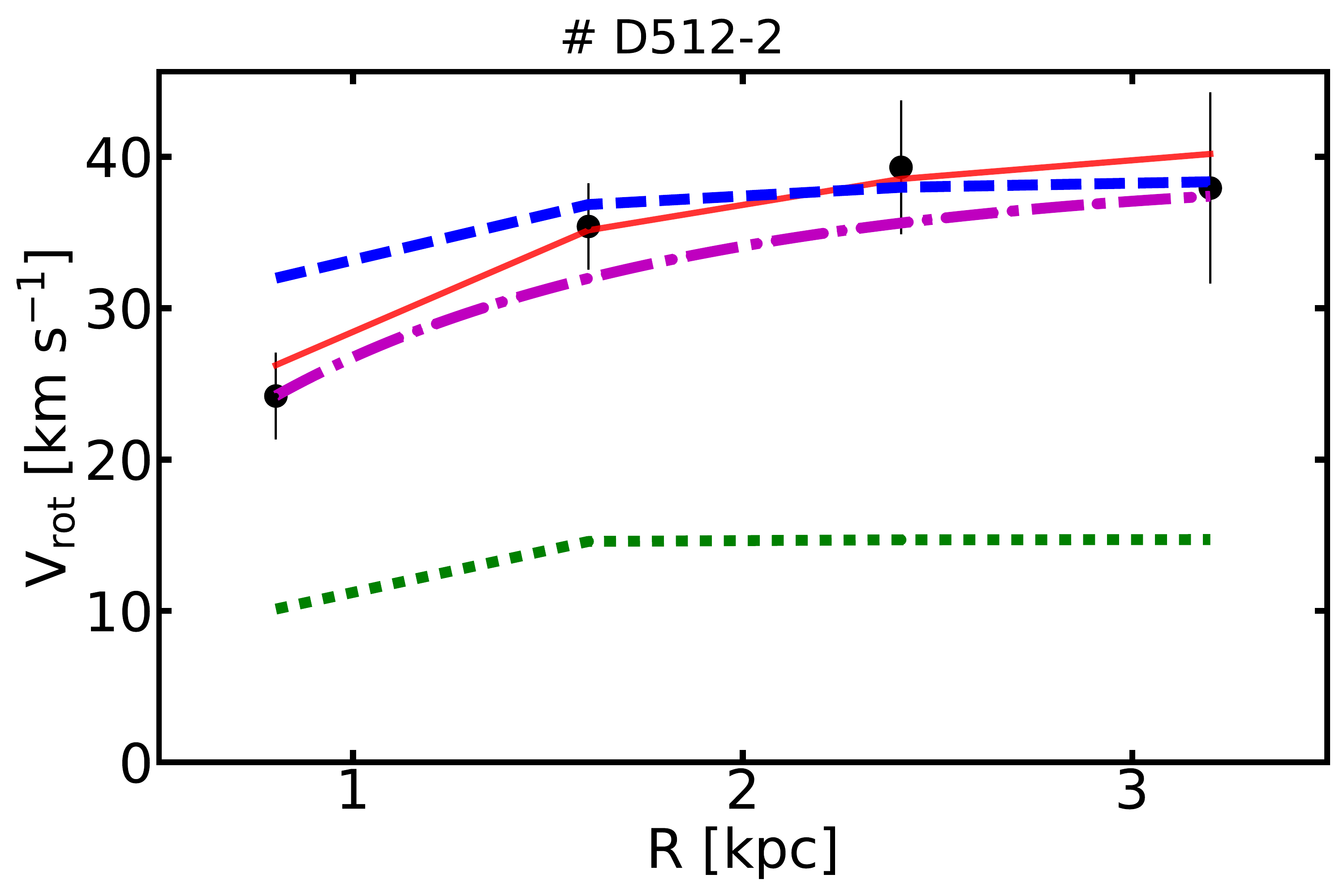}

\includegraphics[width=.33\textwidth]{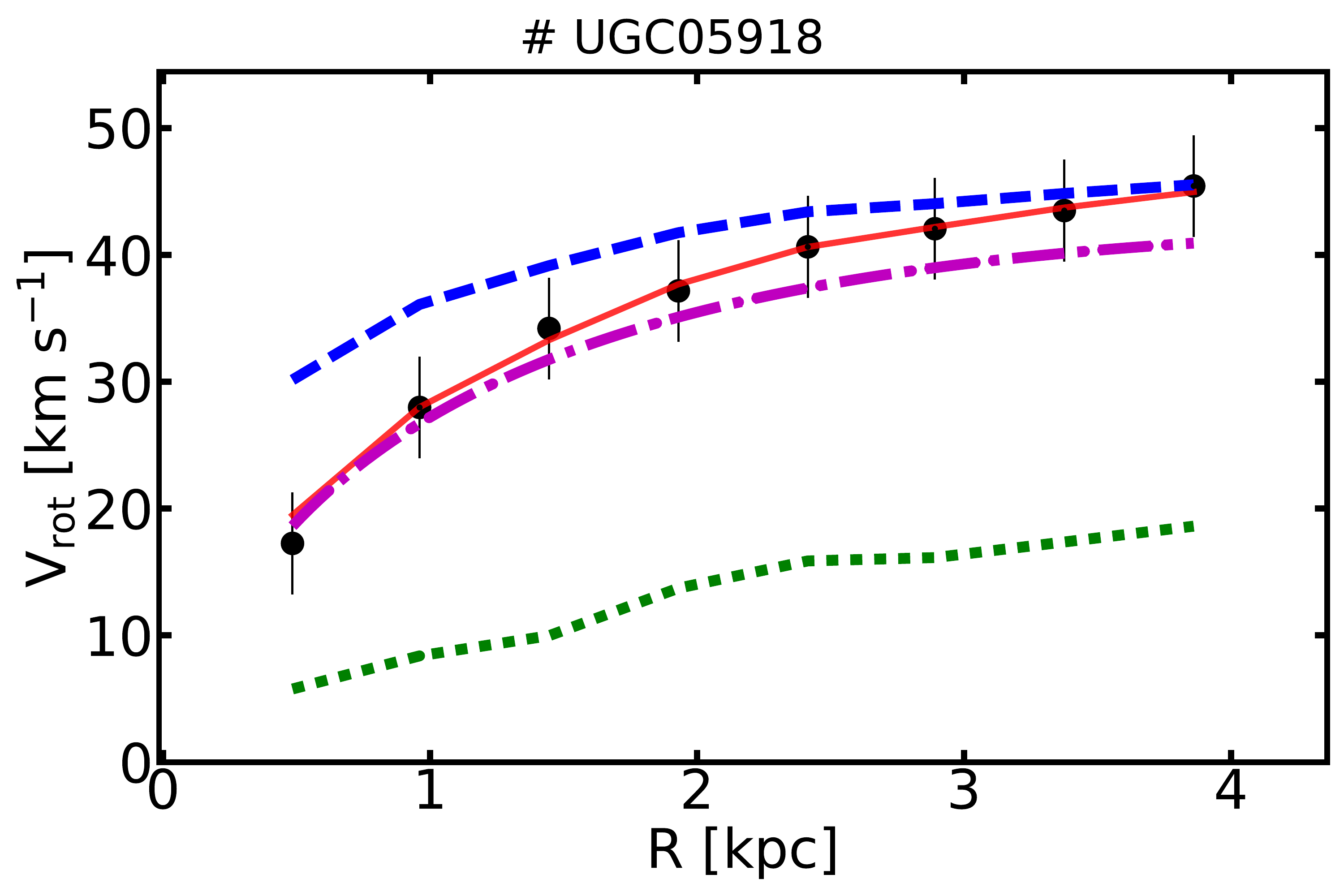}\hfill
\includegraphics[width=.33\textwidth]{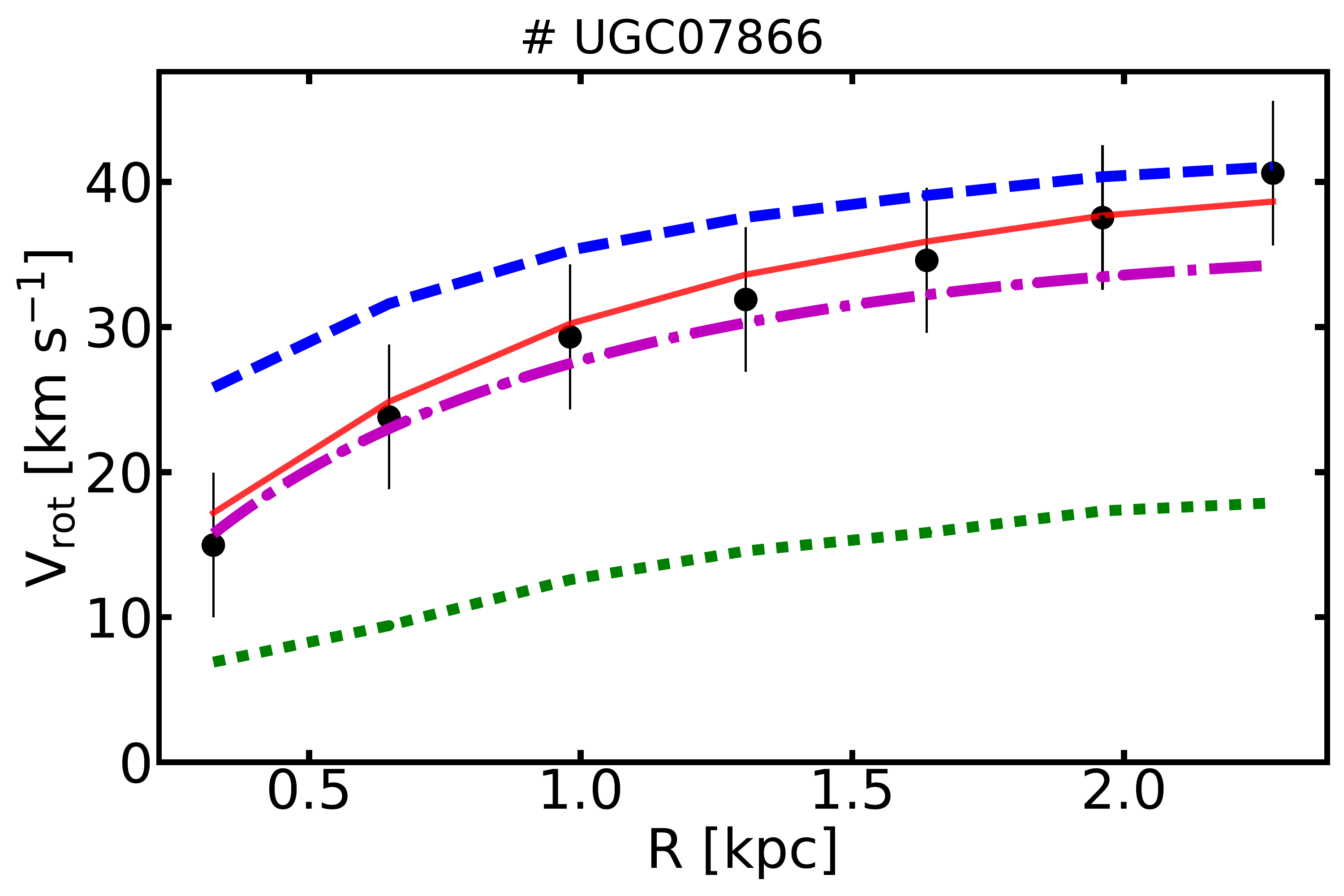}\hfill
\includegraphics[width=.33\textwidth]{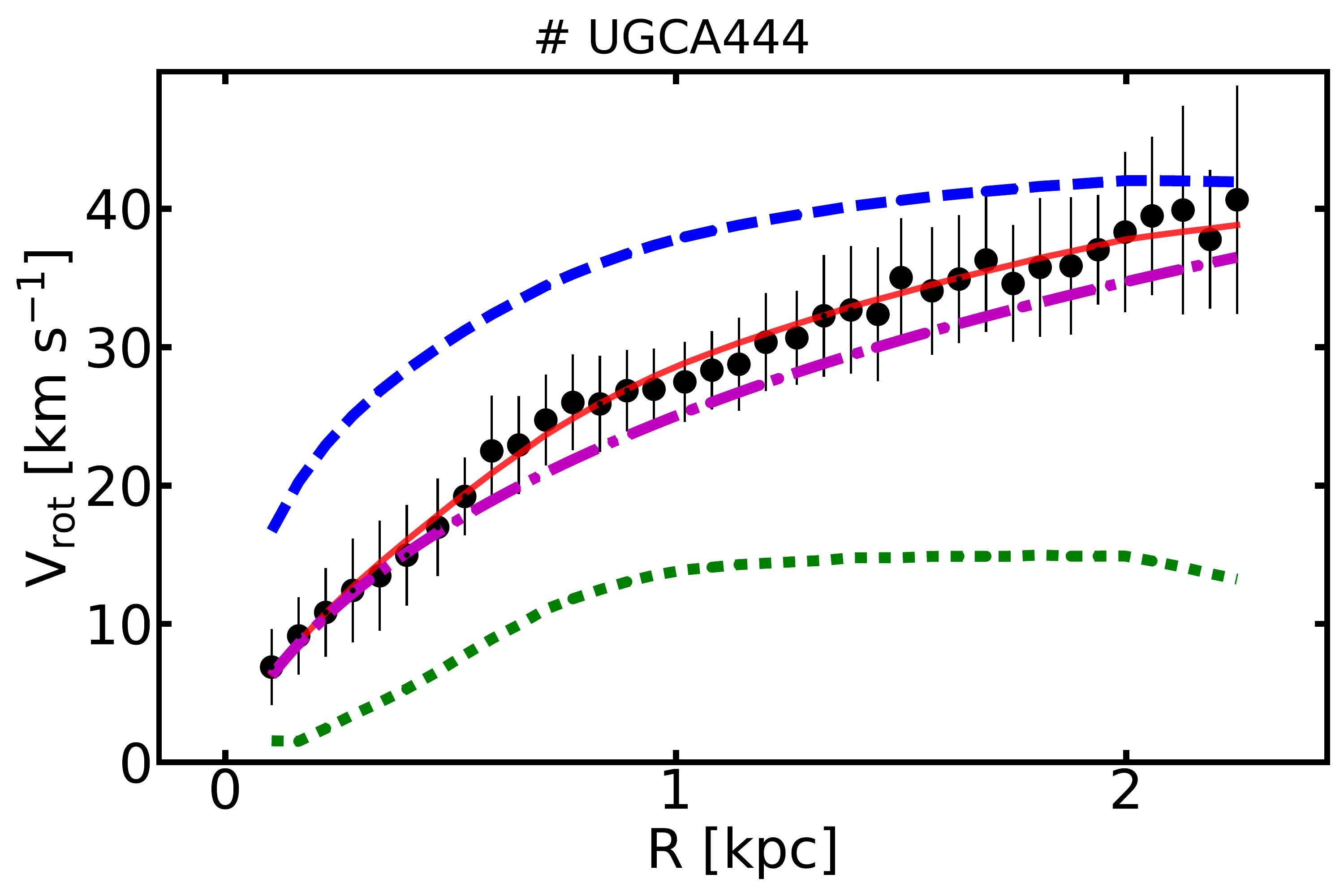}

\includegraphics[width=.33\textwidth]{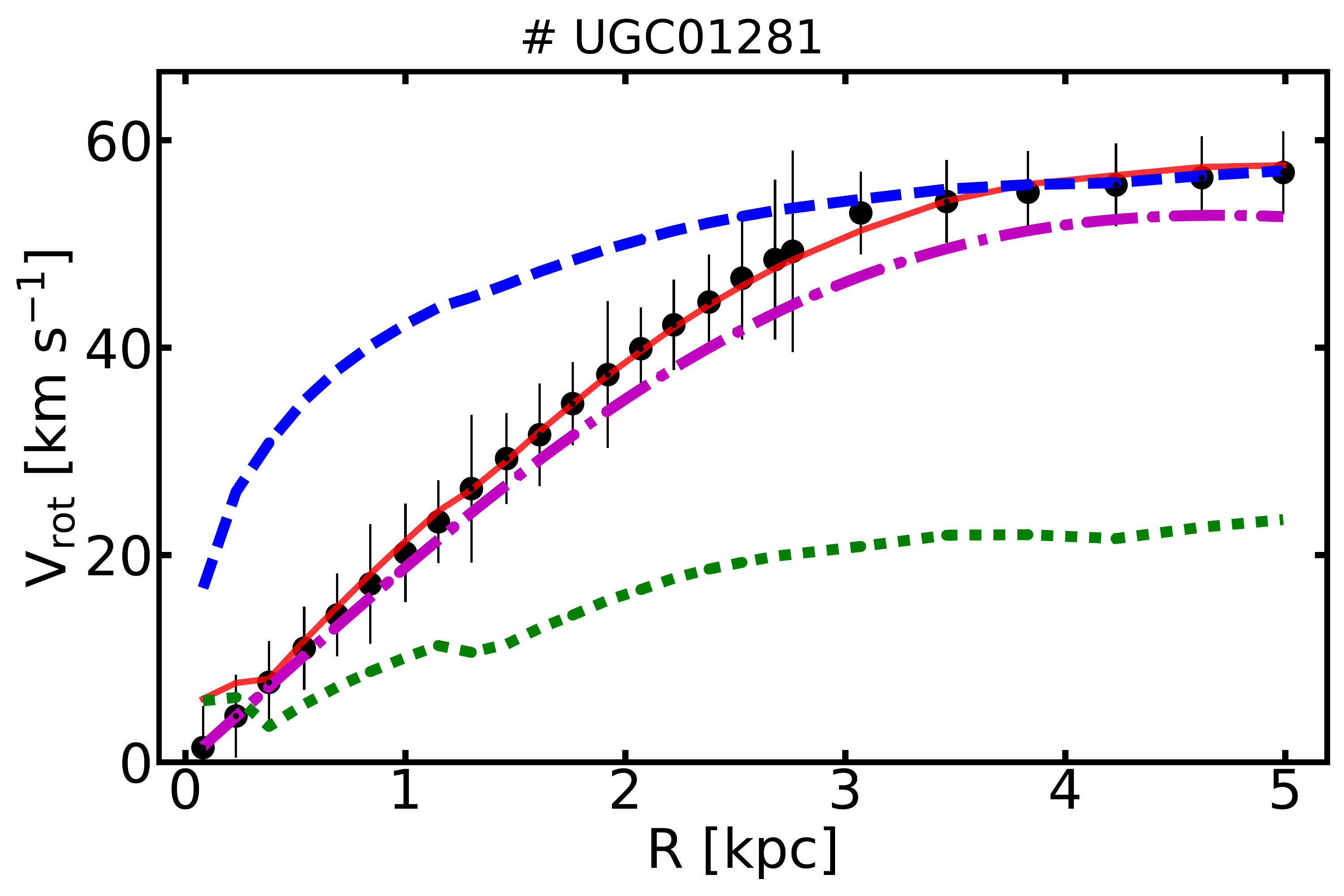}\hfill
\includegraphics[width=.33\textwidth]{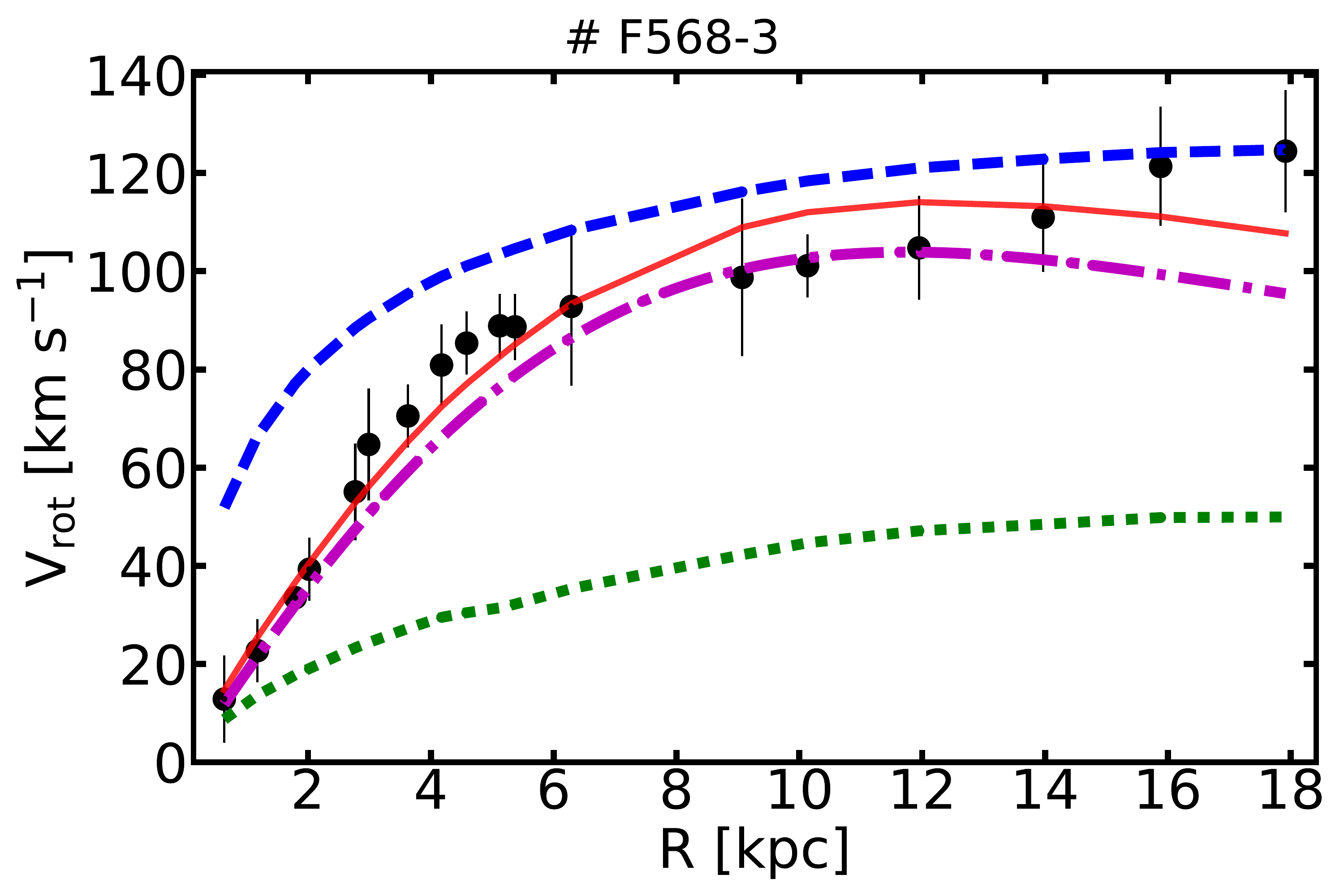}\hfill
\includegraphics[width=.33\textwidth]{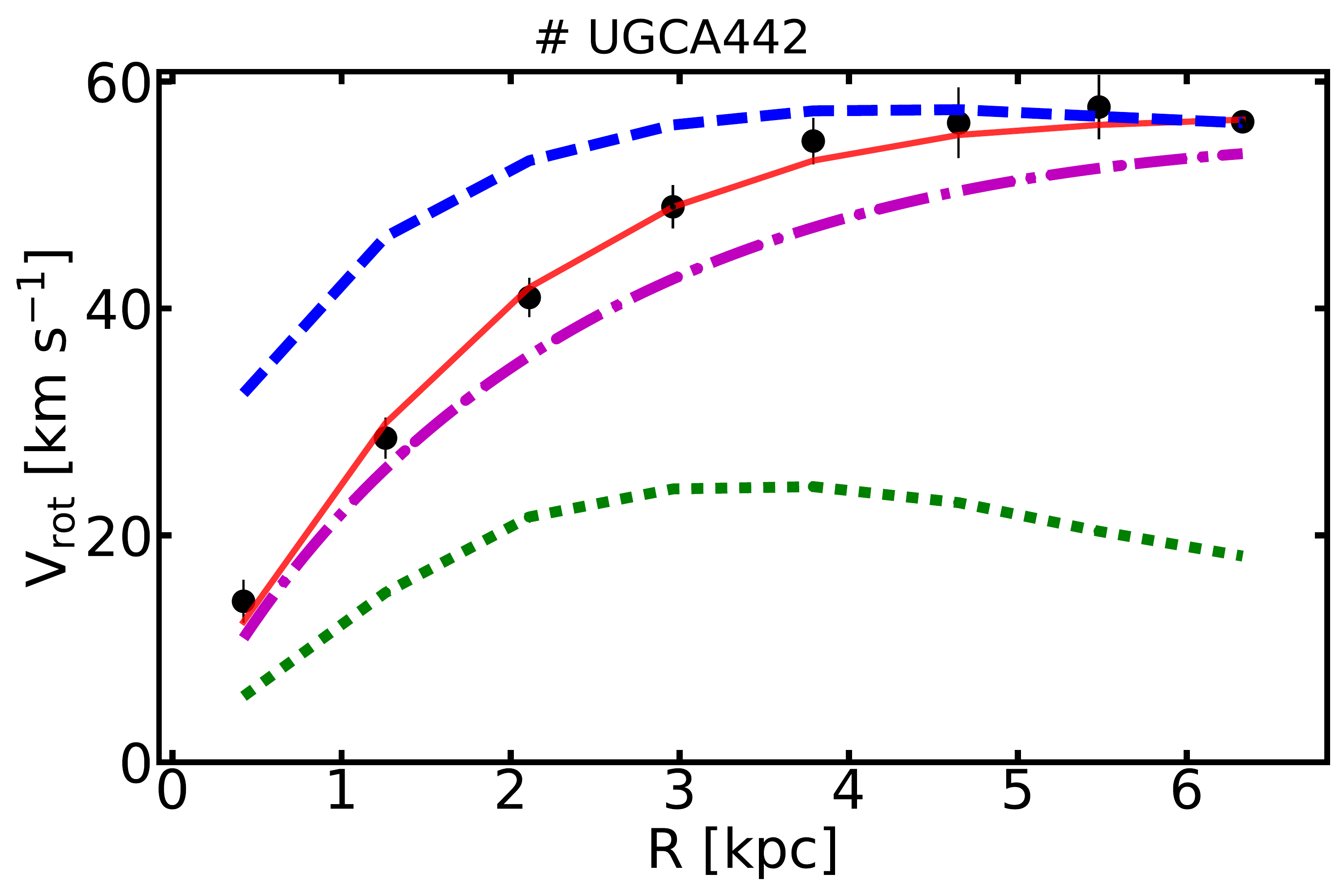}

\includegraphics[width=.33\textwidth]{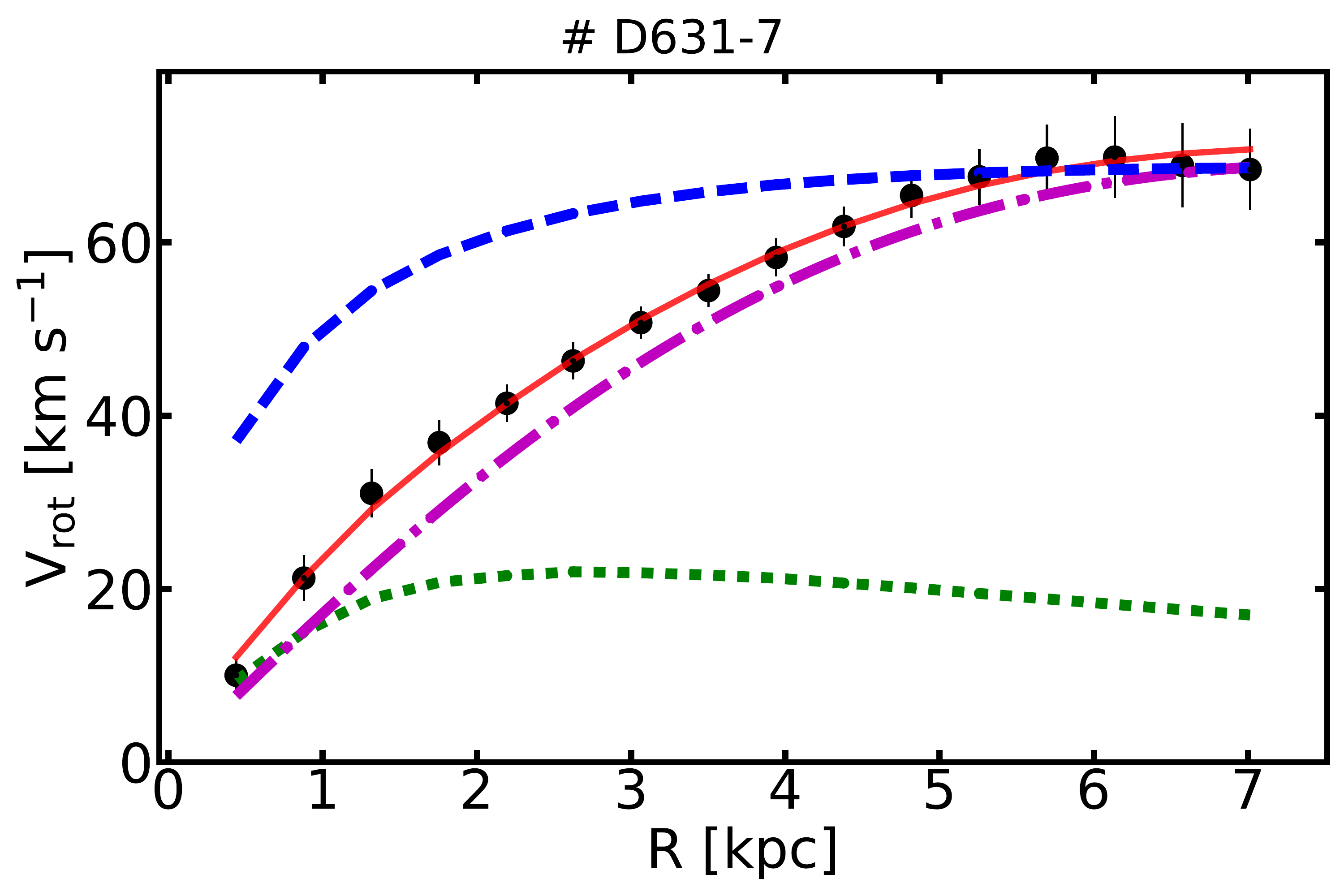}\hfill
\includegraphics[width=.33\textwidth]{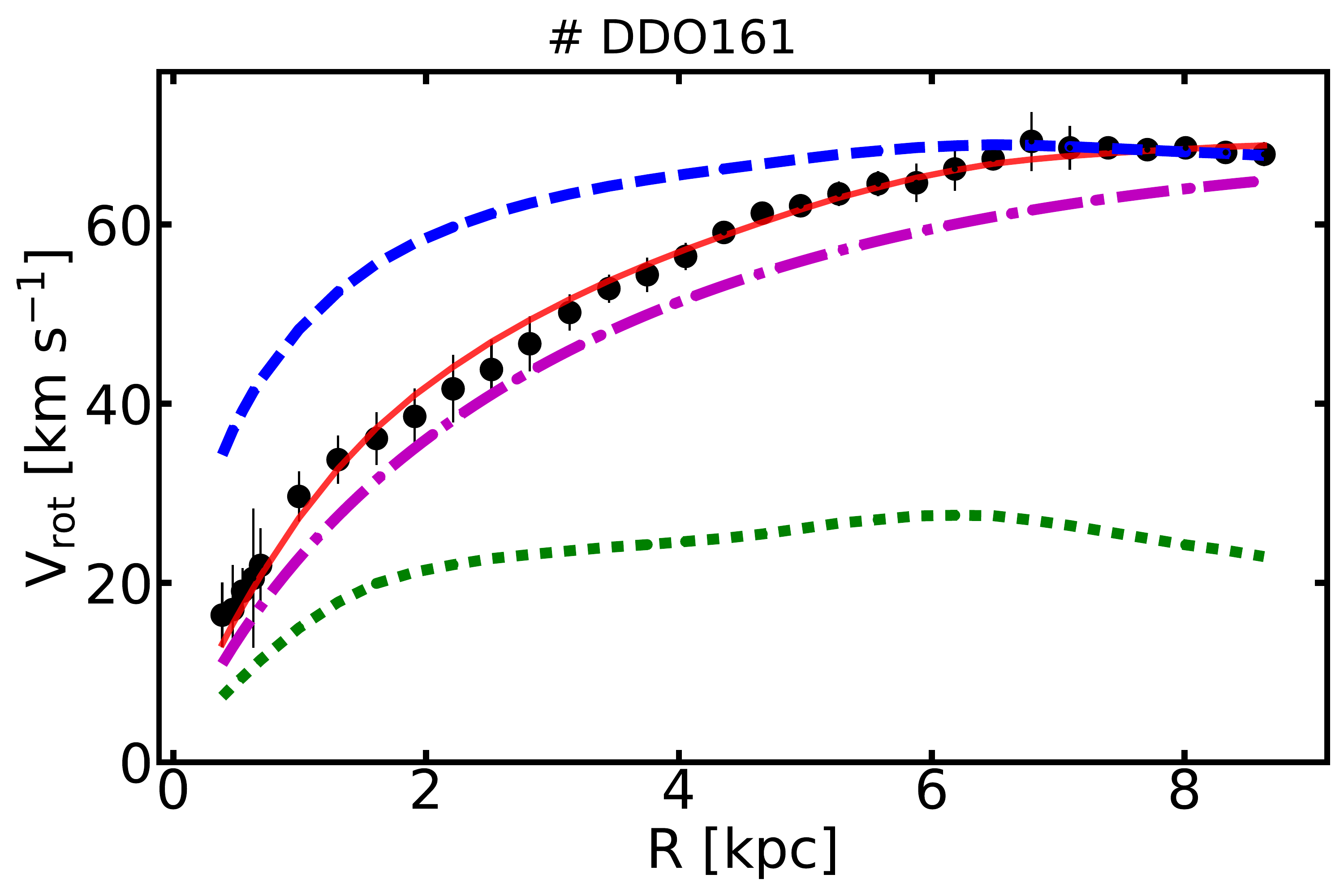}\hfill
\includegraphics[width=.33\textwidth]{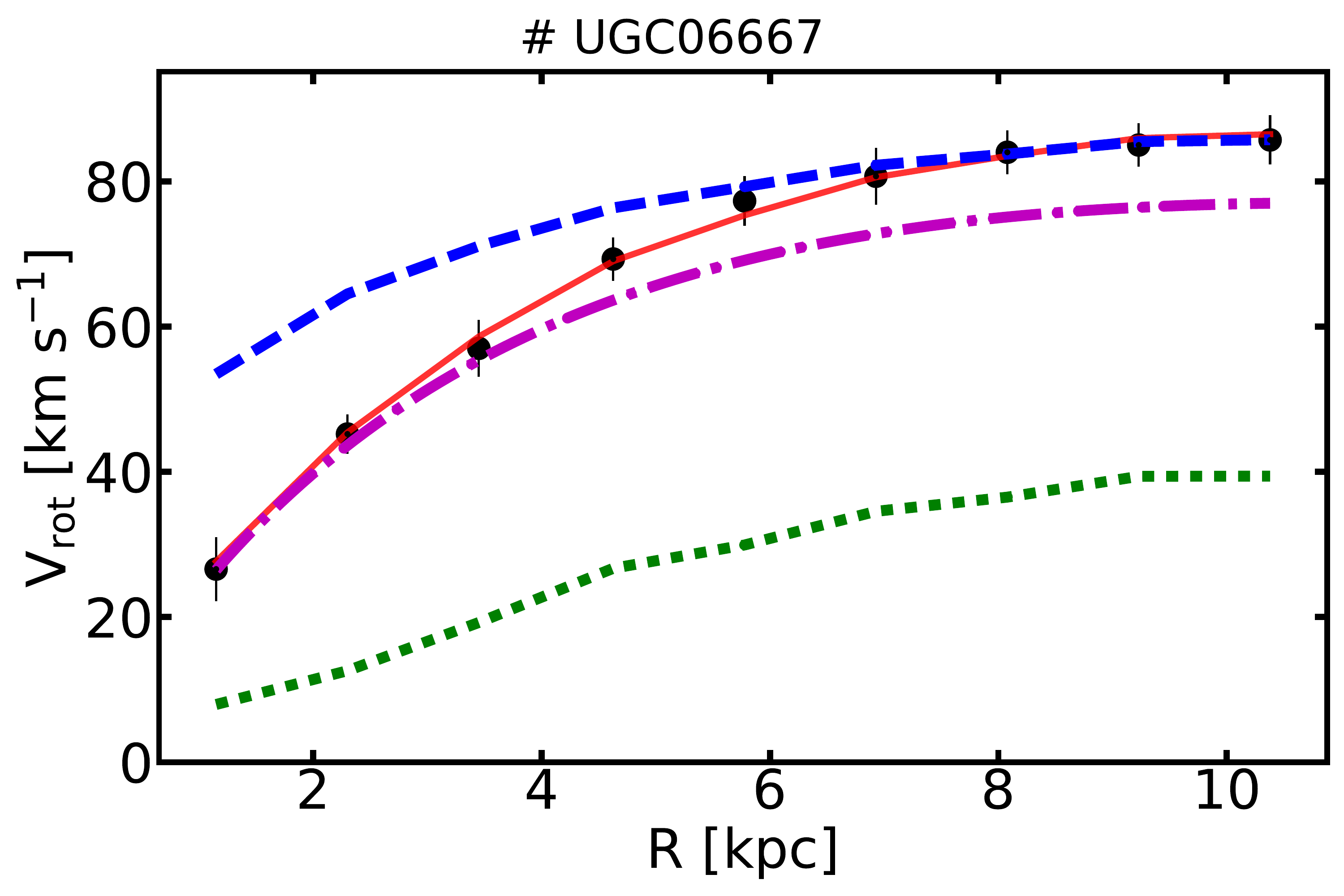}

\includegraphics[width=.33\textwidth]{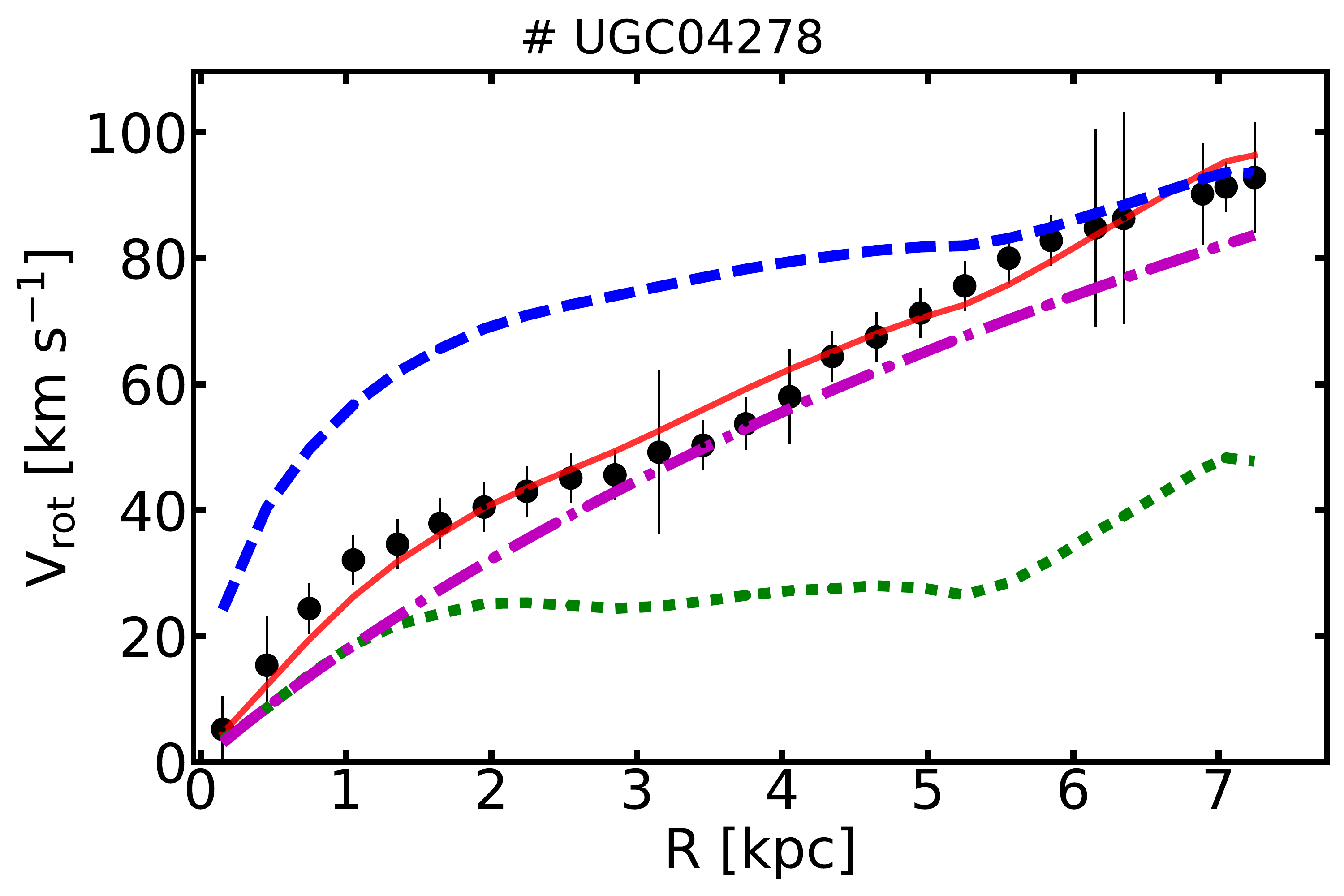}\hfill
\includegraphics[width=.33\textwidth]{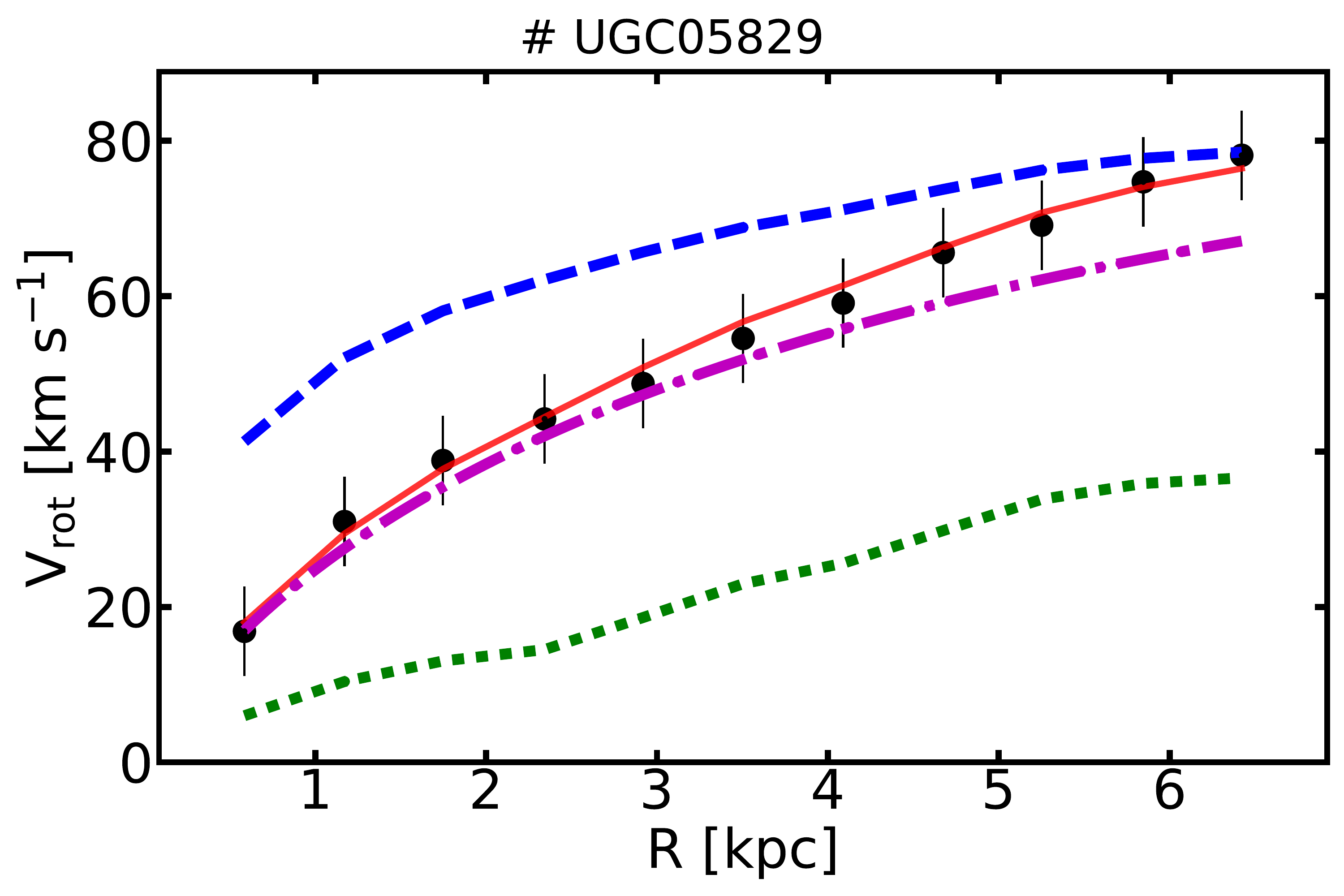}\hfill
\includegraphics[width=.33\textwidth]{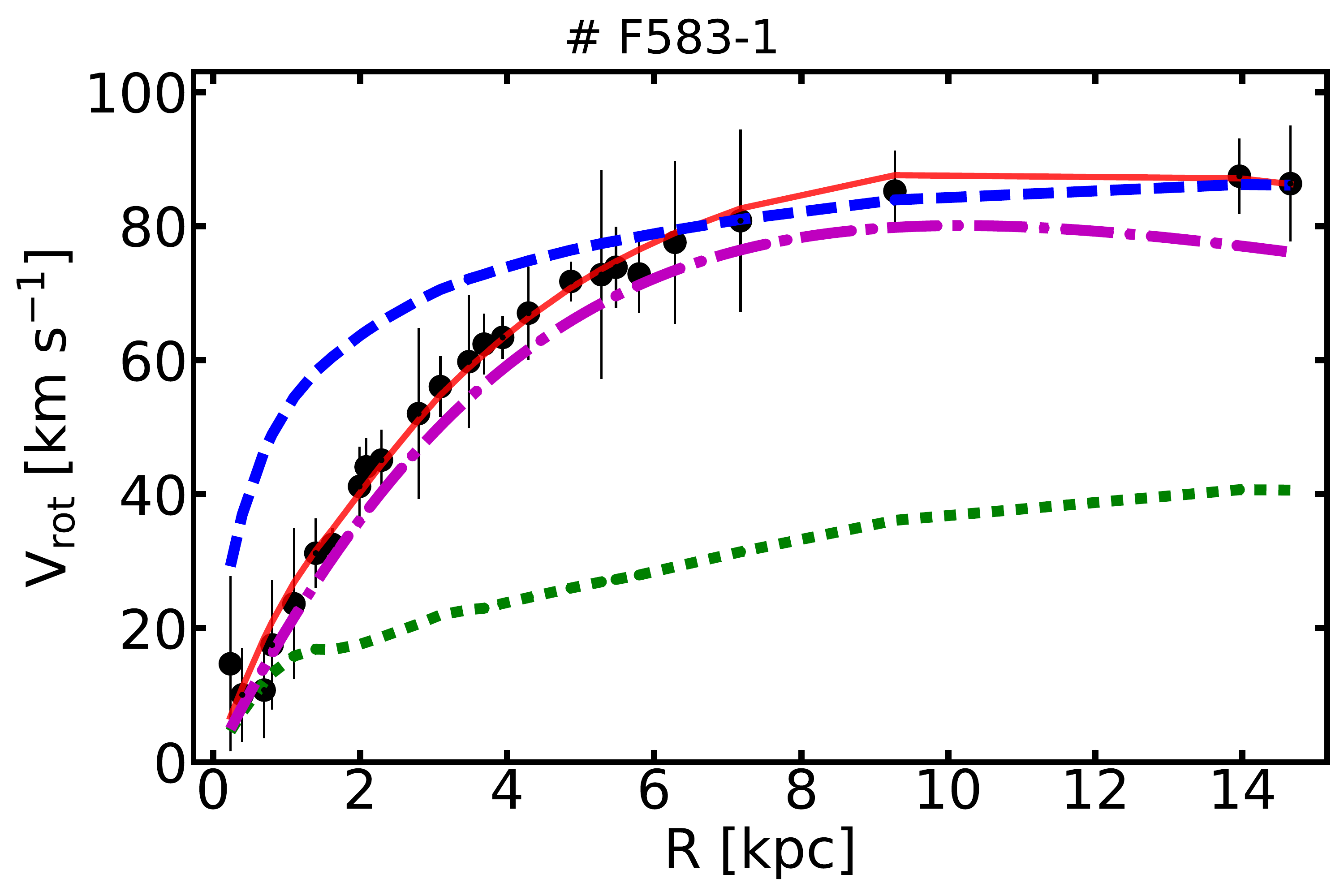}

\includegraphics[width=.33\textwidth]{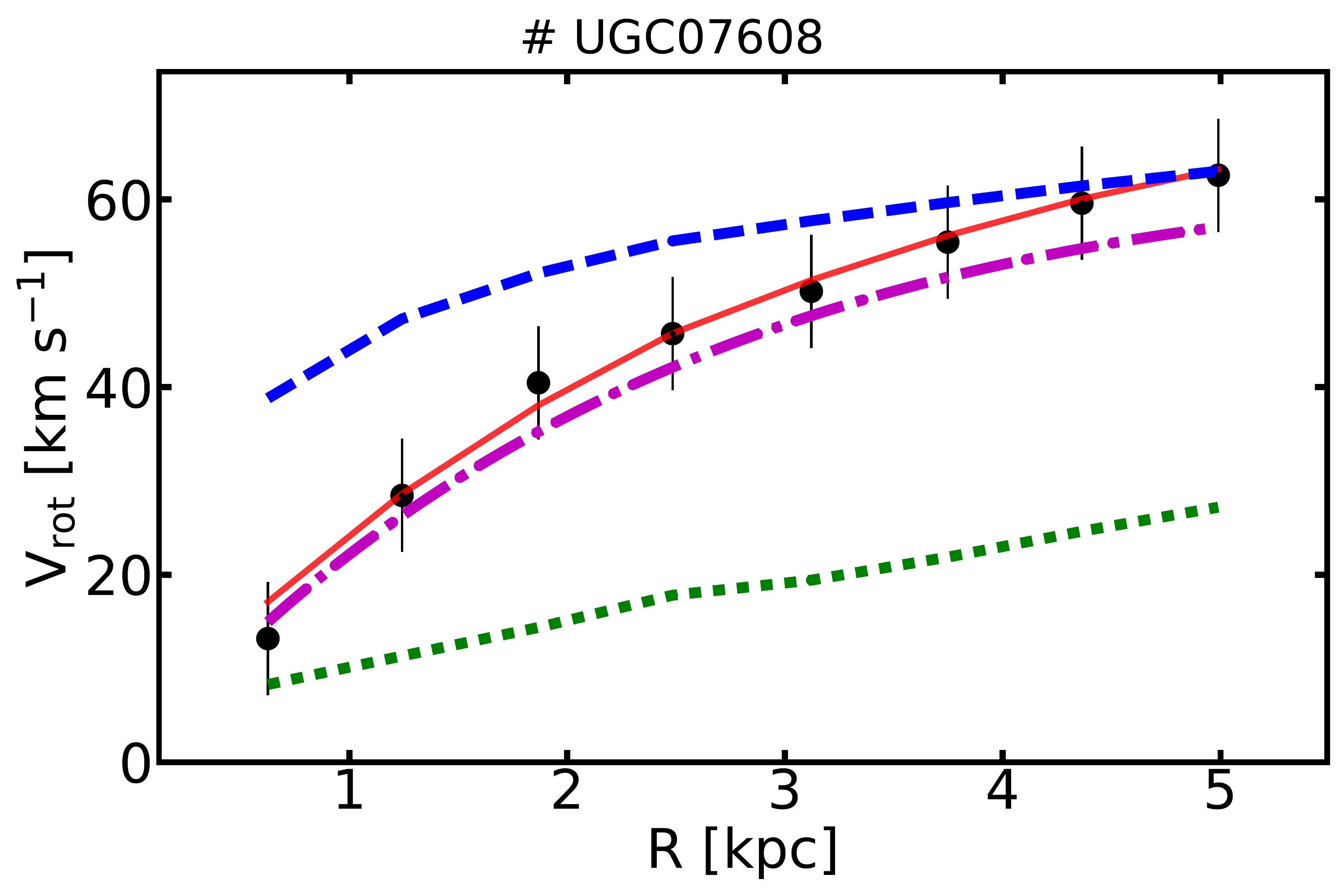}\hfill
\includegraphics[width=.33\textwidth]{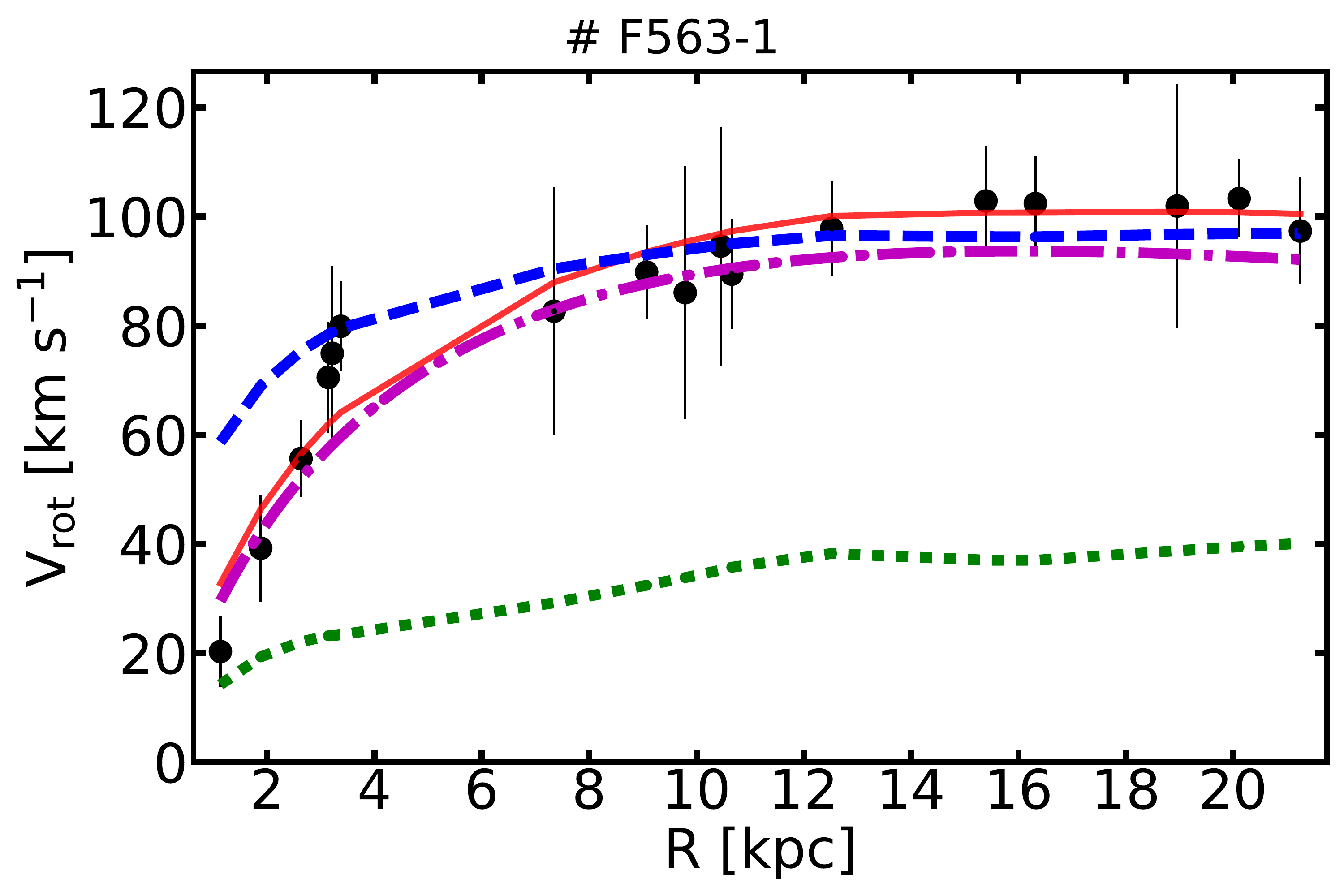}\hfill
\includegraphics[width=.33\textwidth]{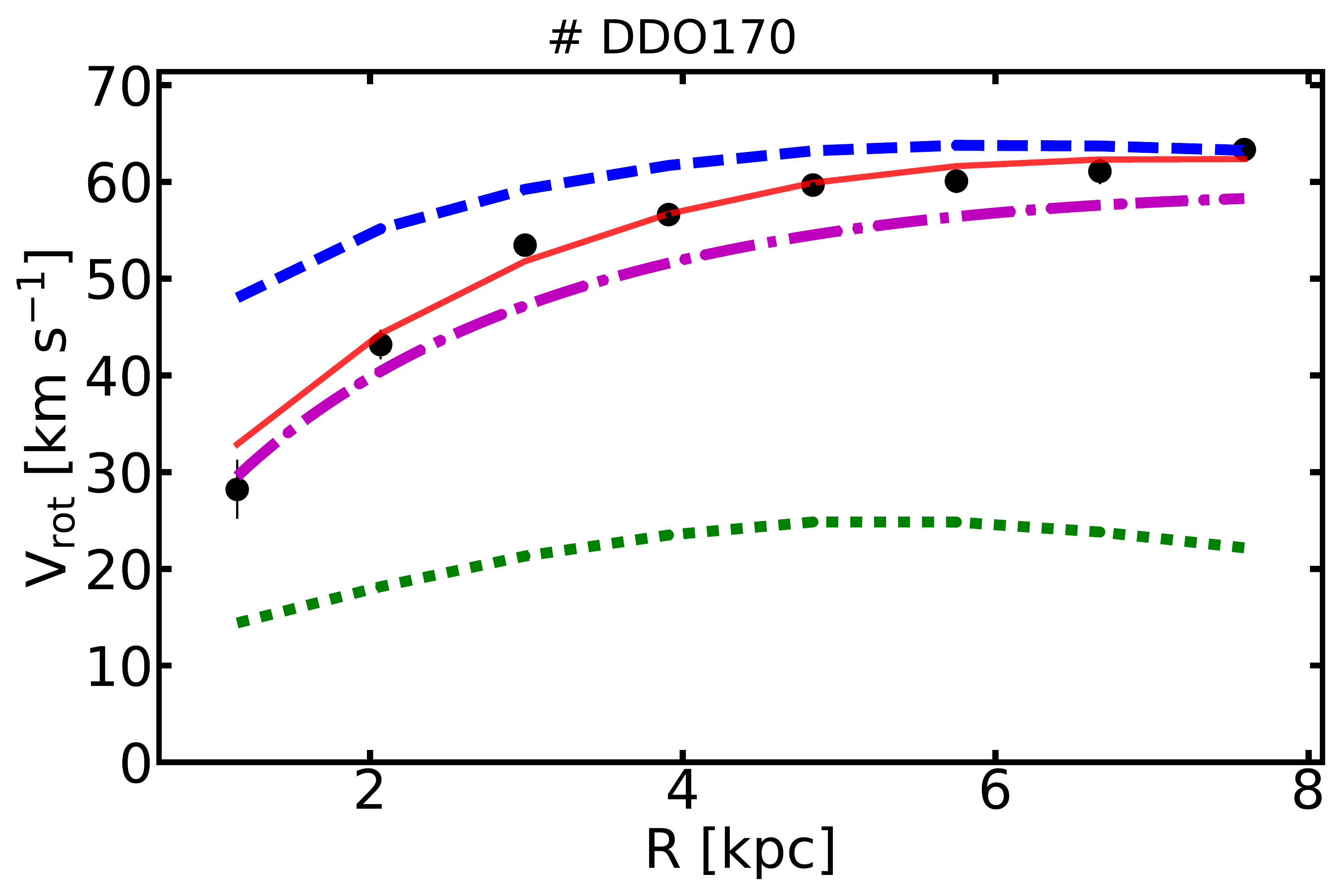}
\caption{Continued.}
\end{figure*}

\begin{figure*}
\centering
\ContinuedFloat

\includegraphics[width=.33\textwidth]{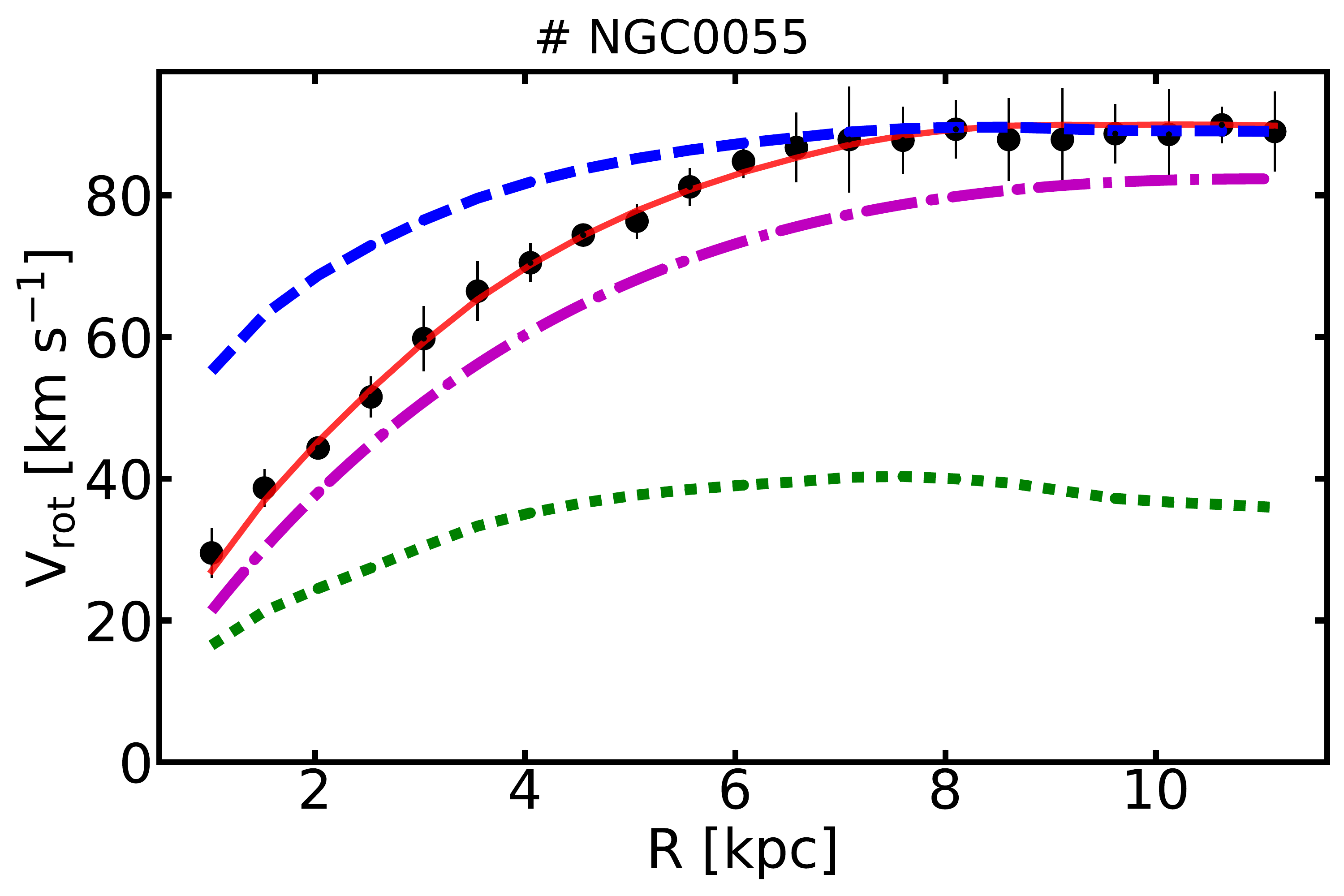}\hfill
\includegraphics[width=.33\textwidth]{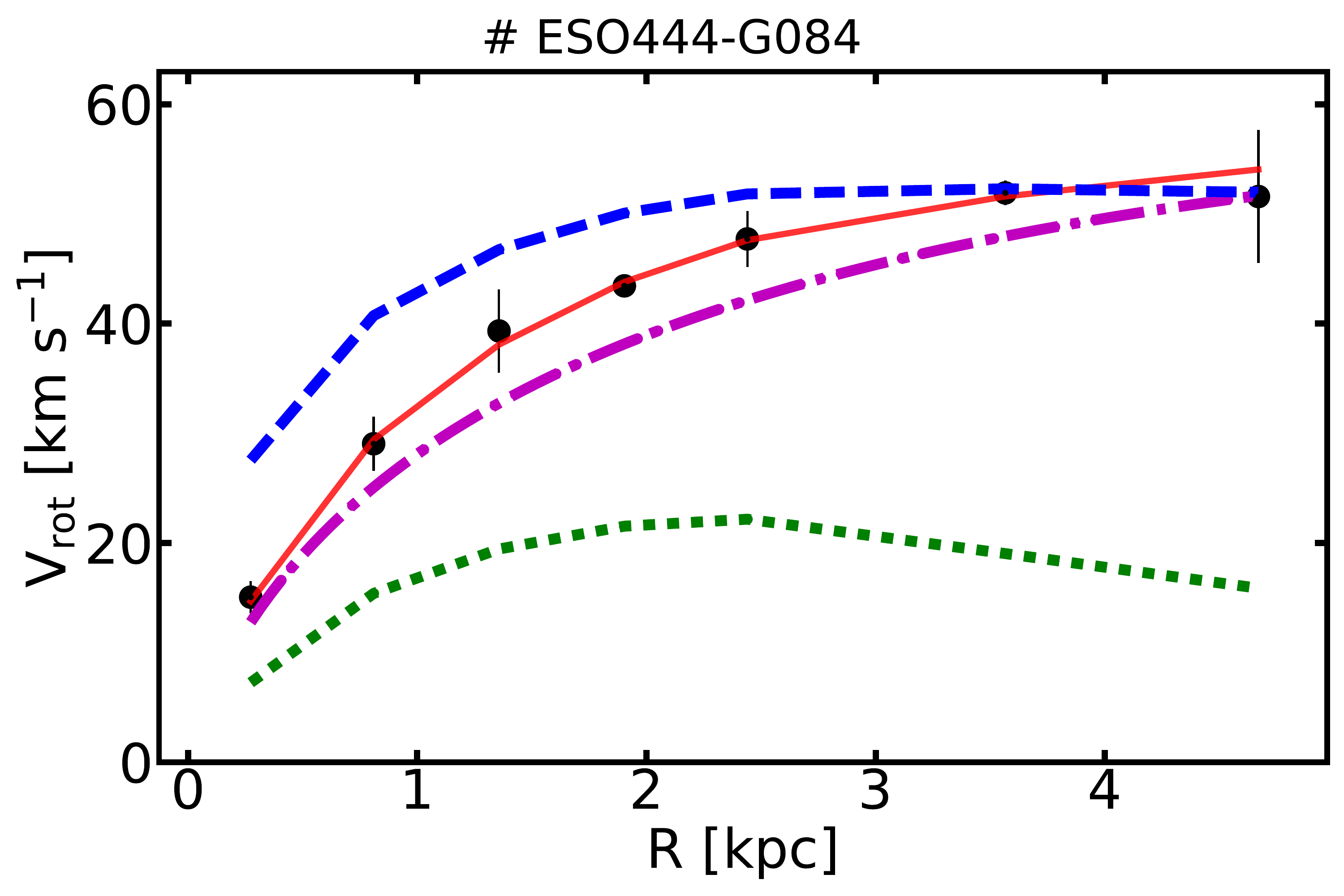}\hfill
\includegraphics[width=.33\textwidth]{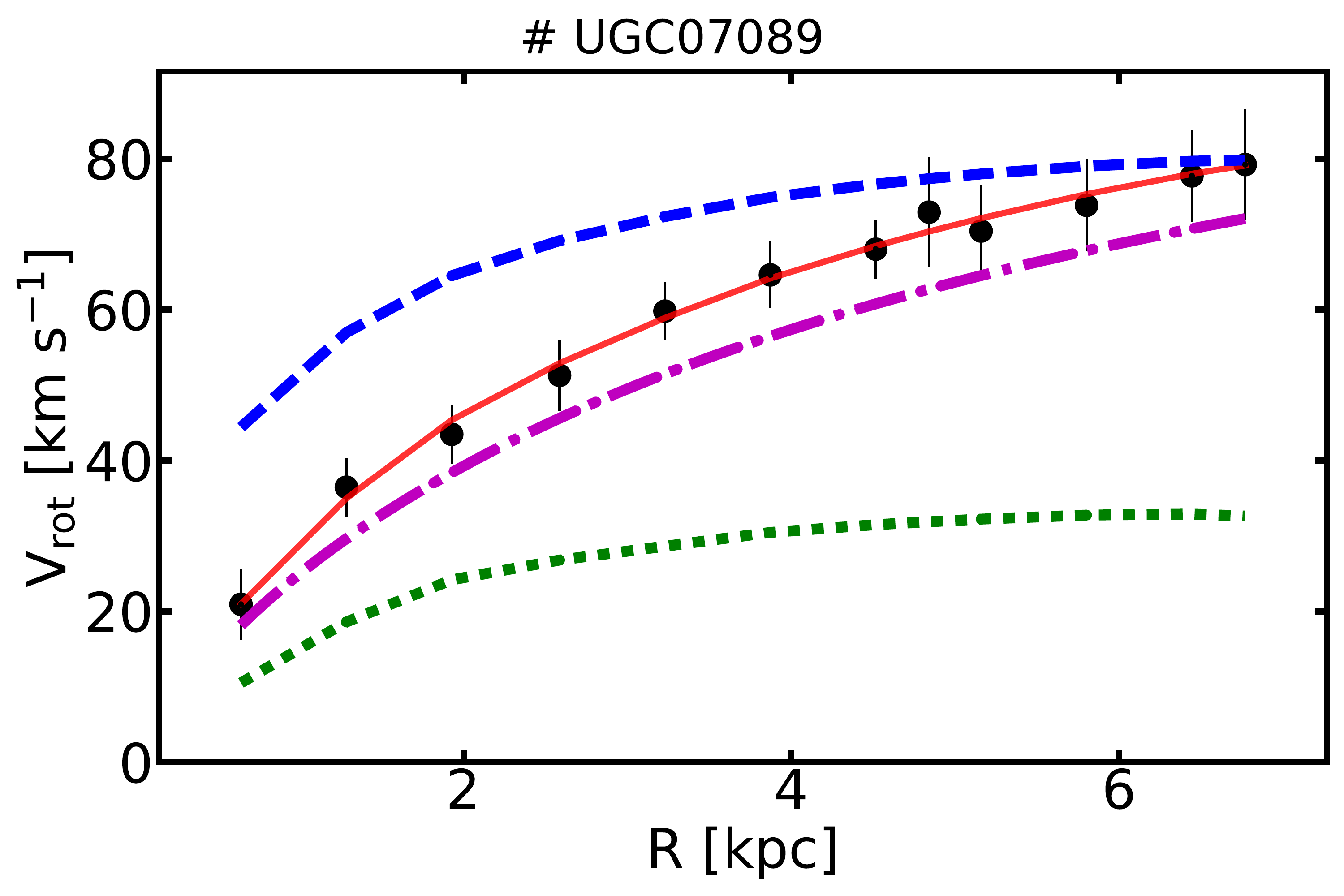}

\includegraphics[width=.33\textwidth]{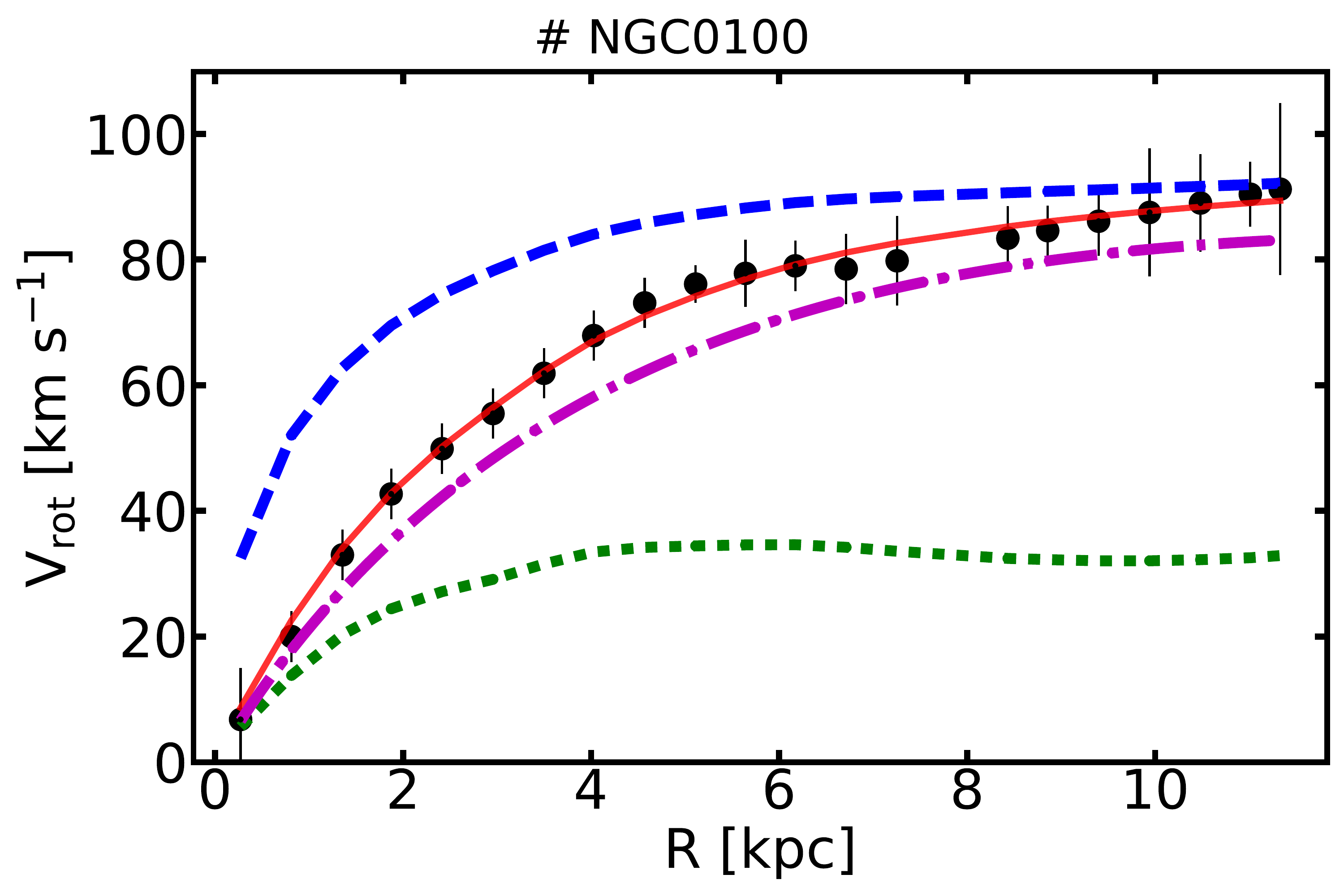}\hfill
\includegraphics[width=.33\textwidth]{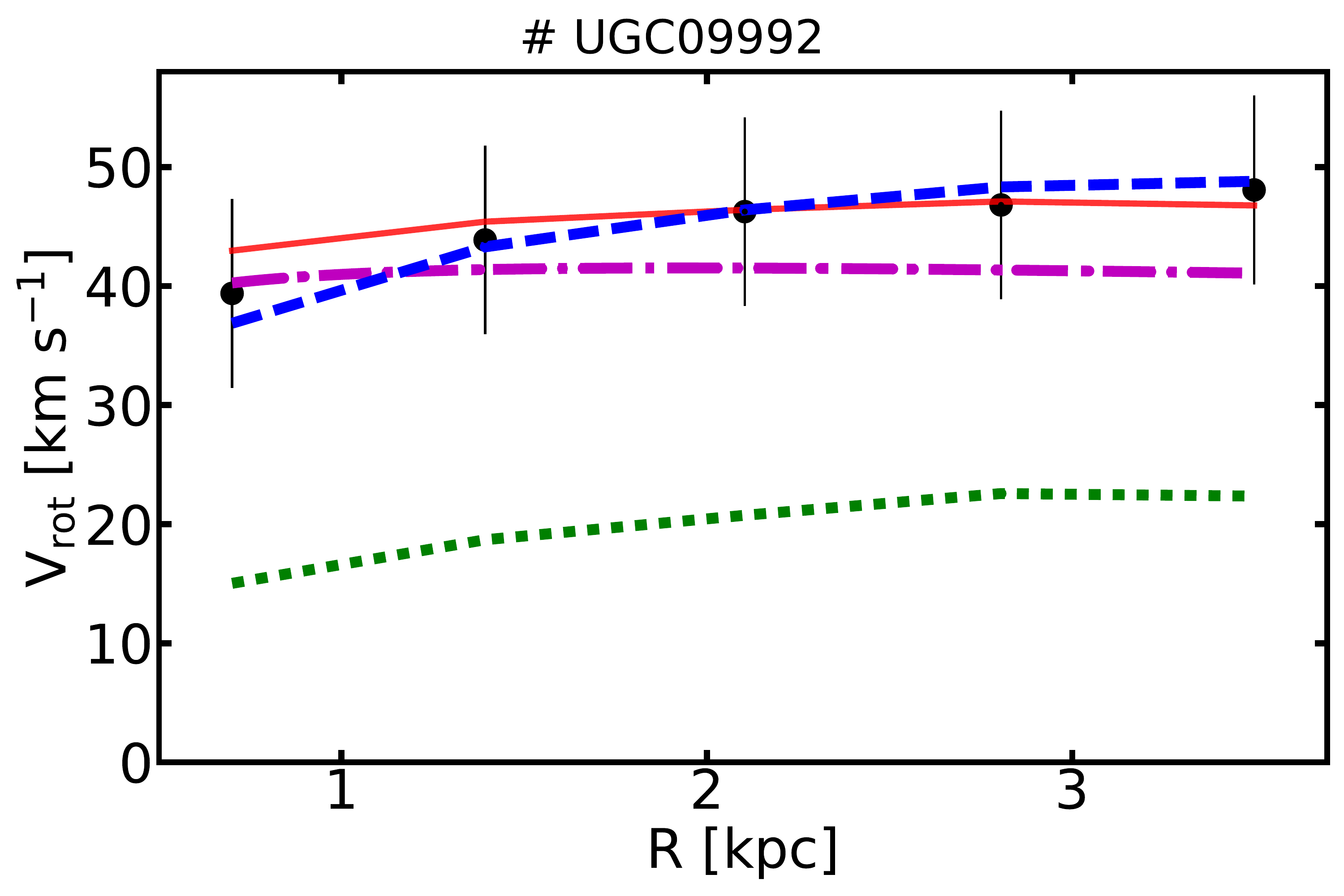}\hfill
\includegraphics[width=.33\textwidth]{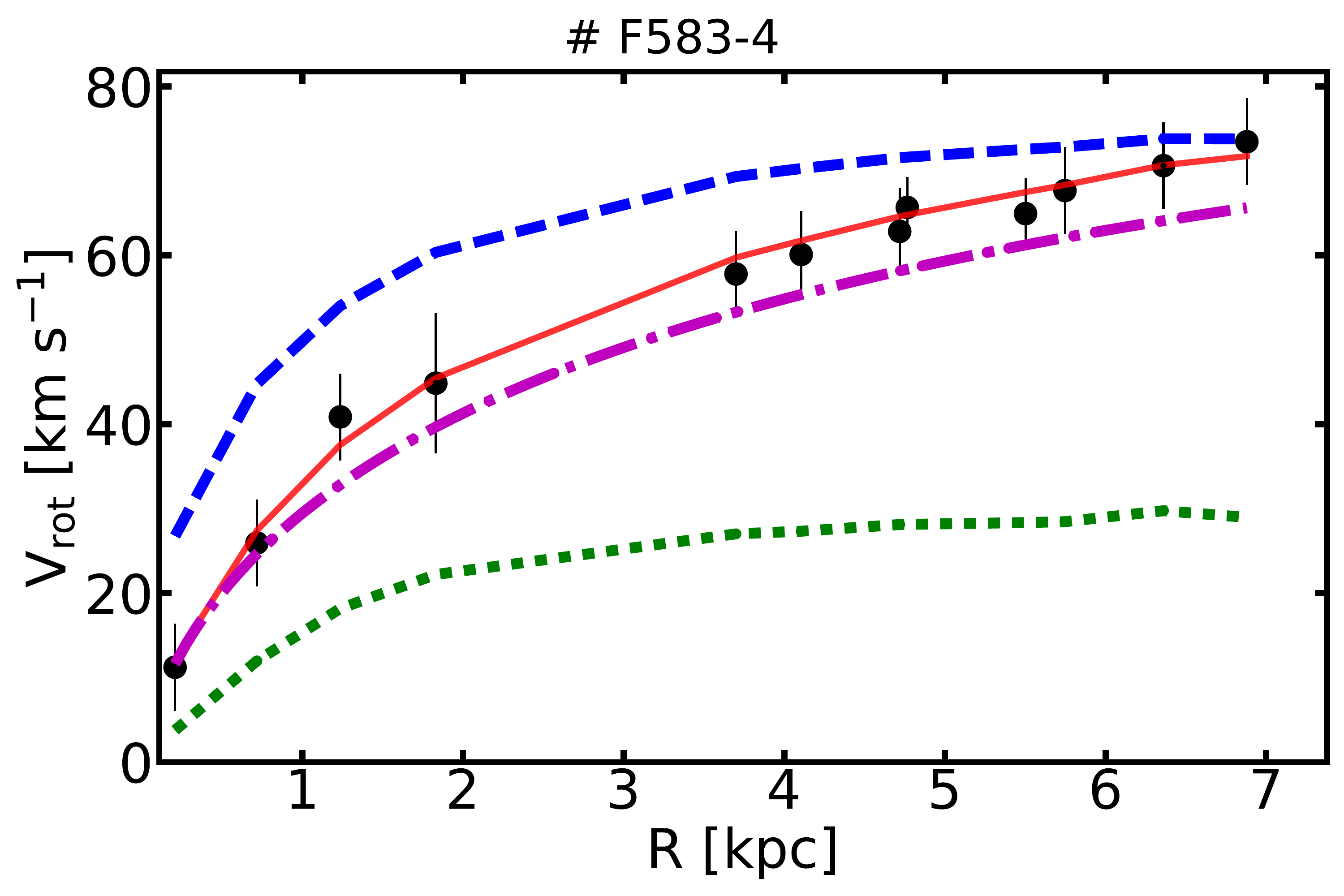}

\includegraphics[width=.33\textwidth]{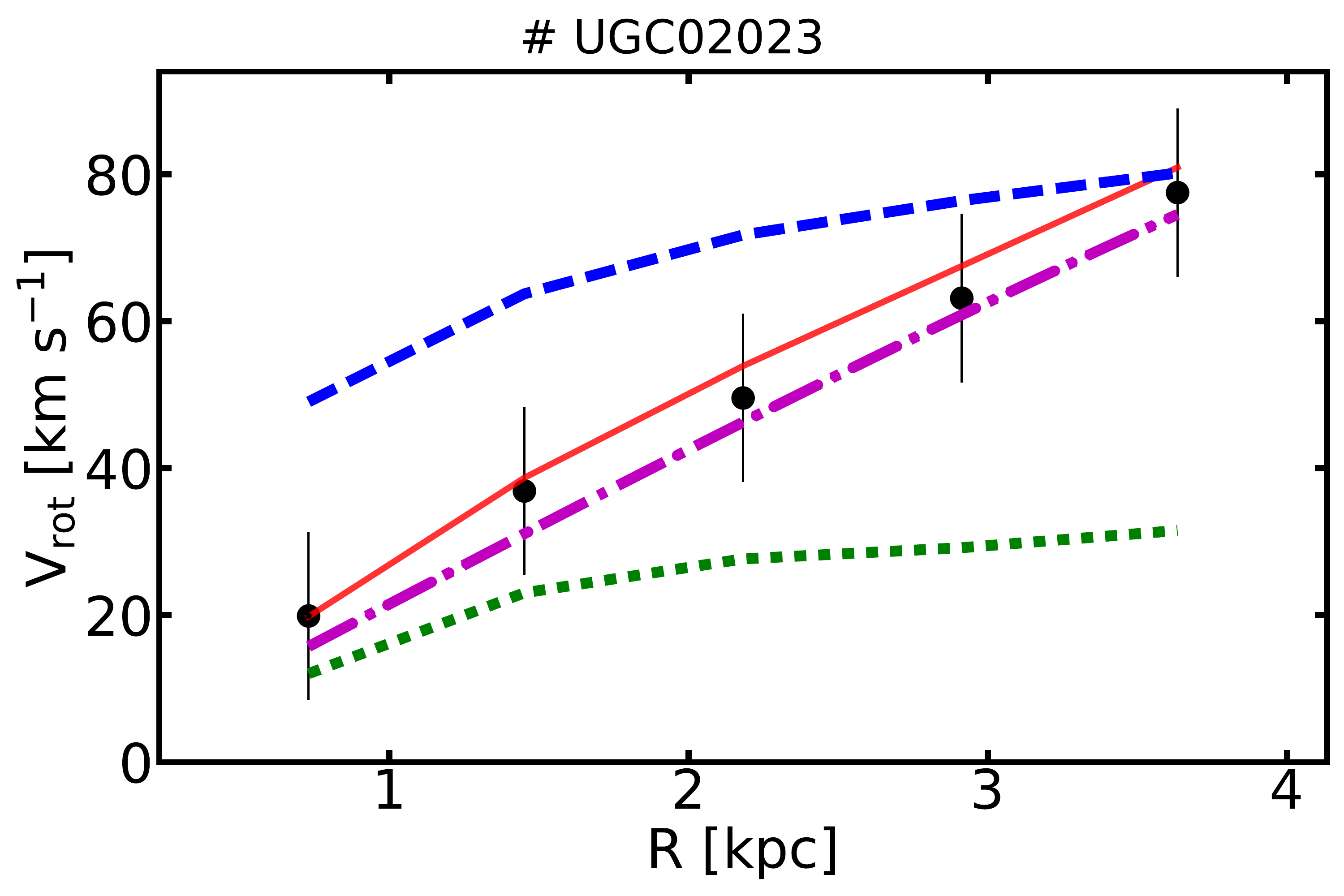}\hfill
\includegraphics[width=.33\textwidth]{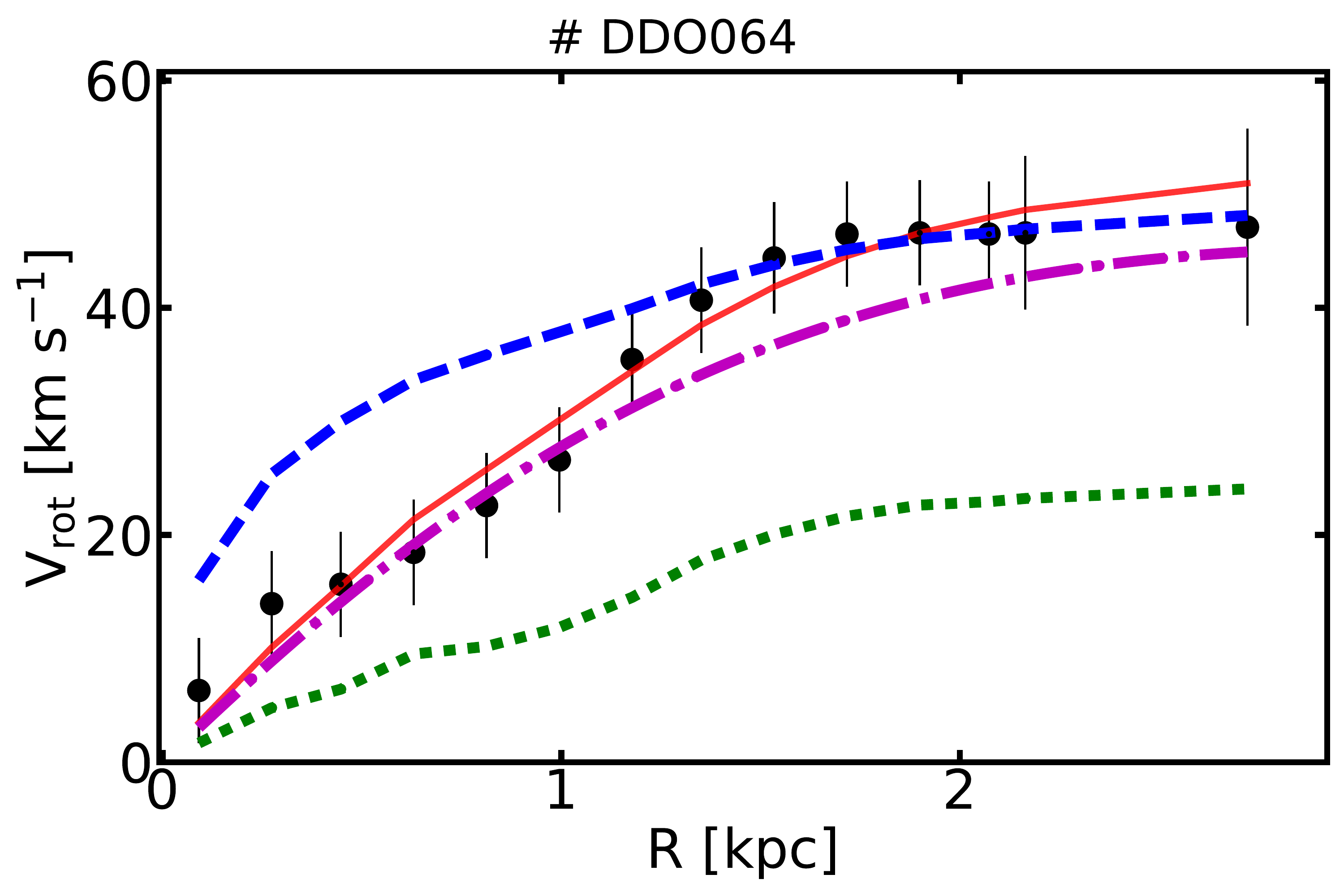}\hfill
\includegraphics[width=.33\textwidth]{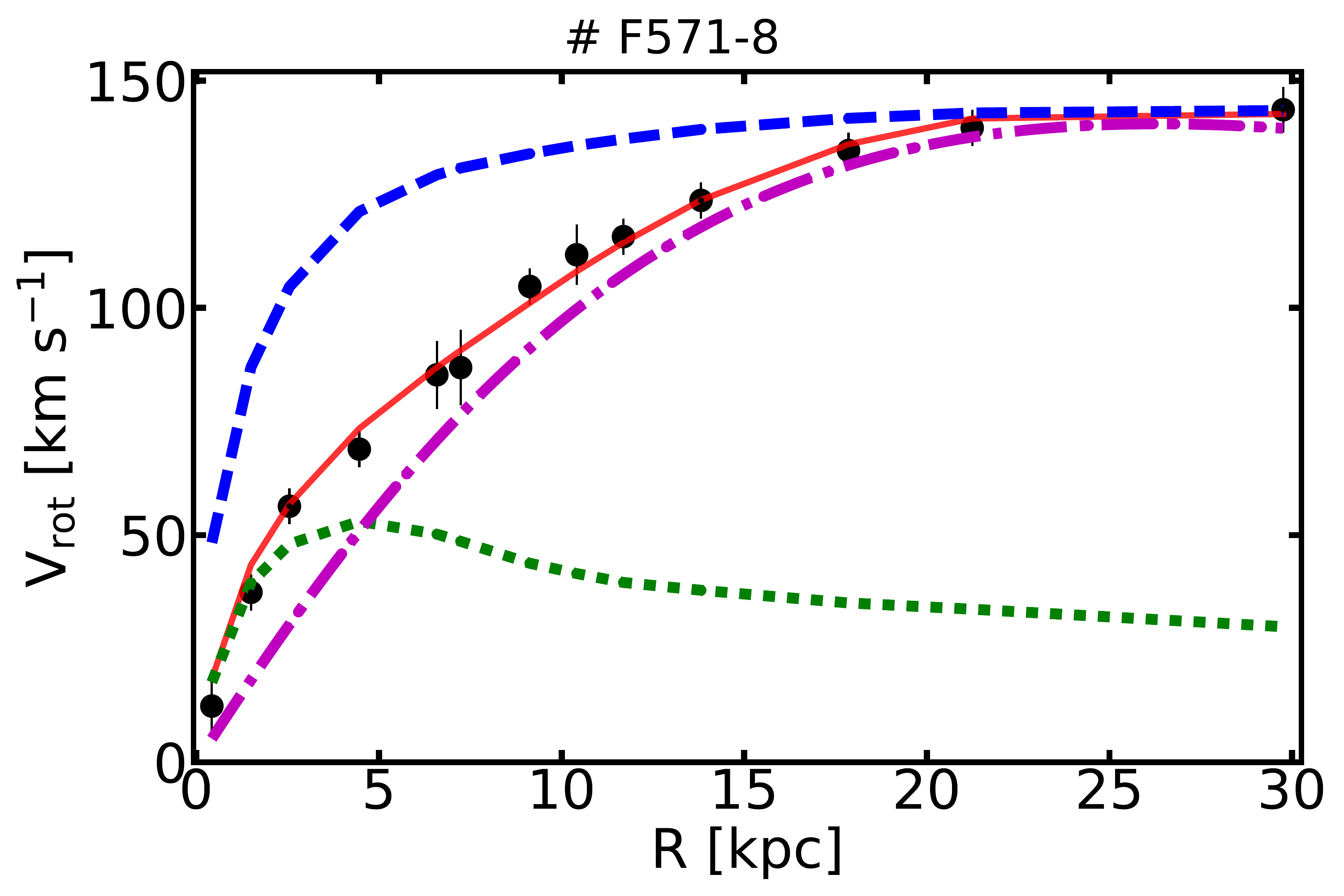}

\includegraphics[width=.33\textwidth]{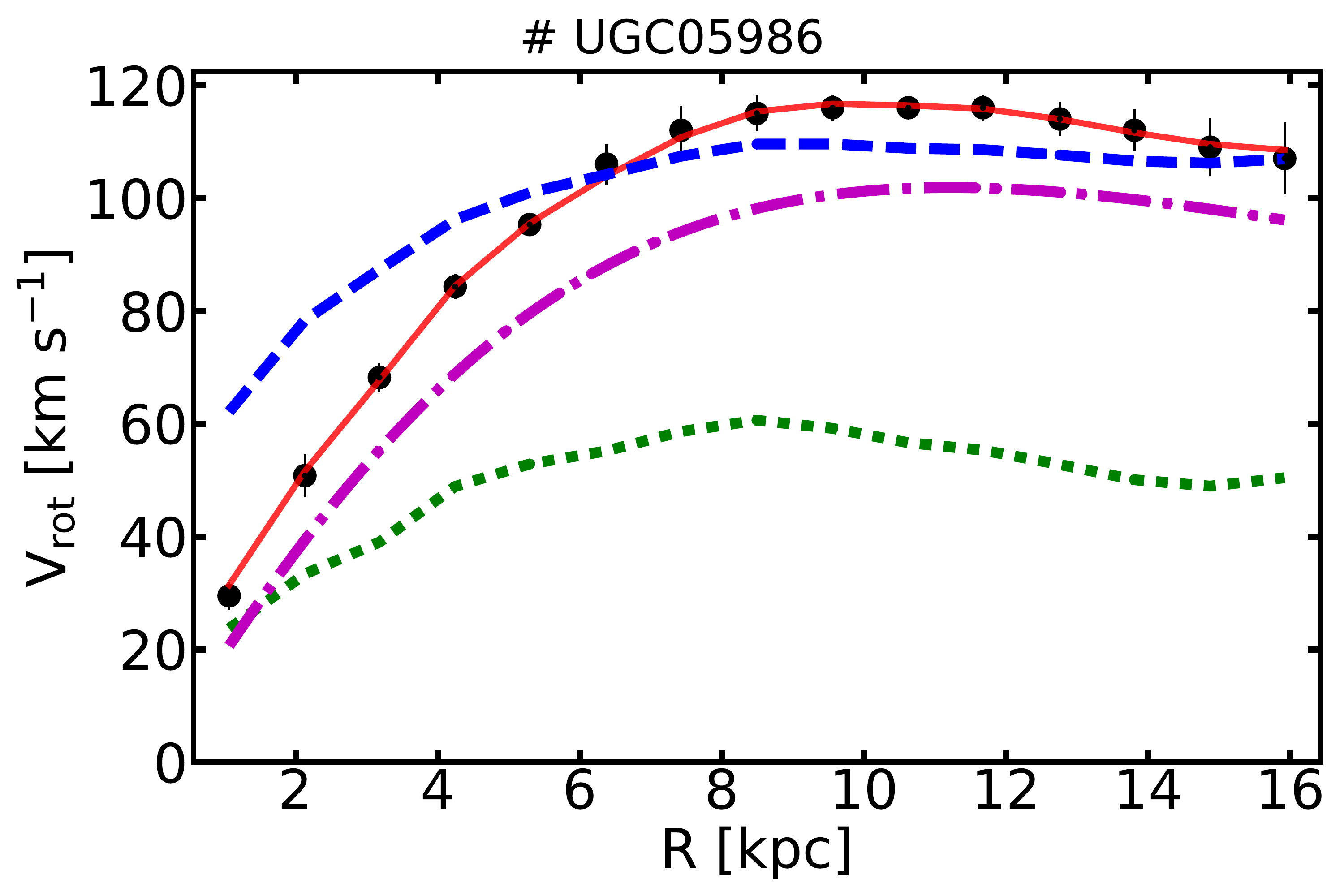}\hfill
\includegraphics[width=.33\textwidth]{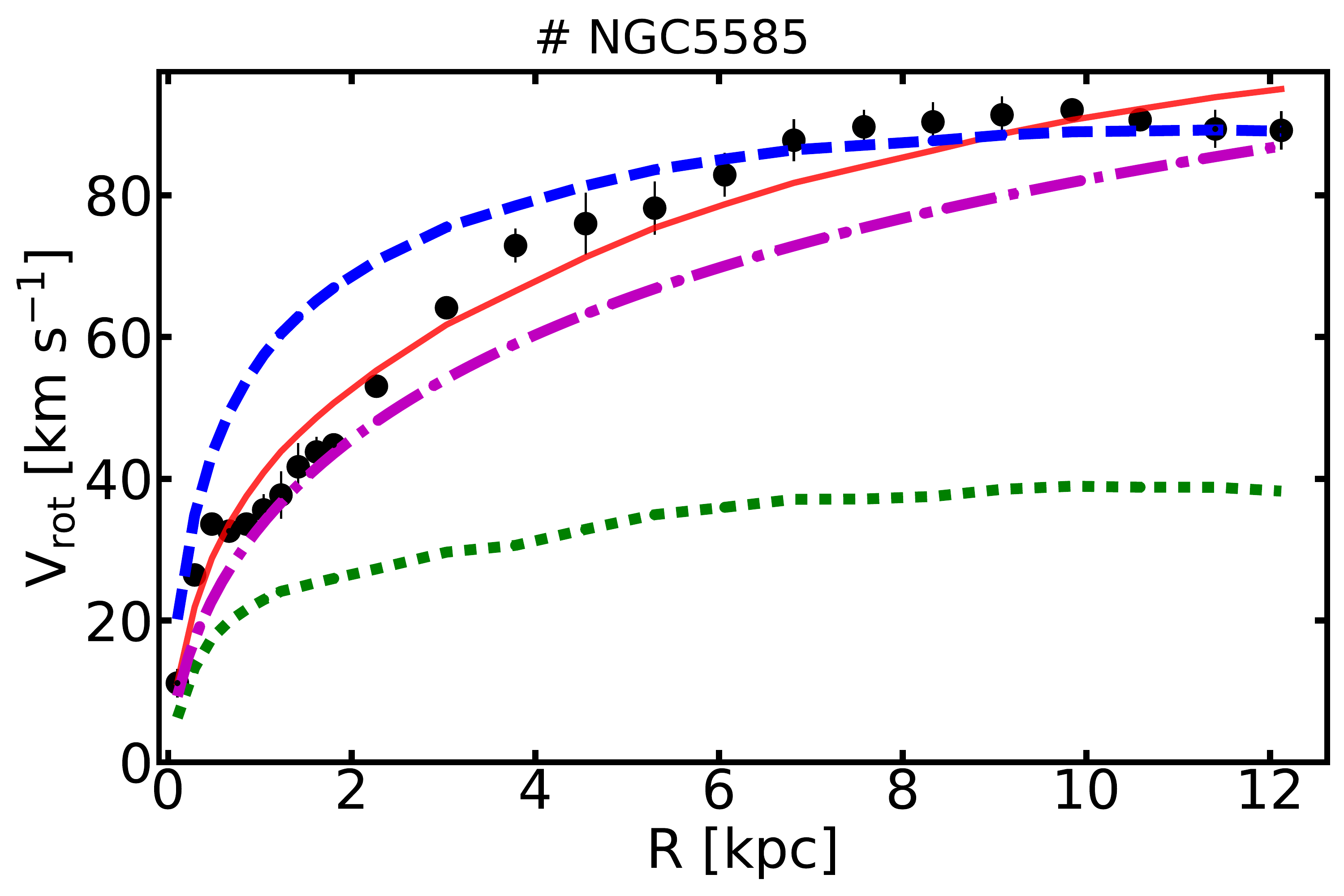}\hfill
\includegraphics[width=.33\textwidth]{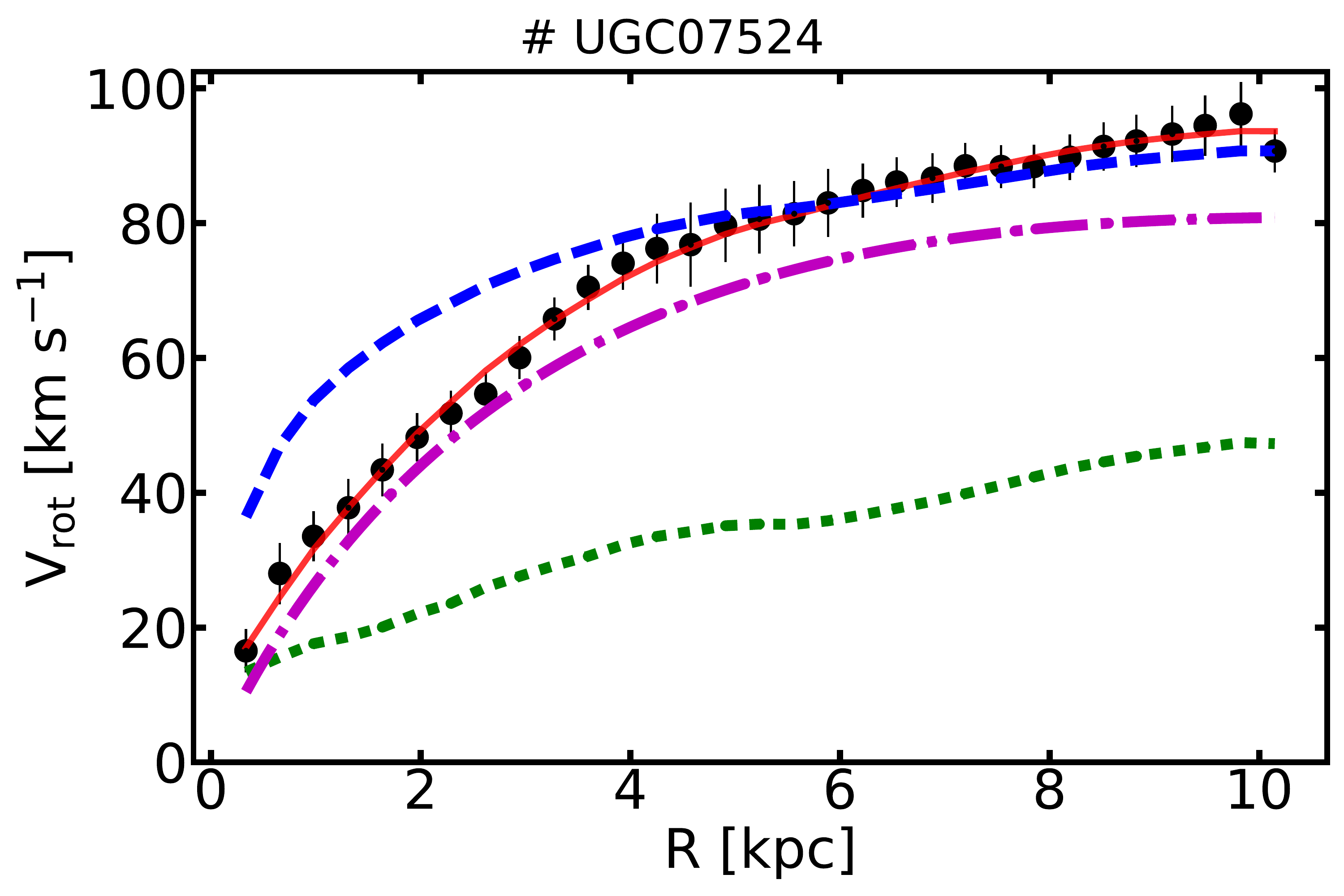}

\includegraphics[width=.33\textwidth]{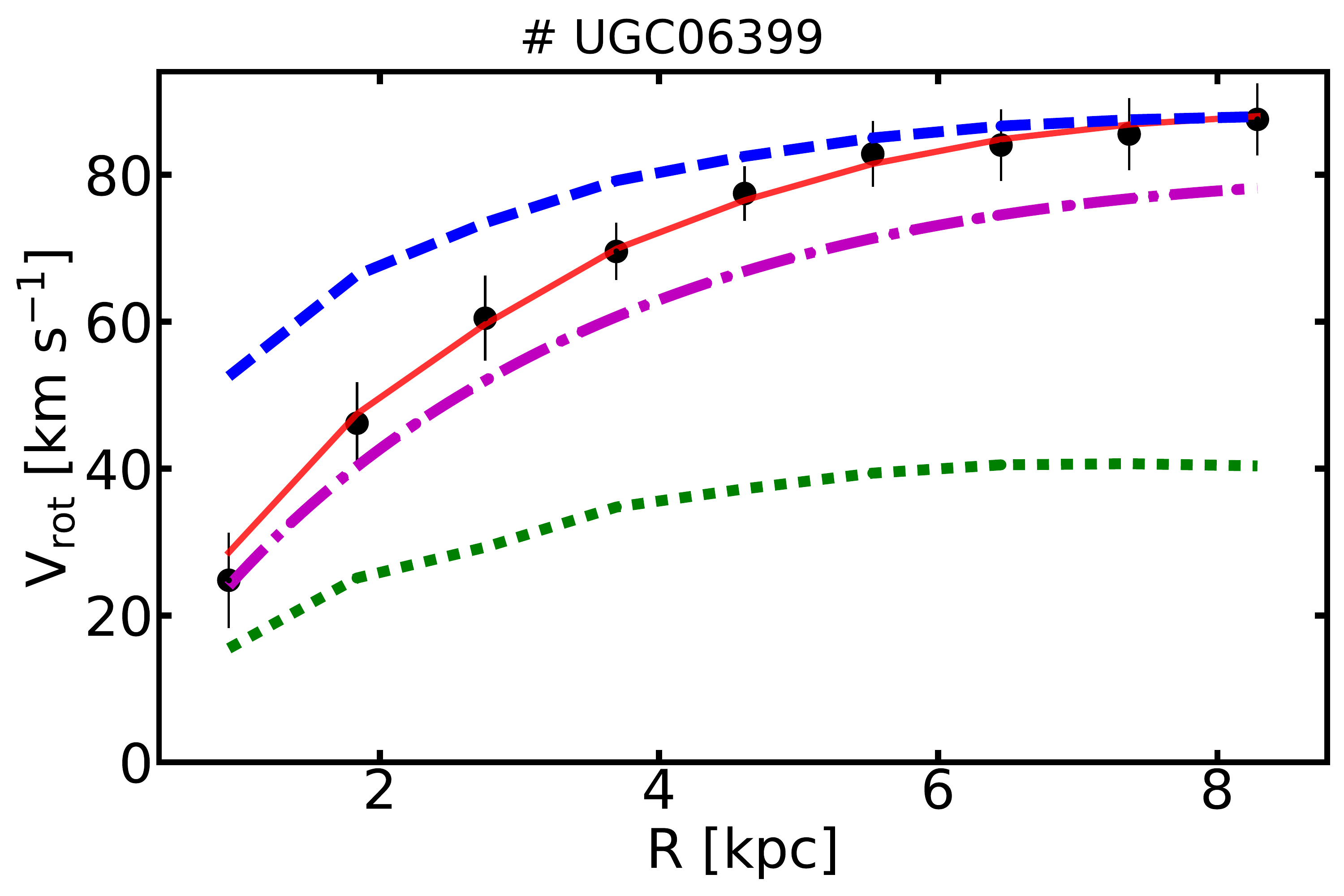}\hfill
\includegraphics[width=.33\textwidth]{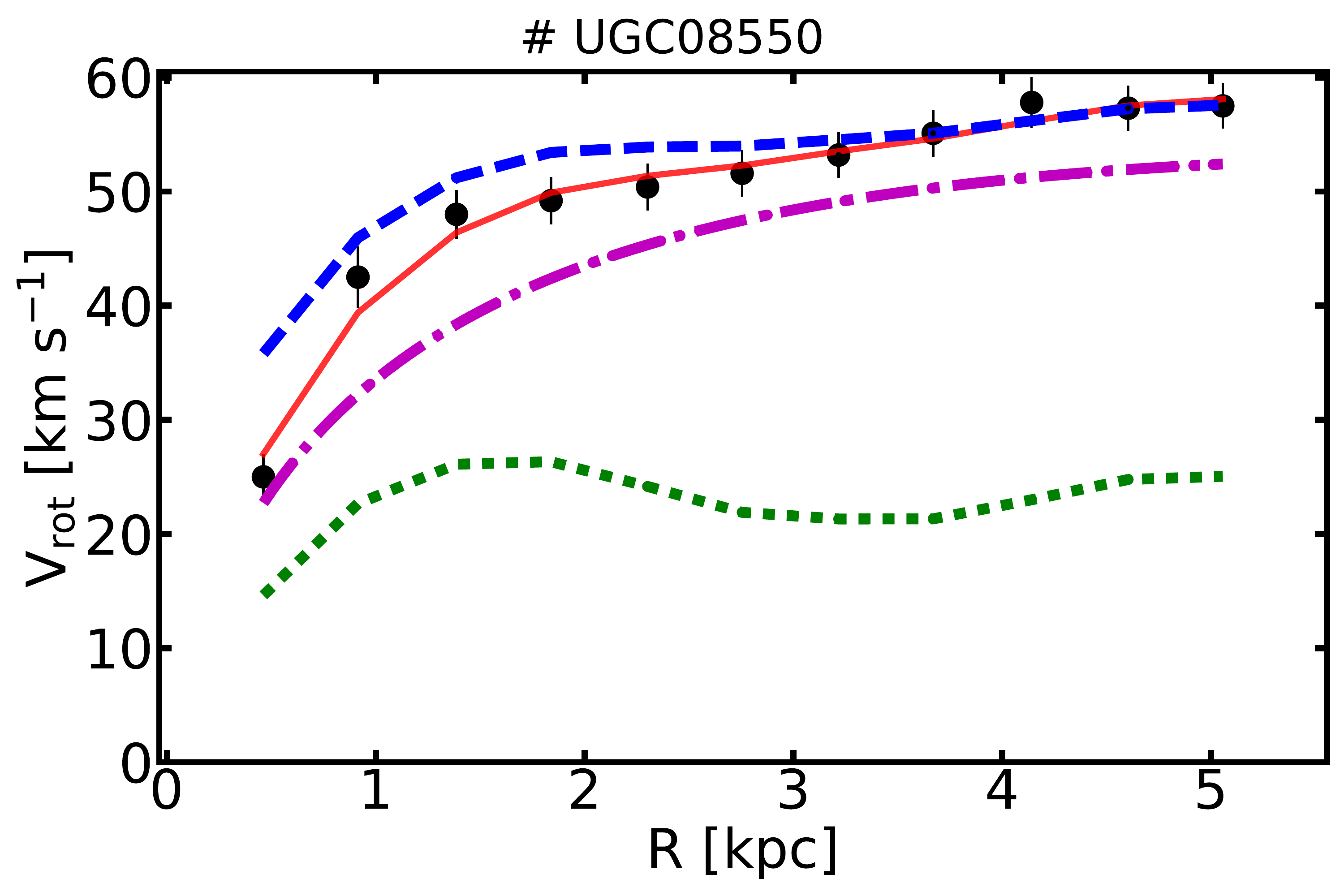}\hfill
\includegraphics[width=.33\textwidth]{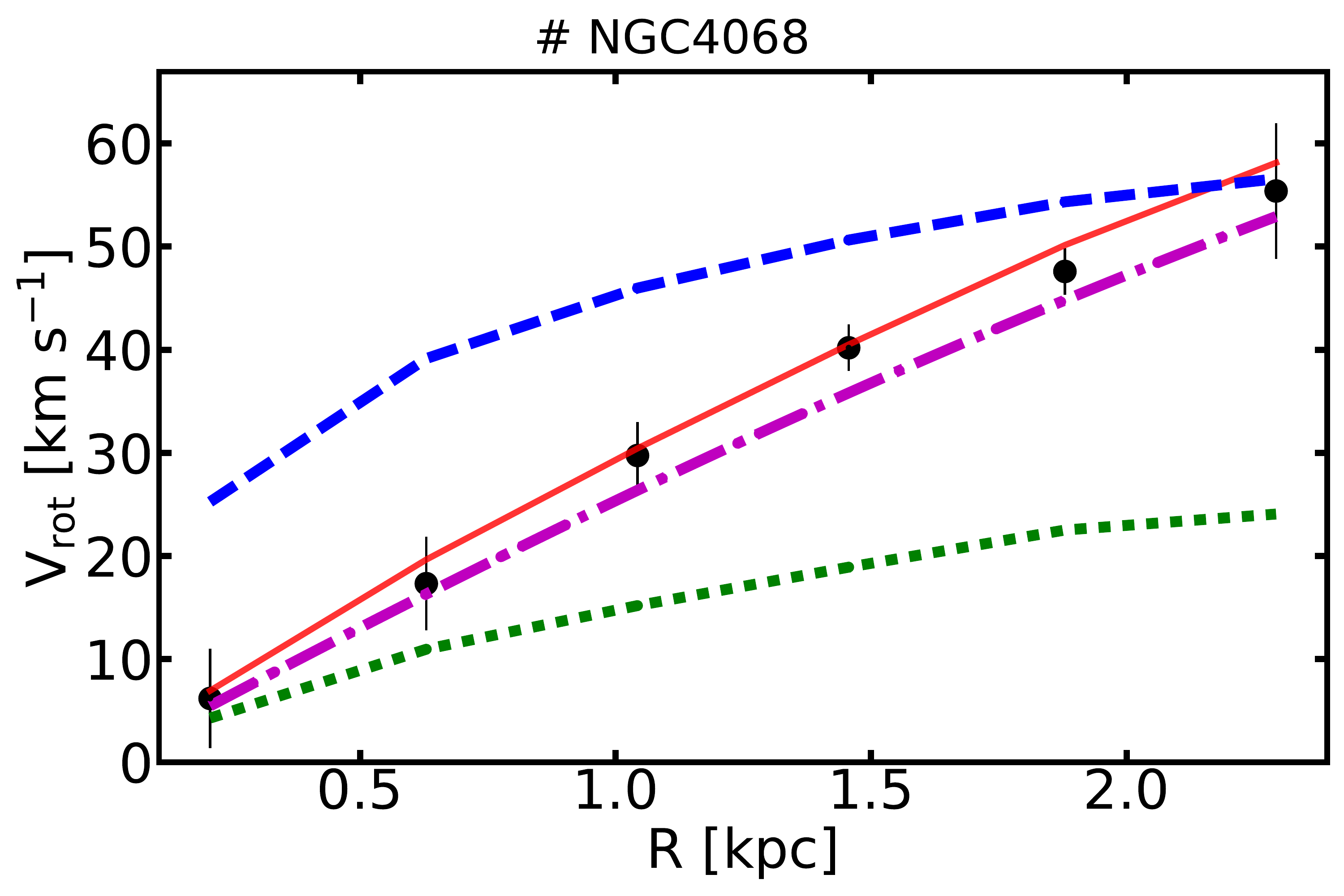}

\includegraphics[width=.33\textwidth]{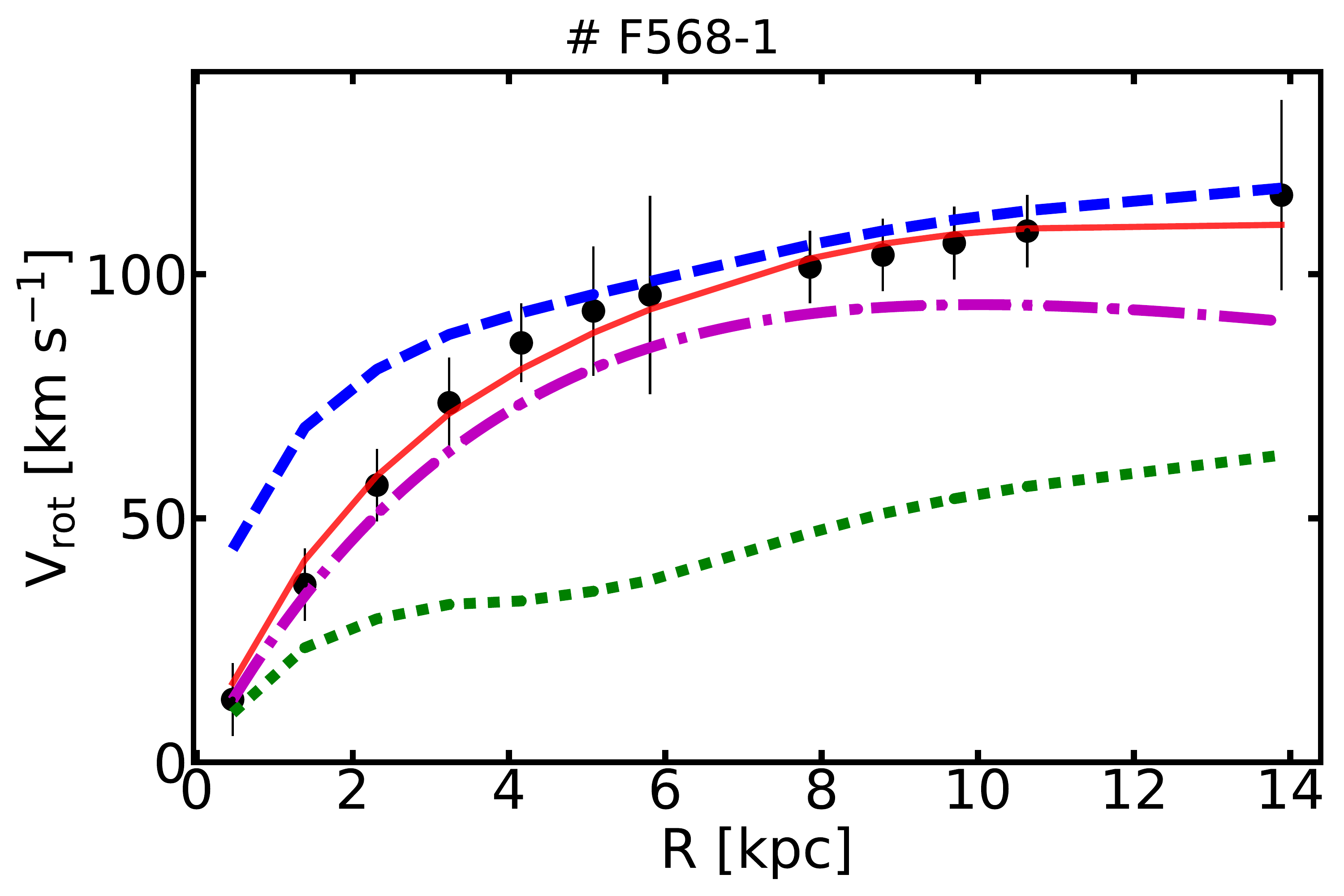}\hfill
\includegraphics[width=.33\textwidth]{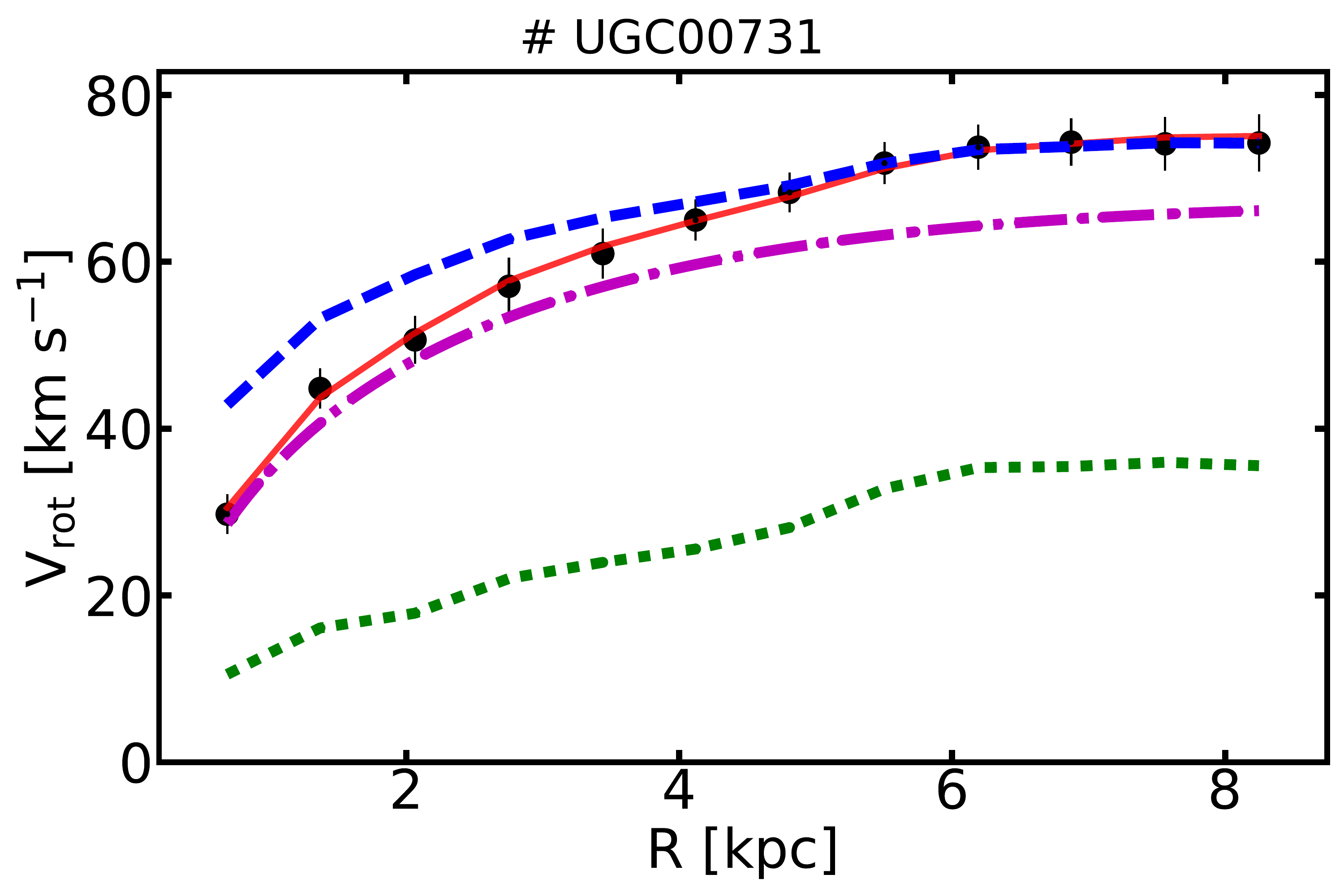}\hfill
\includegraphics[width=.33\textwidth]{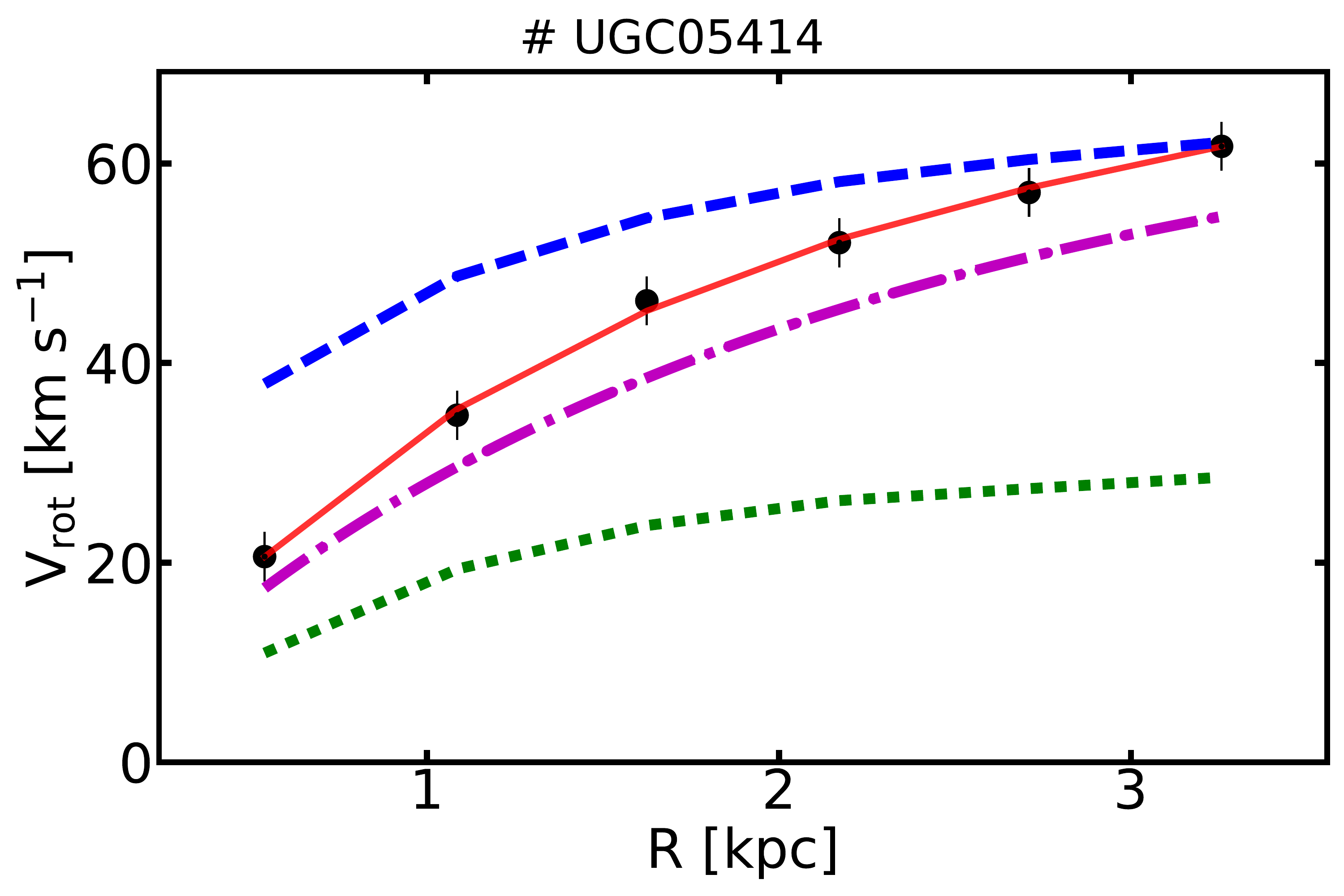}
\caption{Continued.}
\end{figure*}

\begin{figure*}
\centering
\ContinuedFloat
\includegraphics[width=.33\textwidth]{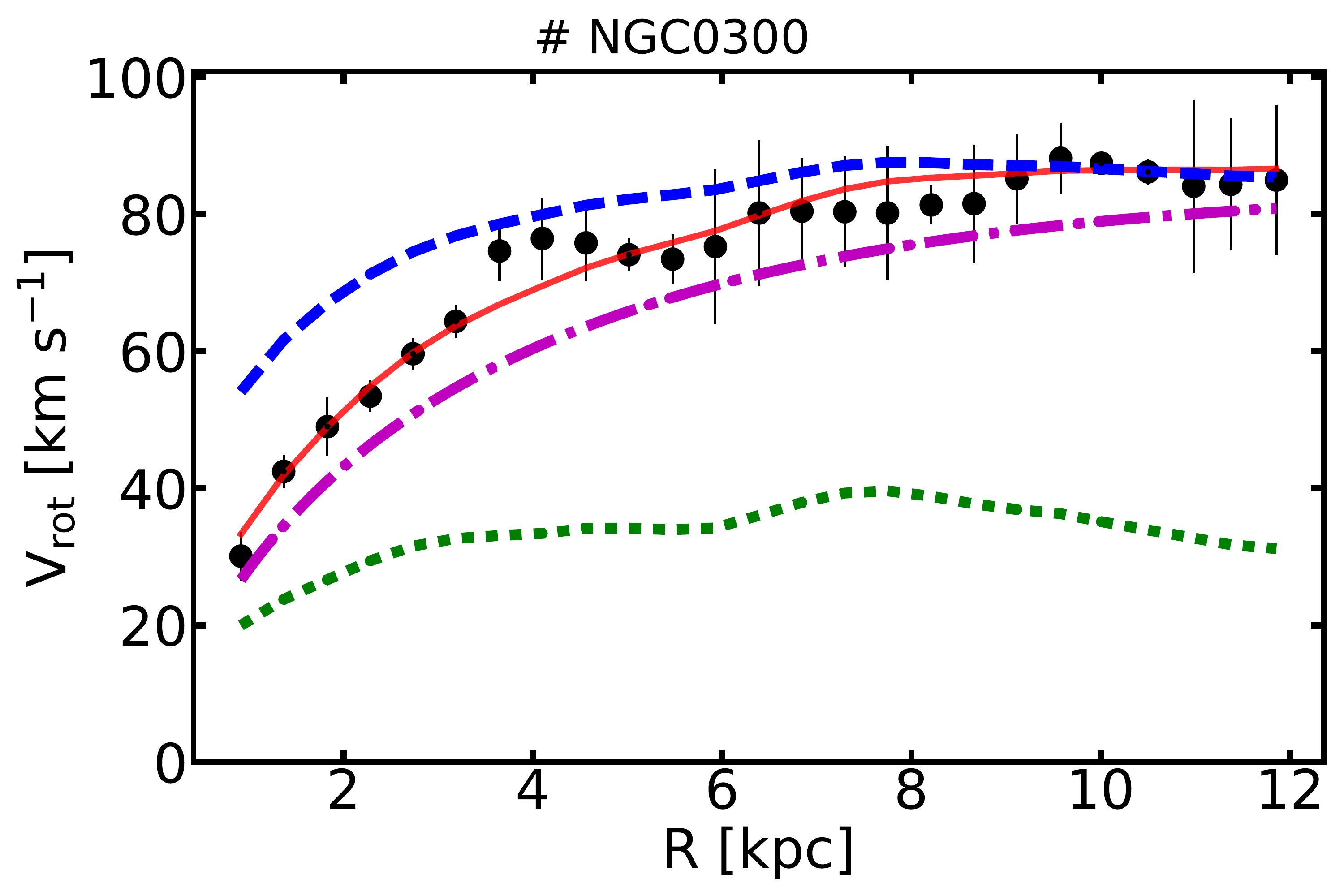}\hfill
\includegraphics[width=.33\textwidth]{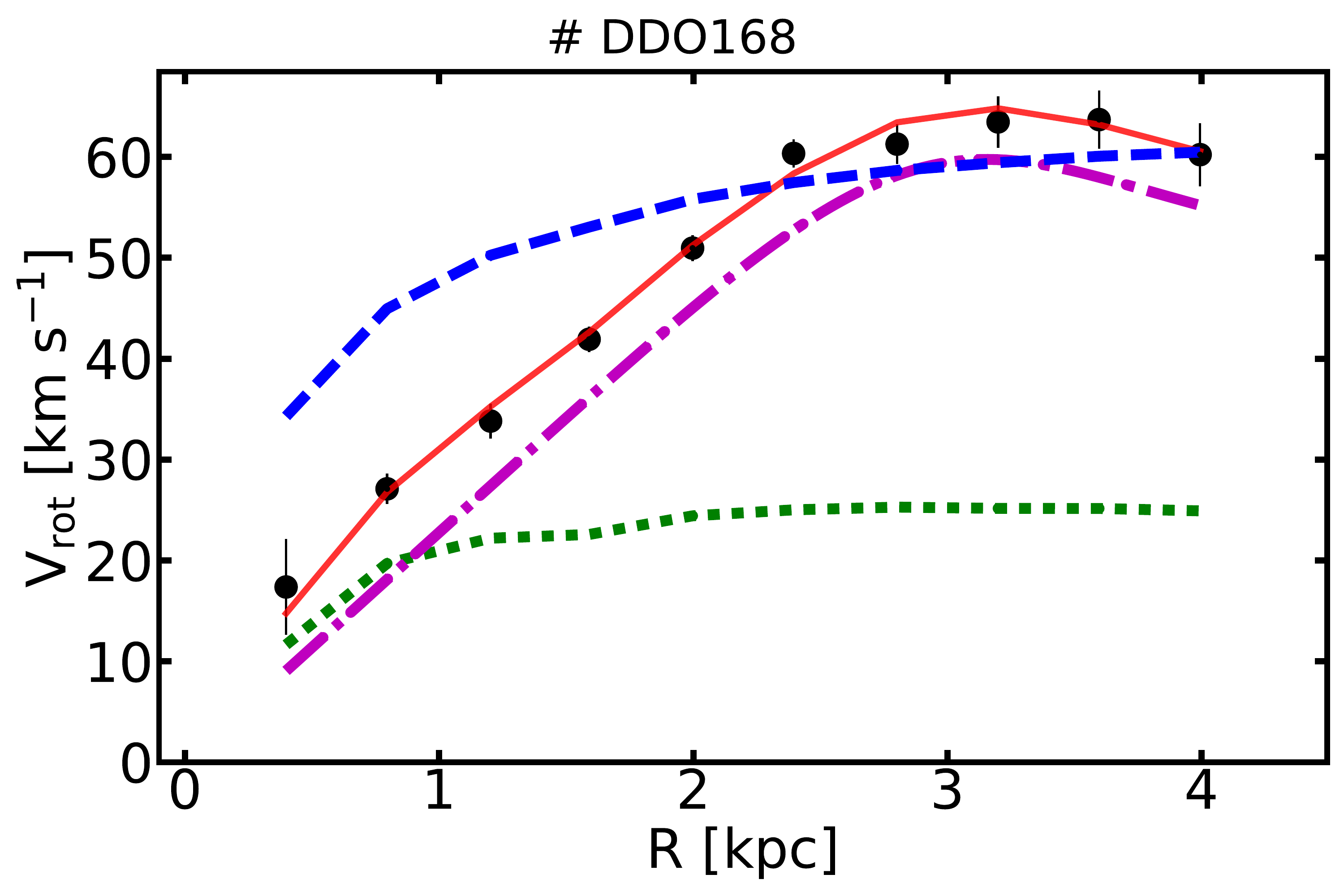}\hfill
\includegraphics[width=.33\textwidth]{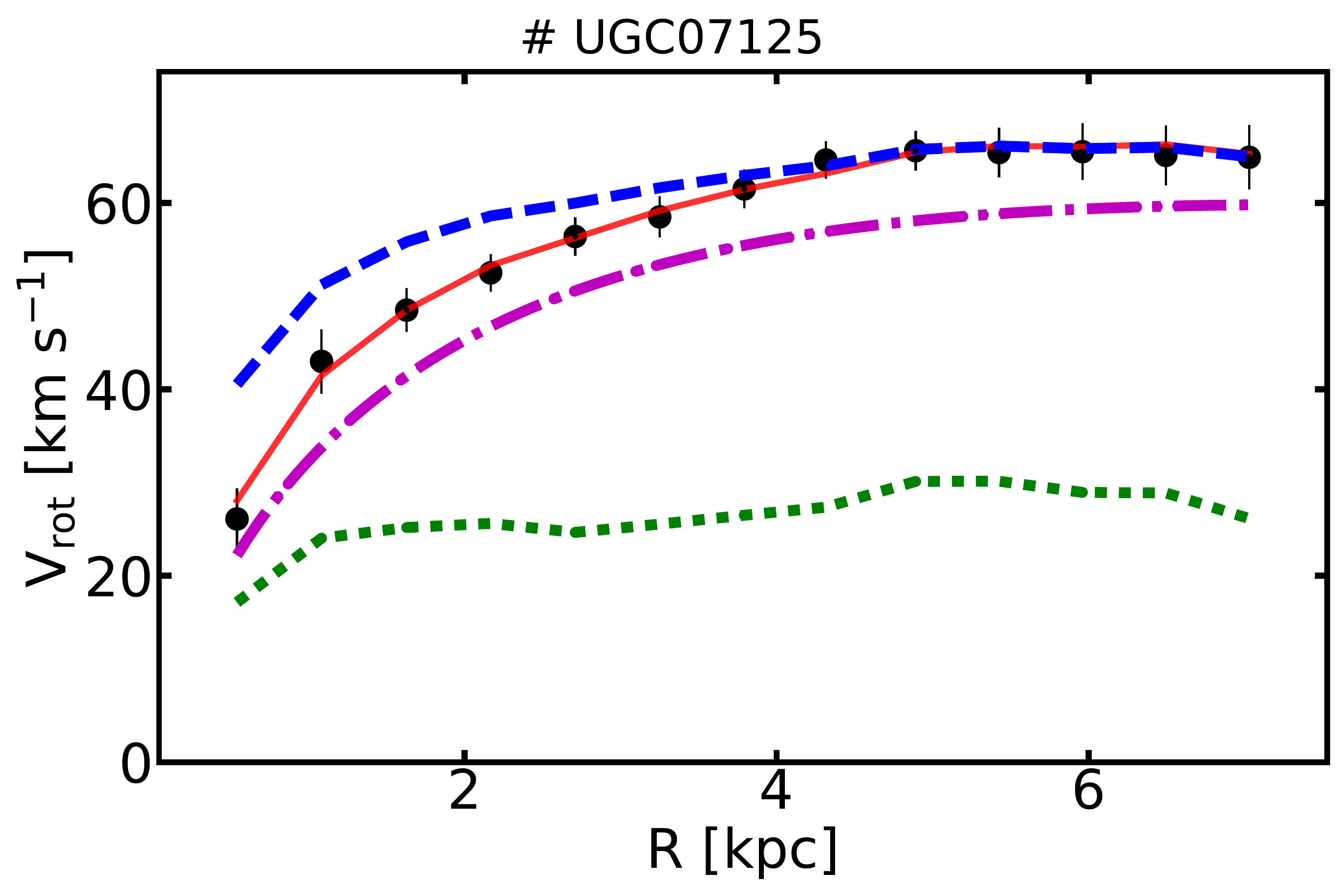}

\includegraphics[width=.33\textwidth]{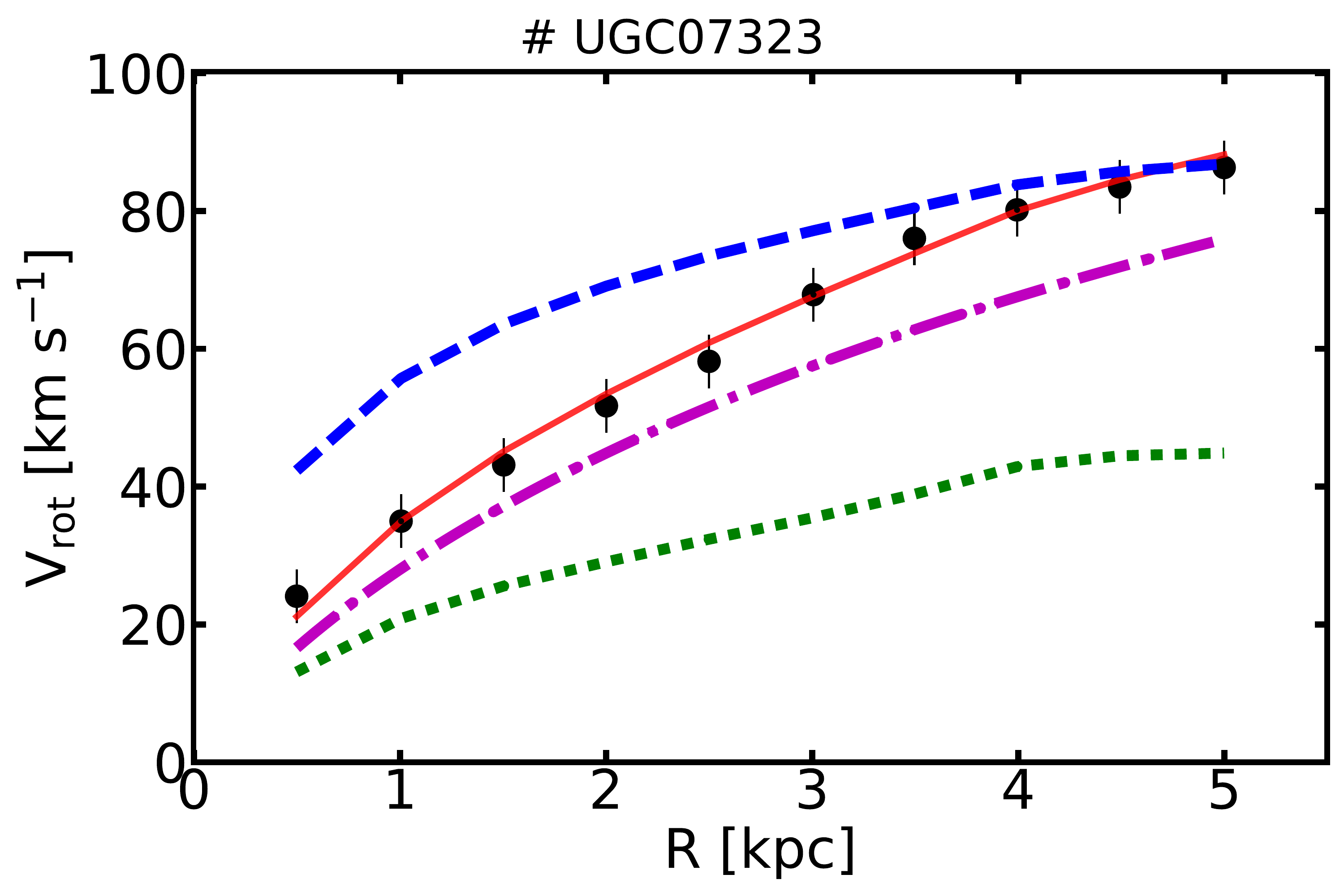}\hfill
\includegraphics[width=.33\textwidth]{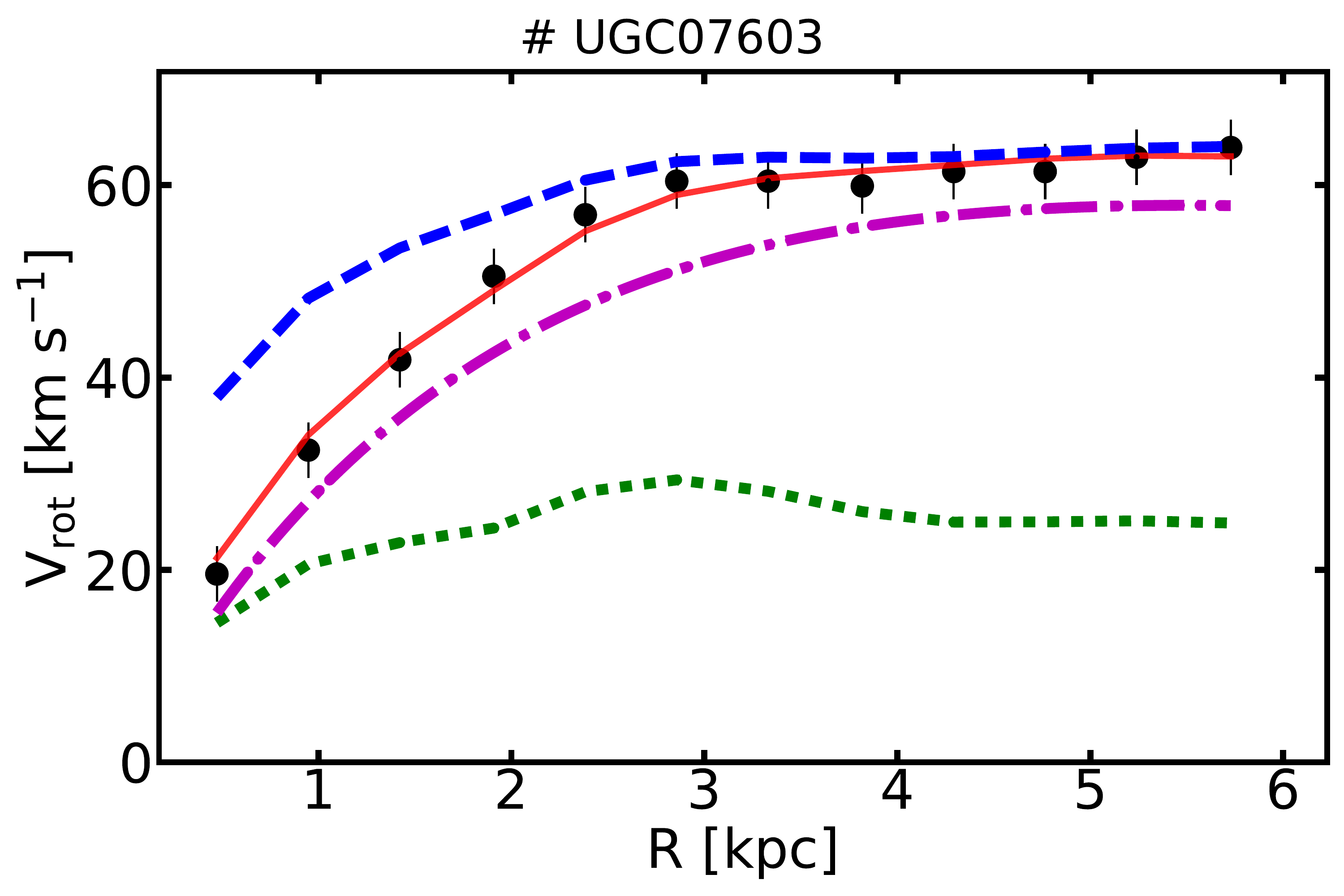}\hfill
\includegraphics[width=.33\textwidth]{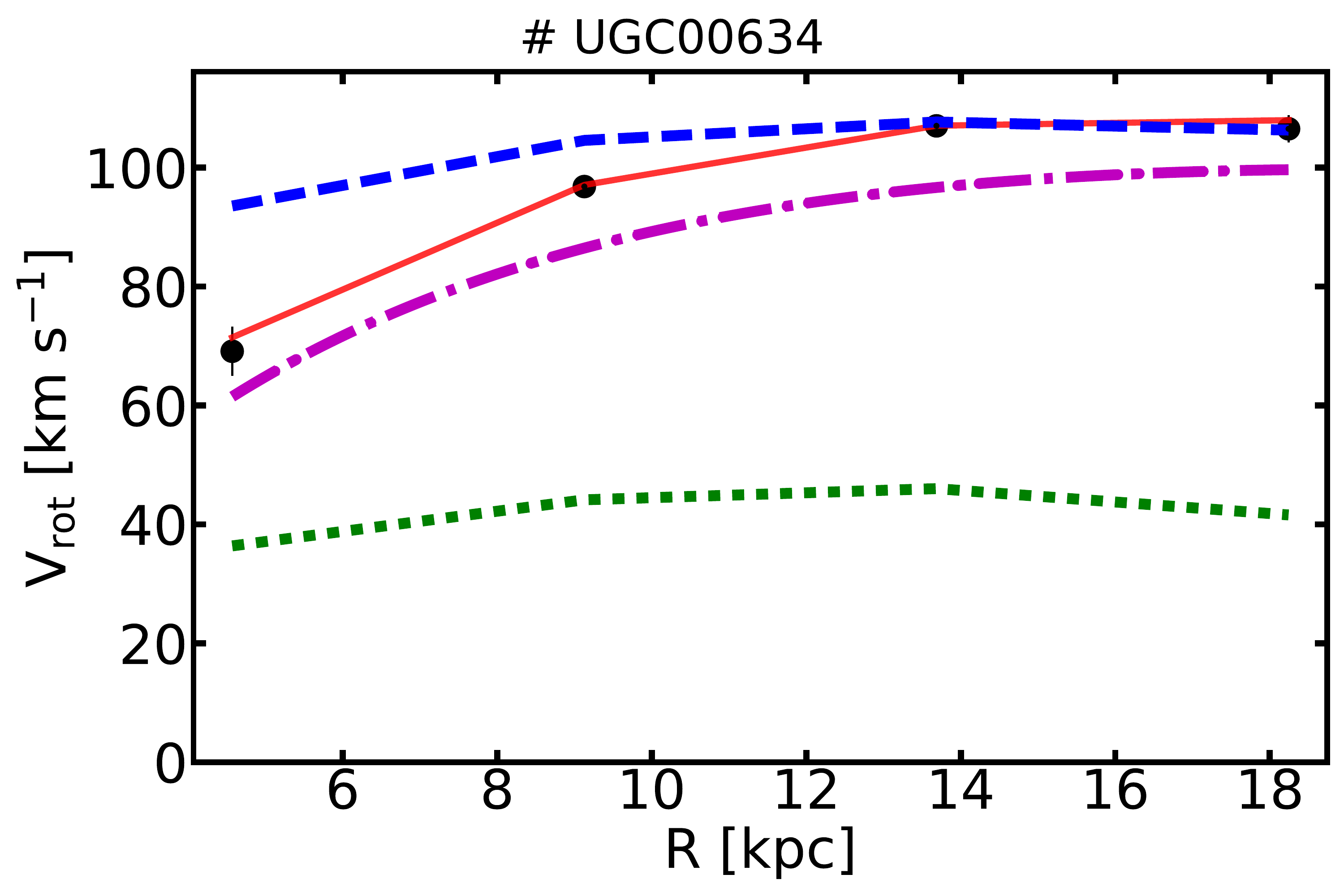}

\includegraphics[width=.33\textwidth]{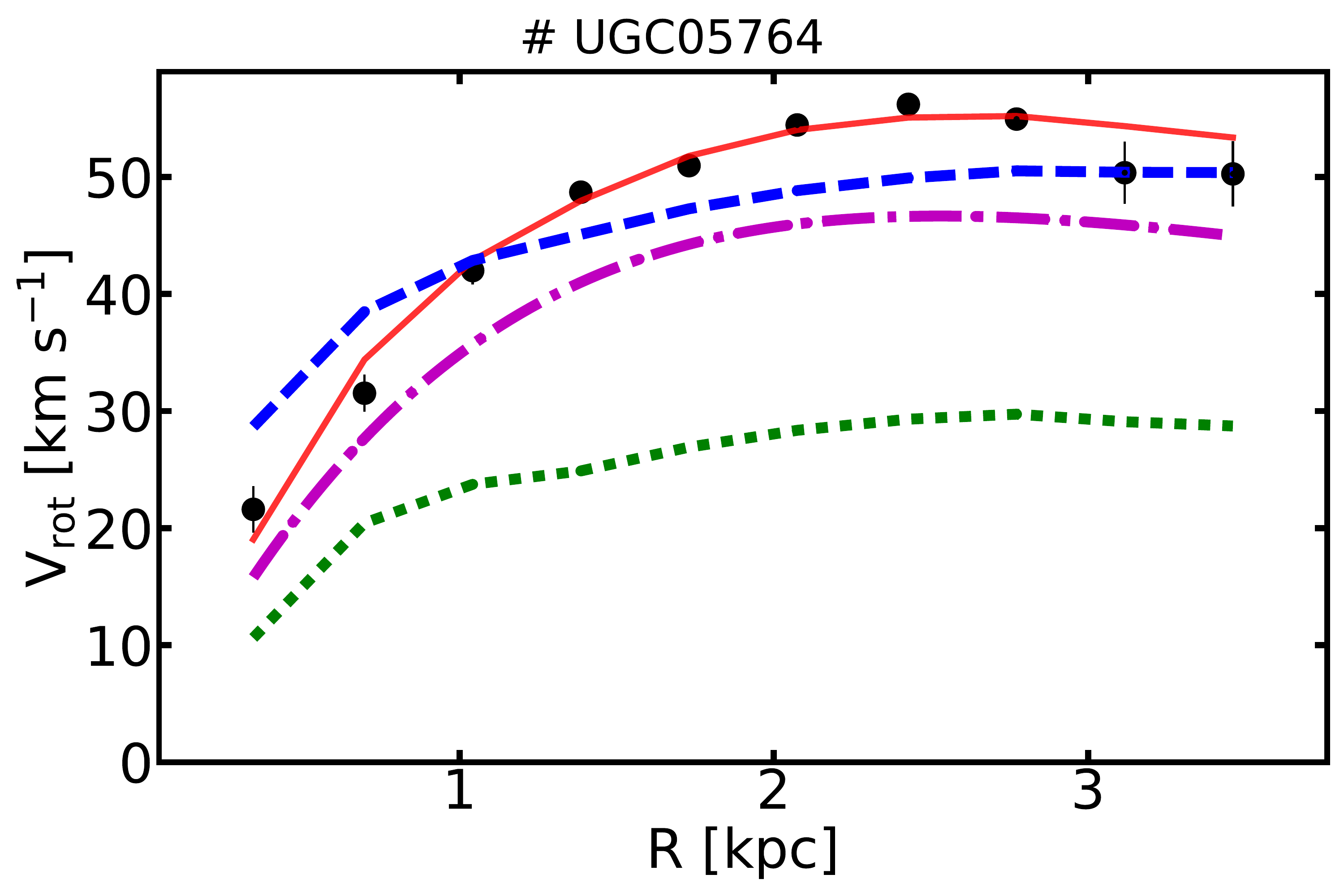}\hfill
\includegraphics[width=.33\textwidth]{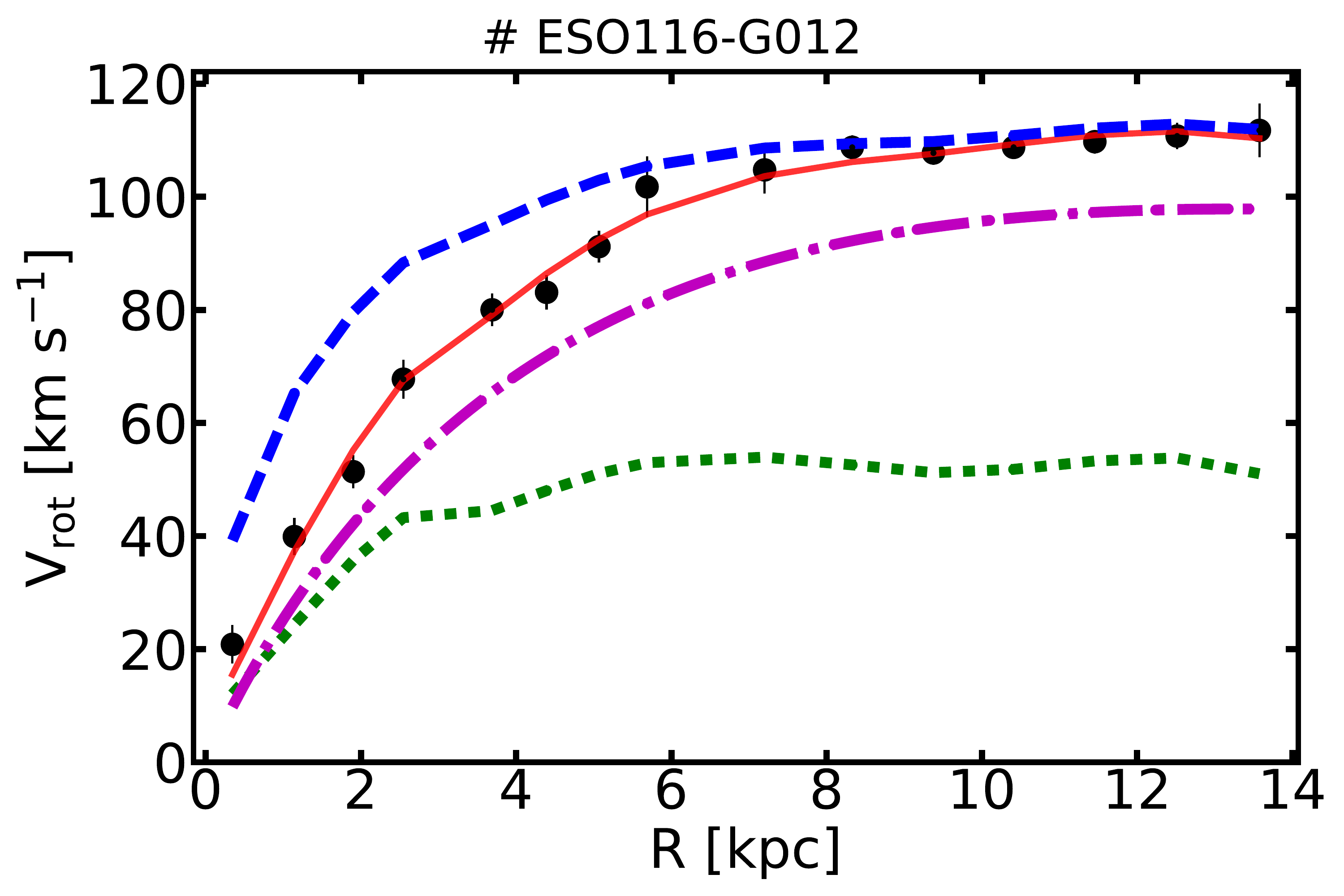}\hfill
\includegraphics[width=.33\textwidth]{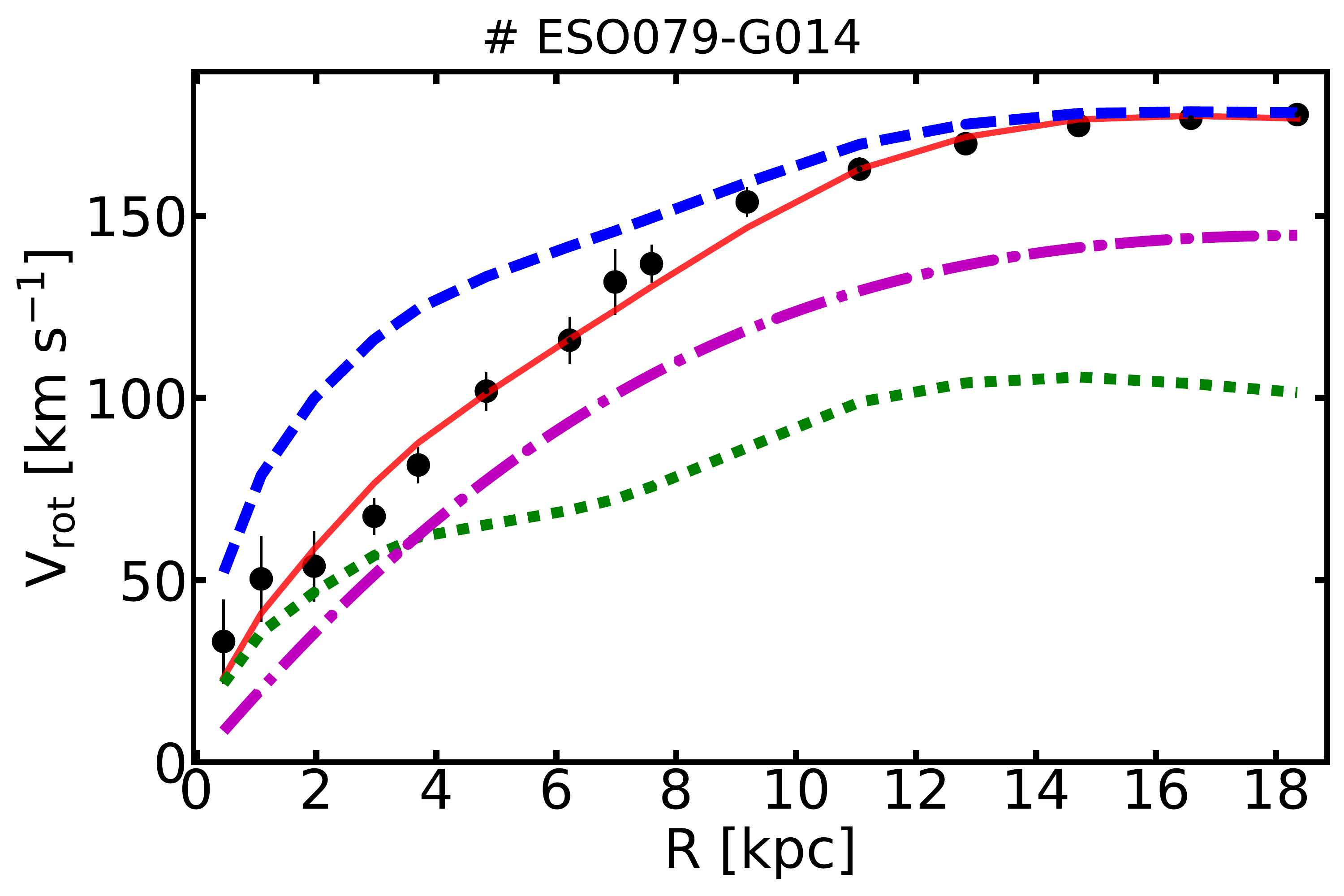}

\includegraphics[width=.33\textwidth]{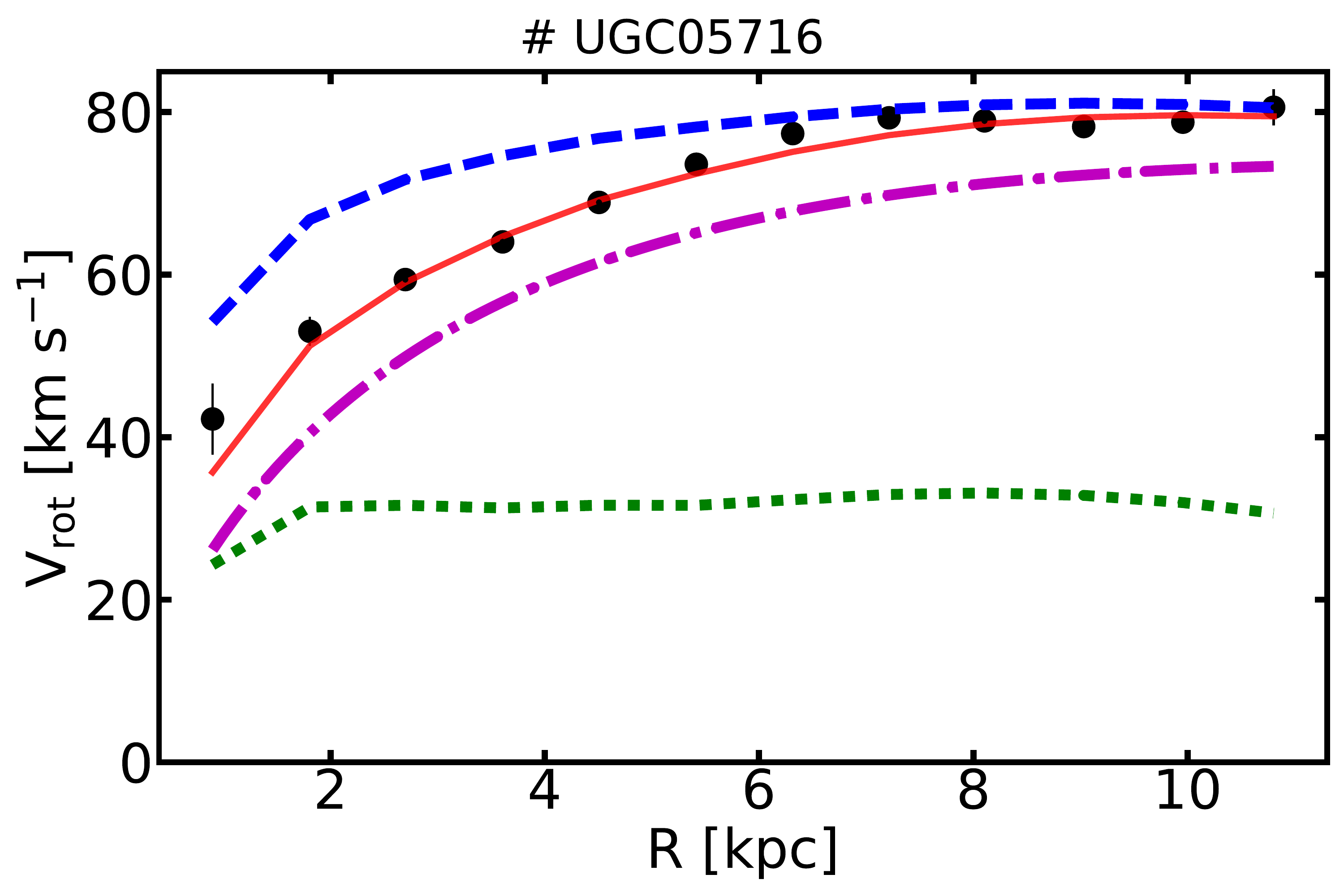}\hfill
\includegraphics[width=.33\textwidth]{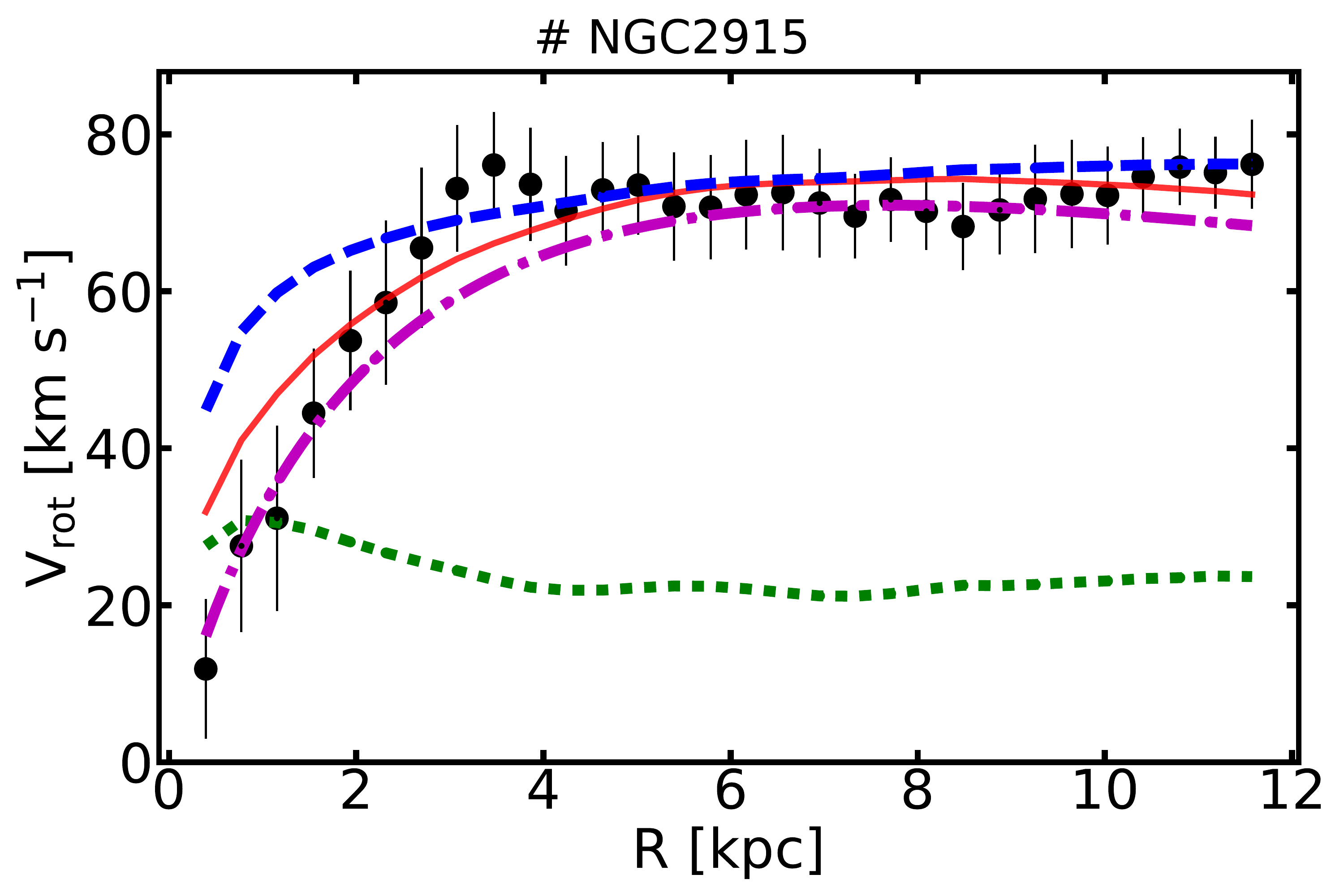}\hfill
\includegraphics[width=.33\textwidth]{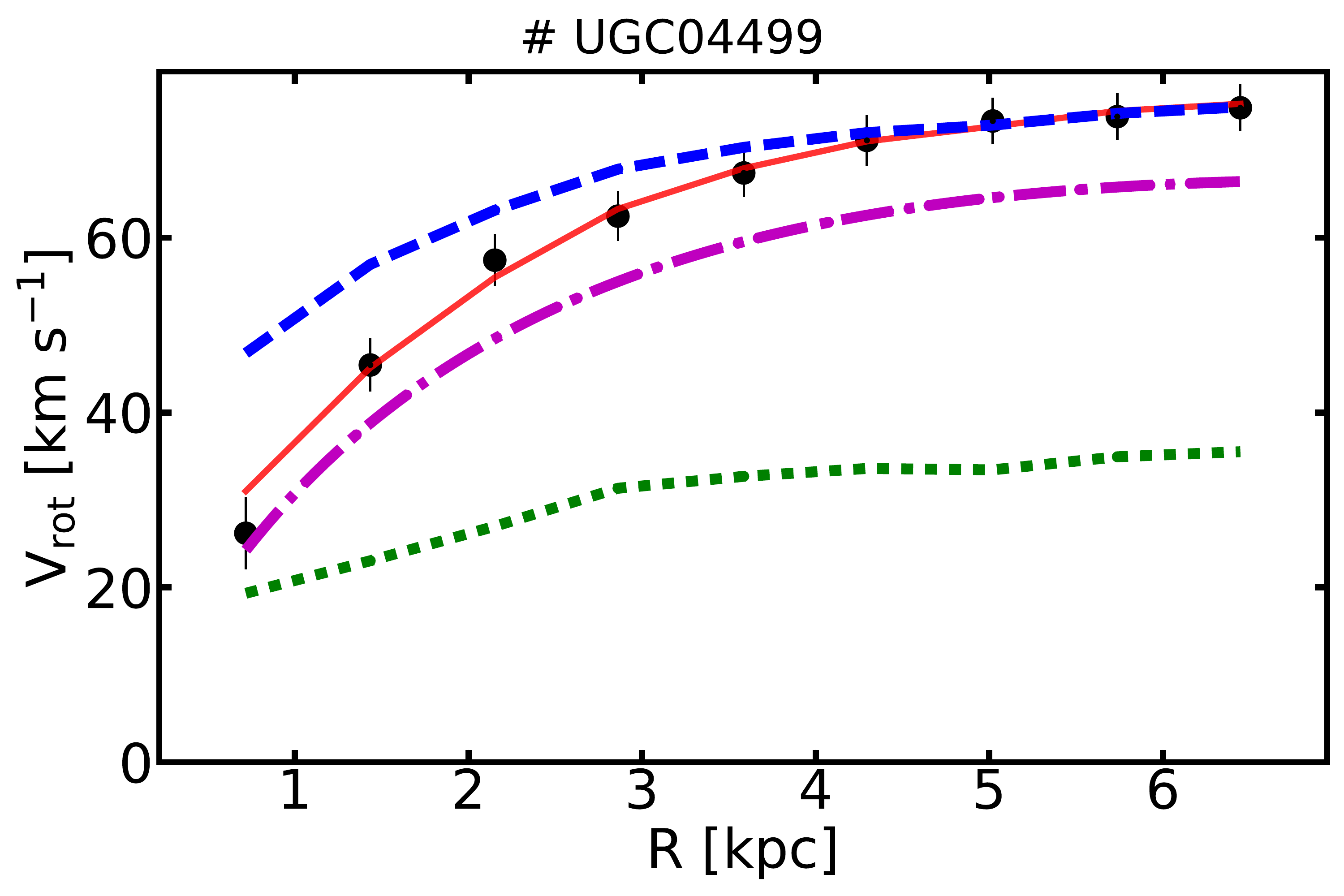}

\includegraphics[width=.33\textwidth]{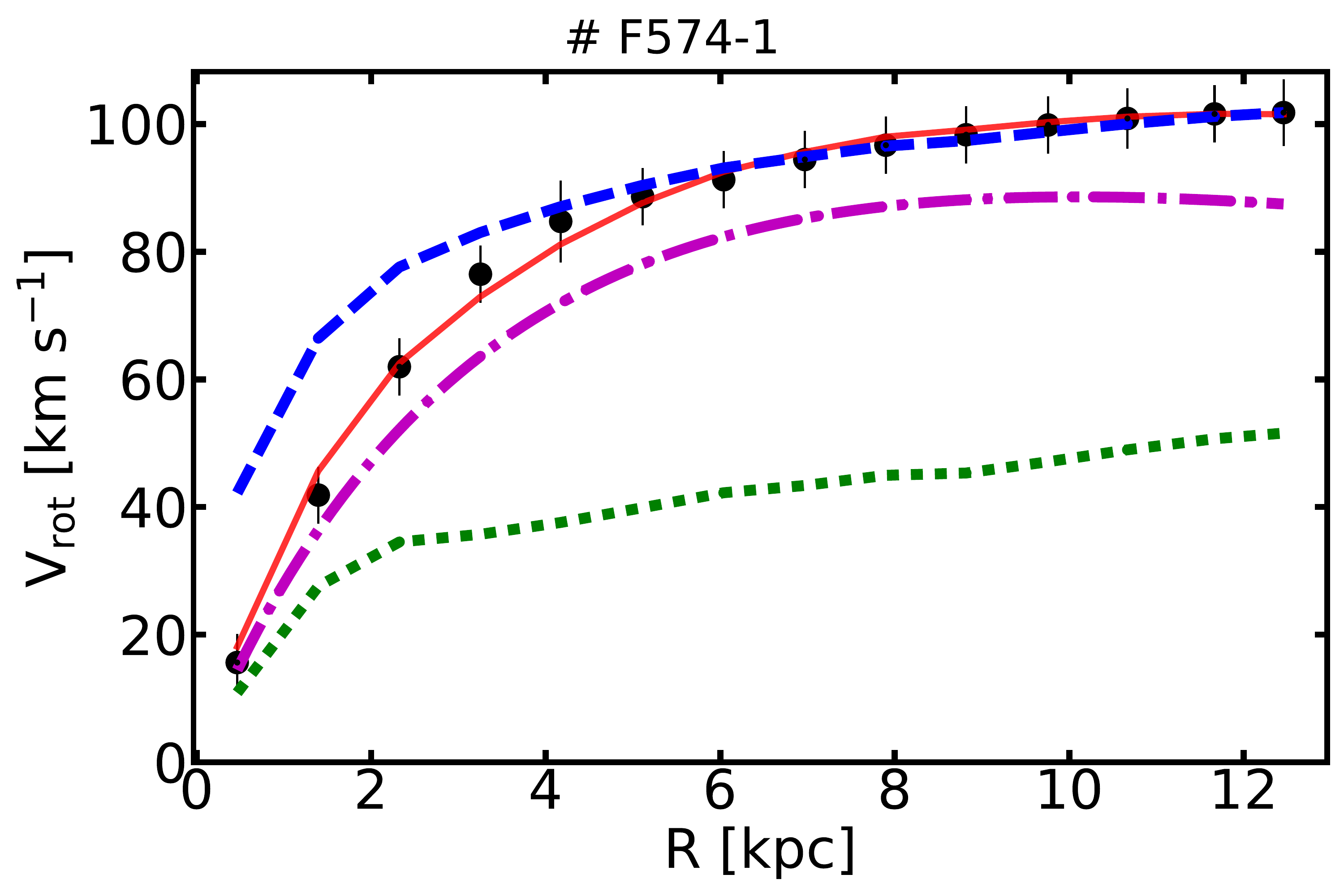}\hfill
\includegraphics[width=.33\textwidth]{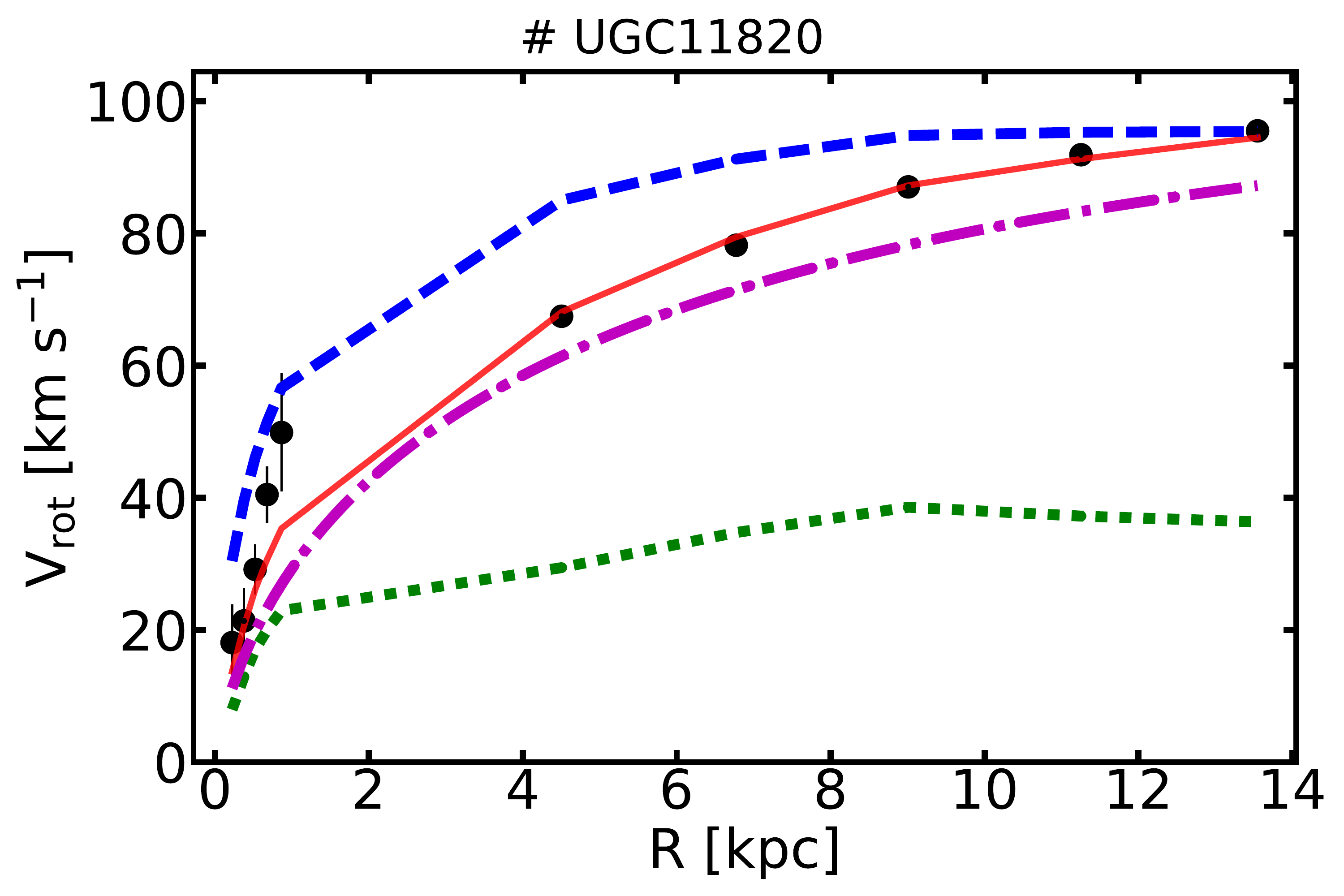}\hfill
\includegraphics[width=.33\textwidth]{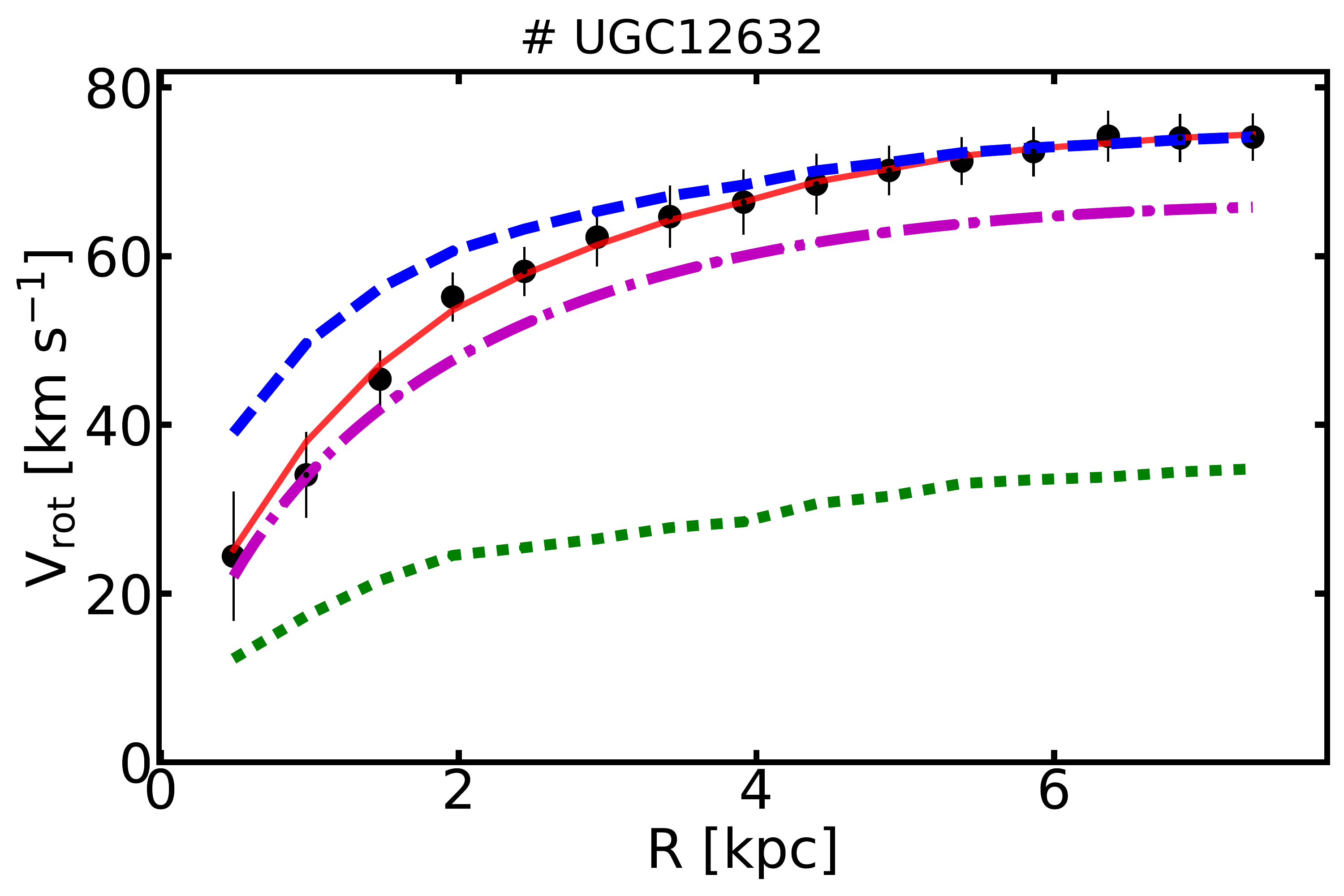}

\includegraphics[width=.33\textwidth]{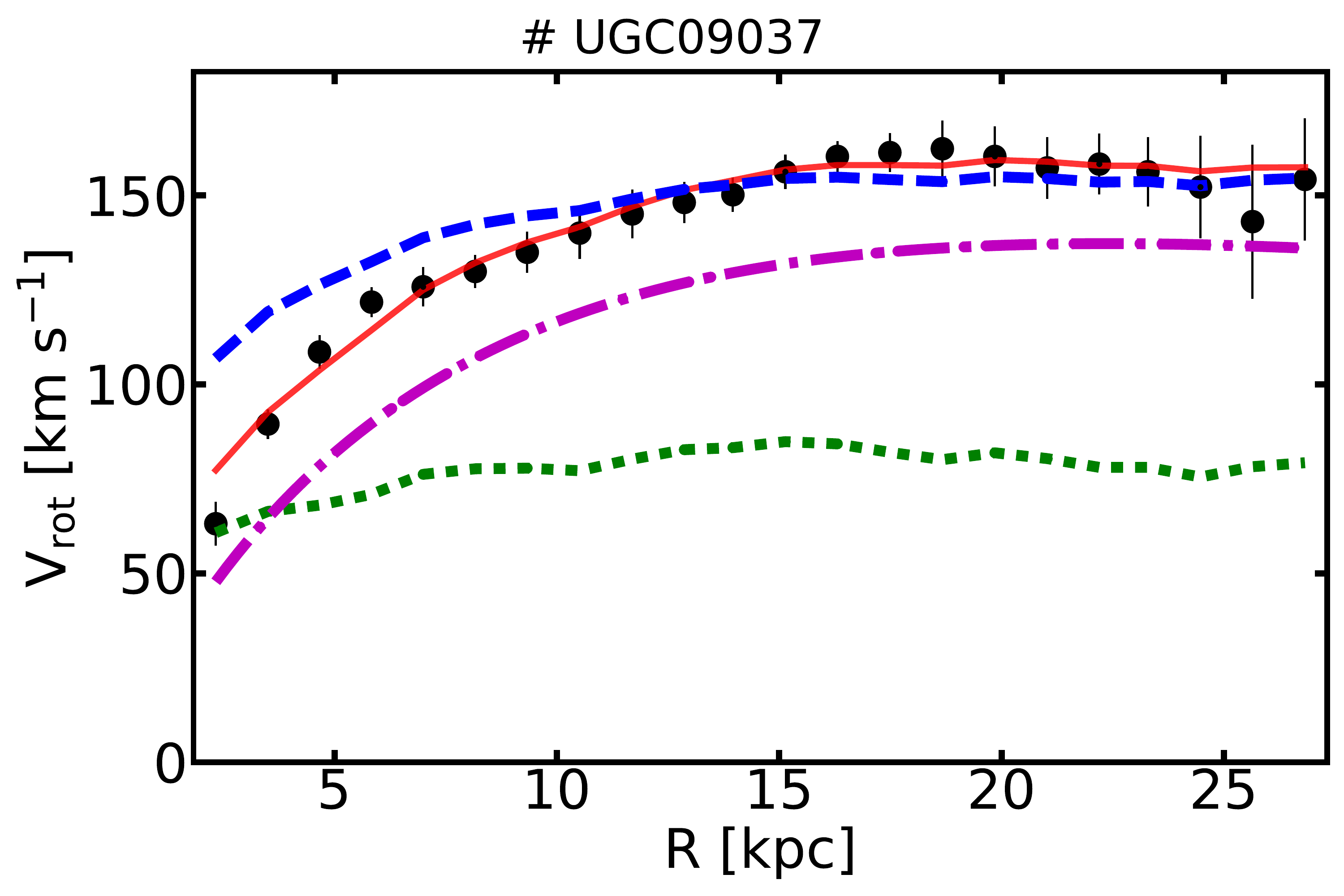}\hfill
\includegraphics[width=.33\textwidth]{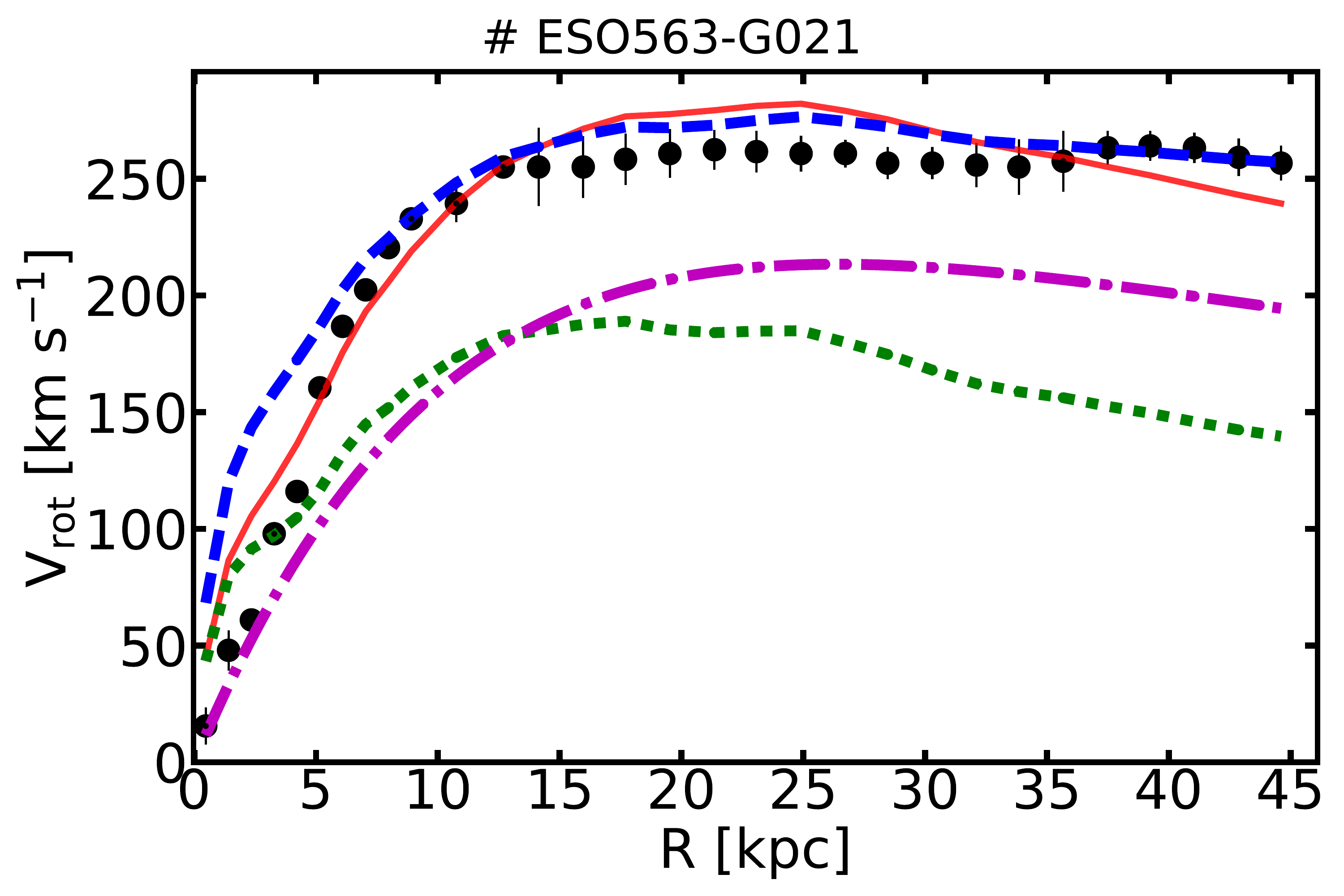}\hfill
\includegraphics[width=.33\textwidth]{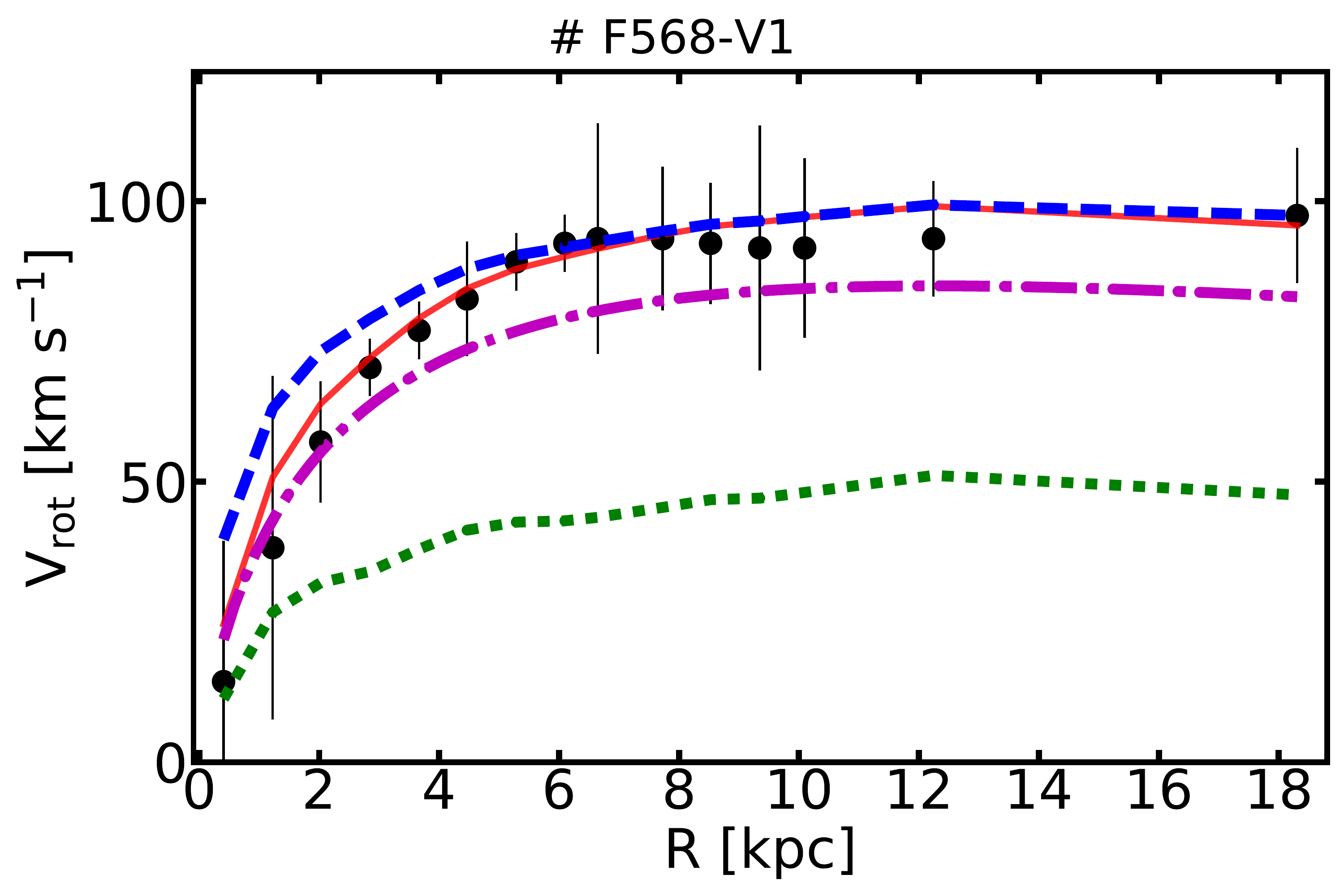}
\caption{Continued.}
\end{figure*}

\begin{figure*}
\centering
\ContinuedFloat
\includegraphics[width=.33\textwidth]{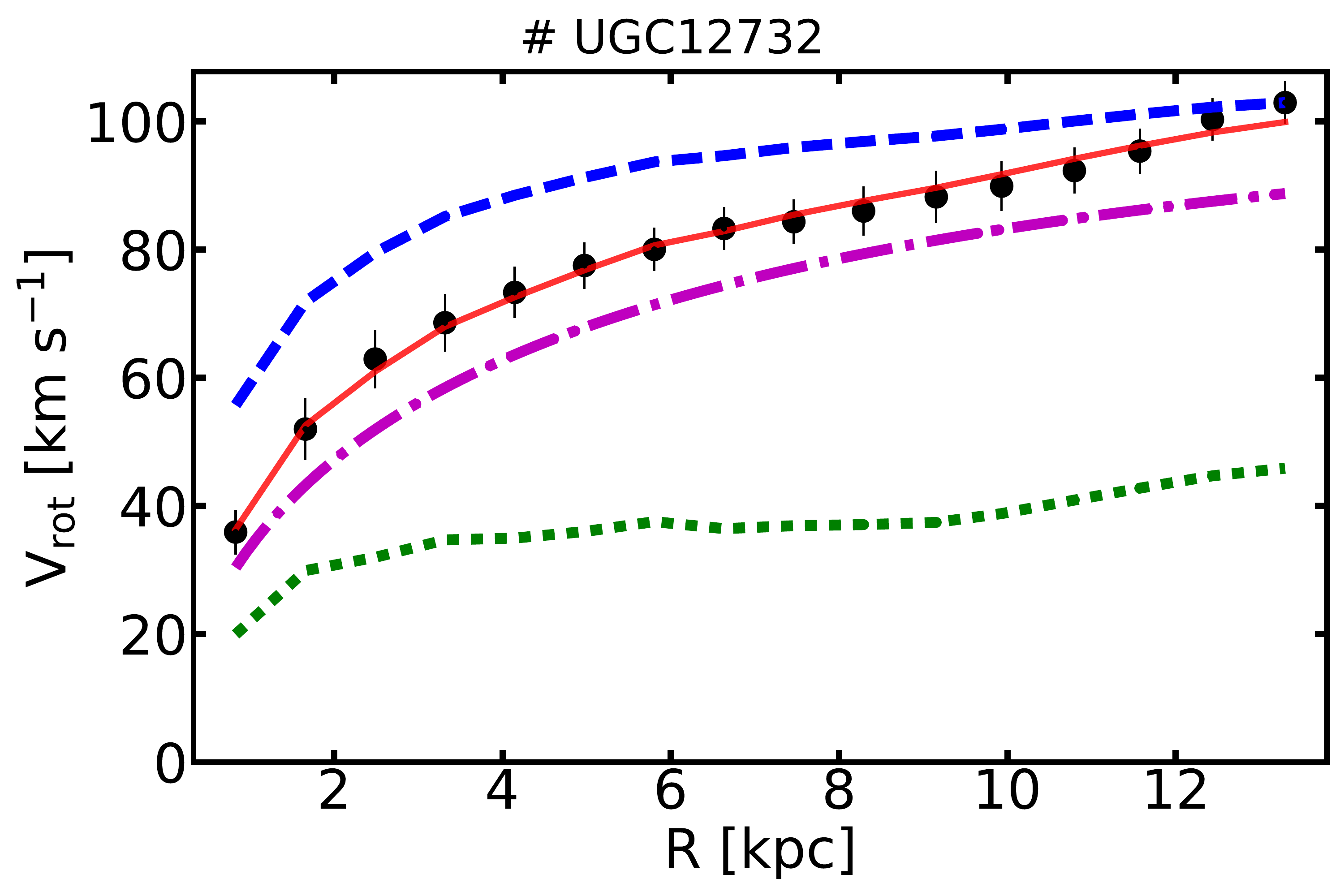}\hfill
\includegraphics[width=.33\textwidth]{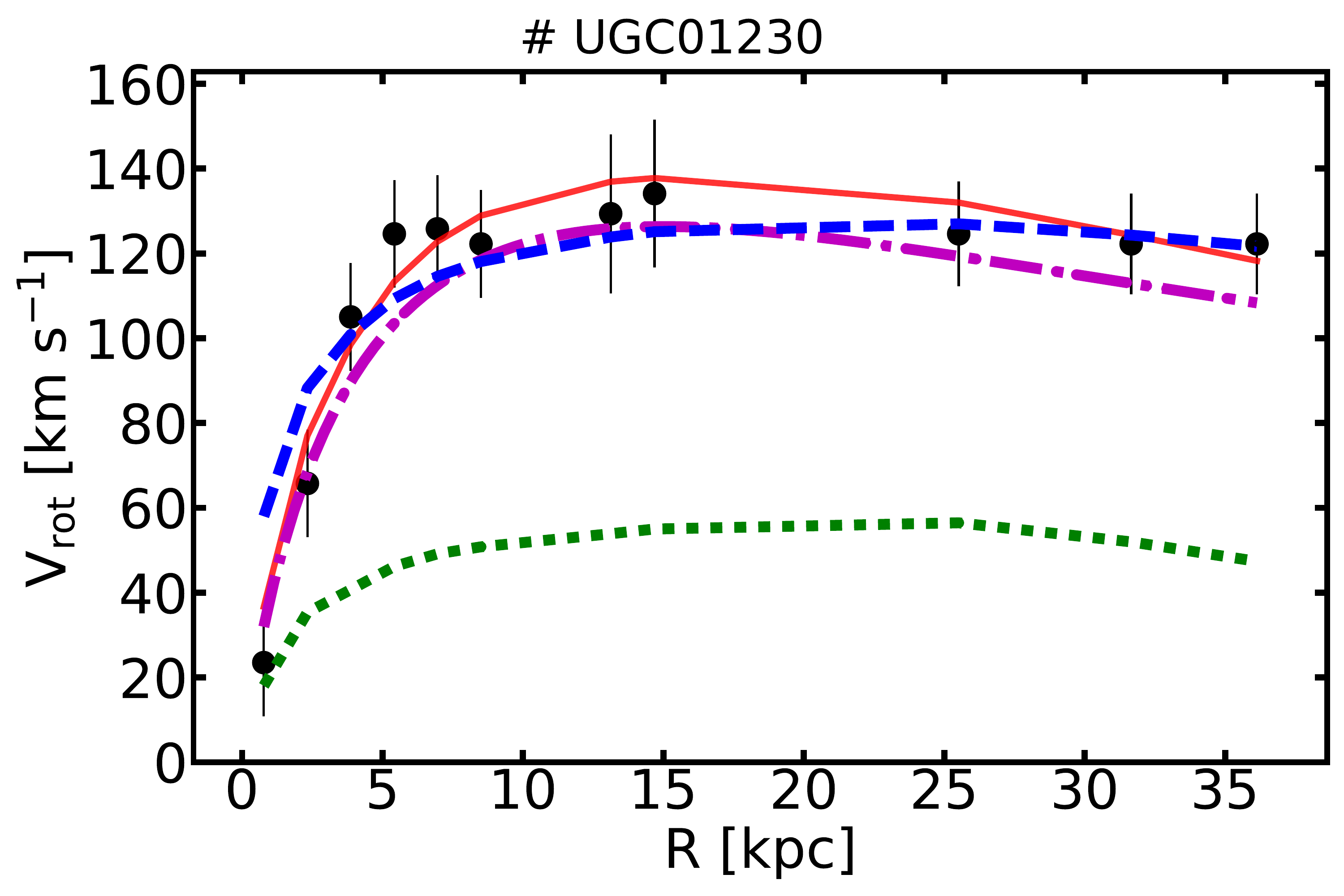}\hfill
\includegraphics[width=.33\textwidth]{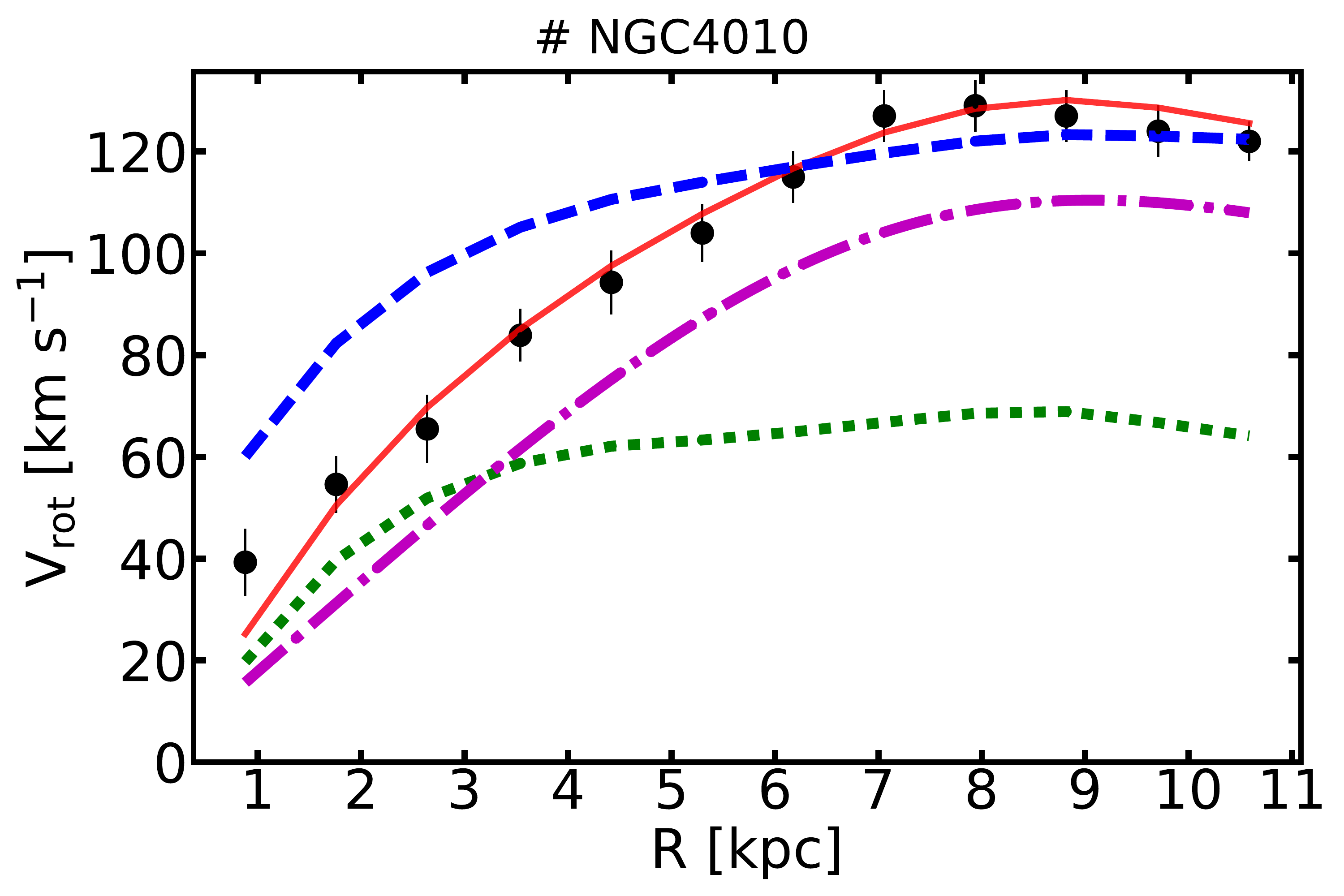}

\includegraphics[width=.33\textwidth]{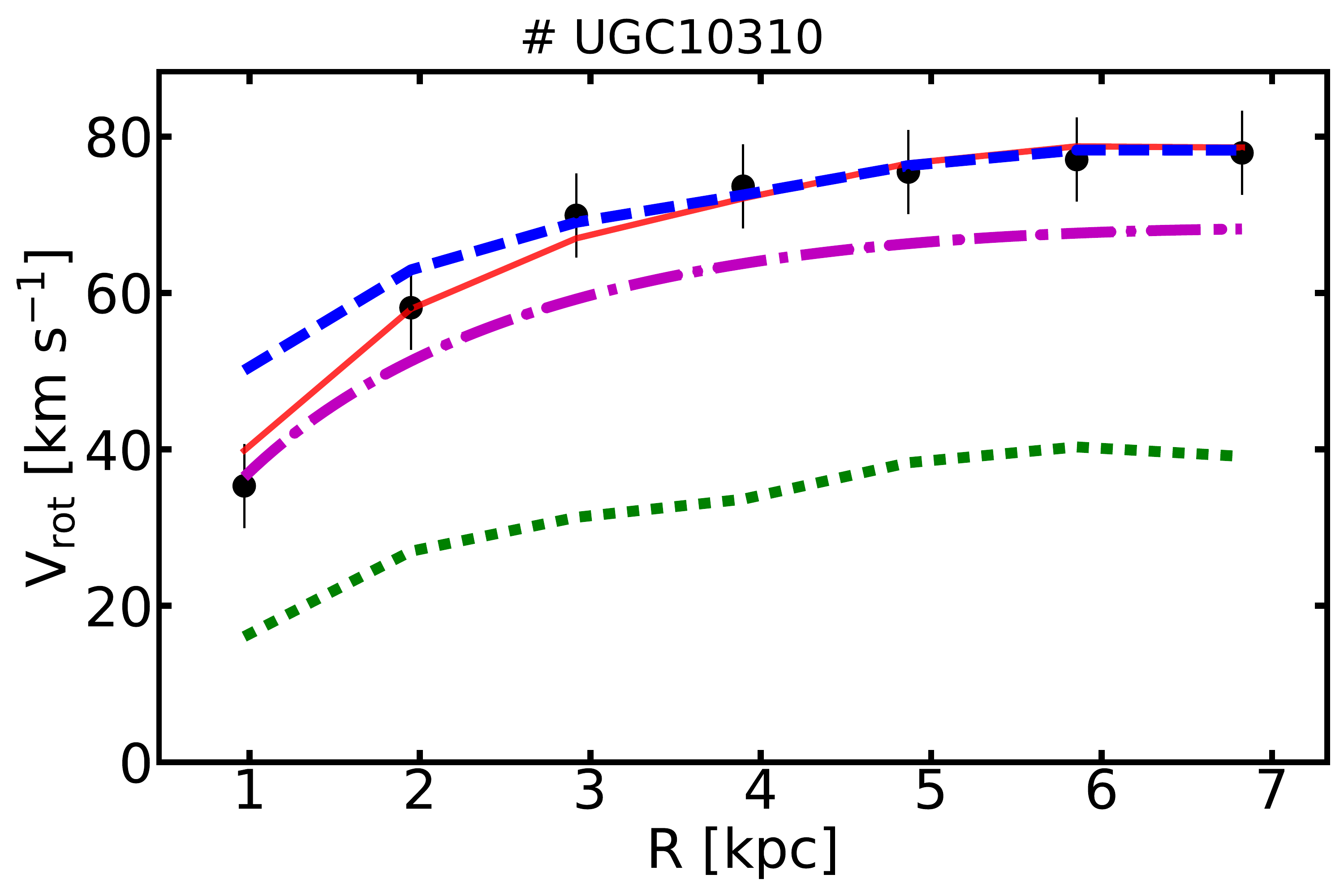}\hfill
\includegraphics[width=.33\textwidth]{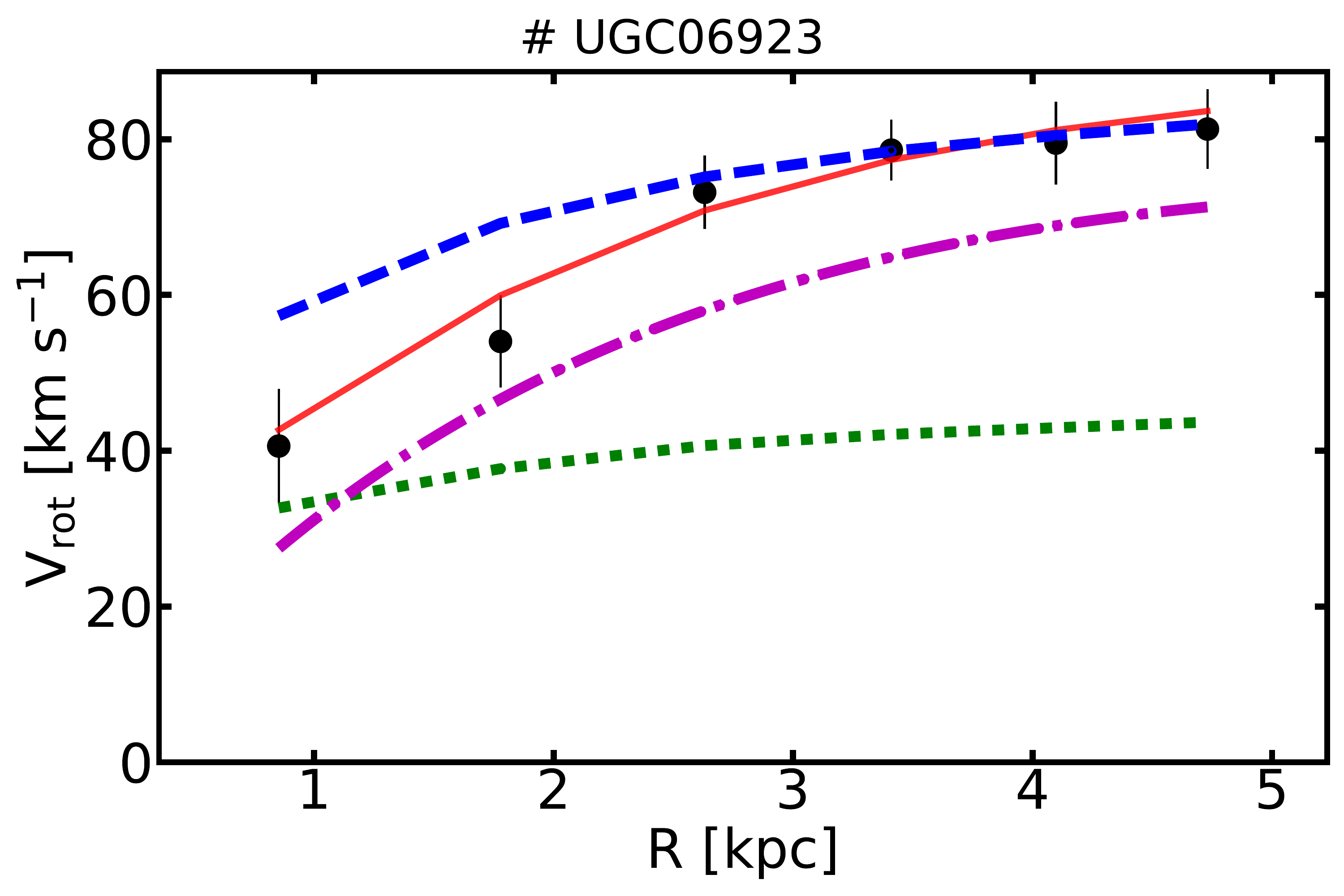}\hfill
\includegraphics[width=.33\textwidth]{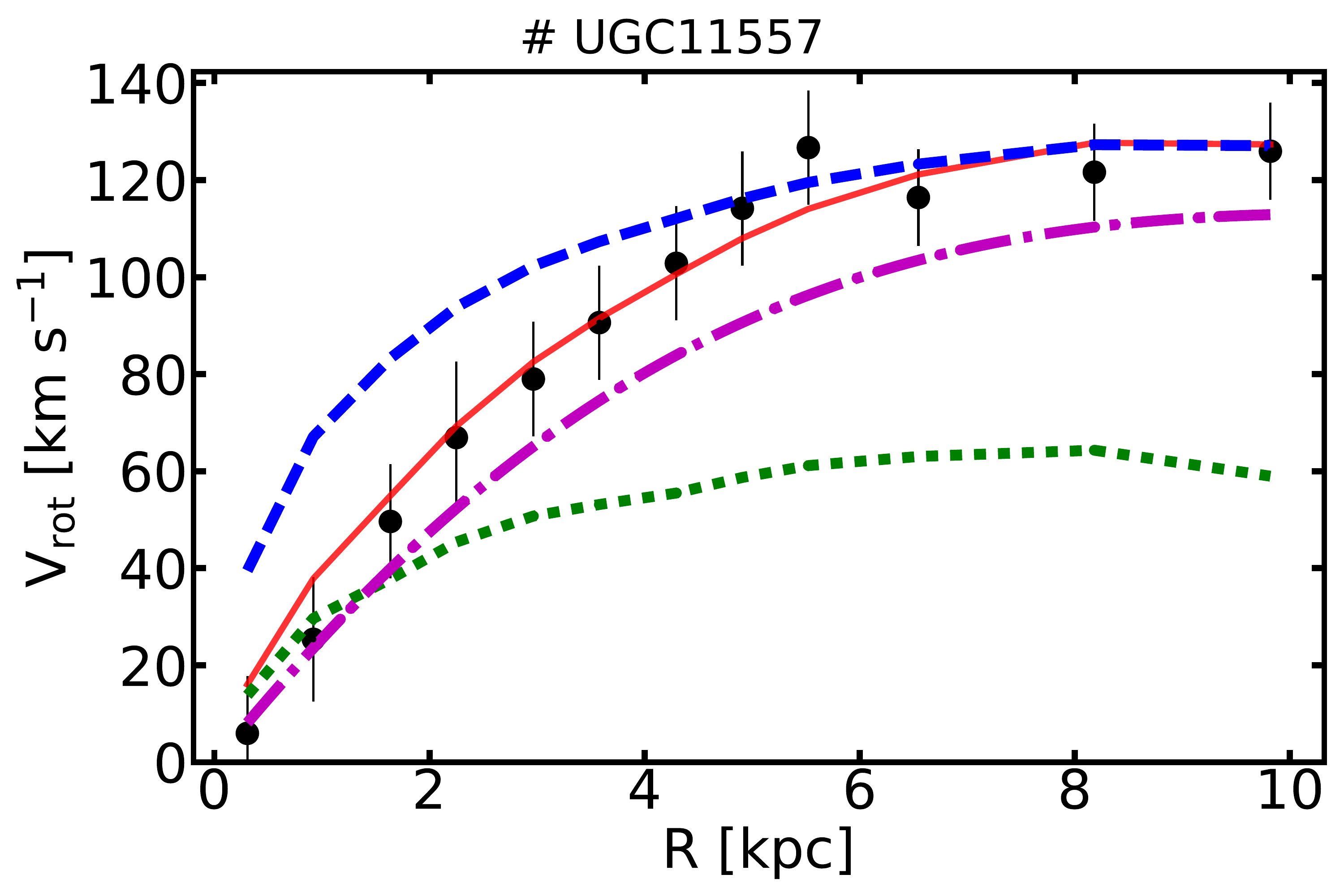}

\includegraphics[width=.33\textwidth]{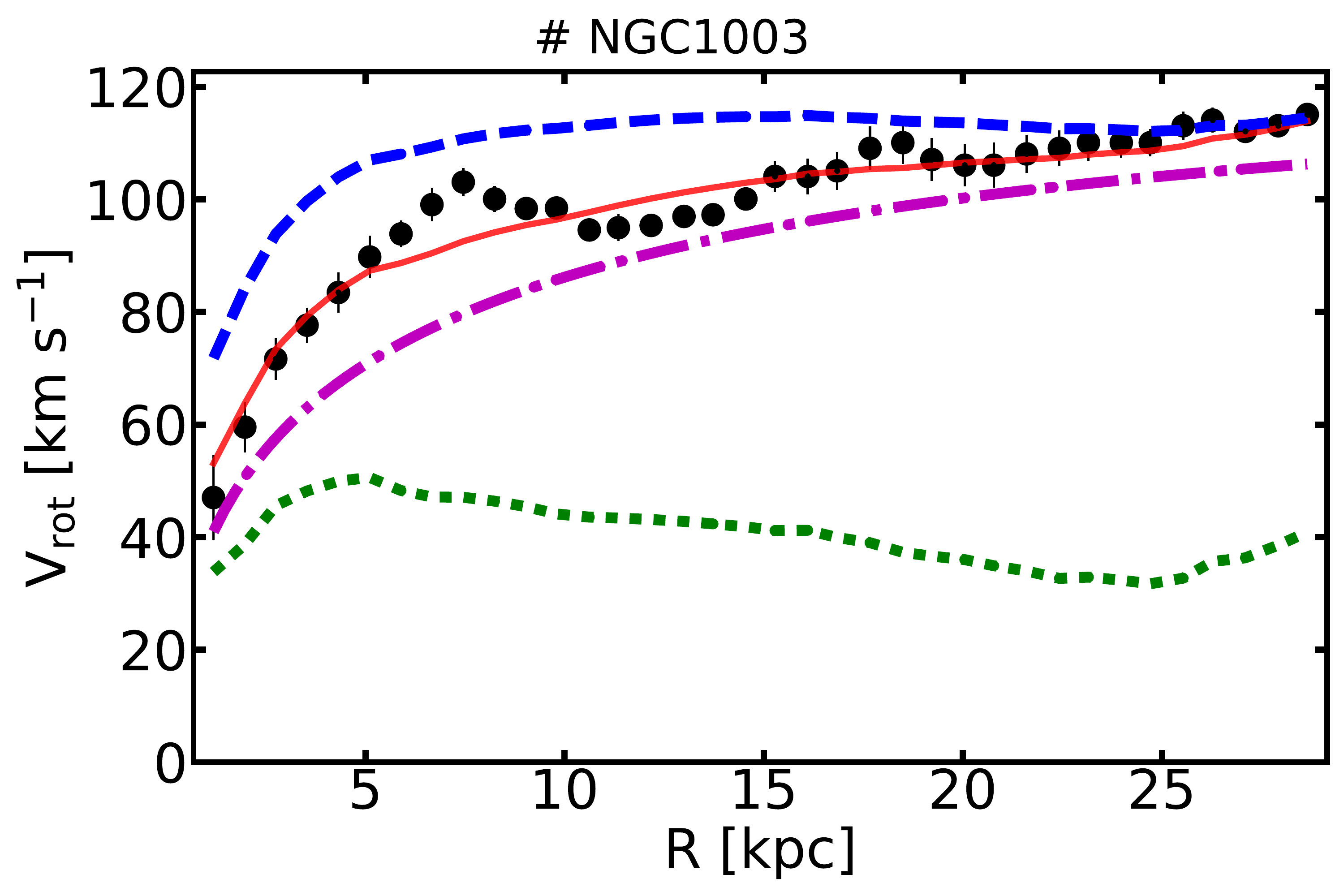}\hfill
\includegraphics[width=.33\textwidth]{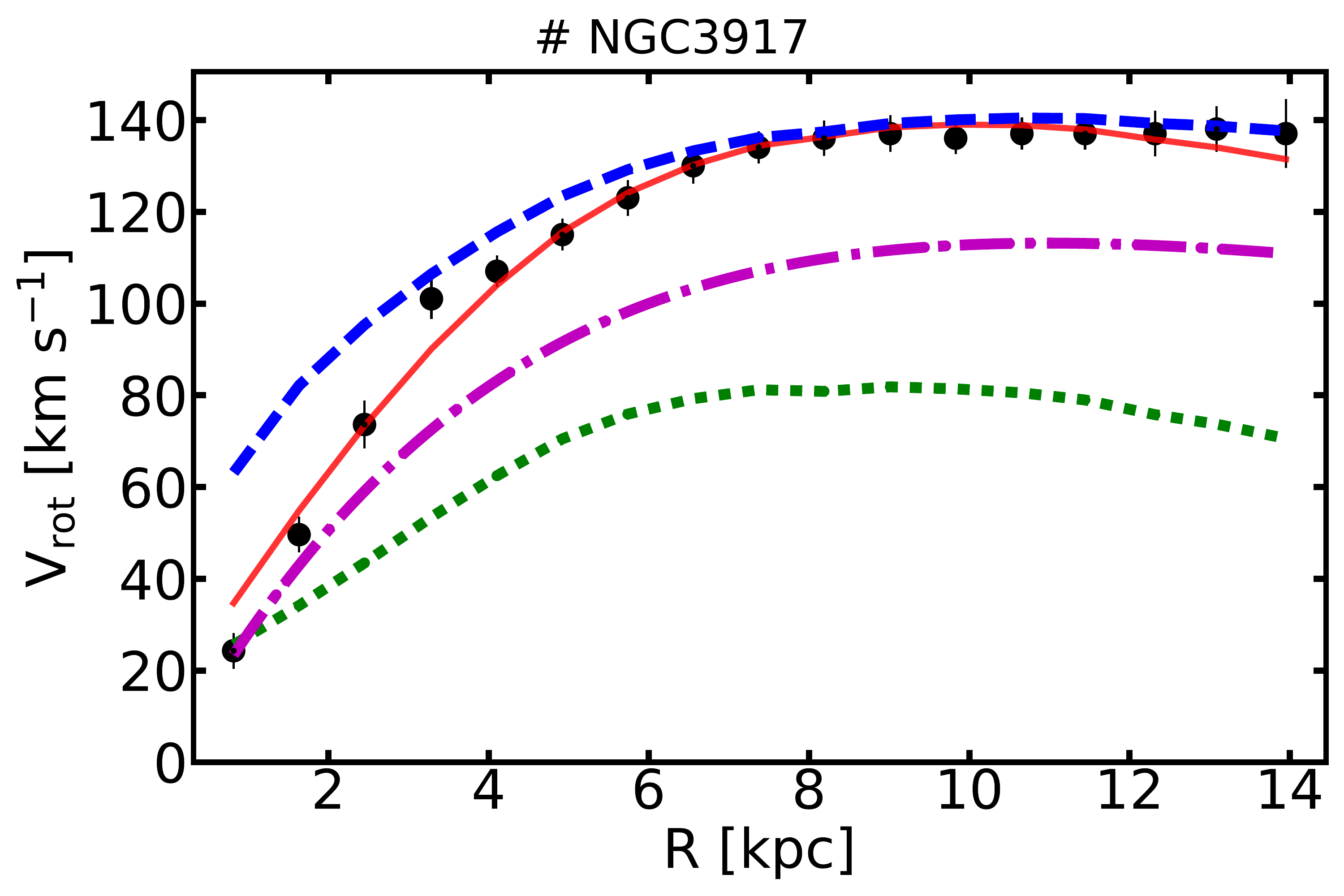}\hfill
\includegraphics[width=.33\textwidth]{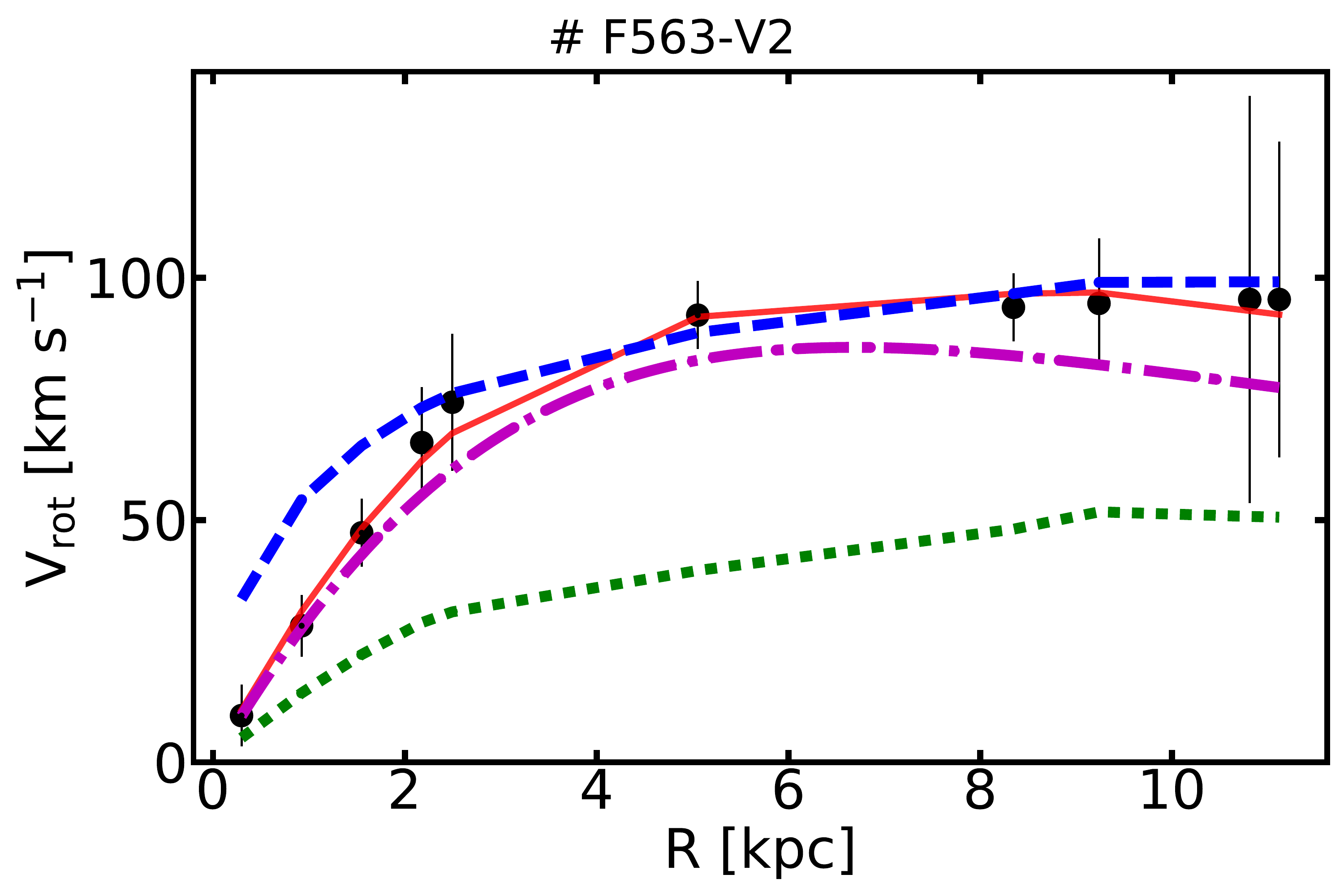}

\includegraphics[width=.33\textwidth]{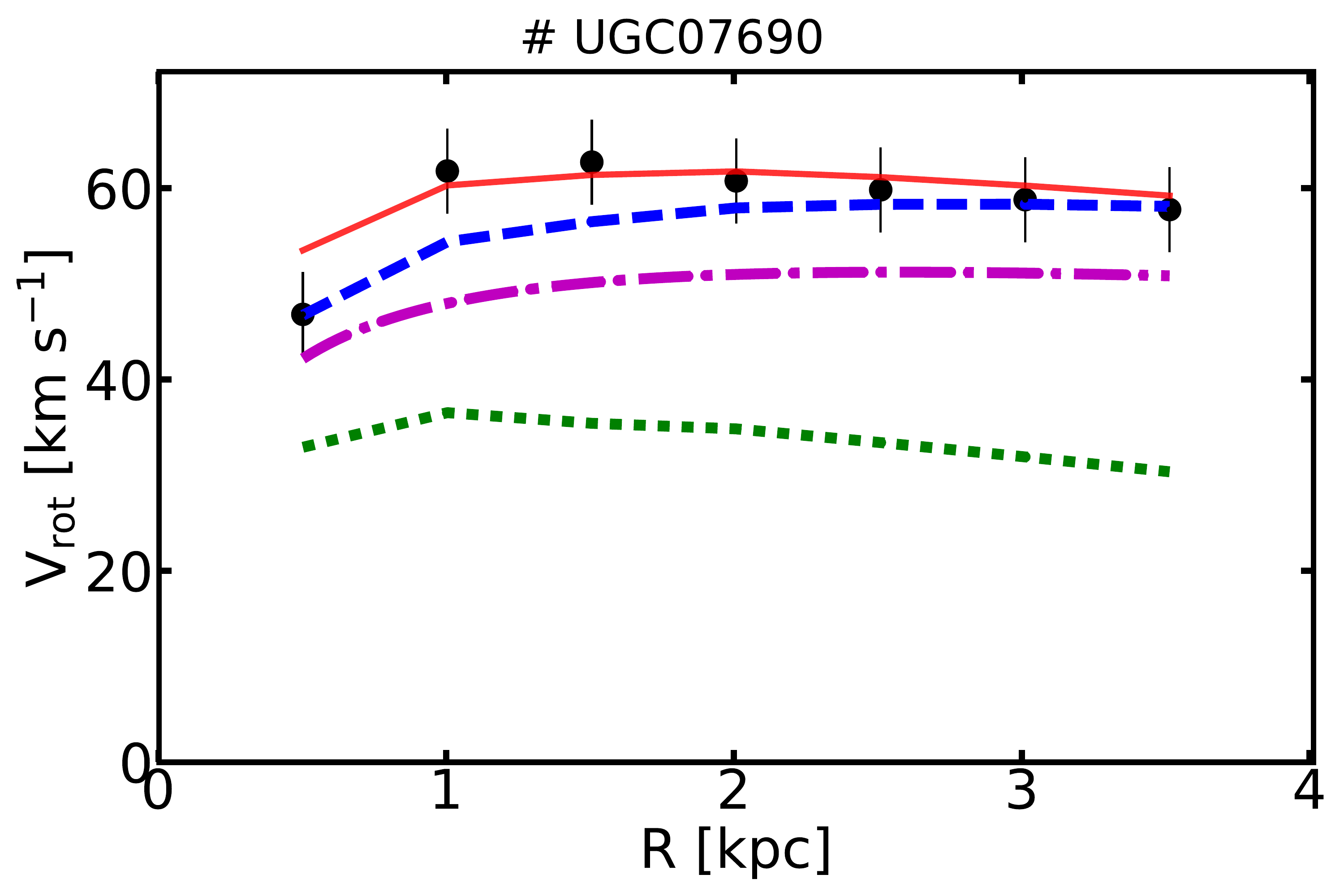}\hfill
\includegraphics[width=.33\textwidth]{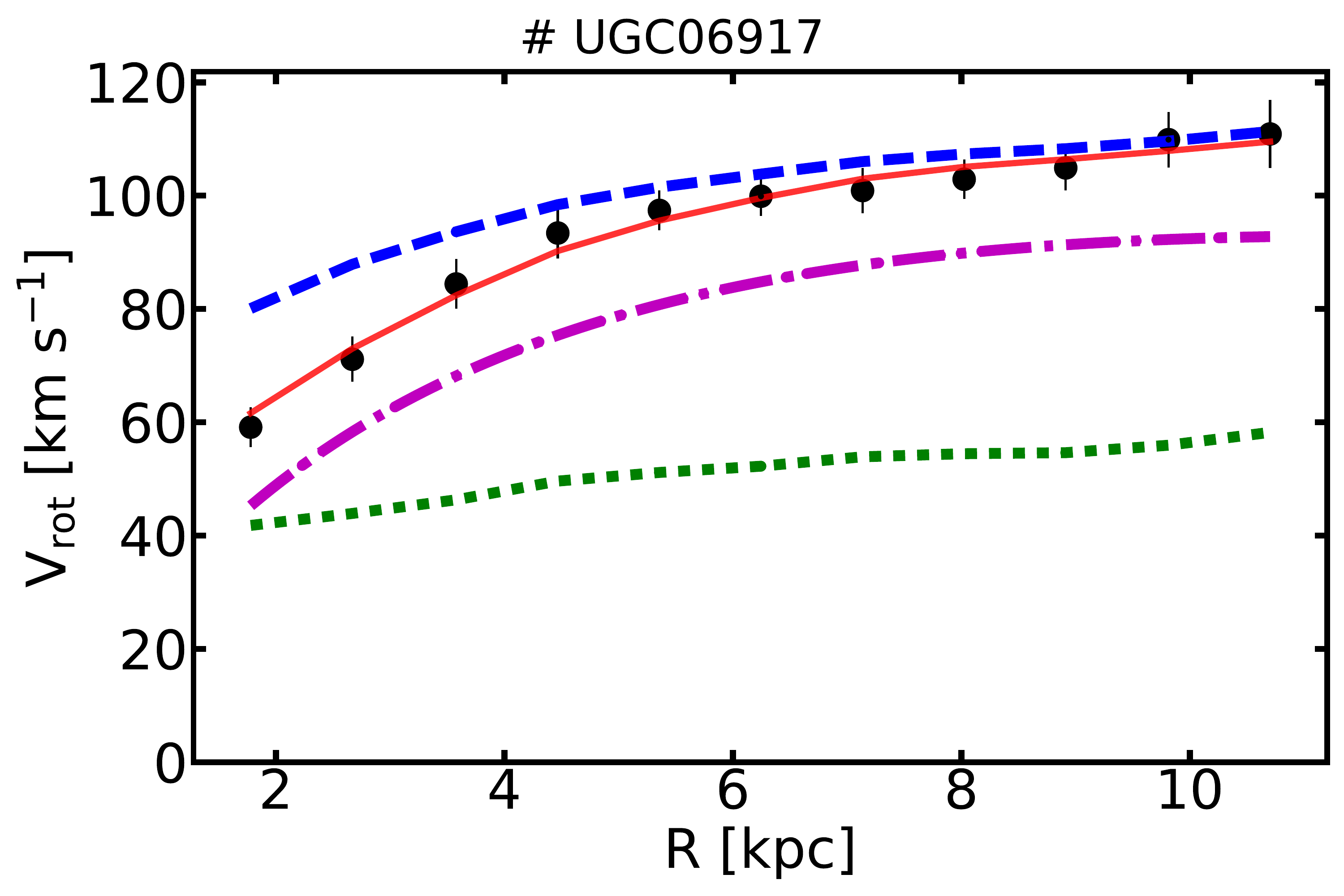}\hfill
\includegraphics[width=.33\textwidth]{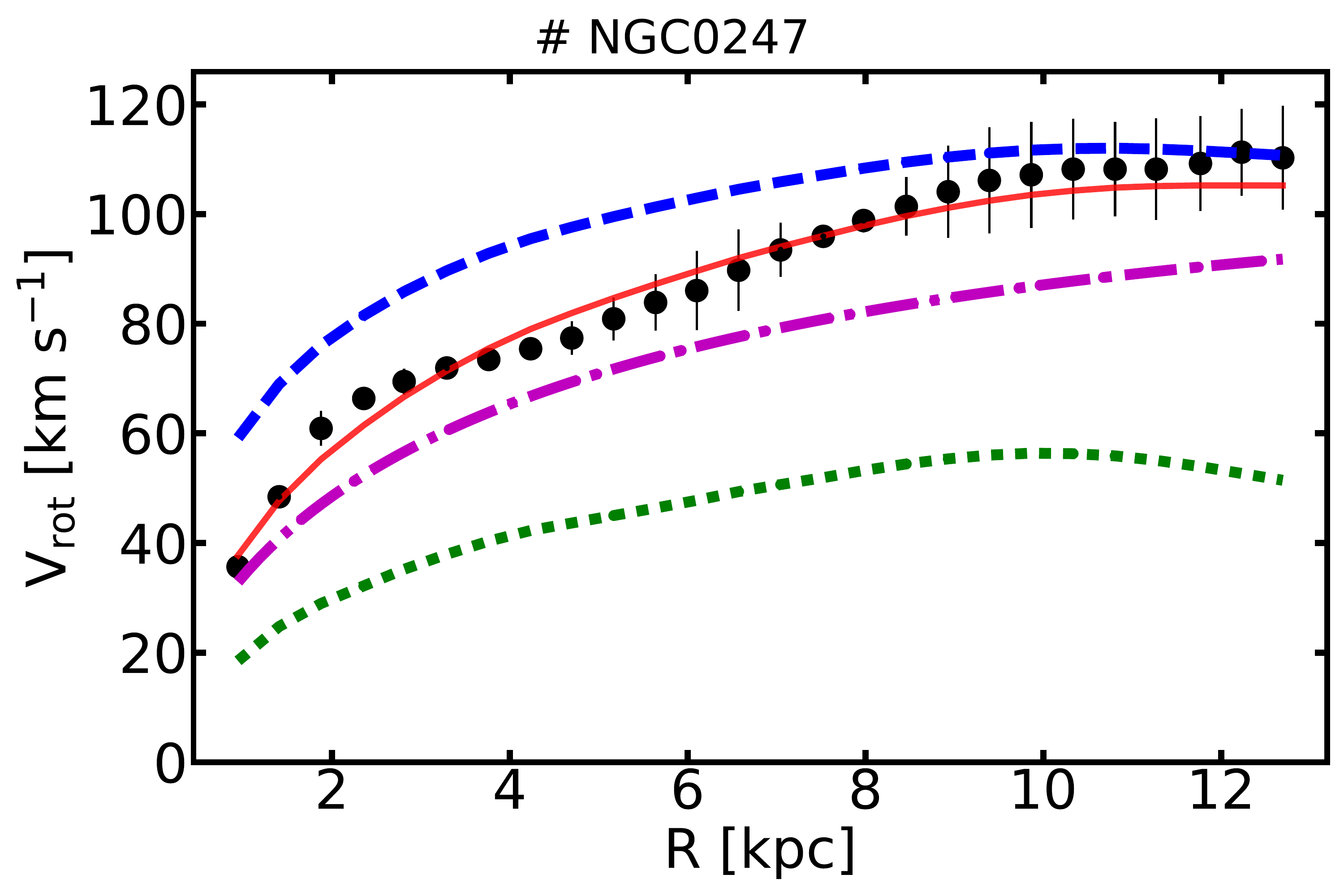}

\includegraphics[width=.33\textwidth]{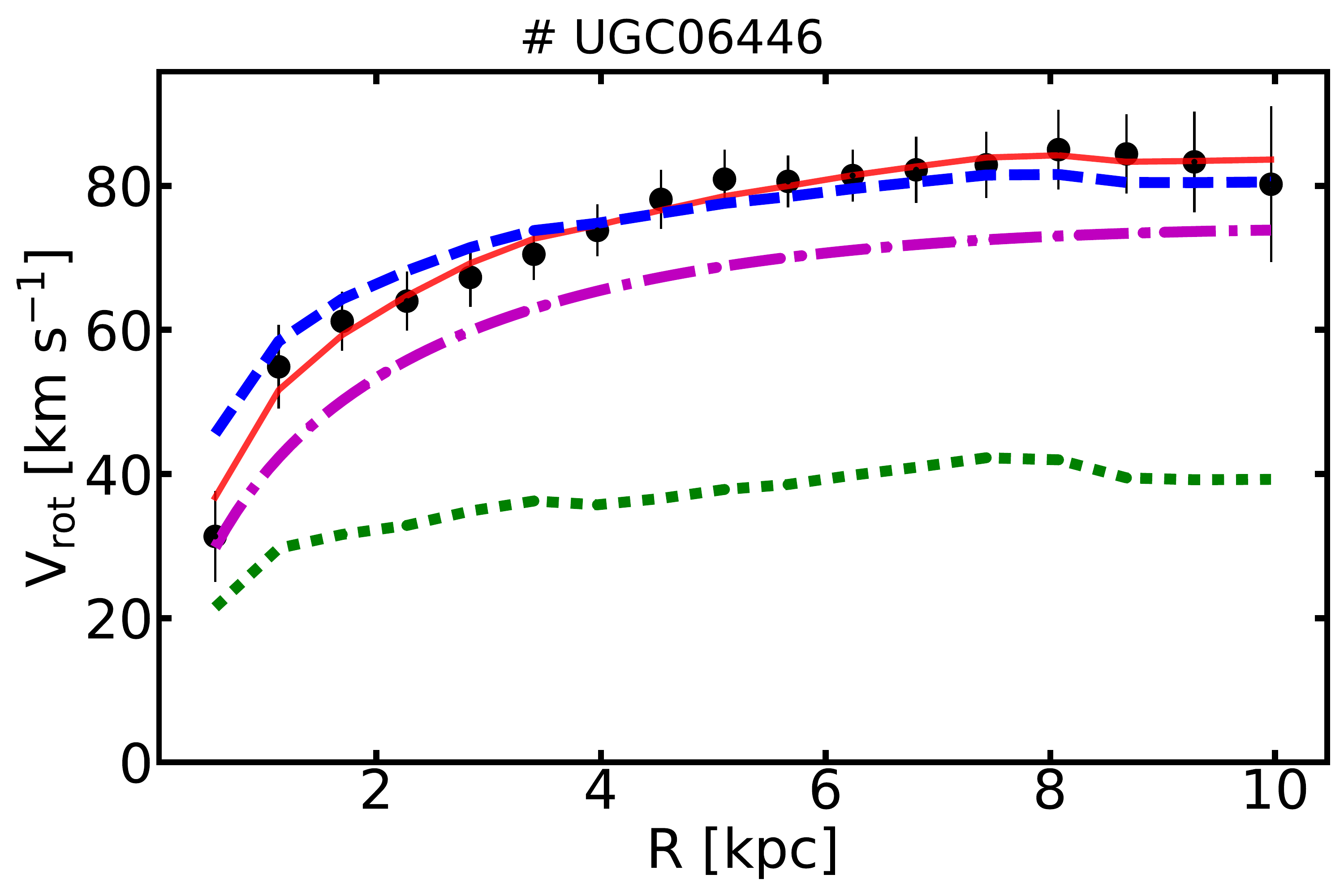}\hfill
\includegraphics[width=.33\textwidth]{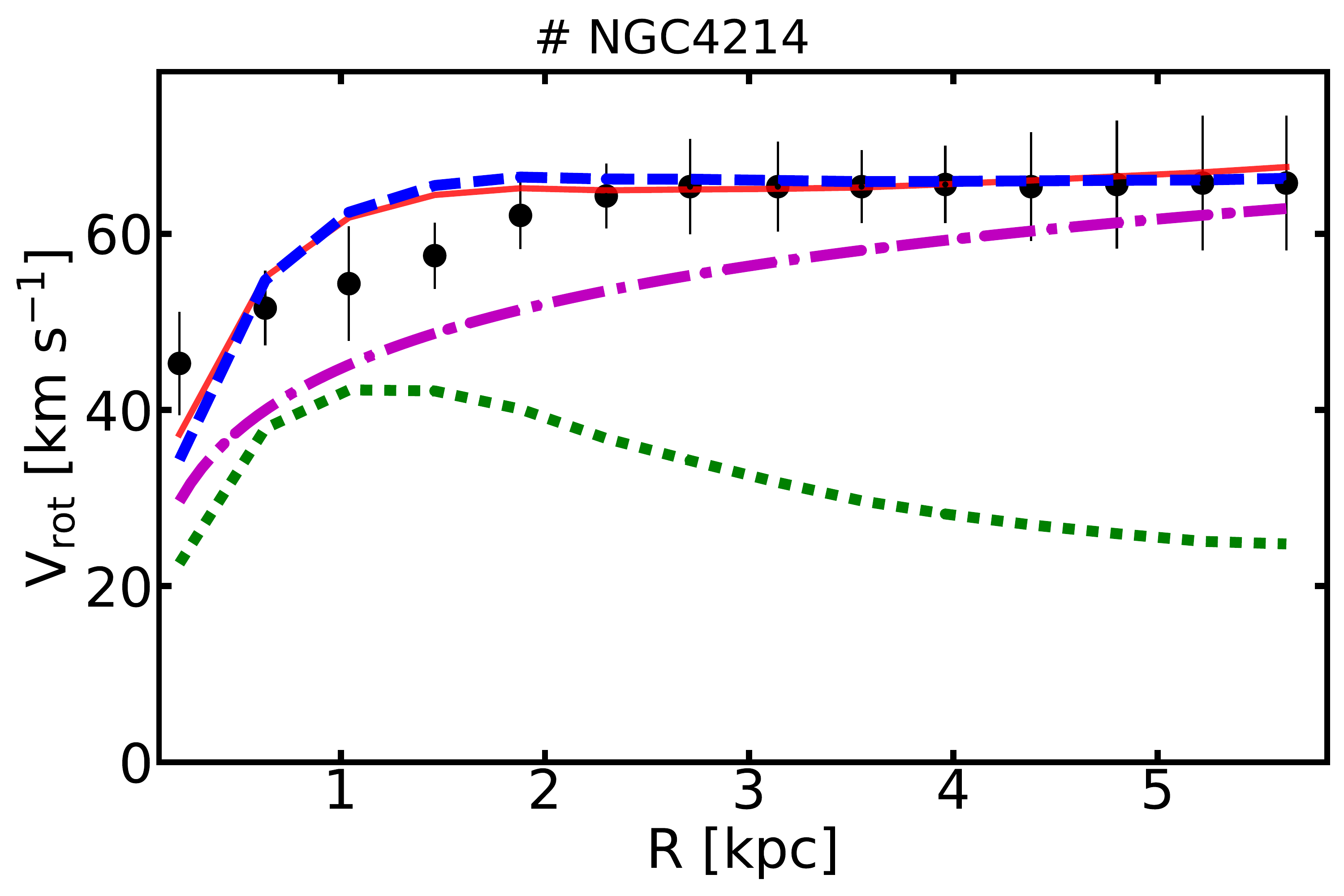}\hfill
\includegraphics[width=.33\textwidth]{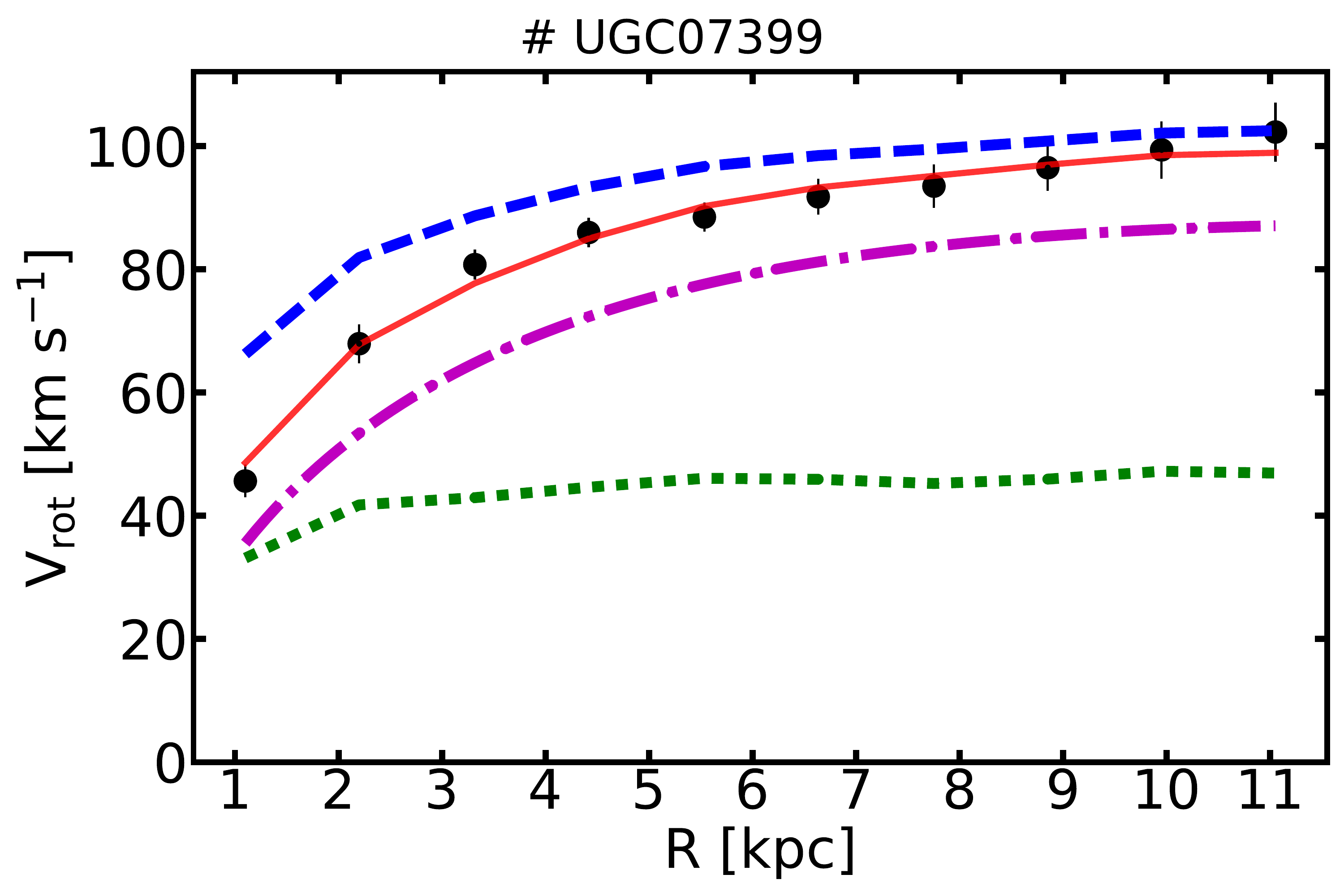}

\includegraphics[width=.33\textwidth]{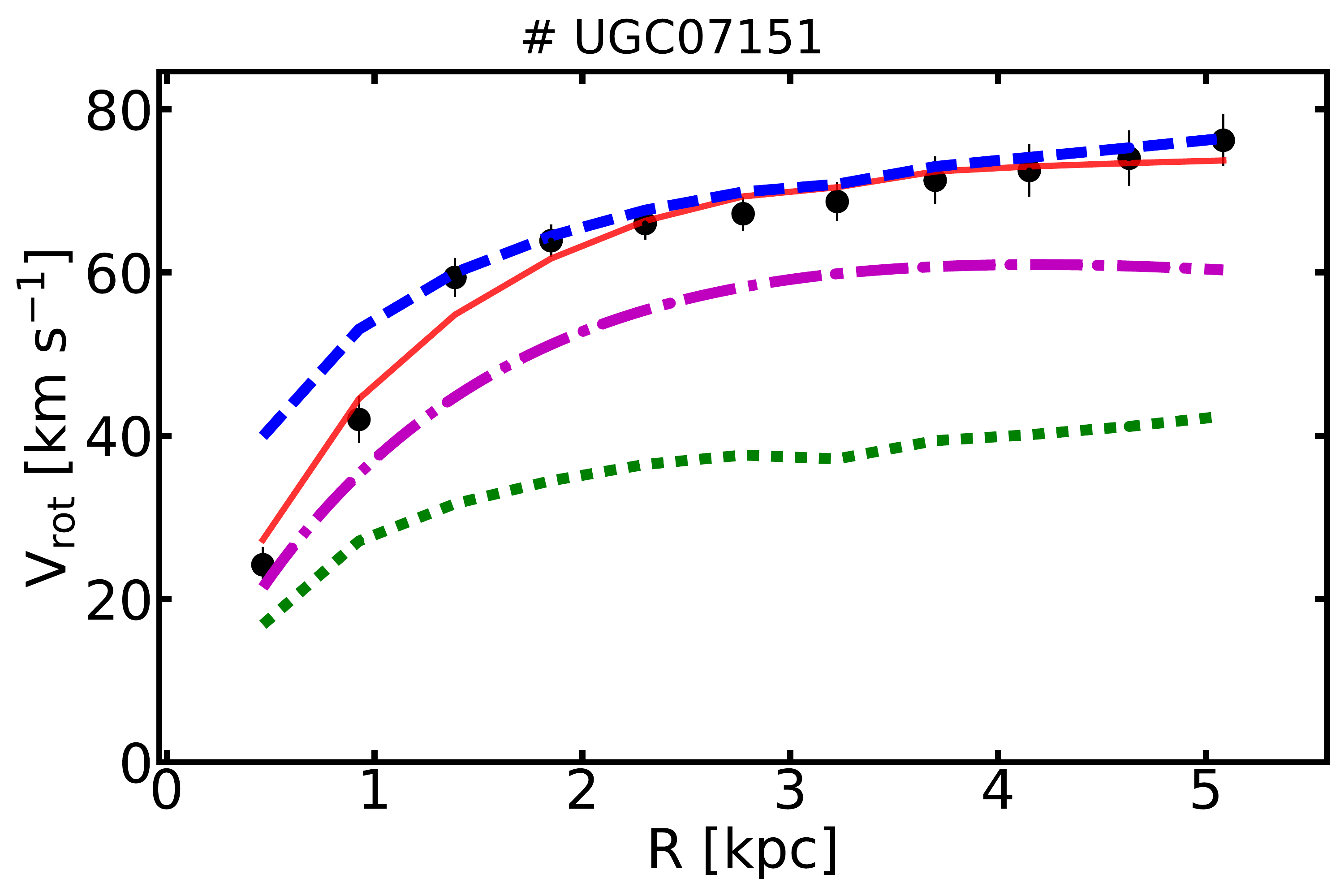}\hfill
\includegraphics[width=.33\textwidth]{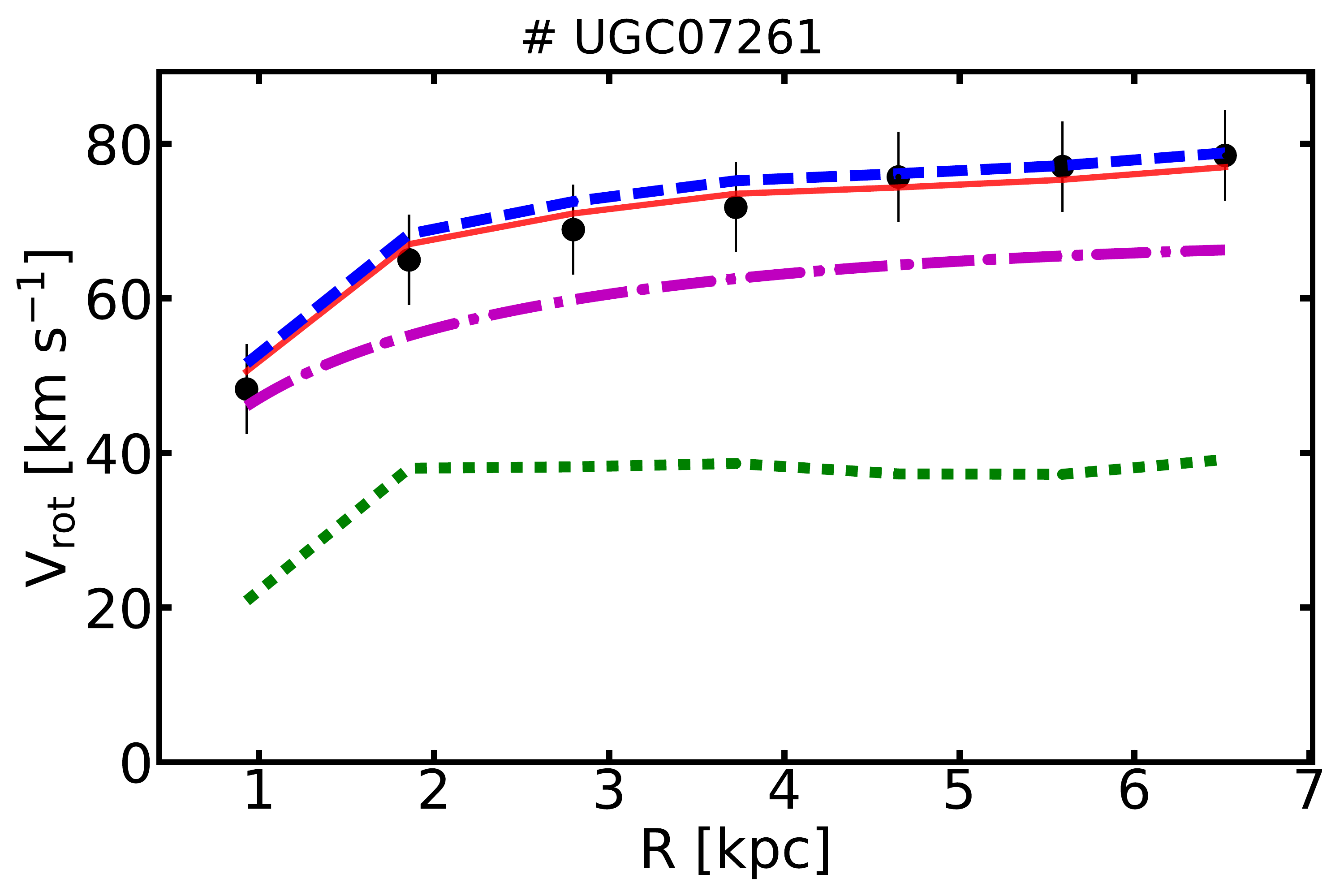}\hfill
\includegraphics[width=.33\textwidth]{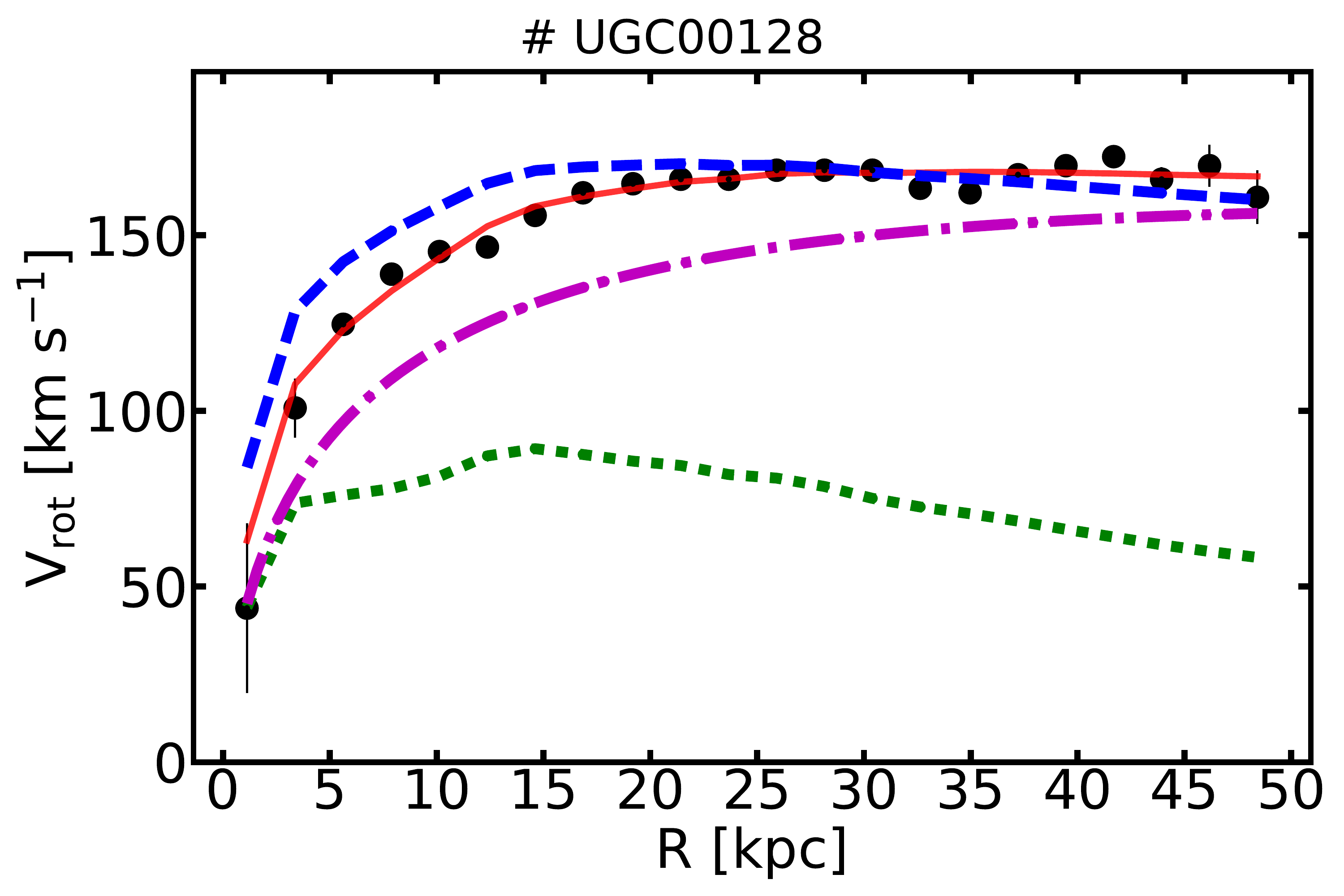}
\caption{Continued.}
\end{figure*}

\begin{figure*}
\centering
\ContinuedFloat
\includegraphics[width=.33\textwidth]{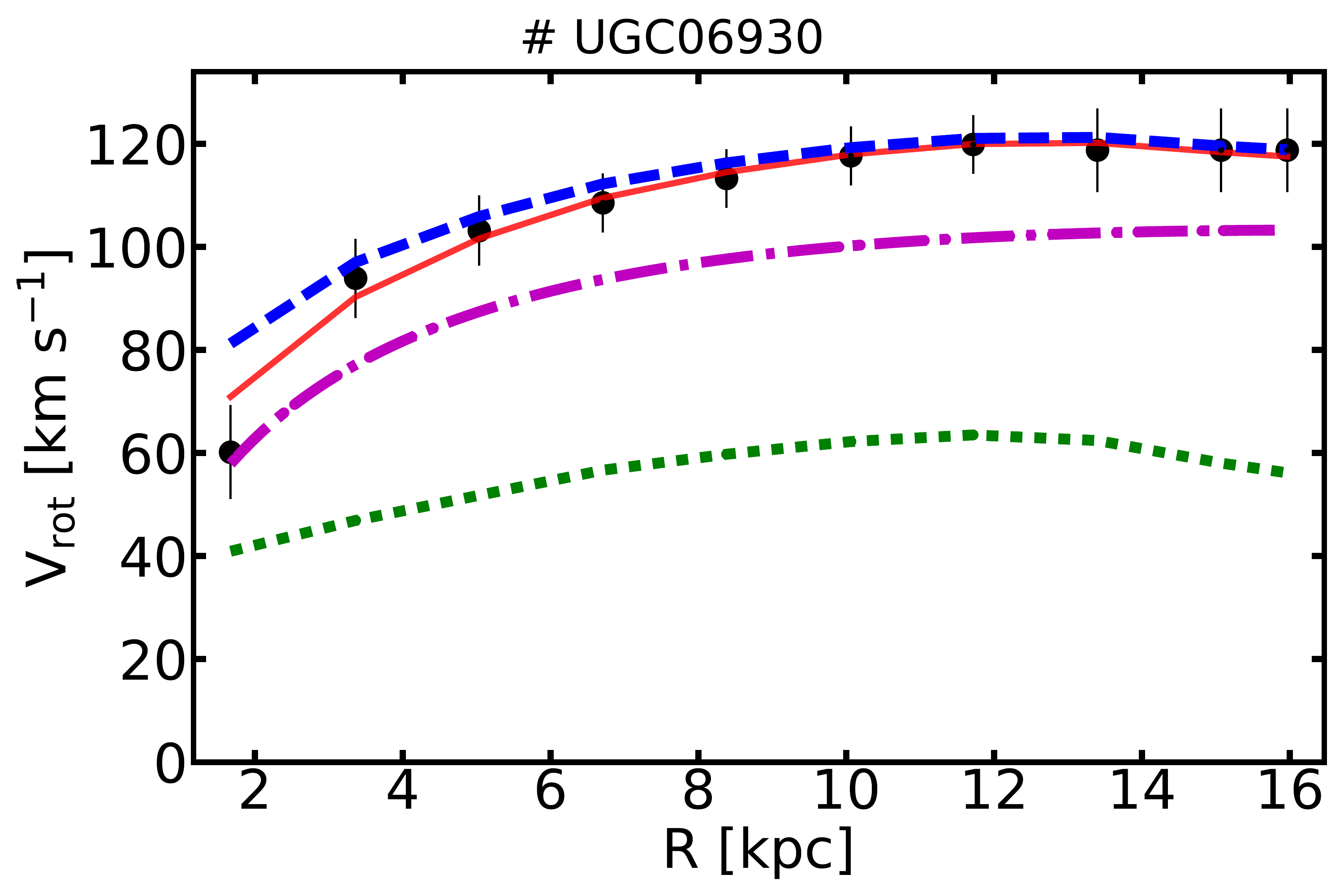}\hfill
\includegraphics[width=.33\textwidth]{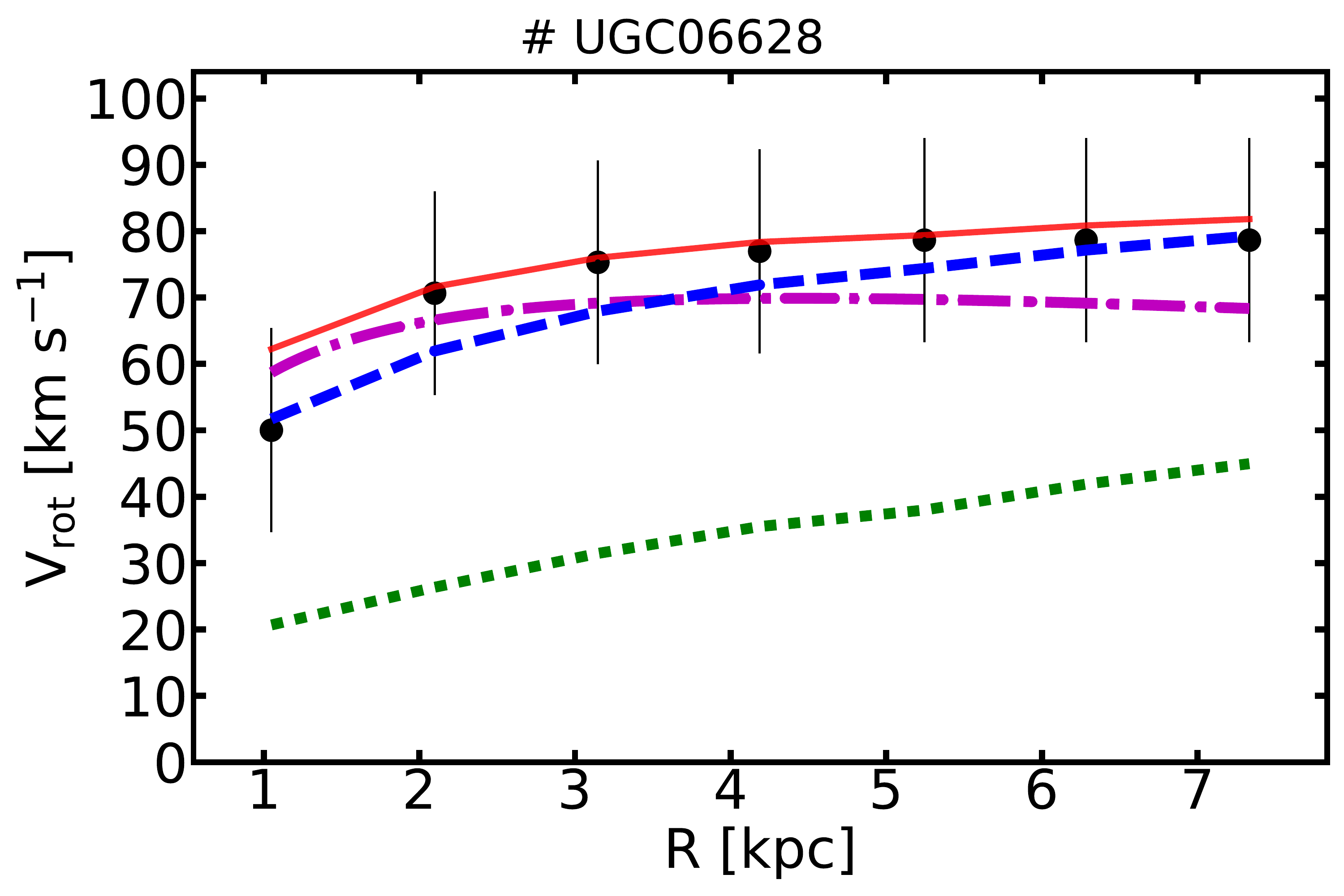}\hfill
\includegraphics[width=.33\textwidth]{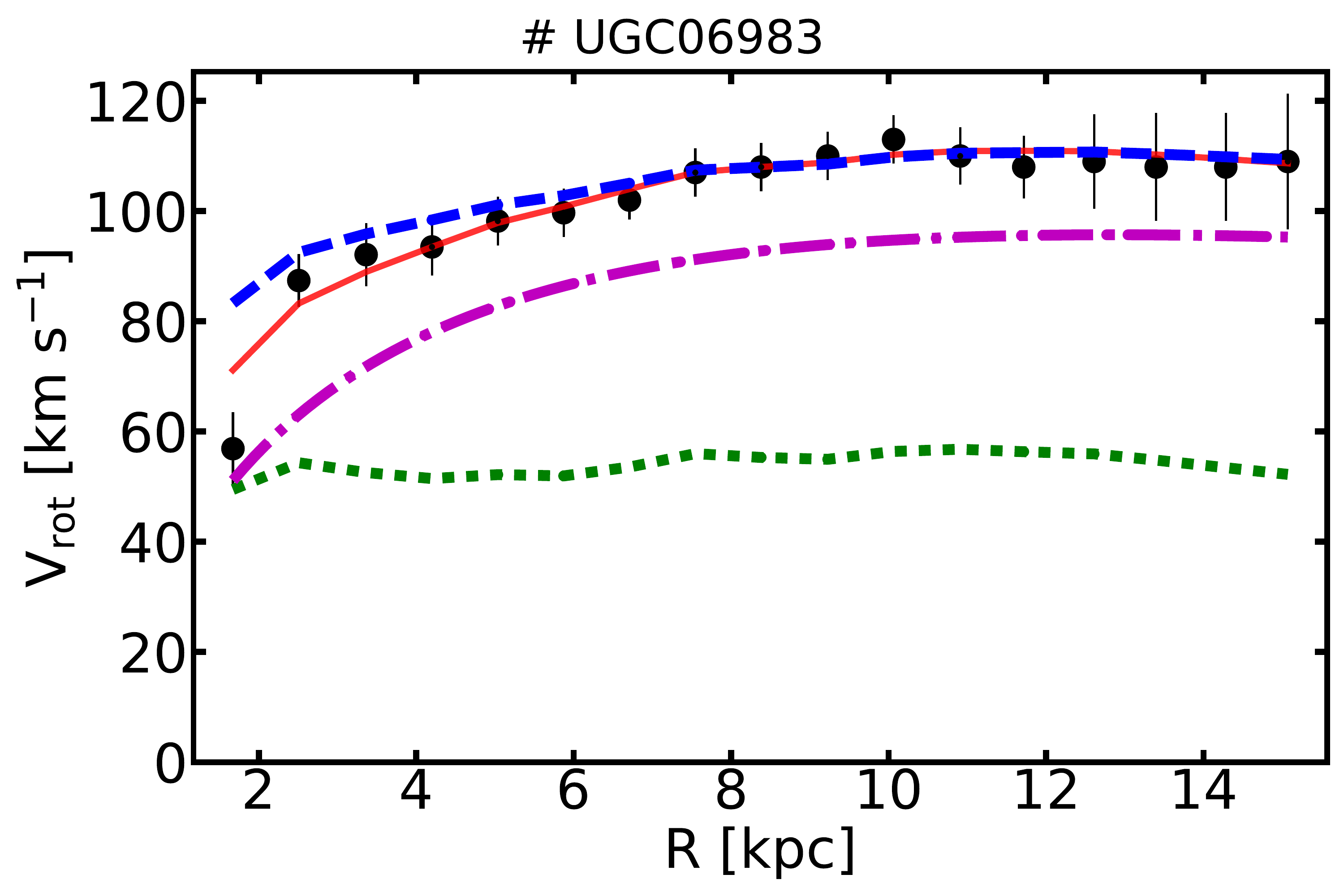}

\includegraphics[width=.33\textwidth]{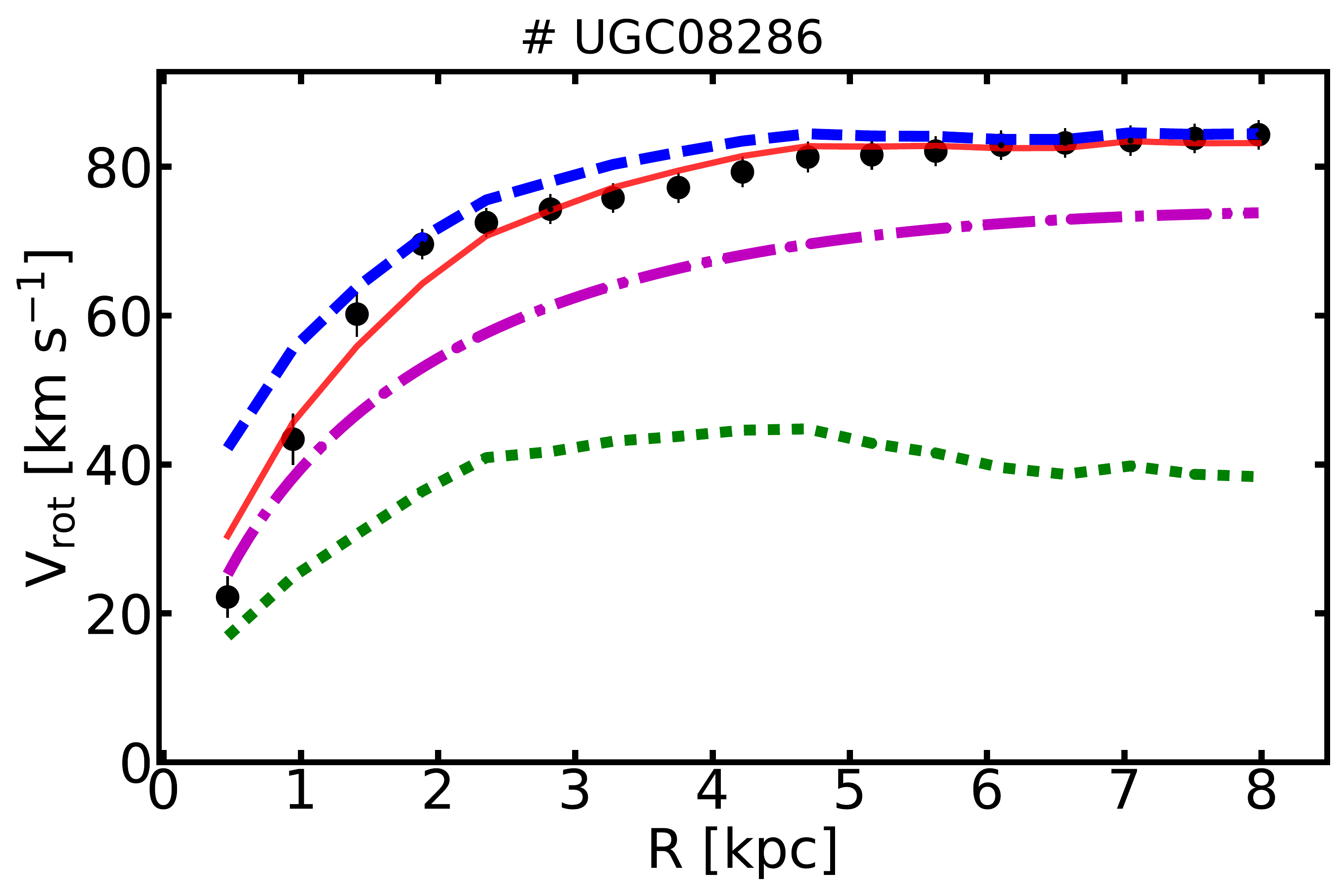}\hfill
\includegraphics[width=.33\textwidth]{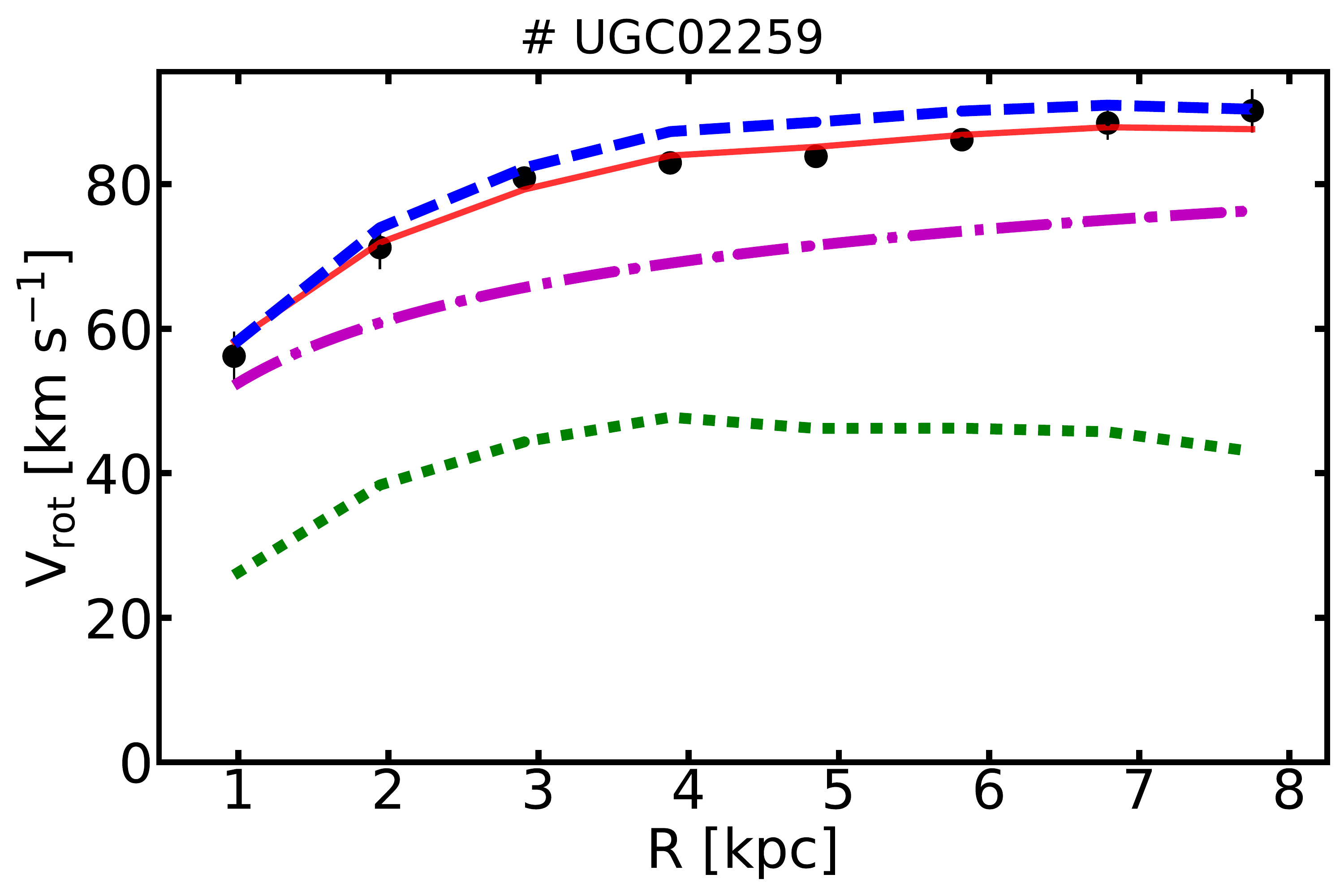}\hfill
\includegraphics[width=.33\textwidth]{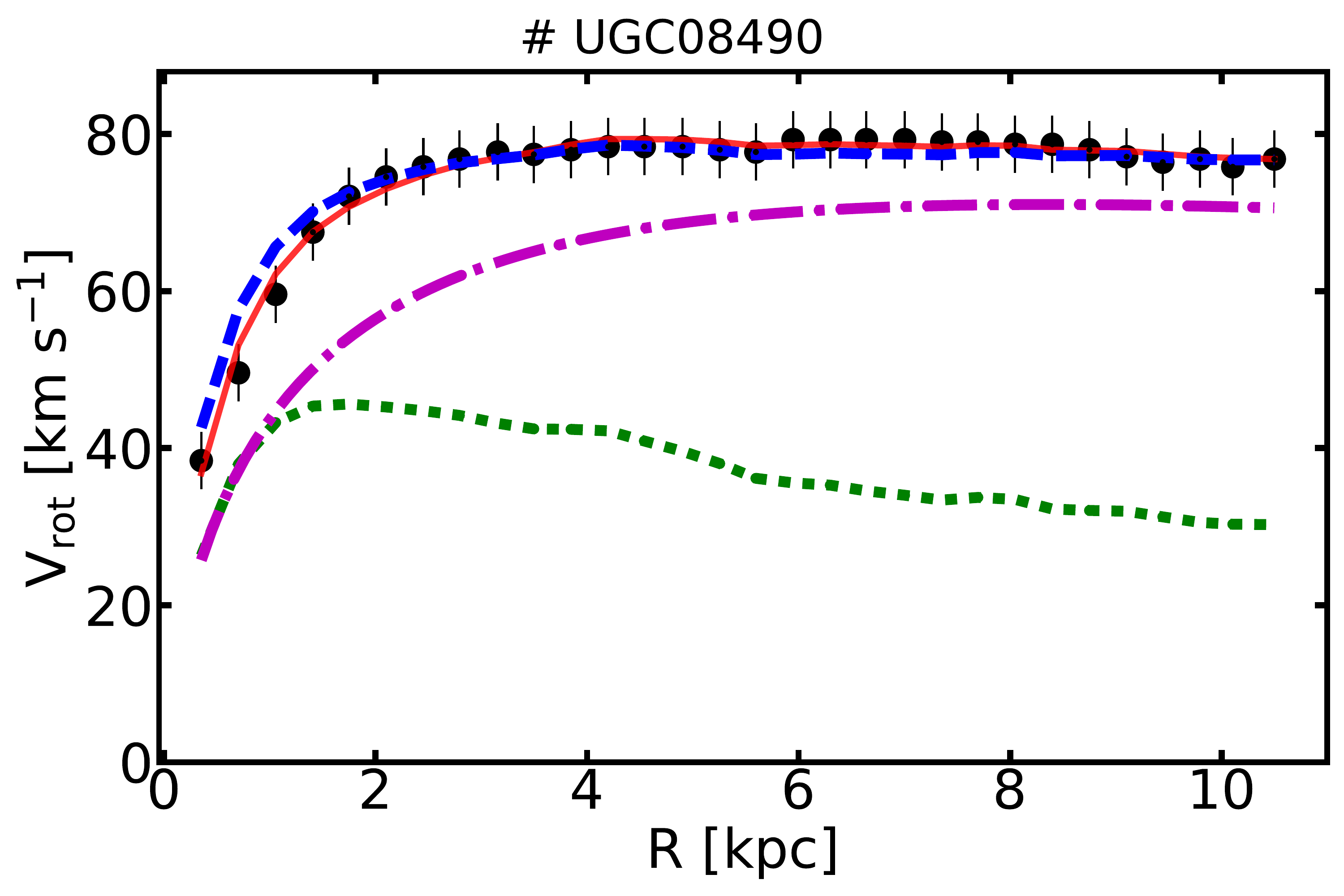}

\includegraphics[width=.33\textwidth]{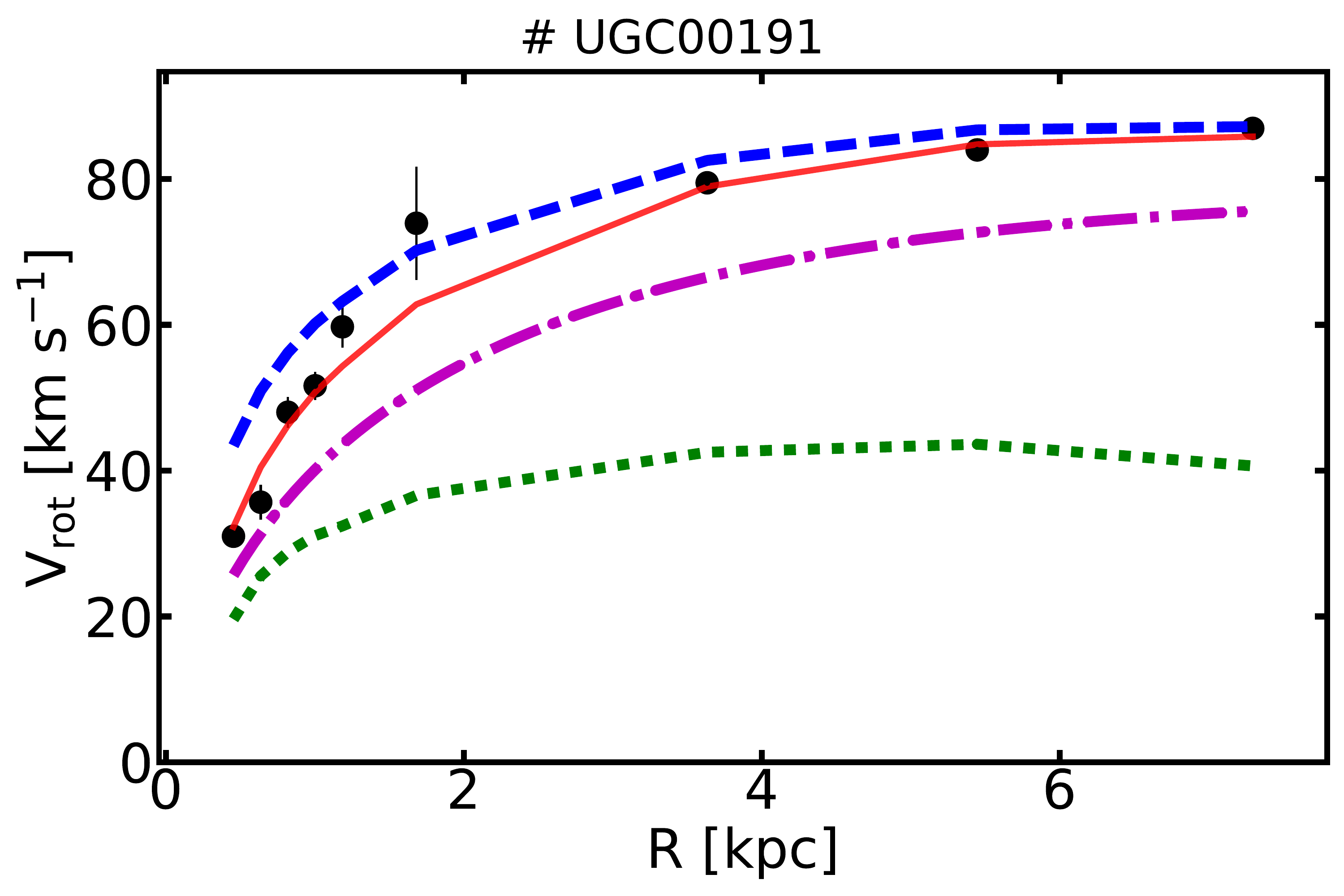}\hfill
\includegraphics[width=.33\textwidth]{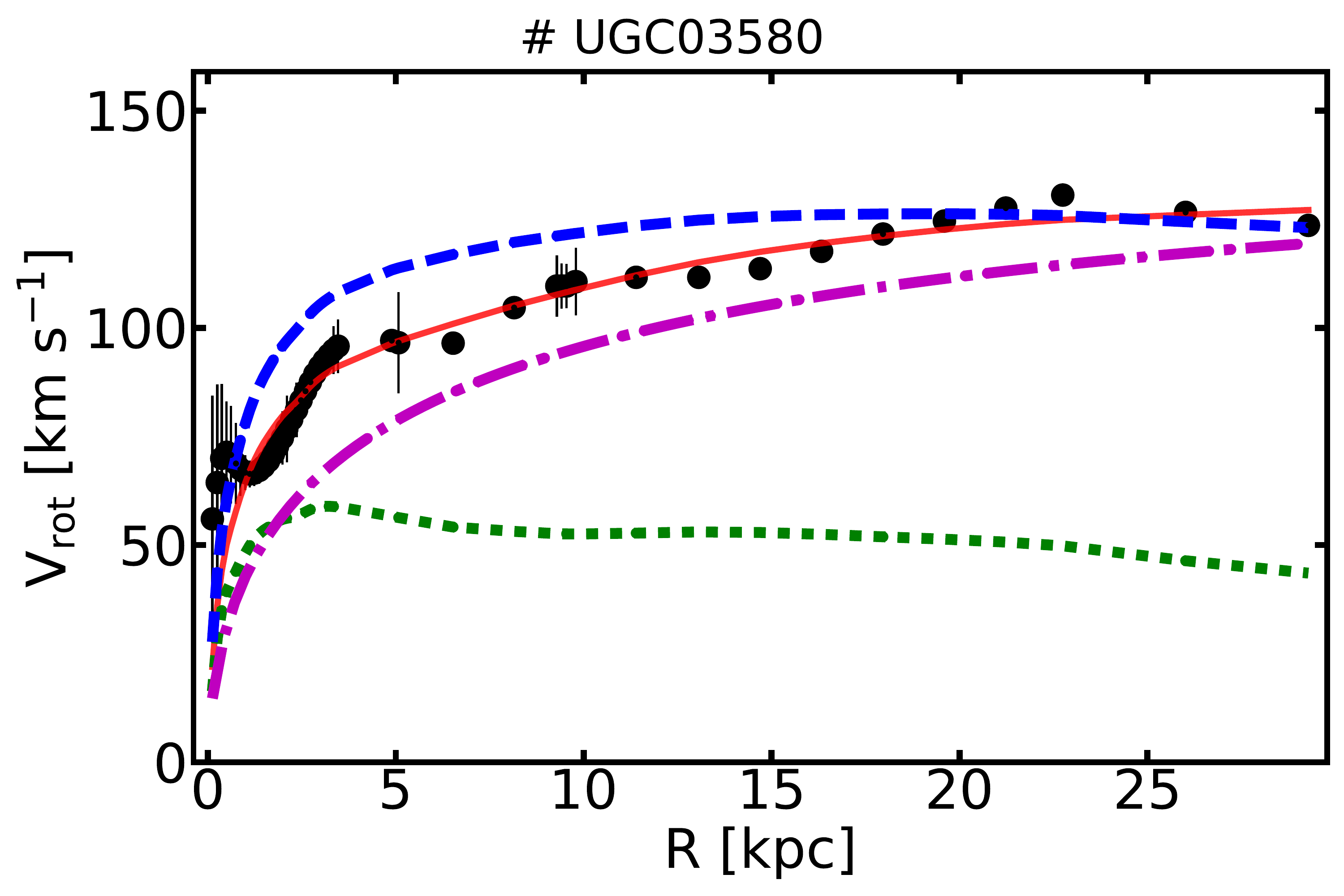}\hfill
\includegraphics[width=.33\textwidth]{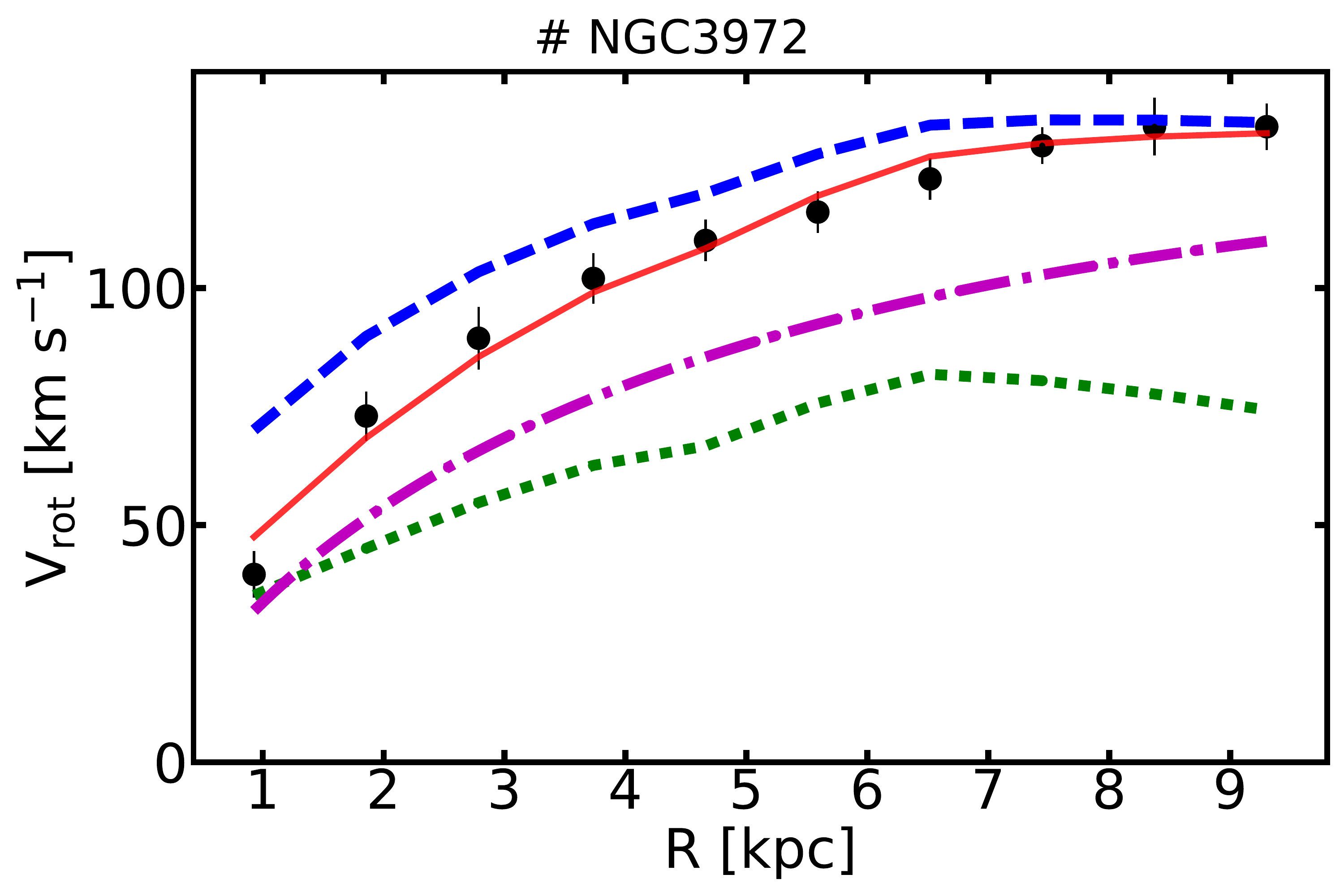}

\includegraphics[width=.33\textwidth]{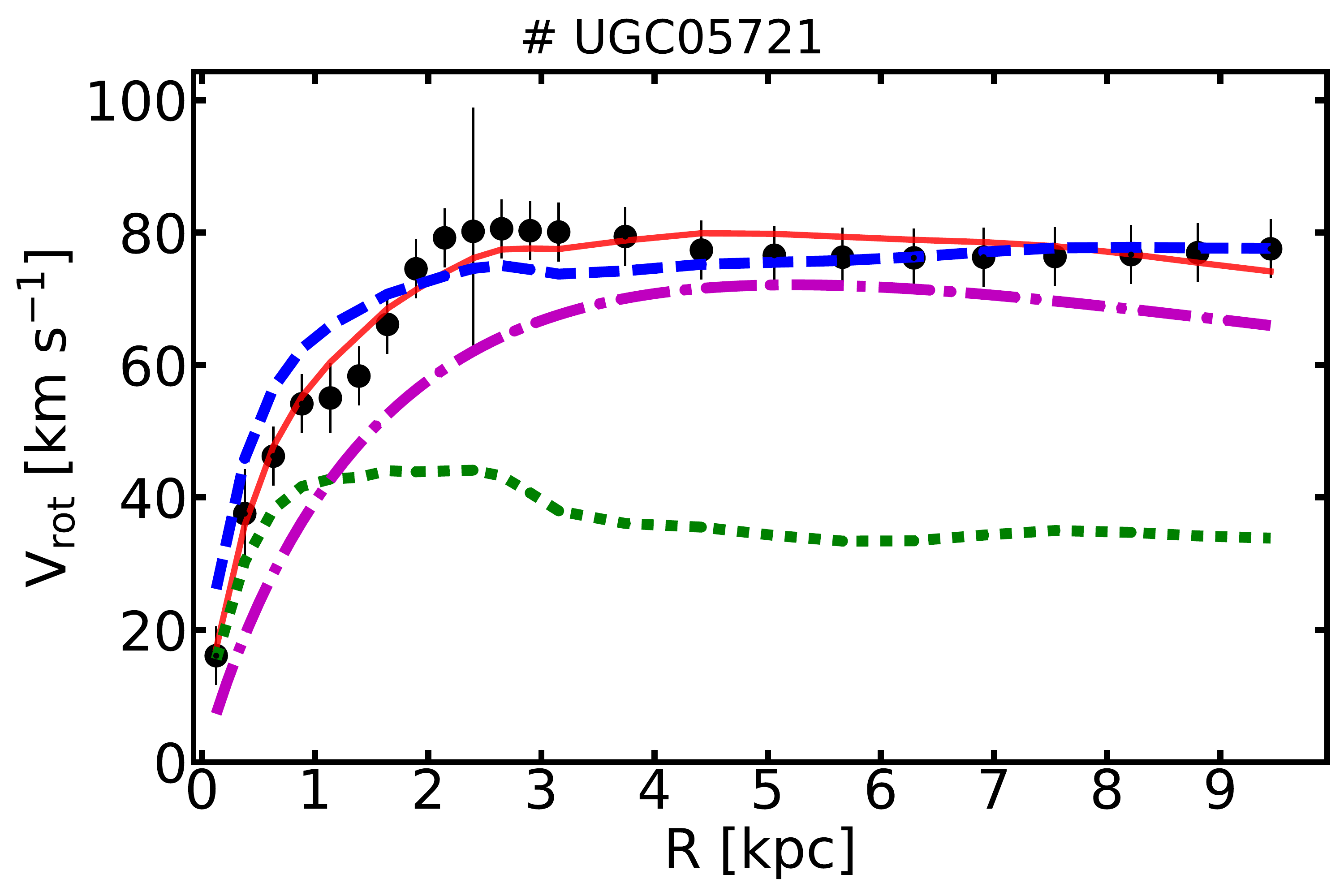}\hfill
\includegraphics[width=.33\textwidth]{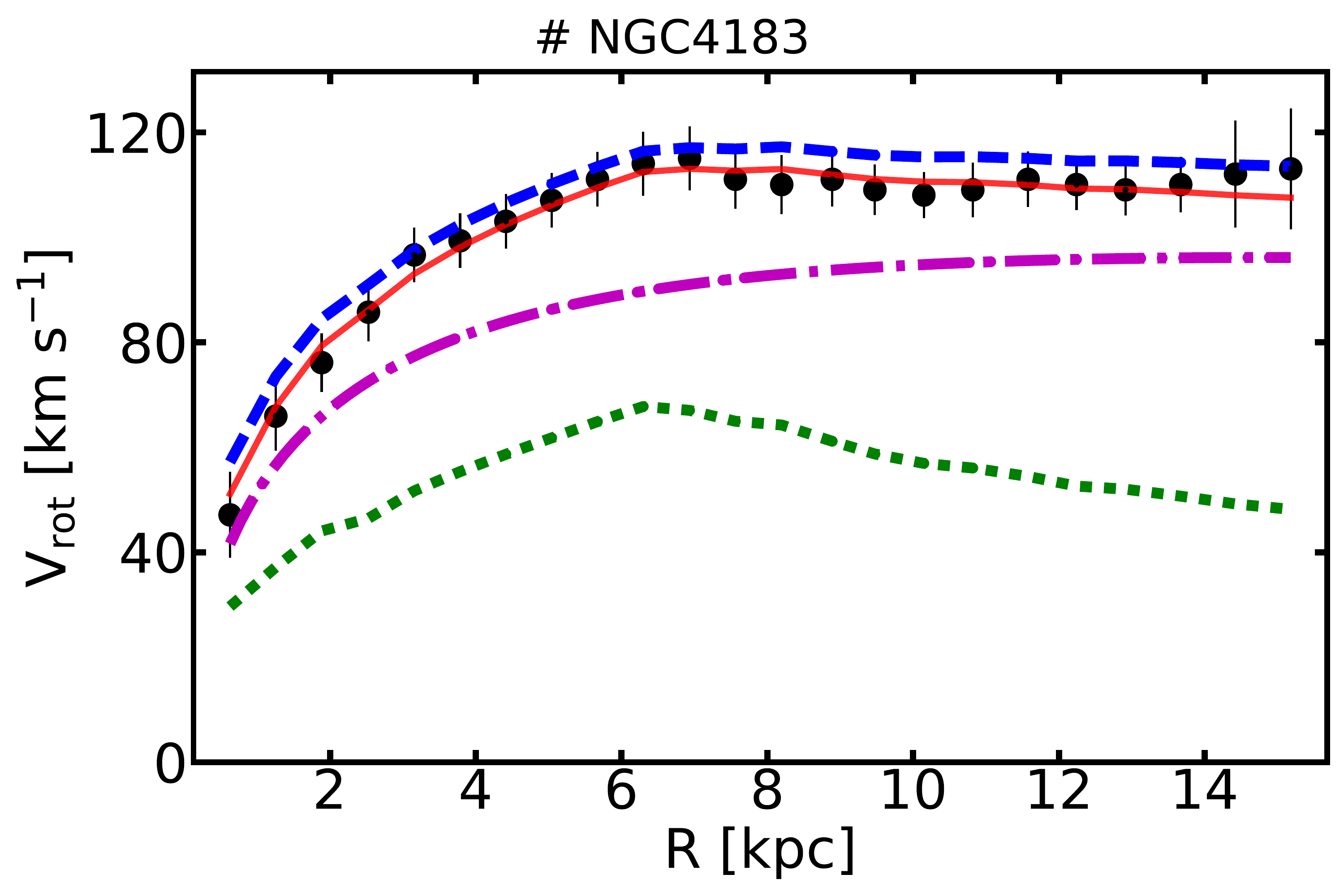}\hfill
\includegraphics[width=.33\textwidth]{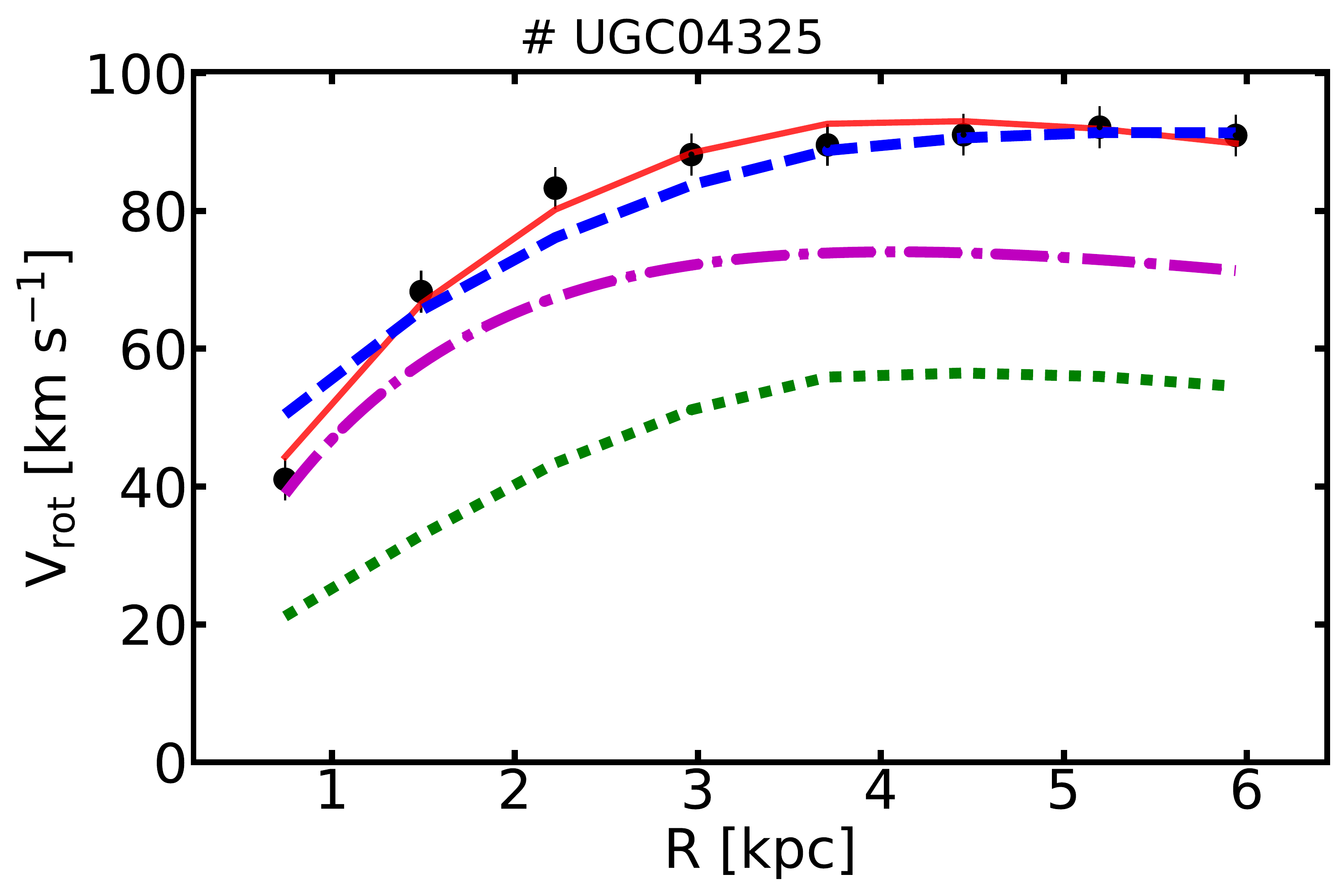}

\includegraphics[width=.33\textwidth]{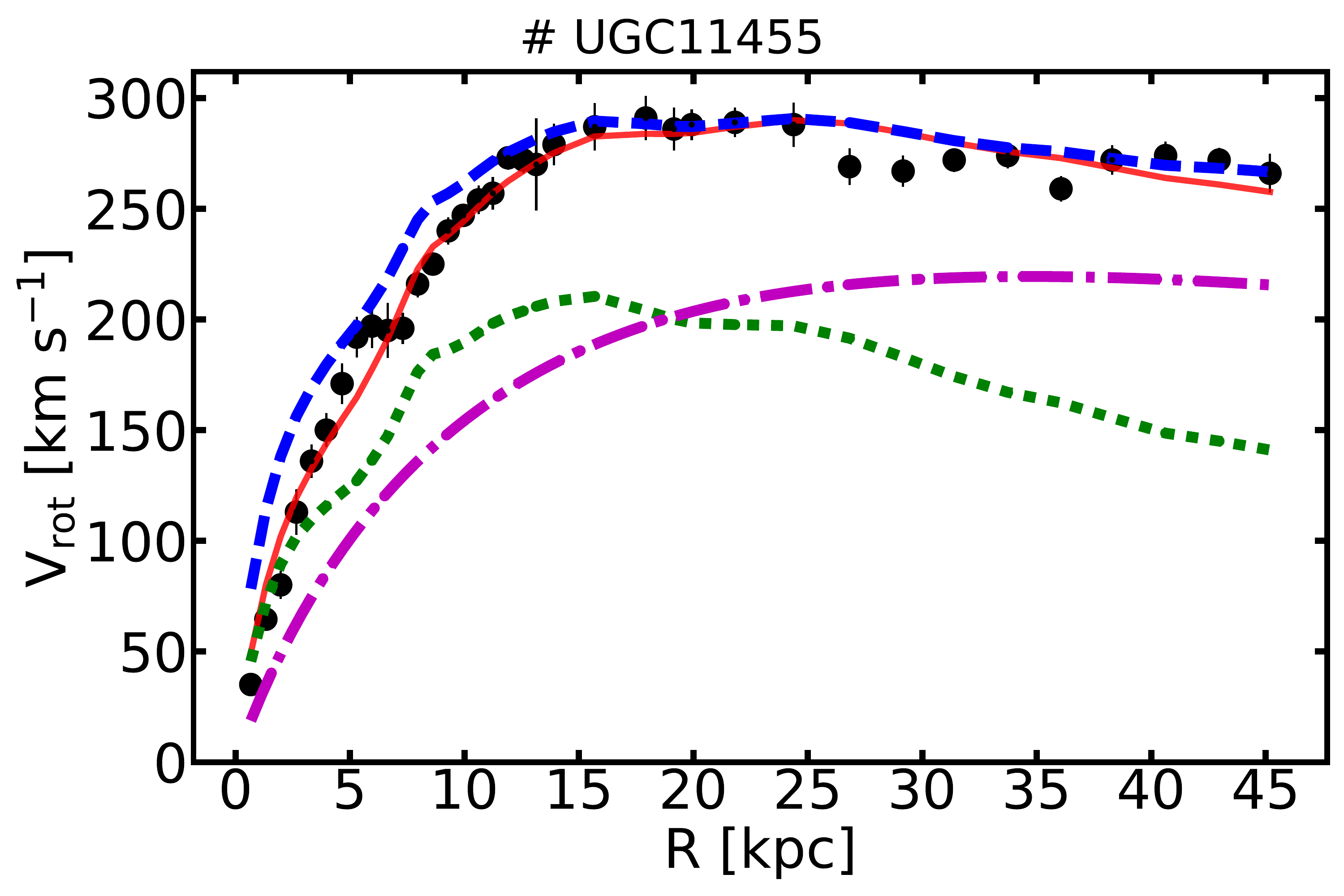}\hfill
\includegraphics[width=.33\textwidth]{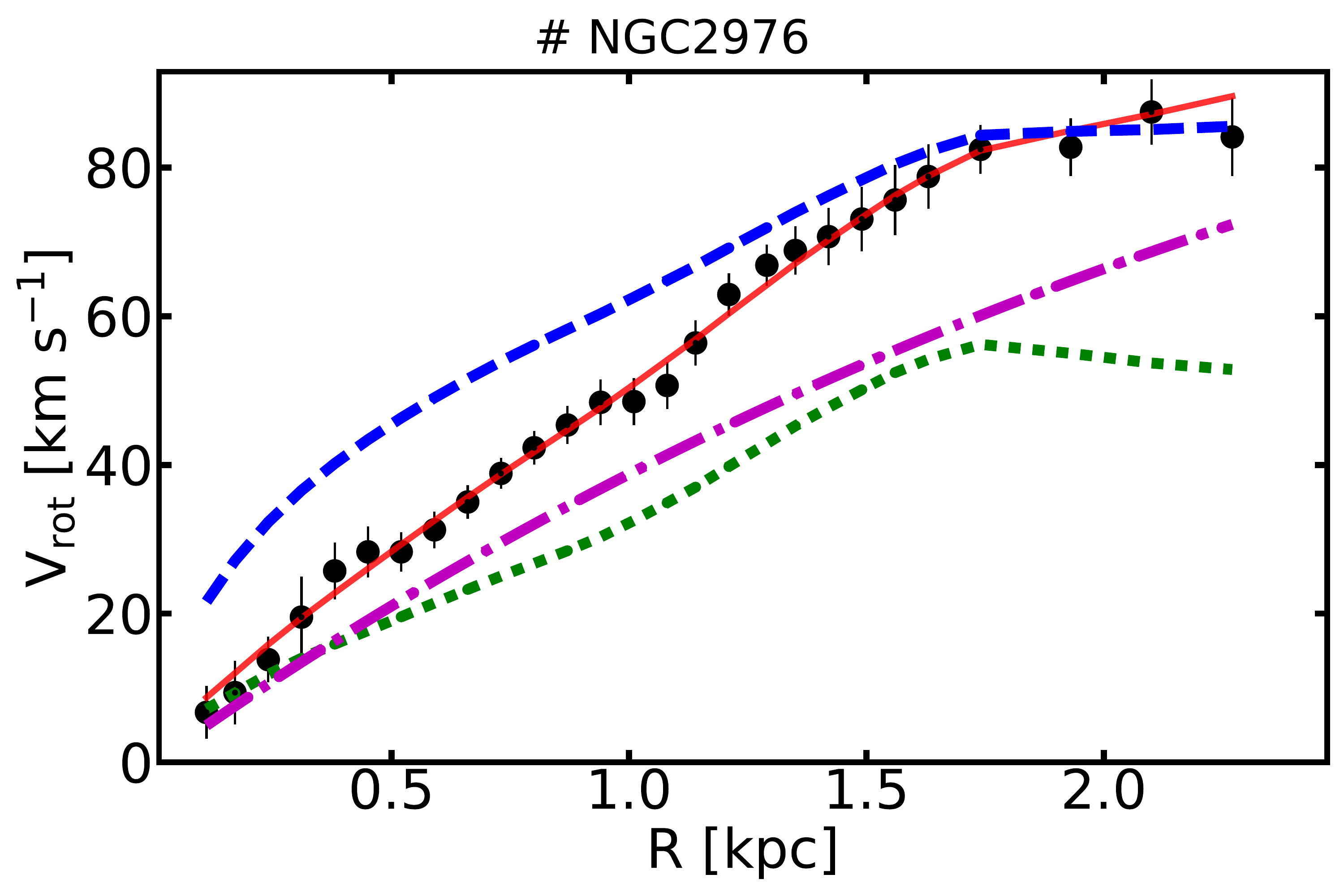}\hfill
\includegraphics[width=.33\textwidth]{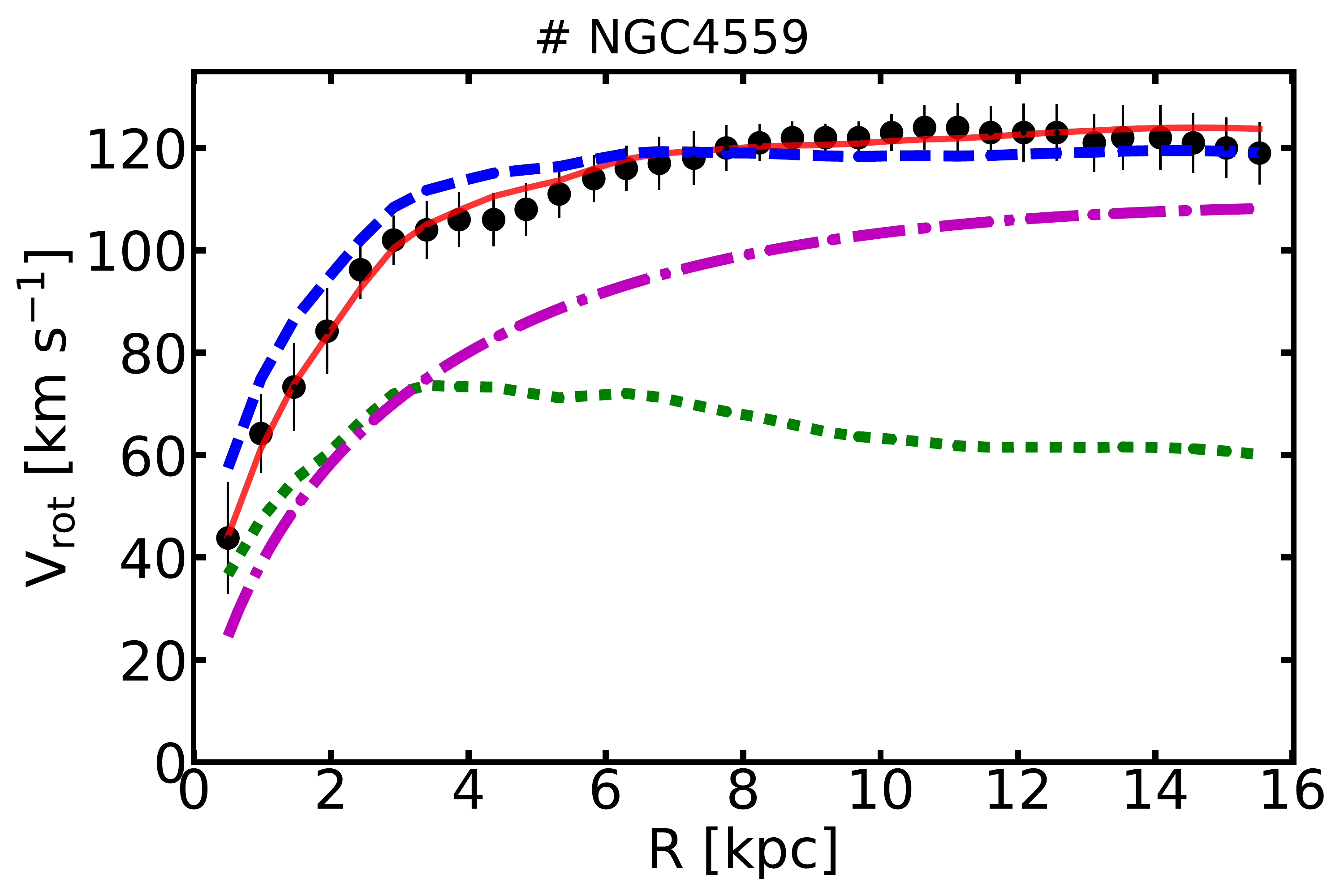}

\includegraphics[width=.33\textwidth]{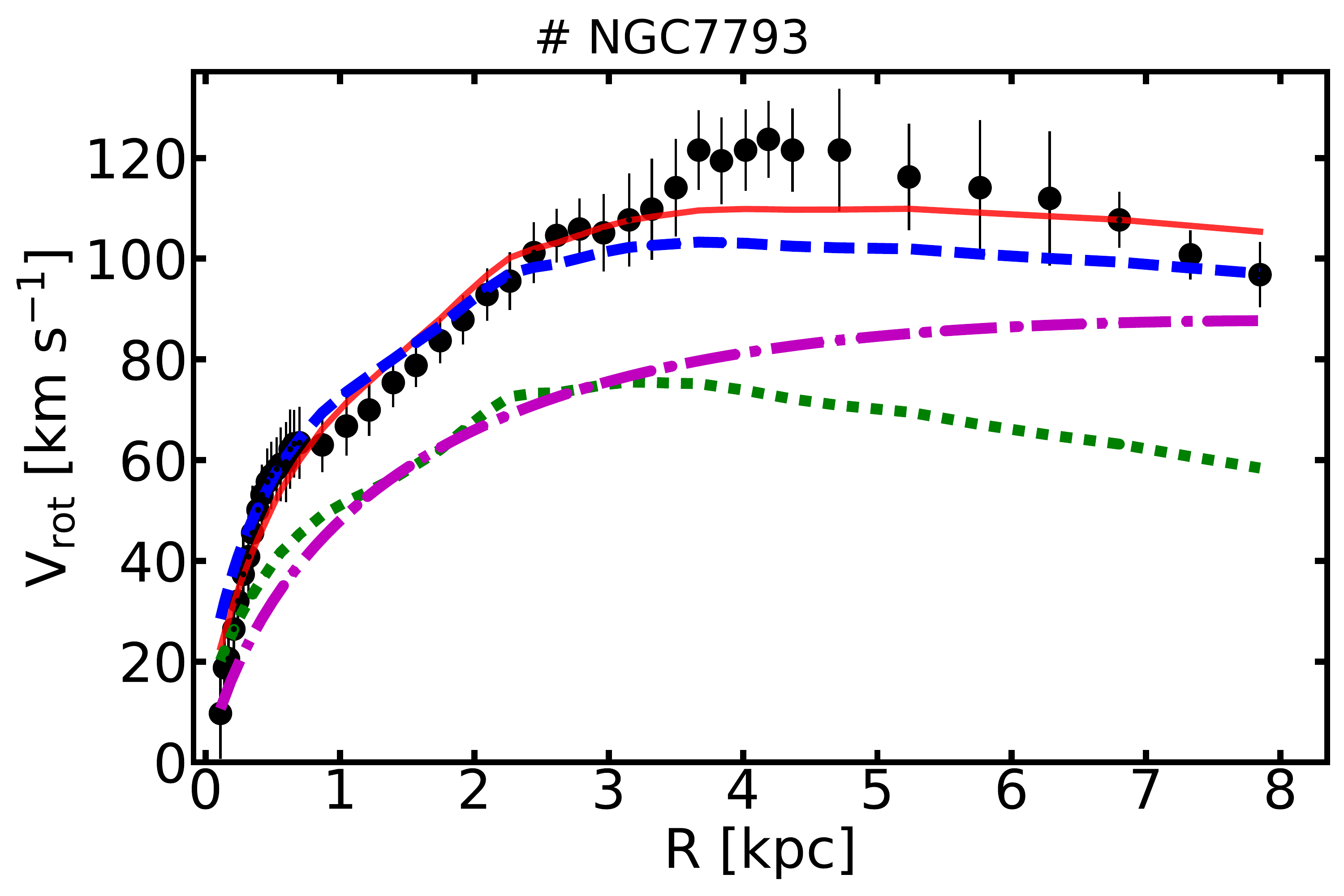}\hfill
\includegraphics[width=.33\textwidth]{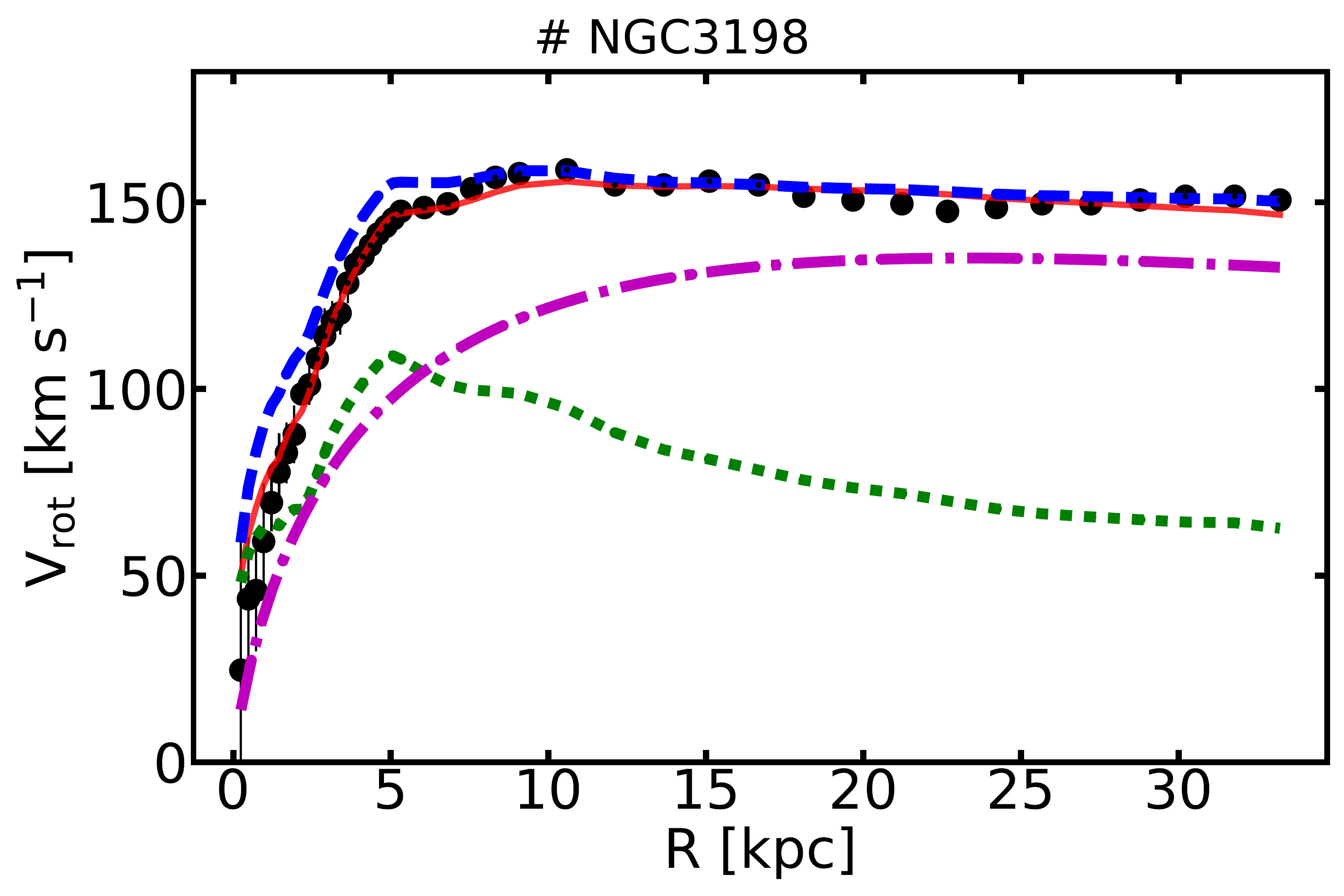}\hfill
\includegraphics[width=.33\textwidth]{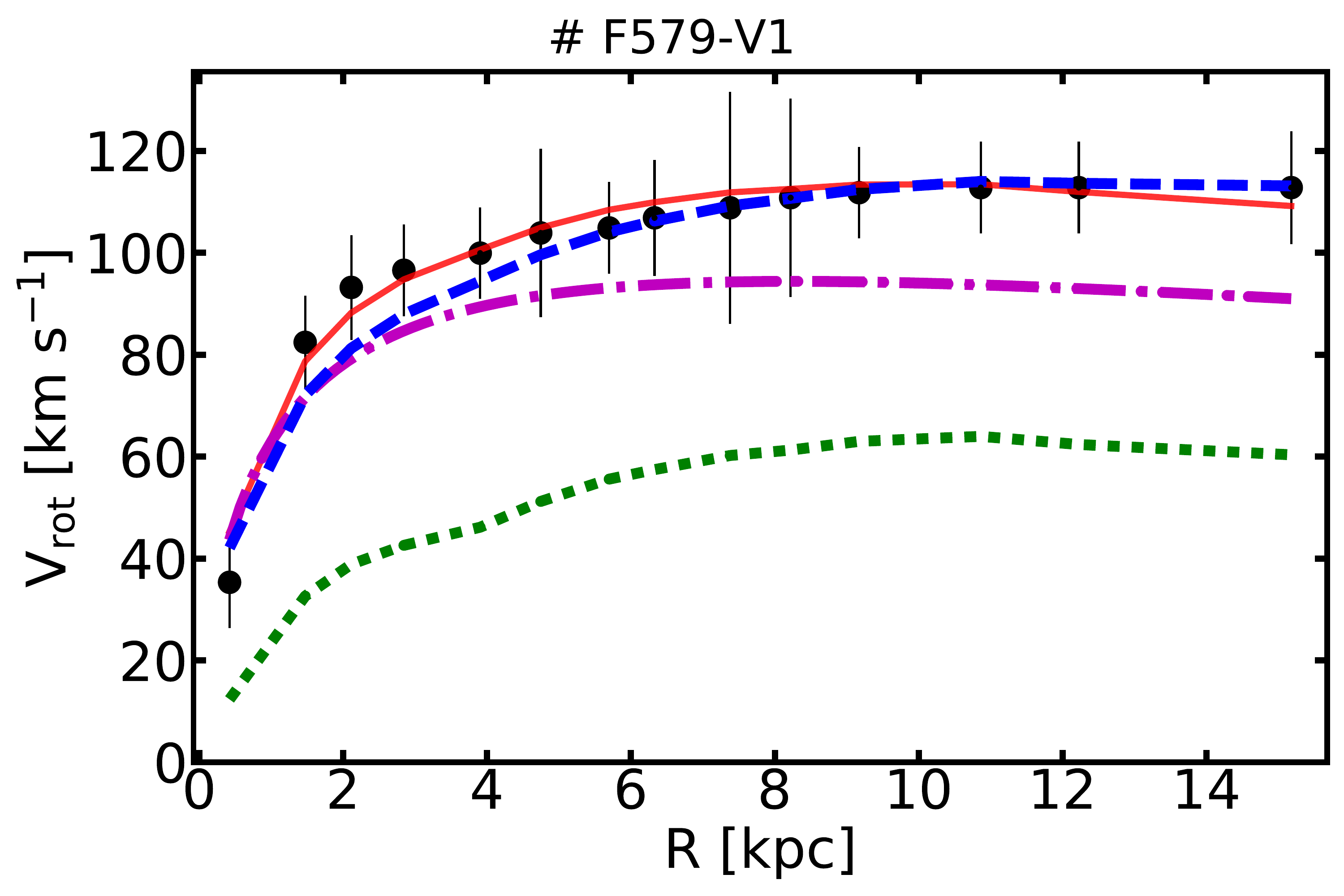}
\caption{Continued.}
\end{figure*}

\begin{figure*}
\centering
\ContinuedFloat
\includegraphics[width=.33\textwidth]{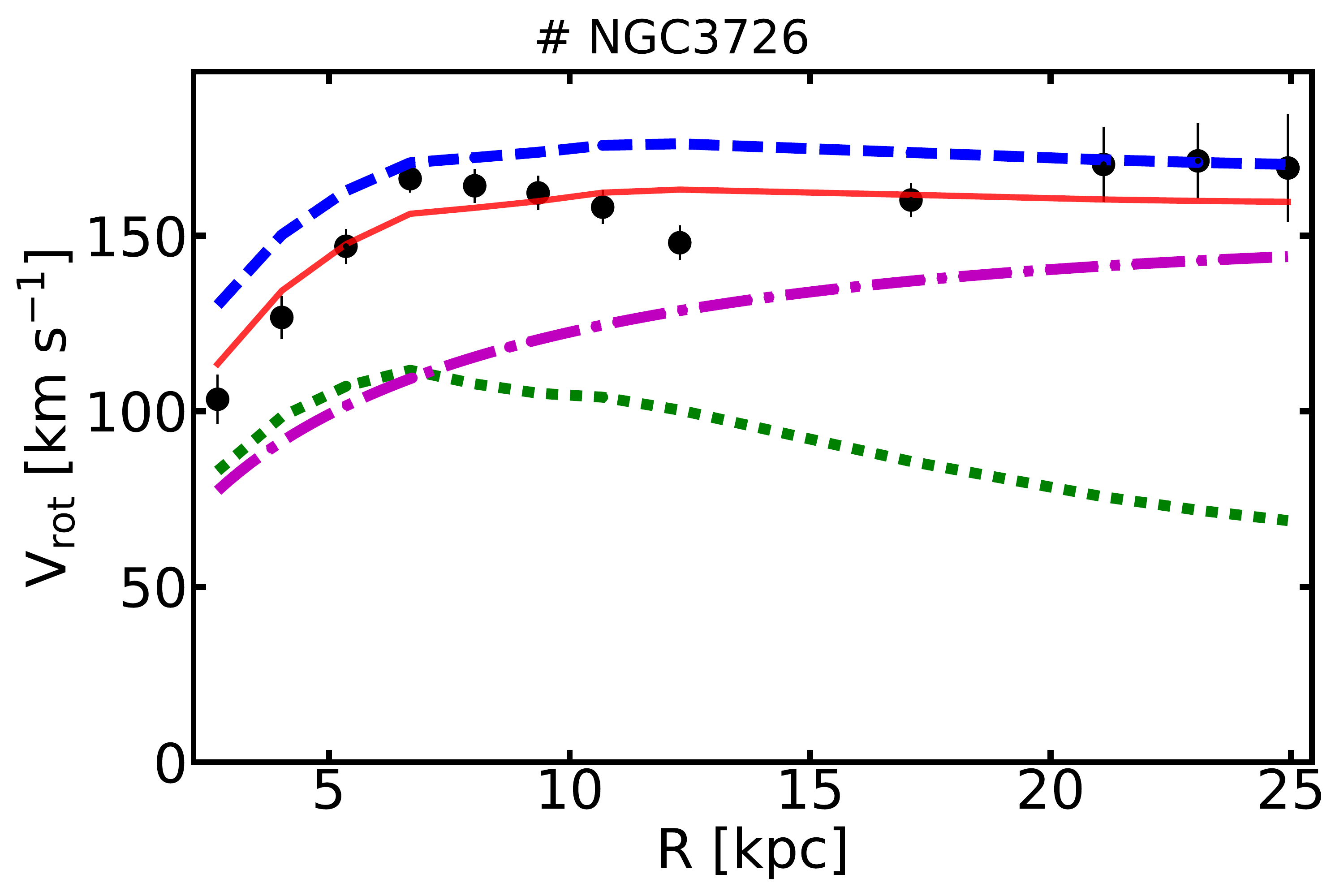}\hfill
\includegraphics[width=.33\textwidth]{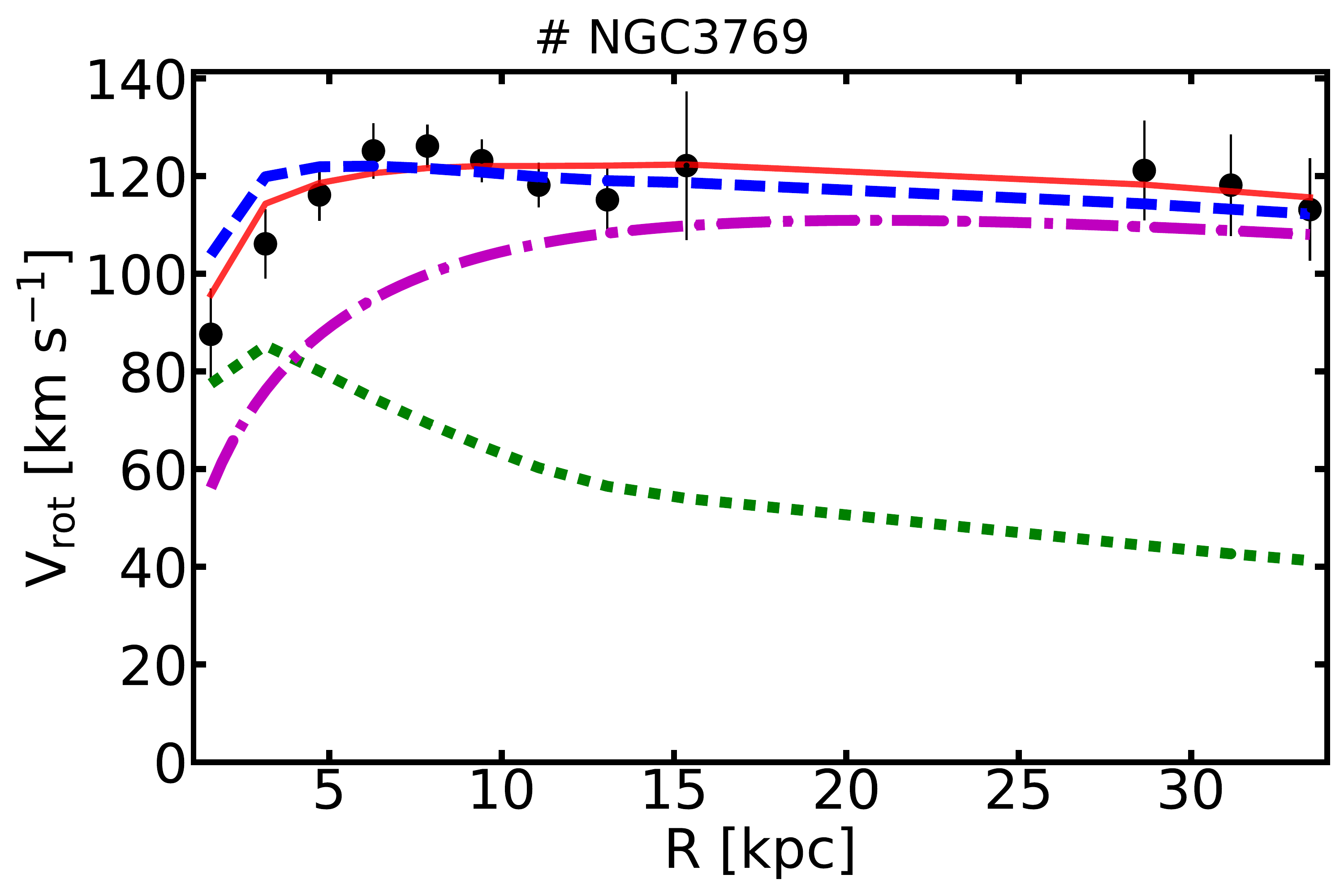}\hfill
\includegraphics[width=.33\textwidth]{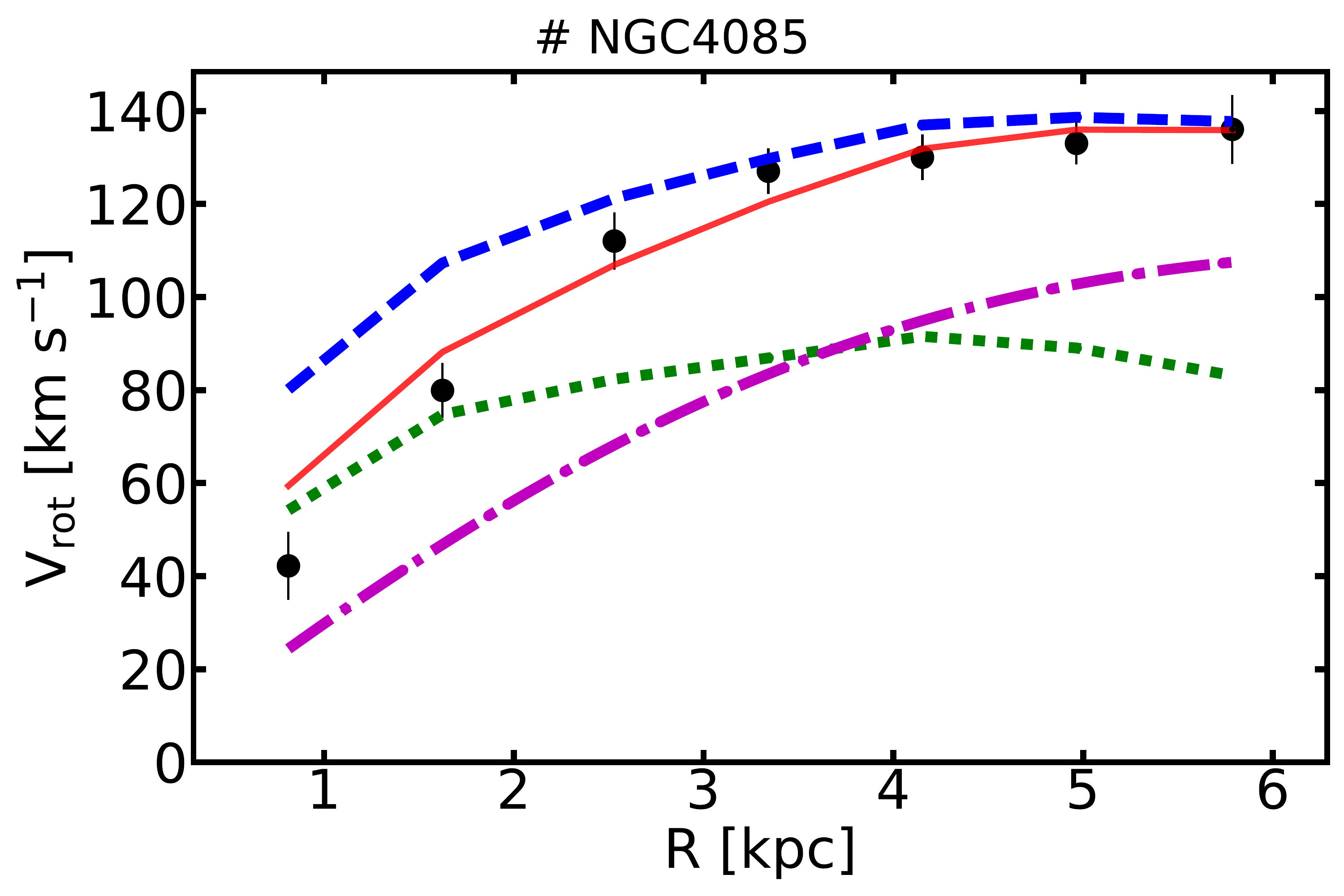}

\includegraphics[width=.33\textwidth]{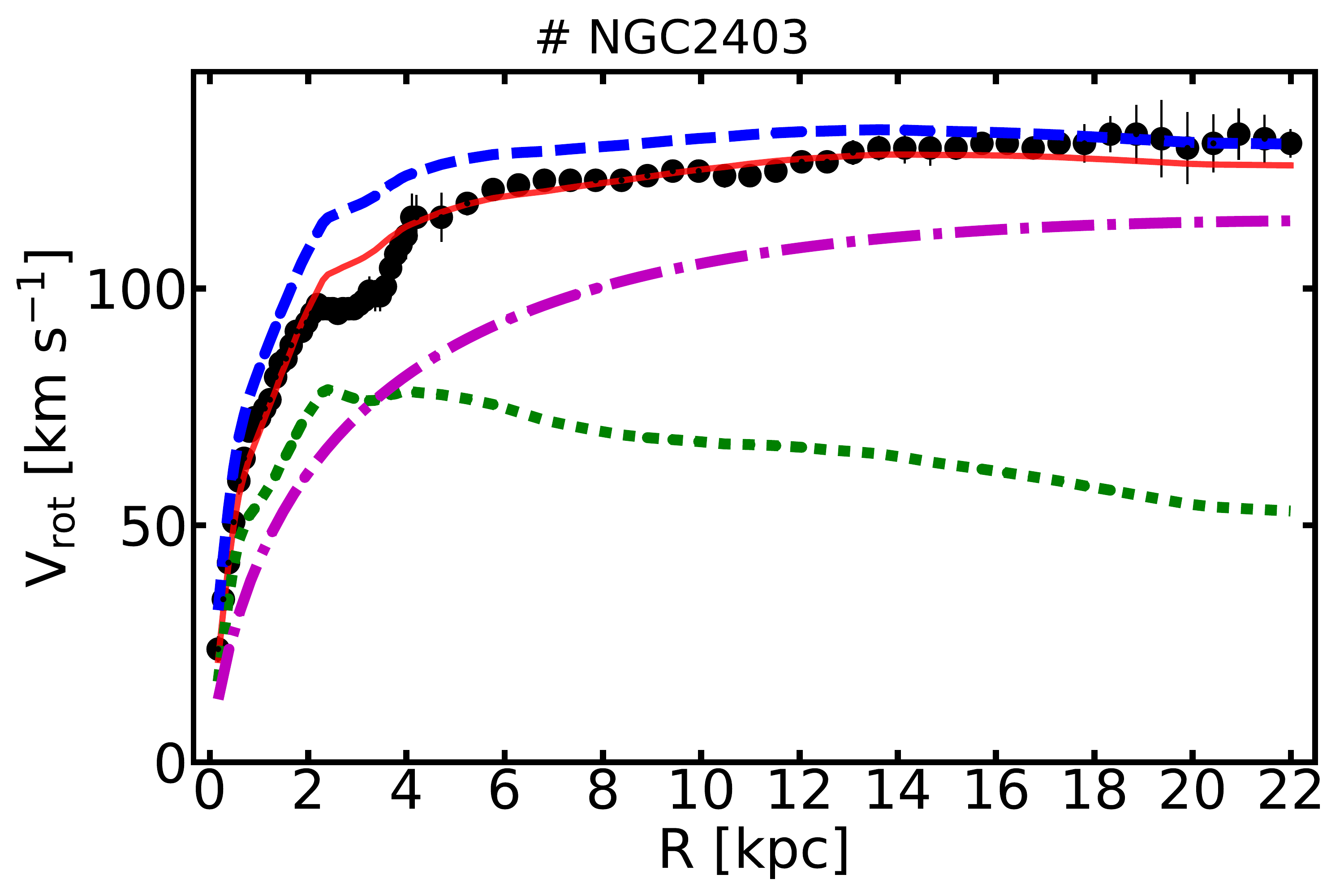}\hfill
\includegraphics[width=.33\textwidth]{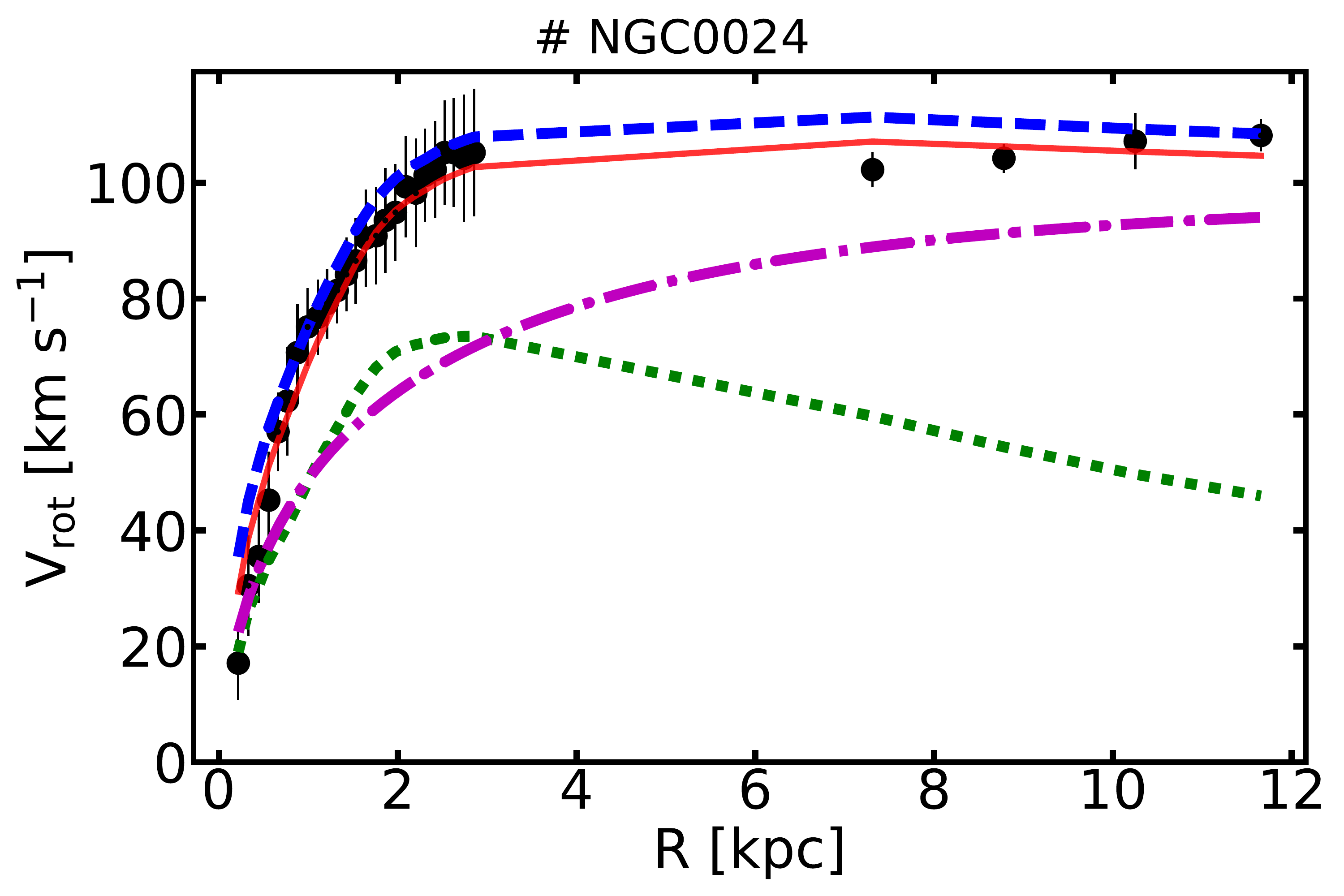}\hfill
\includegraphics[width=.33\textwidth]{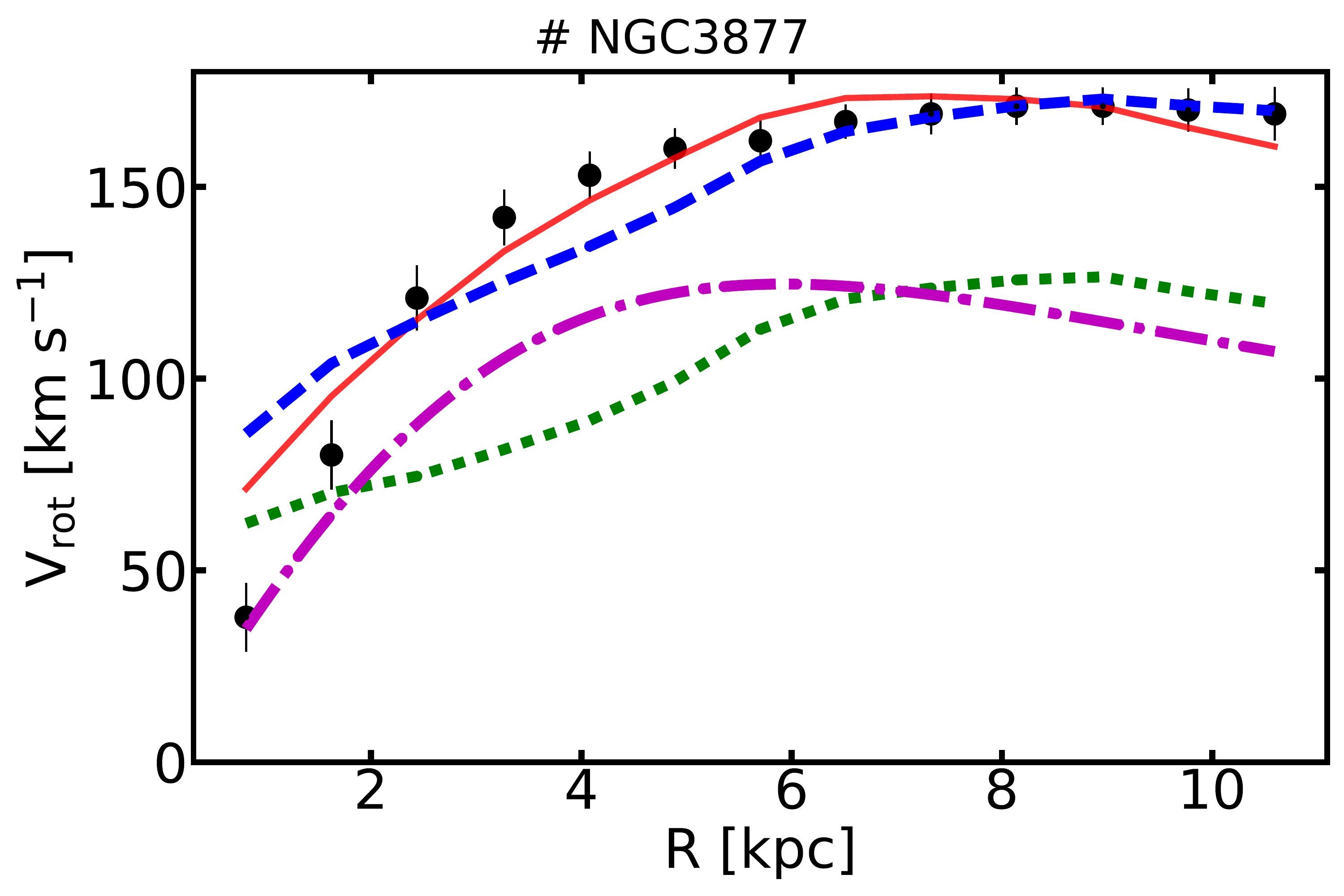}

\includegraphics[width=.33\textwidth]{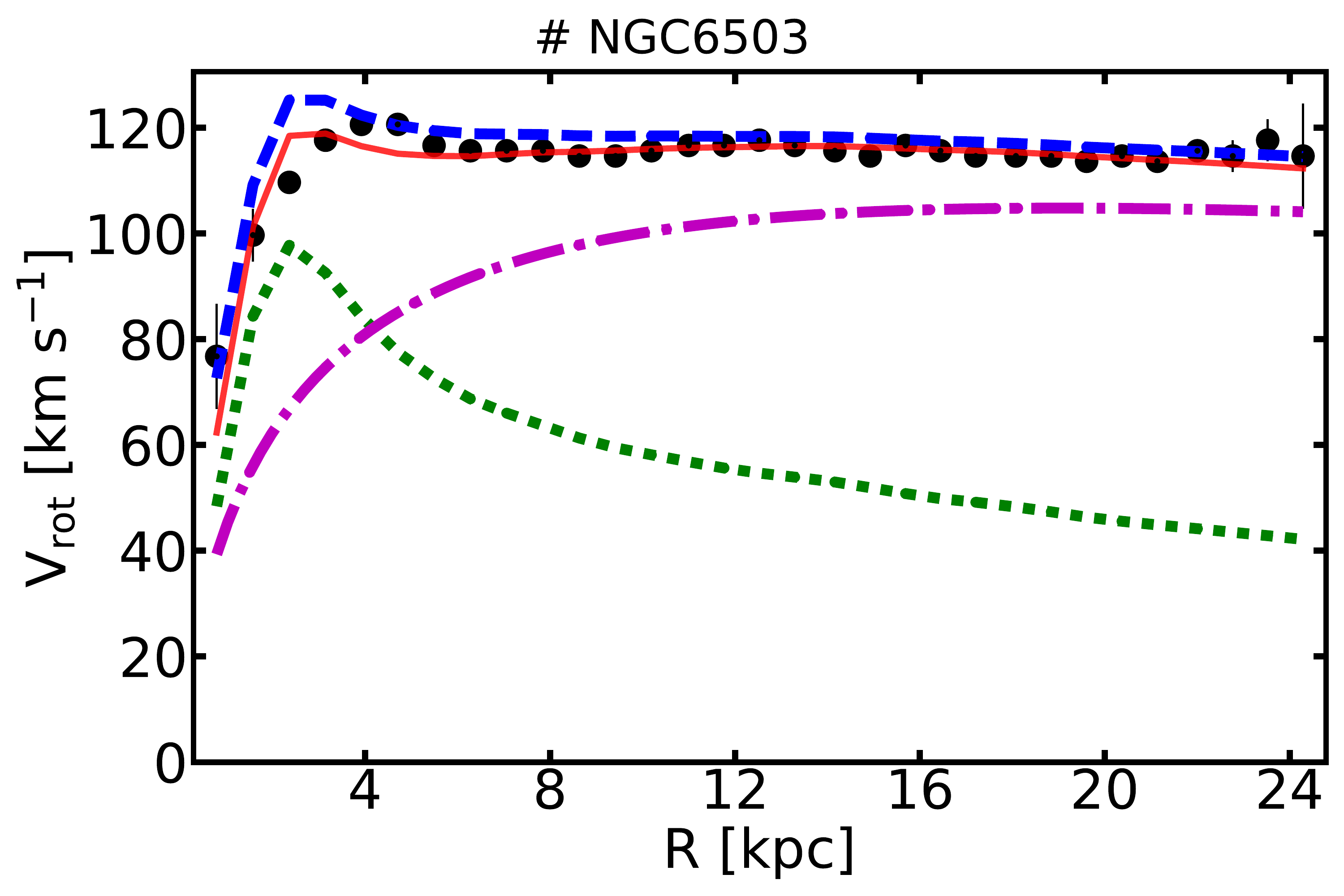}\hfill
\includegraphics[width=.33\textwidth]{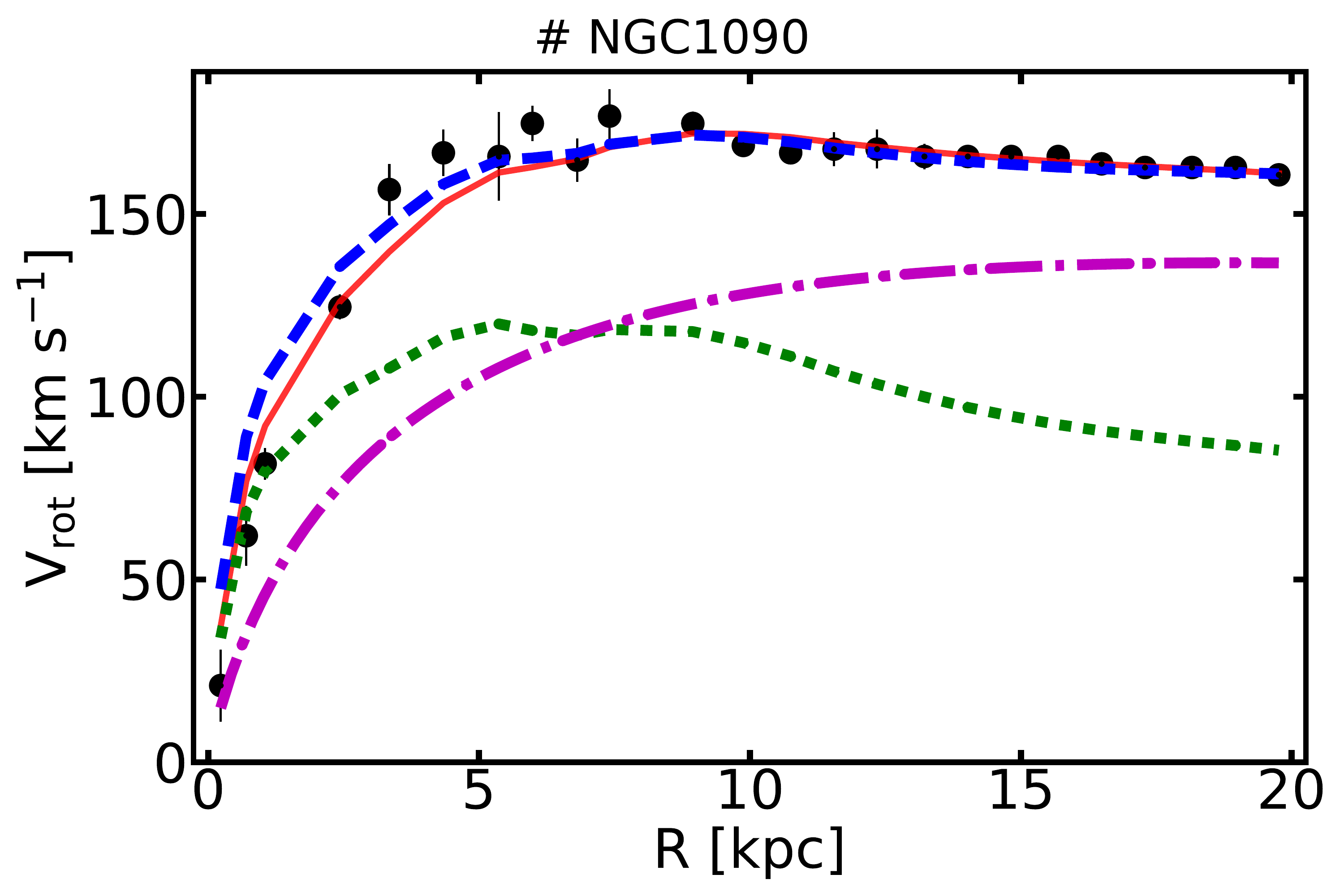}\hfill
\includegraphics[width=.33\textwidth]{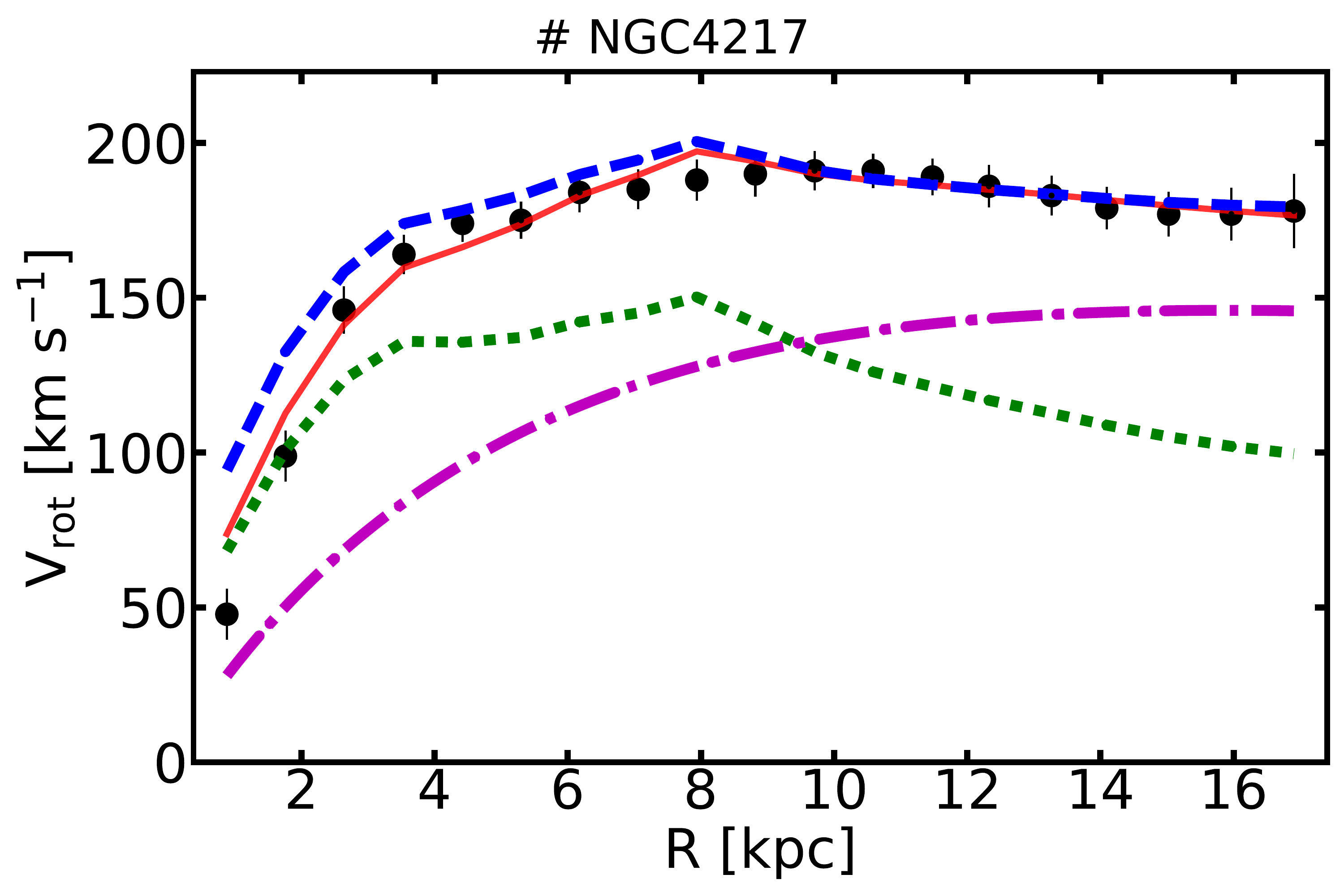}

\includegraphics[width=.33\textwidth]{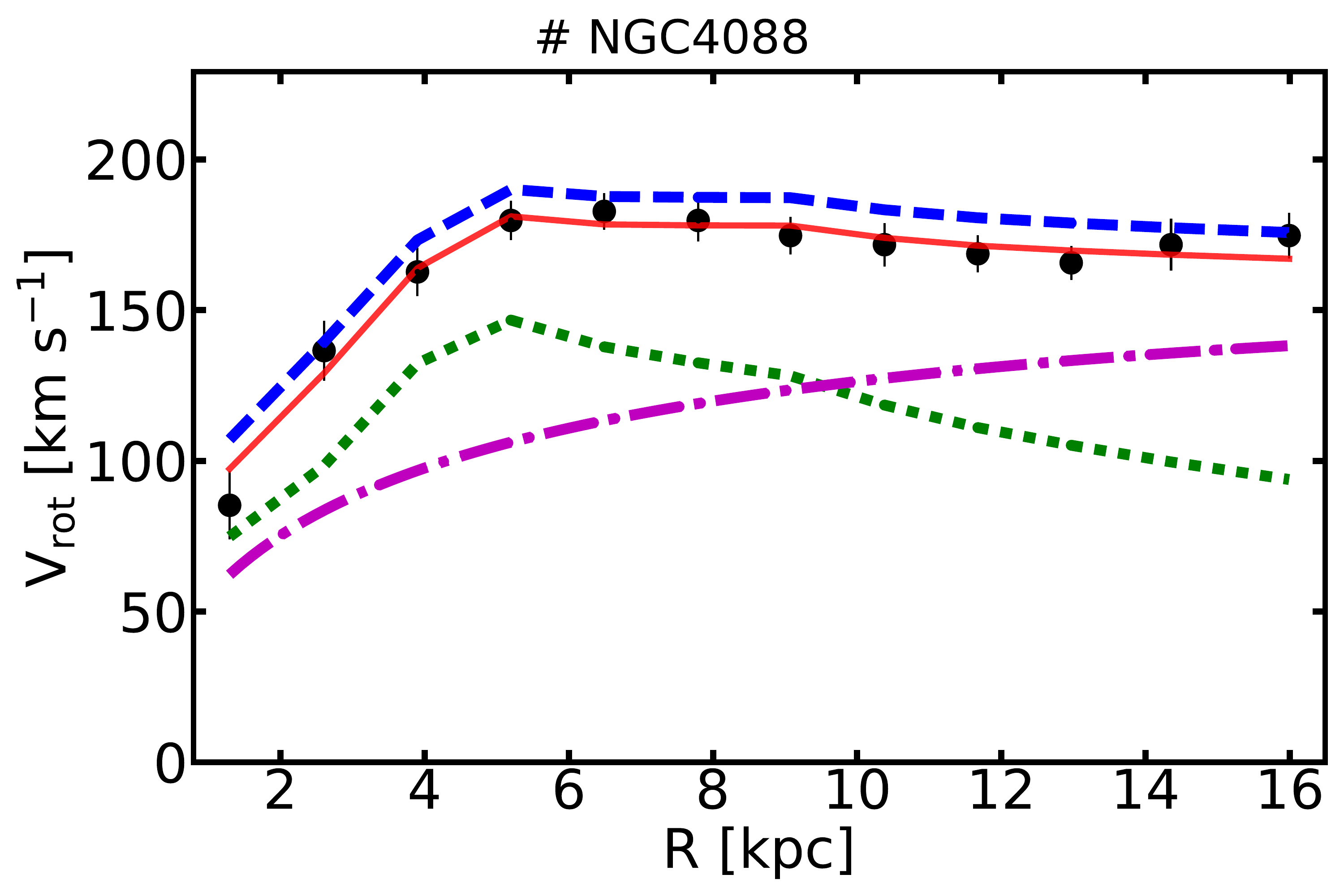}\hfill
\includegraphics[width=.33\textwidth]{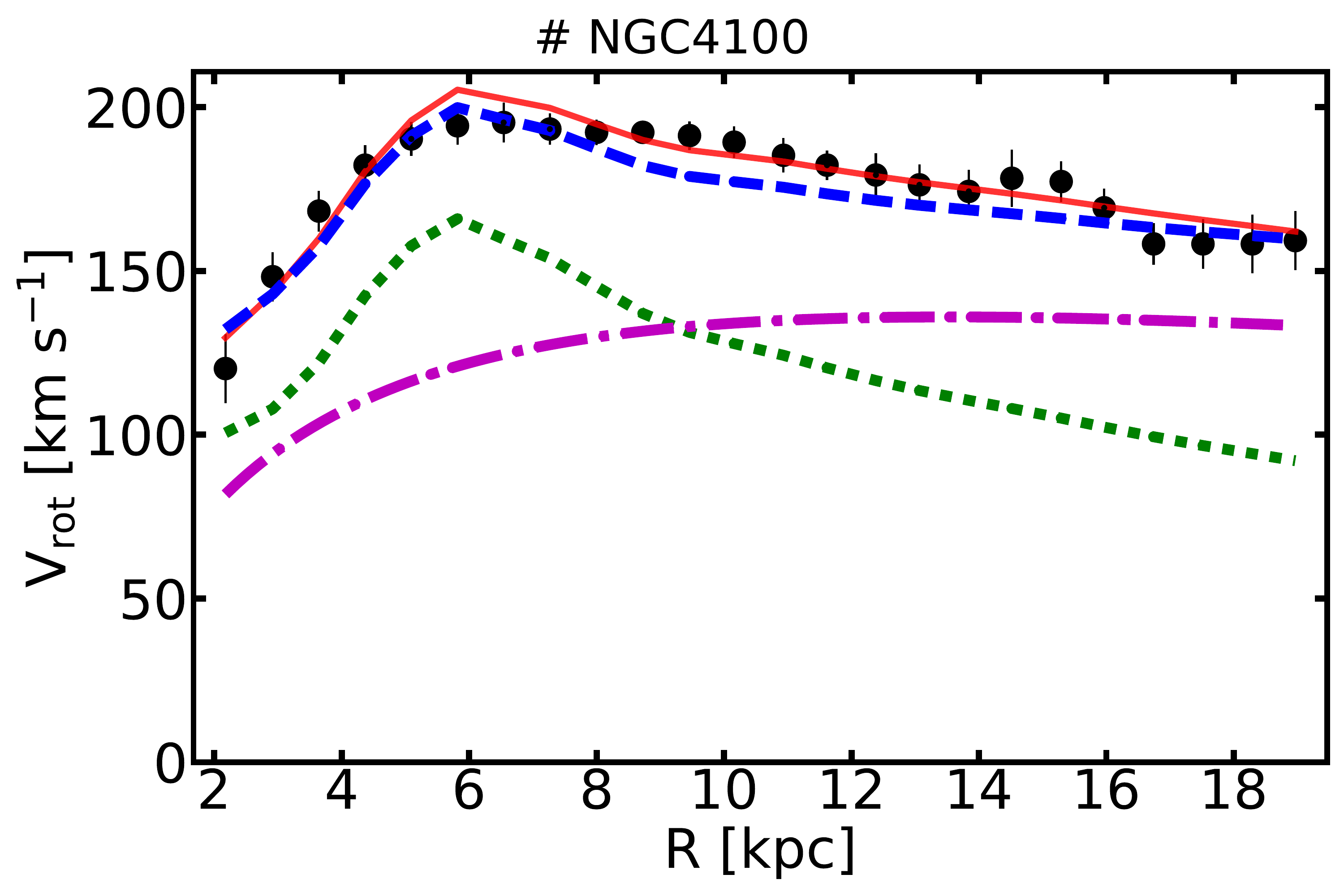}\hfill
\includegraphics[width=.33\textwidth]{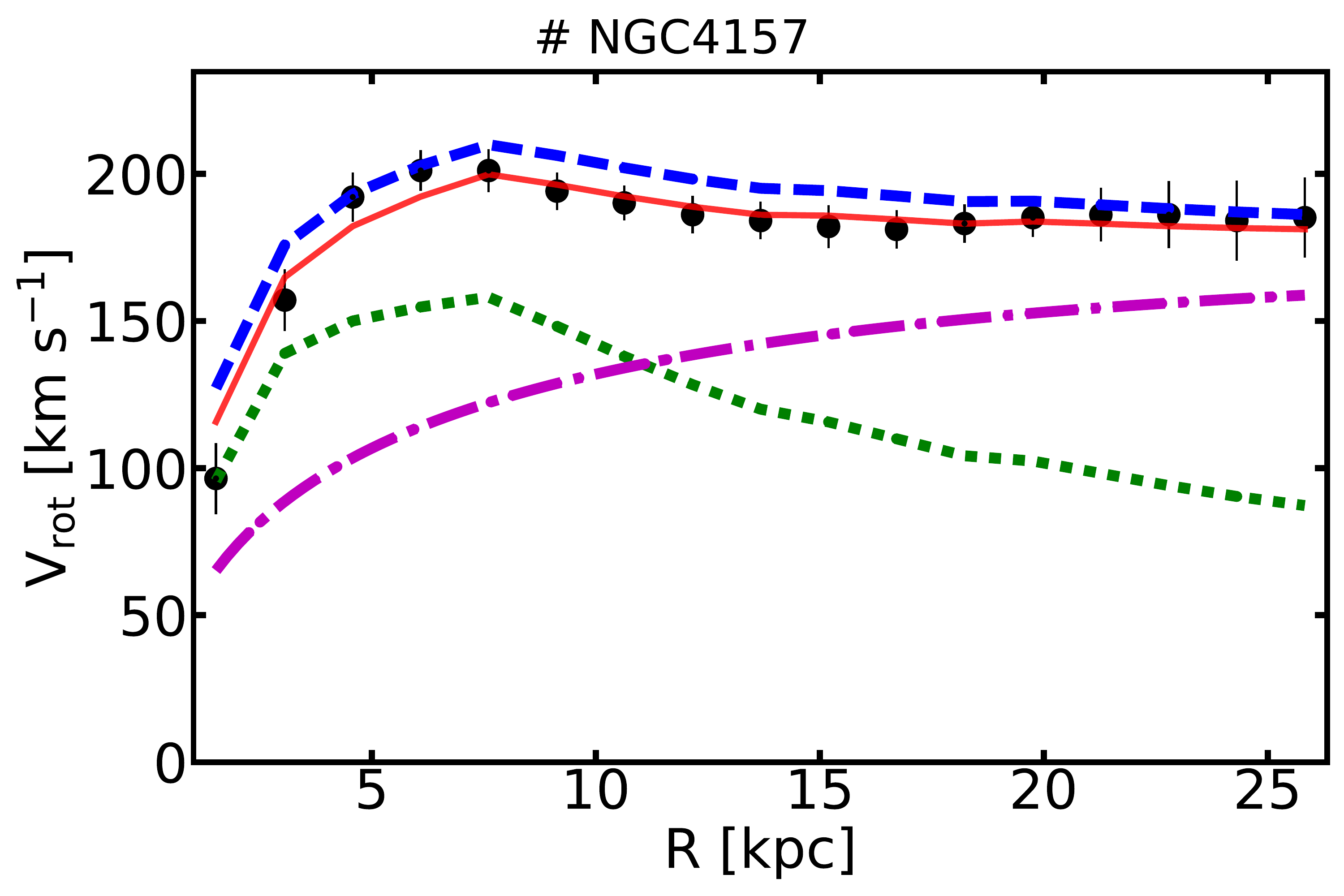}

\includegraphics[width=.33\textwidth]{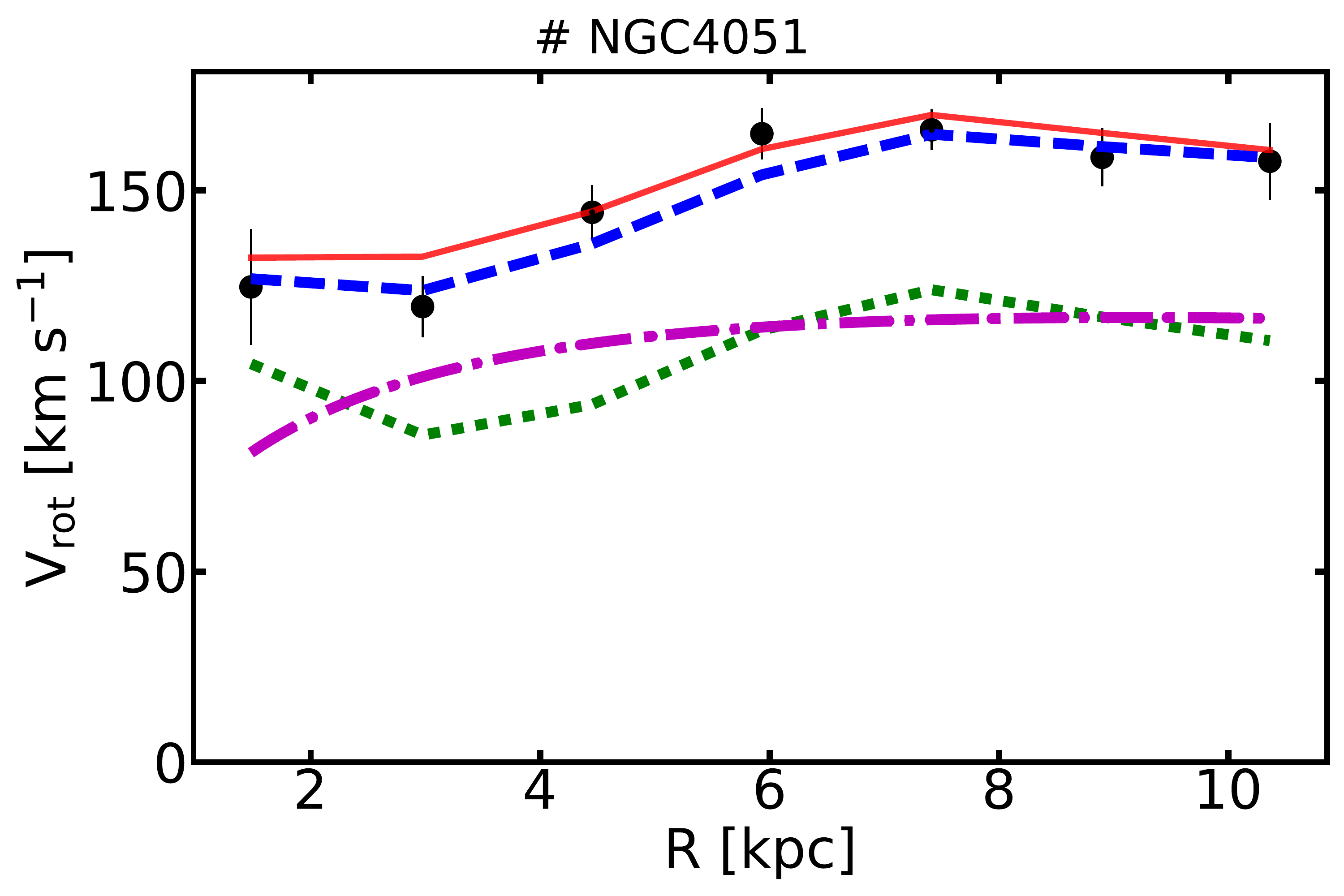}\hfill
\includegraphics[width=.33\textwidth]{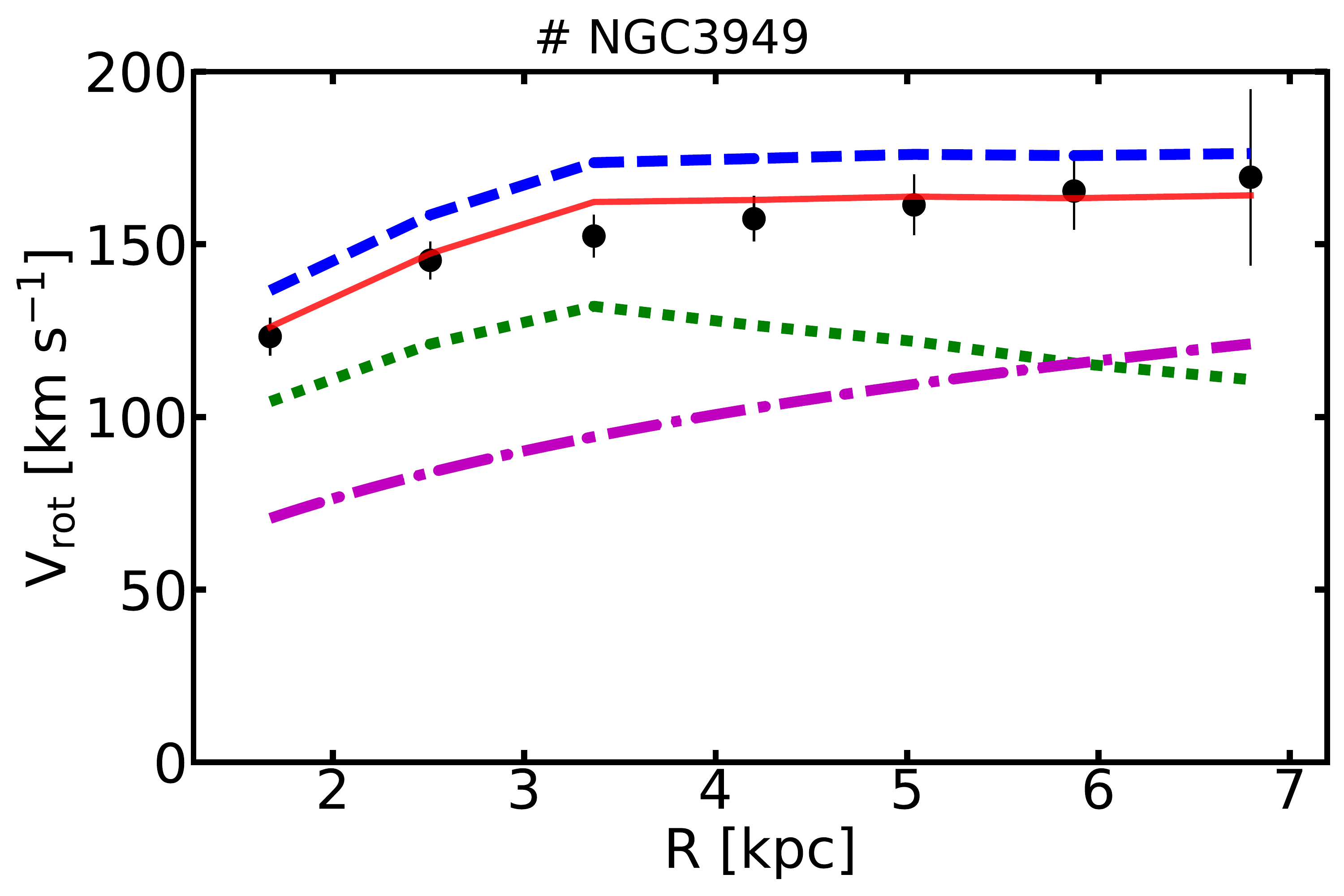}\hfill
\includegraphics[width=.33\textwidth]{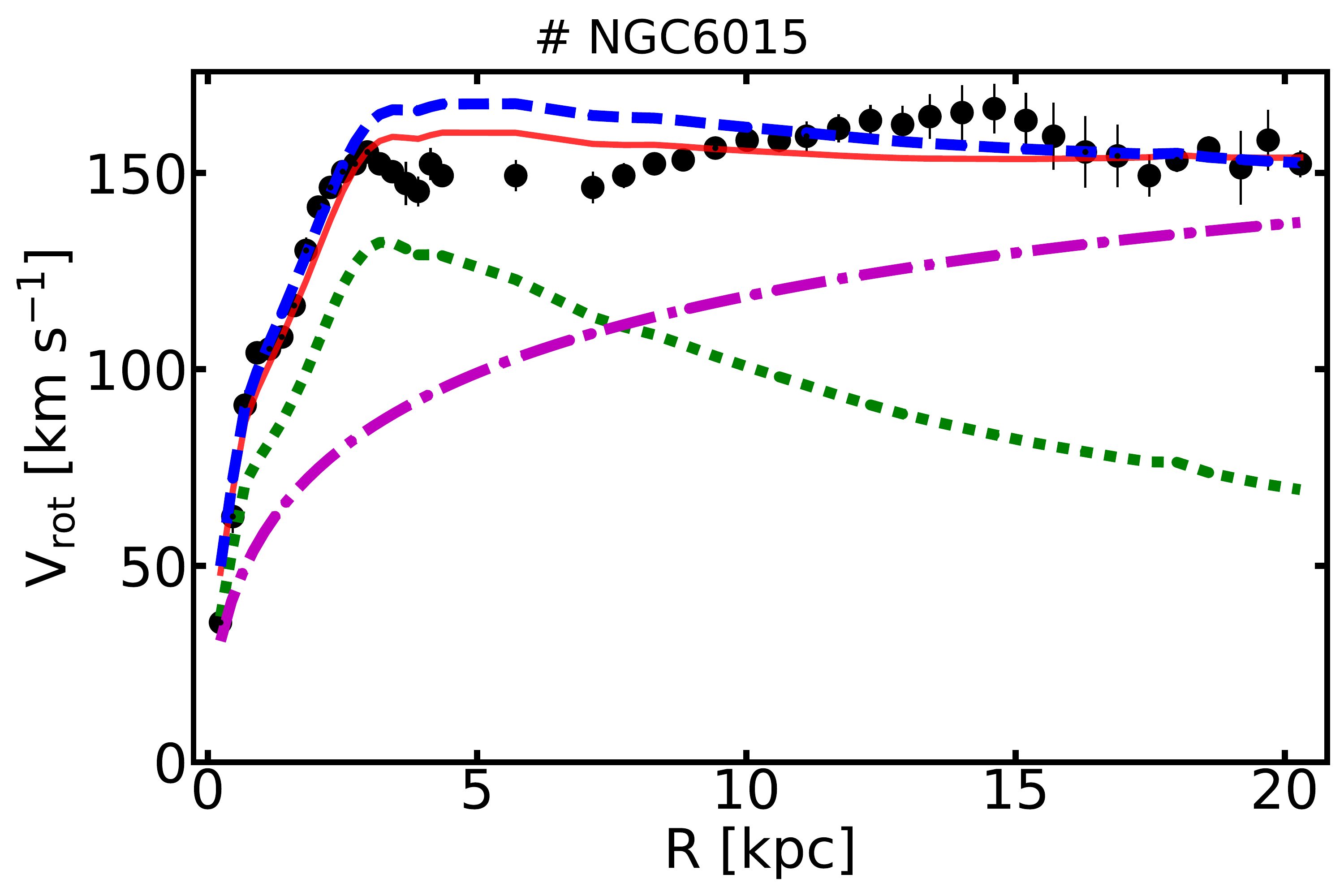}

\includegraphics[width=.33\textwidth]{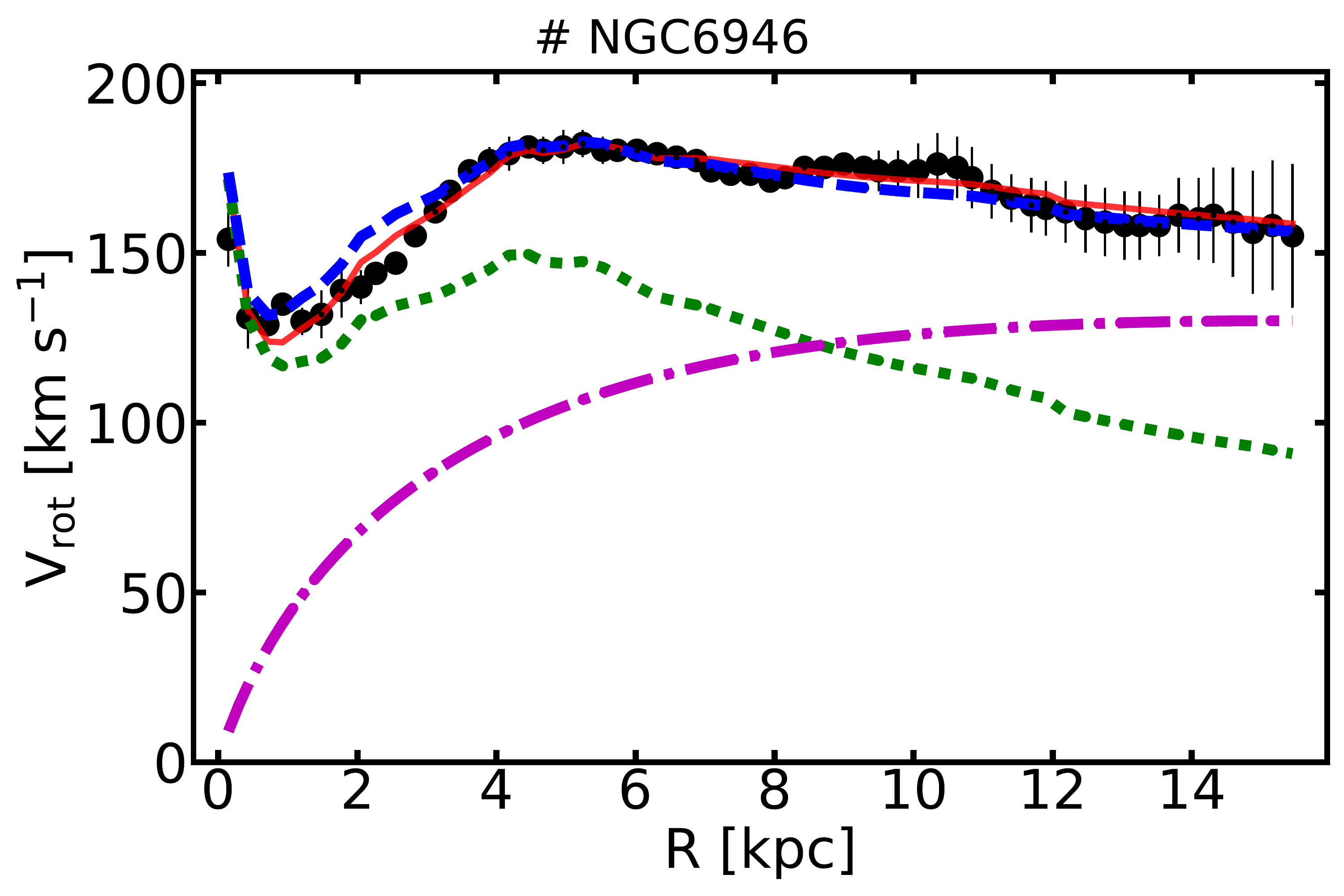}\hfill
\includegraphics[width=.33\textwidth]{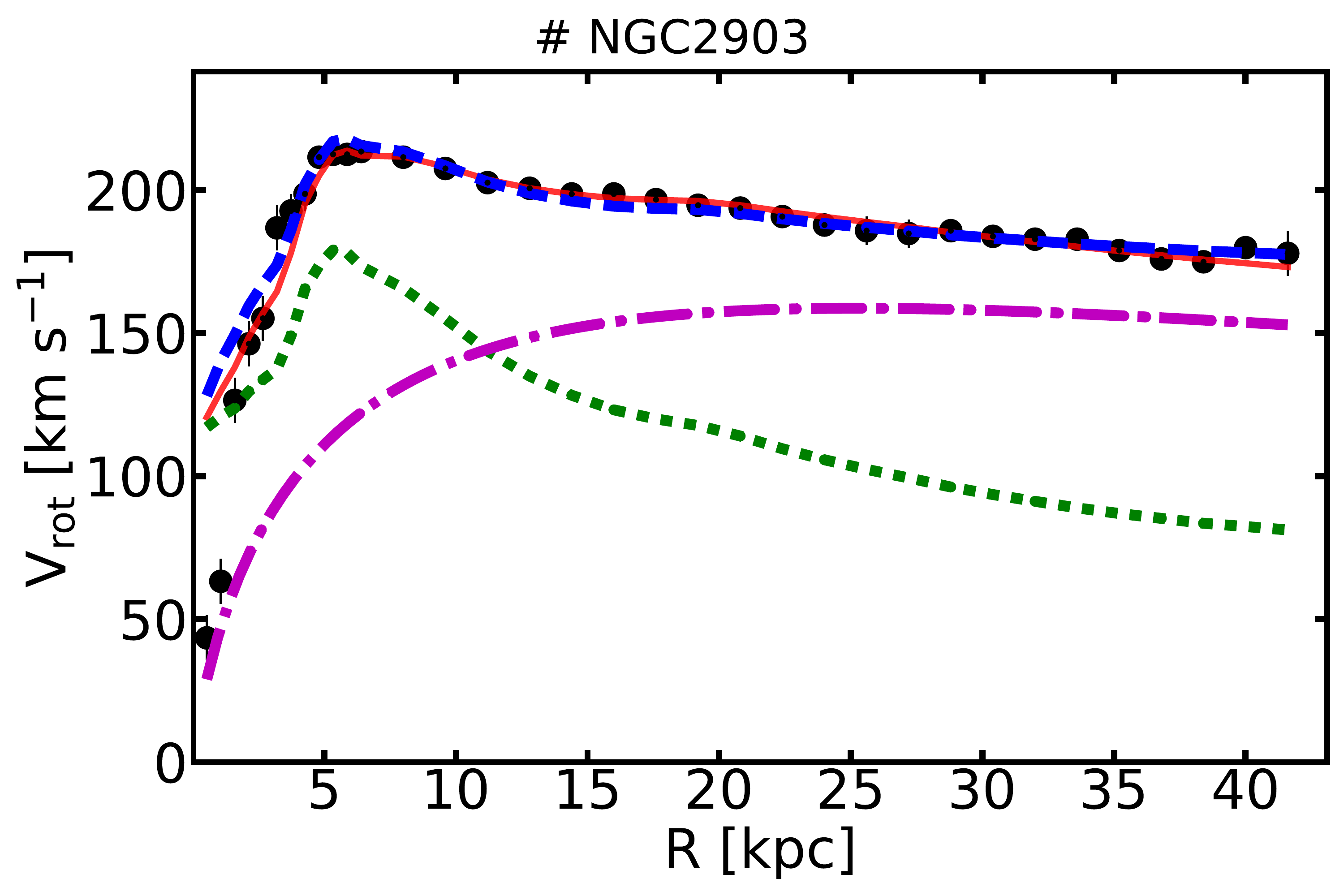}\hfill
\includegraphics[width=.33\textwidth]{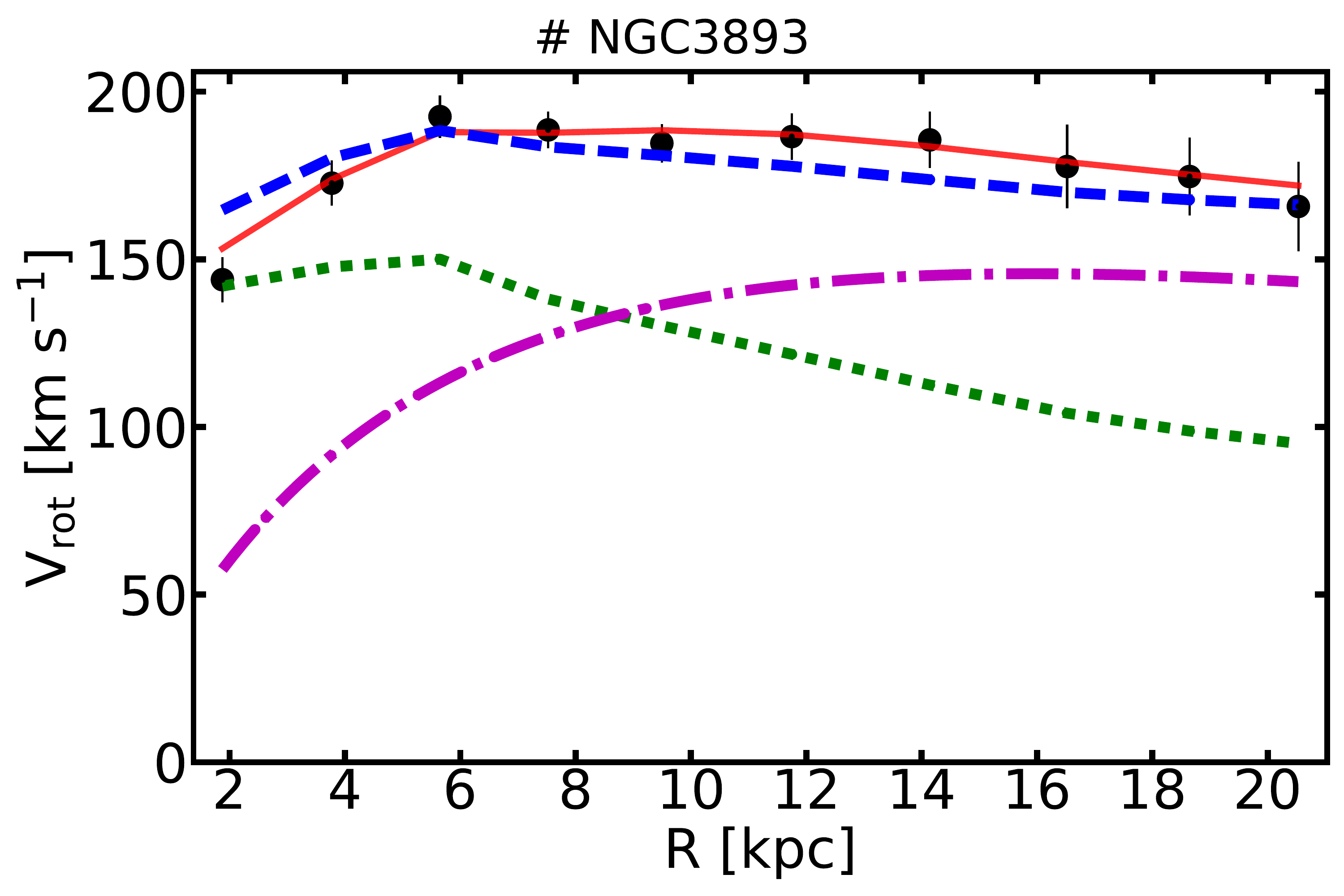}
\caption{Continued.}
\end{figure*}

\begin{figure*}

\centering
\ContinuedFloat
\includegraphics[width=.33\textwidth]{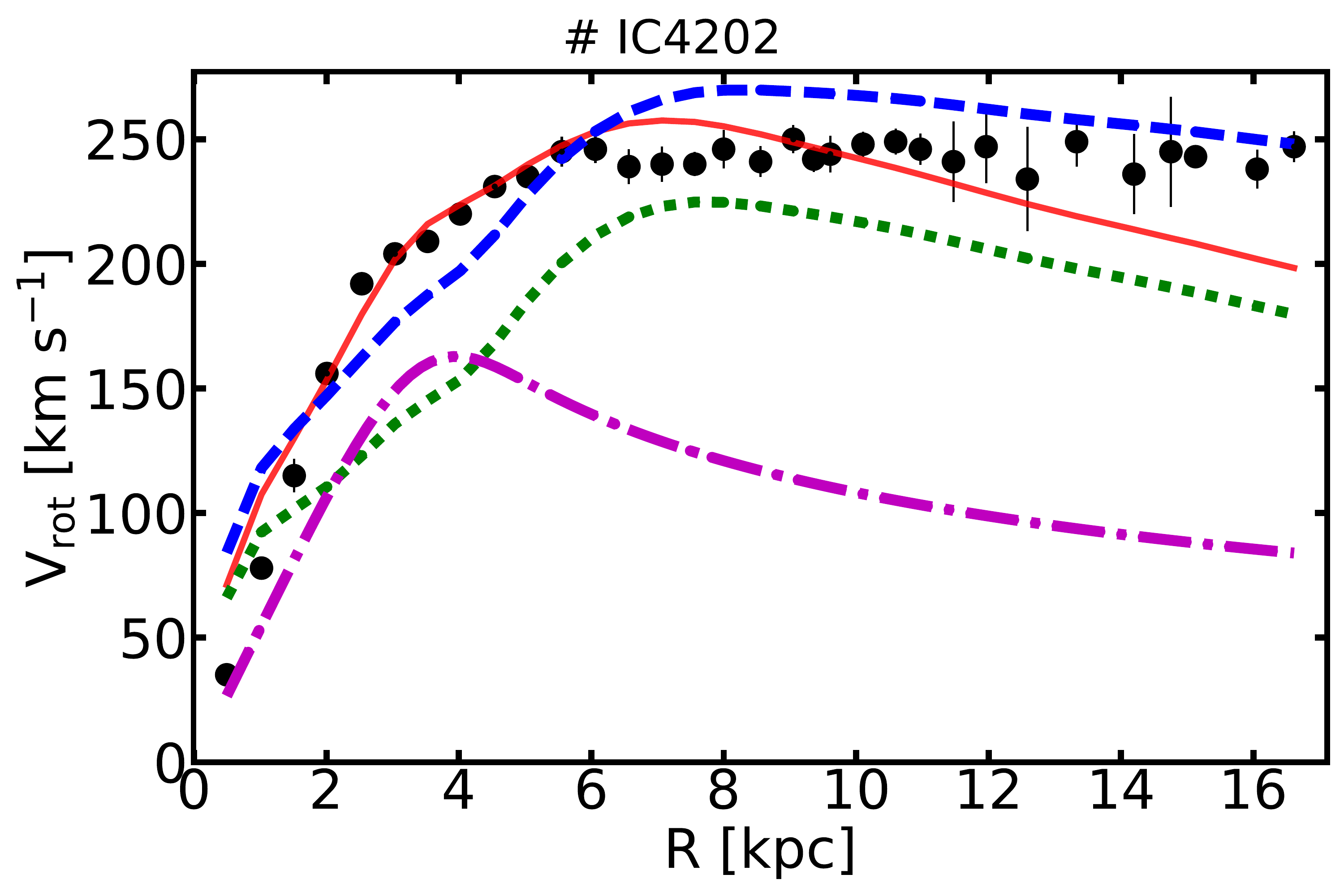}\hfill
\includegraphics[width=.33\textwidth]{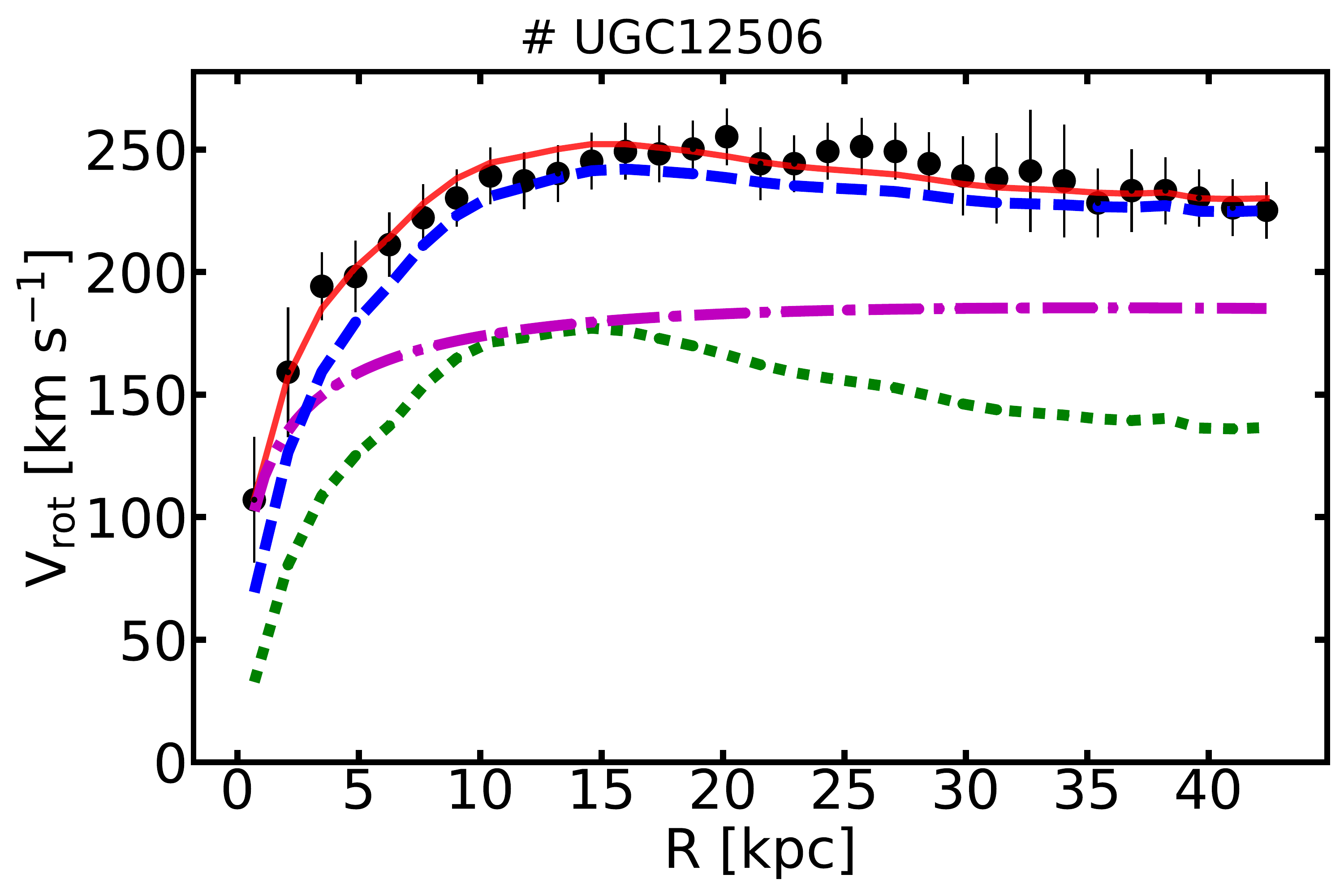}\hfill
\includegraphics[width=.33\textwidth]{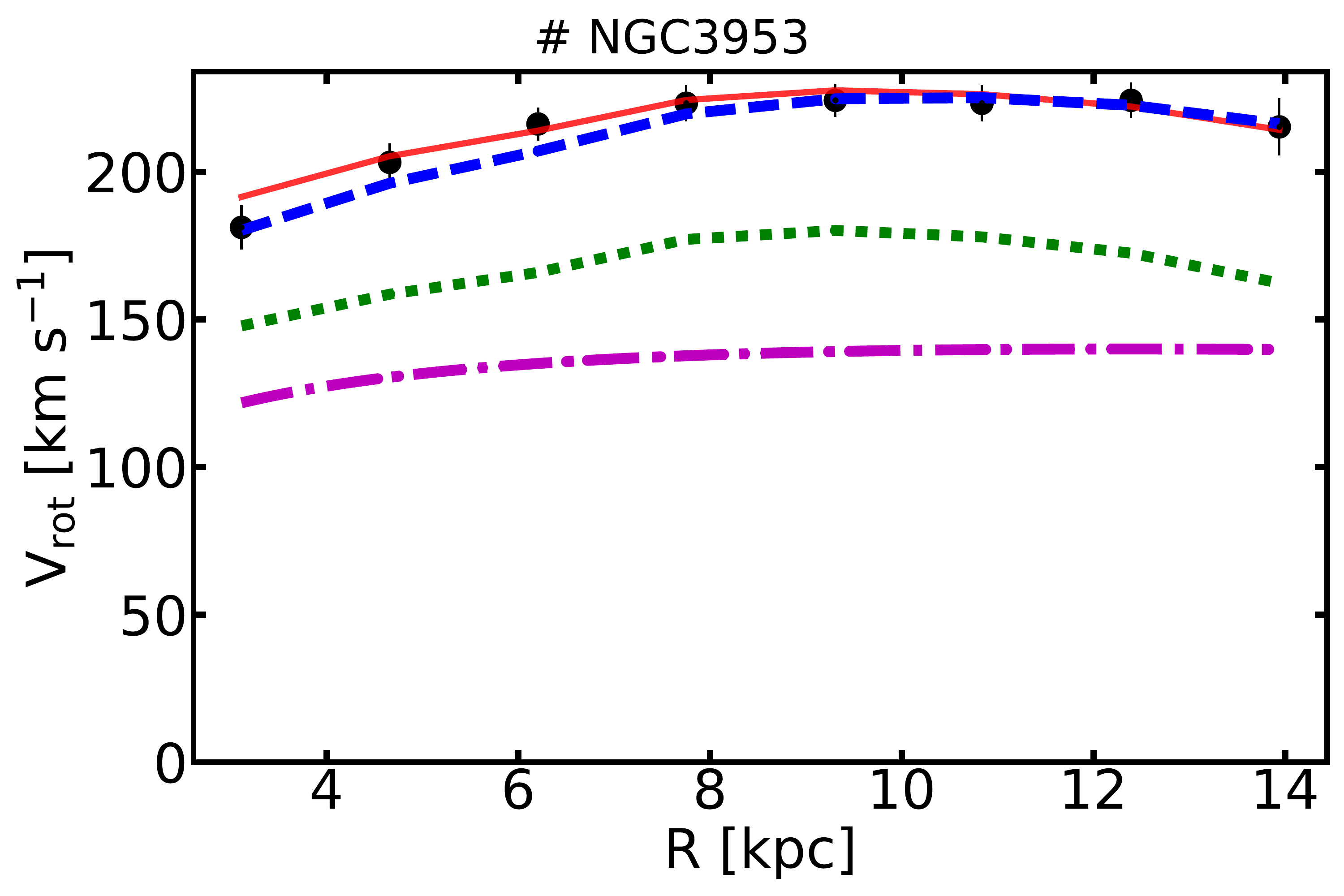}

\includegraphics[width=.33\textwidth]{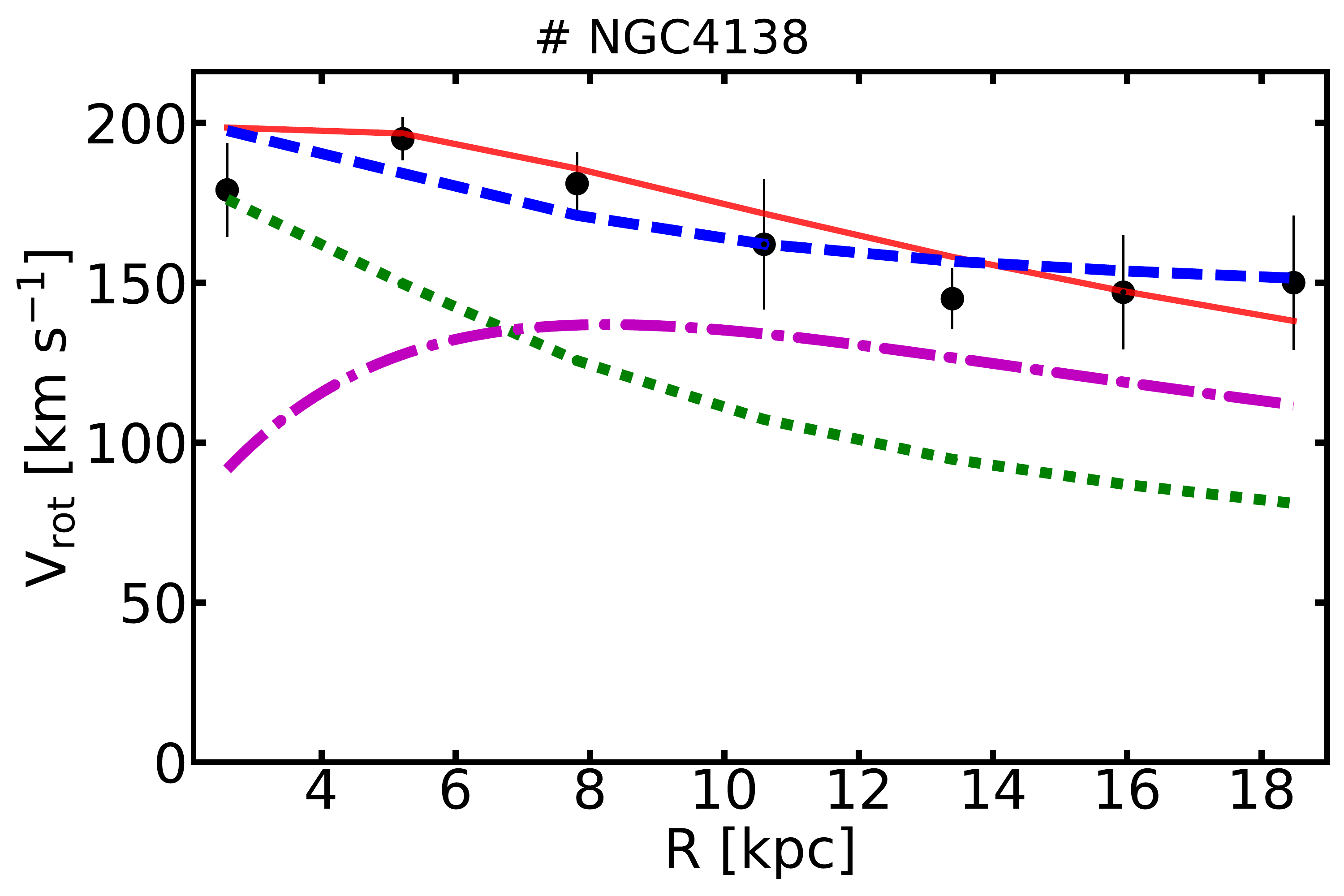}\hfill
\includegraphics[width=.33\textwidth]{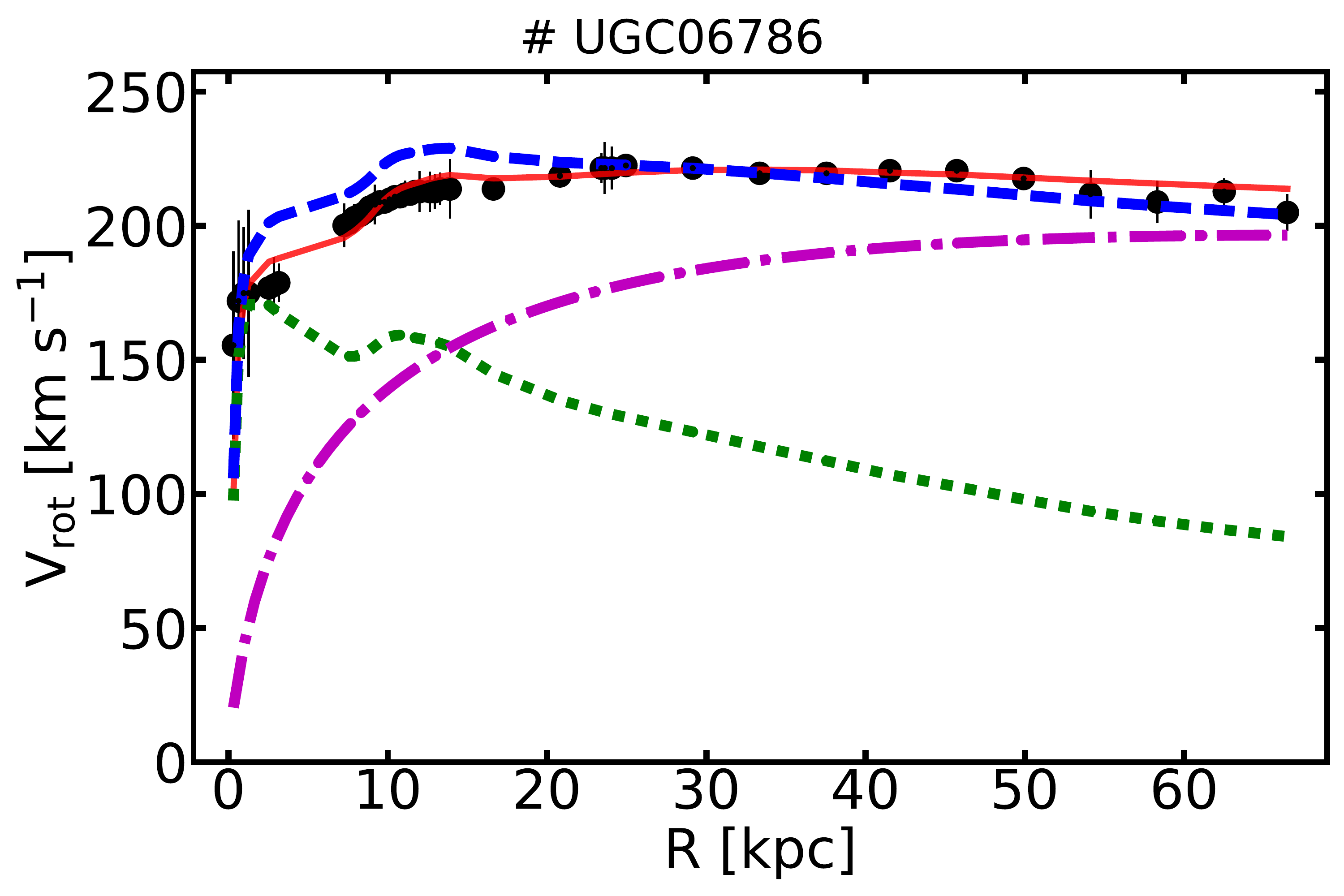}\hfill
\includegraphics[width=.33\textwidth]{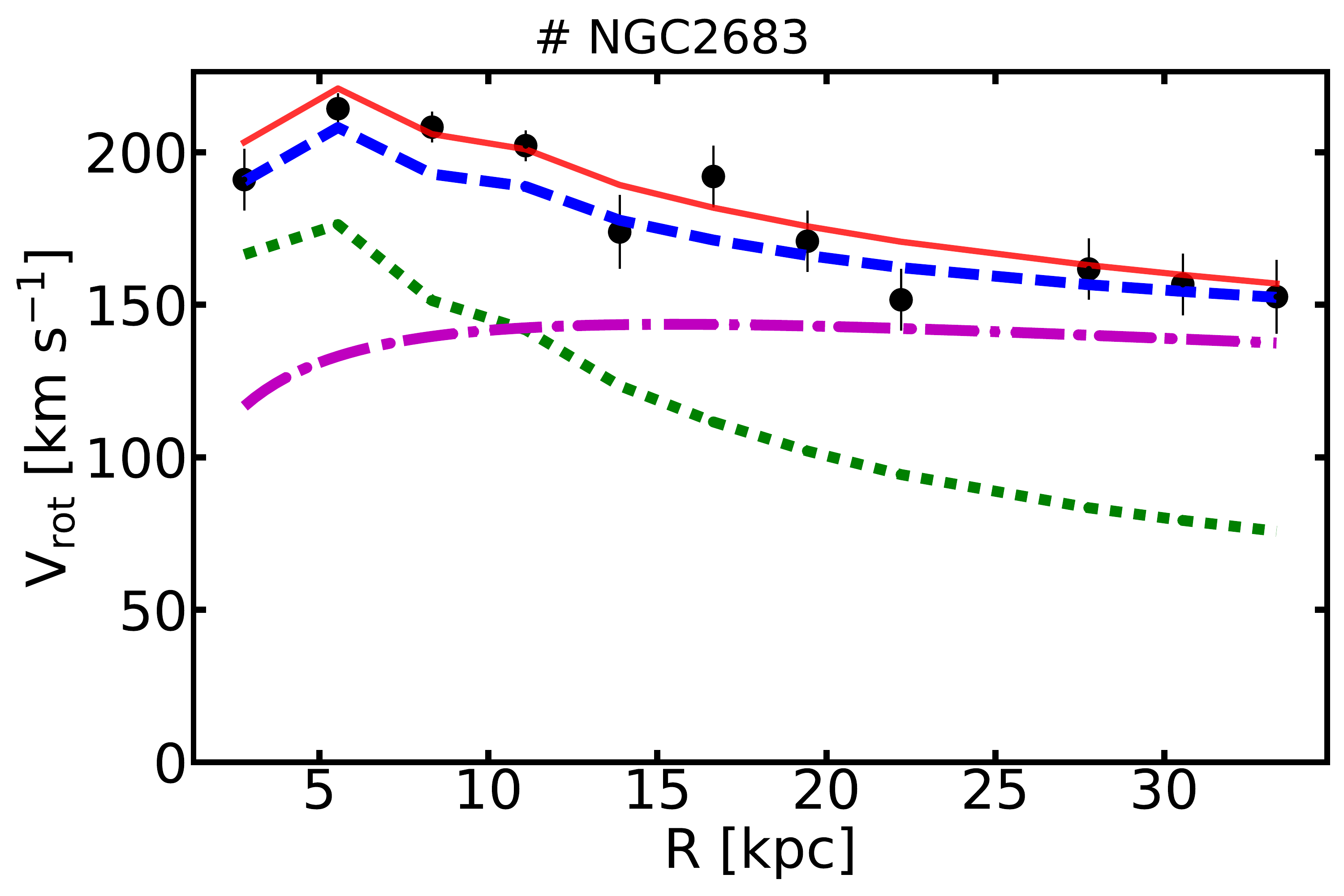}

\includegraphics[width=.33\textwidth]{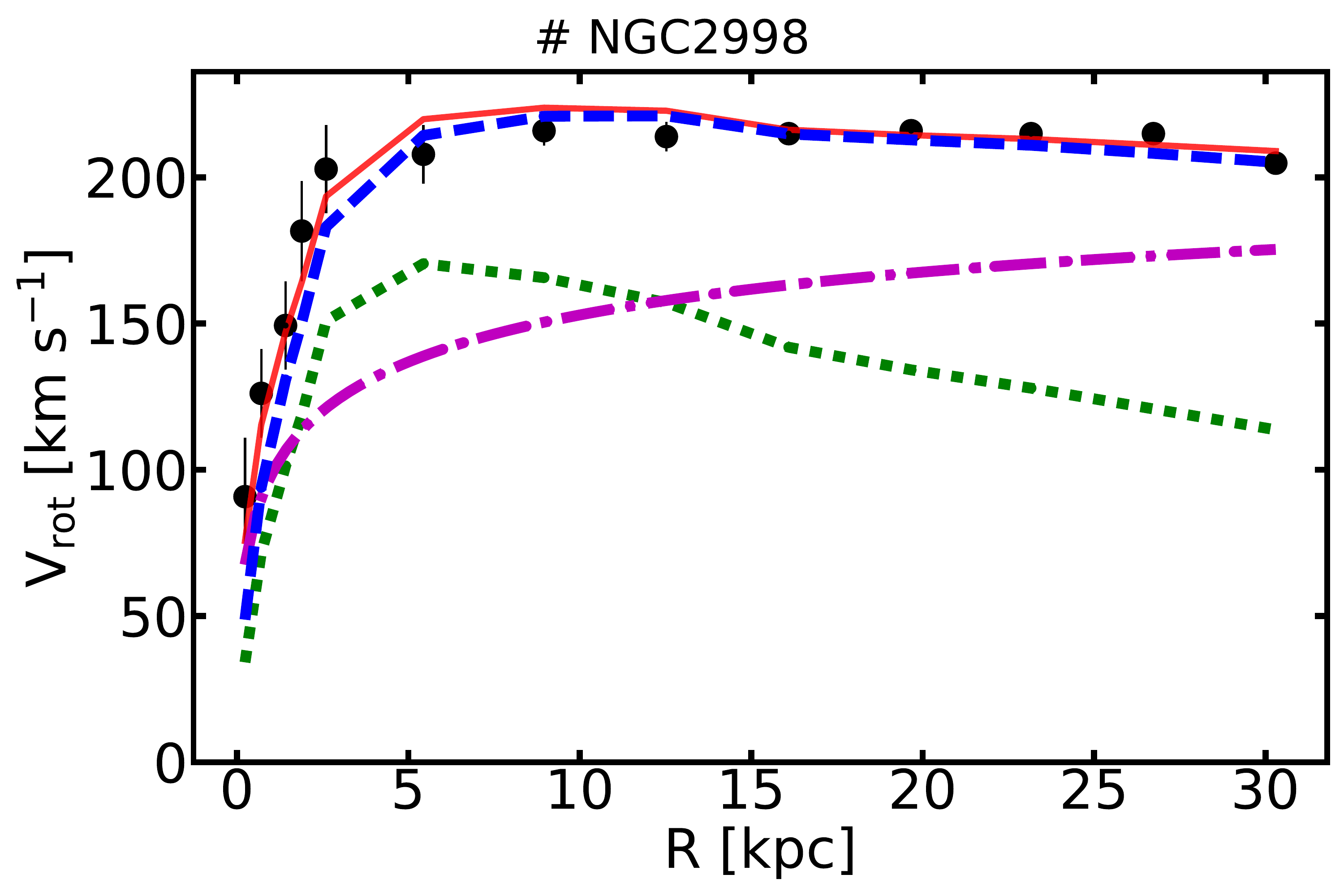}\hfill
\includegraphics[width=.33\textwidth]{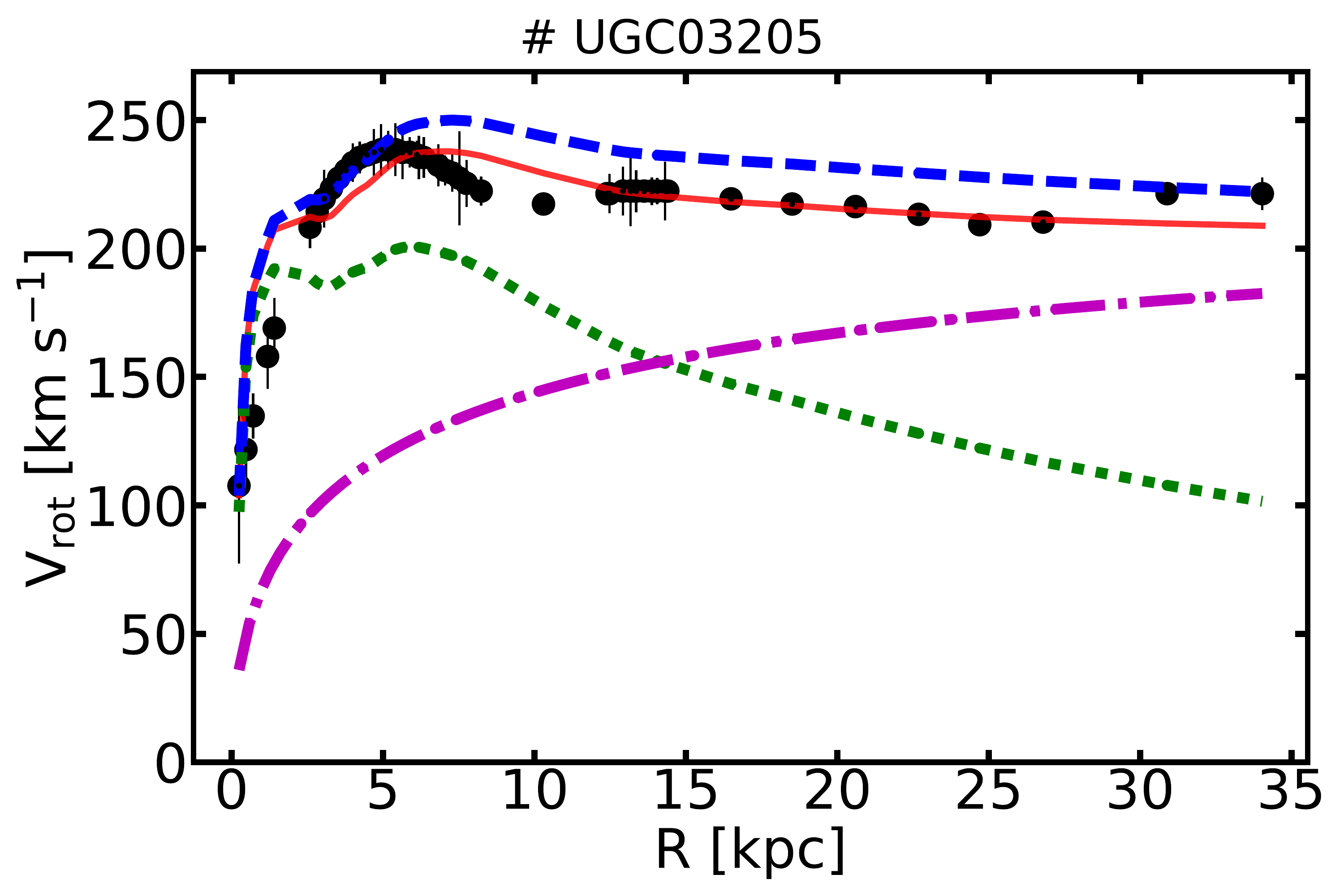}\hfill
\includegraphics[width=.33\textwidth]{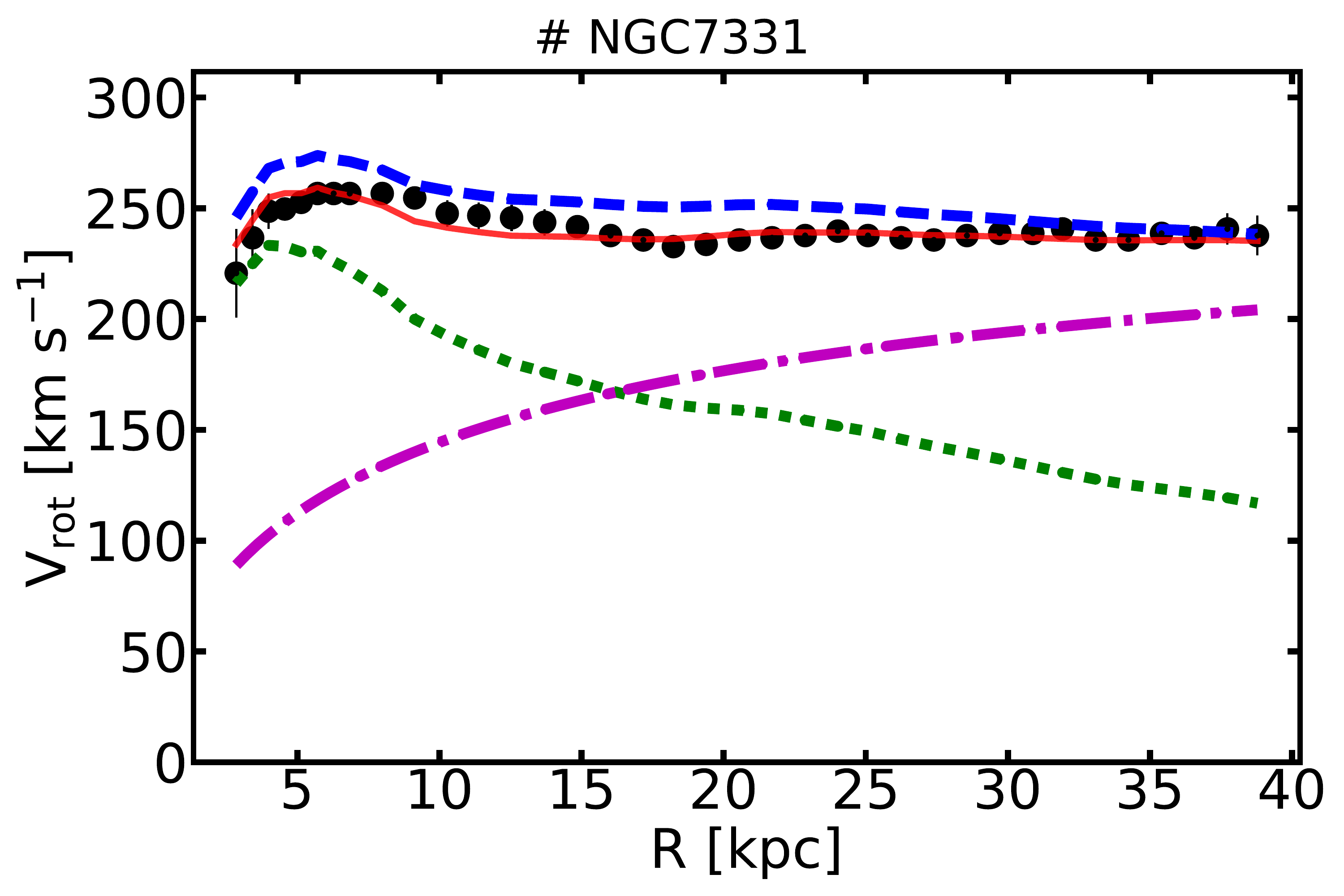}

\includegraphics[width=.33\textwidth]{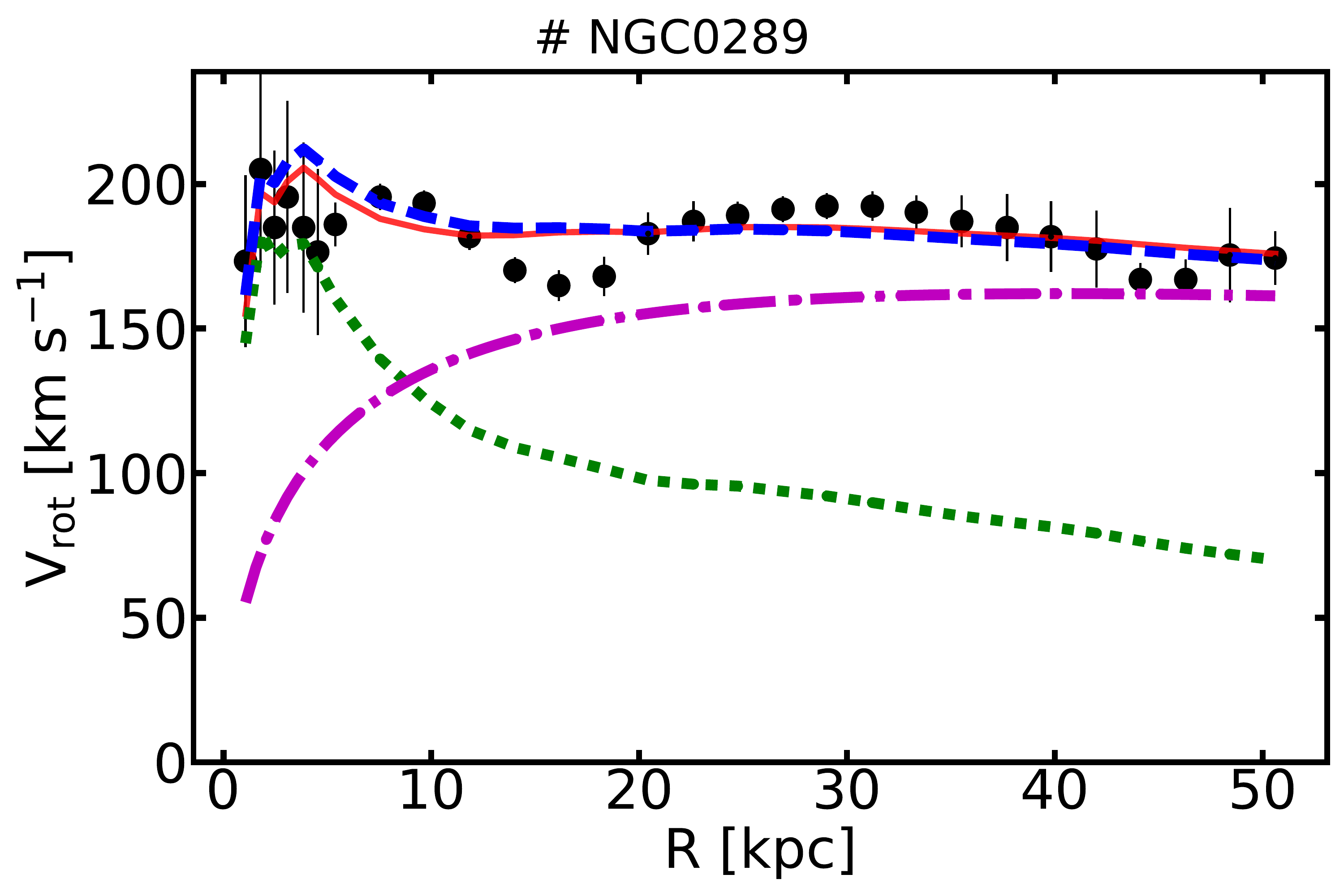}\hfill
\includegraphics[width=.33\textwidth]{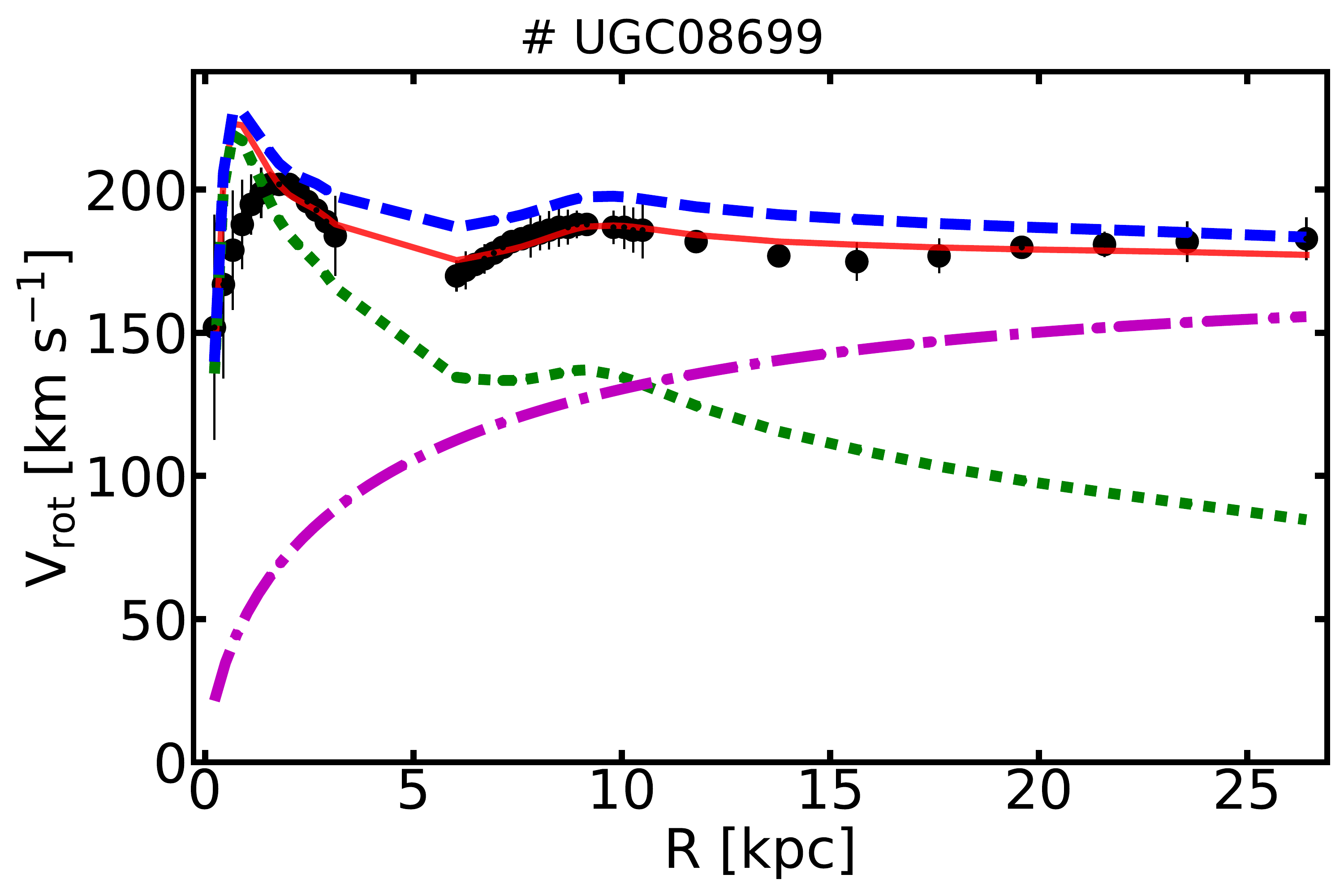}\hfill
\includegraphics[width=.33\textwidth]{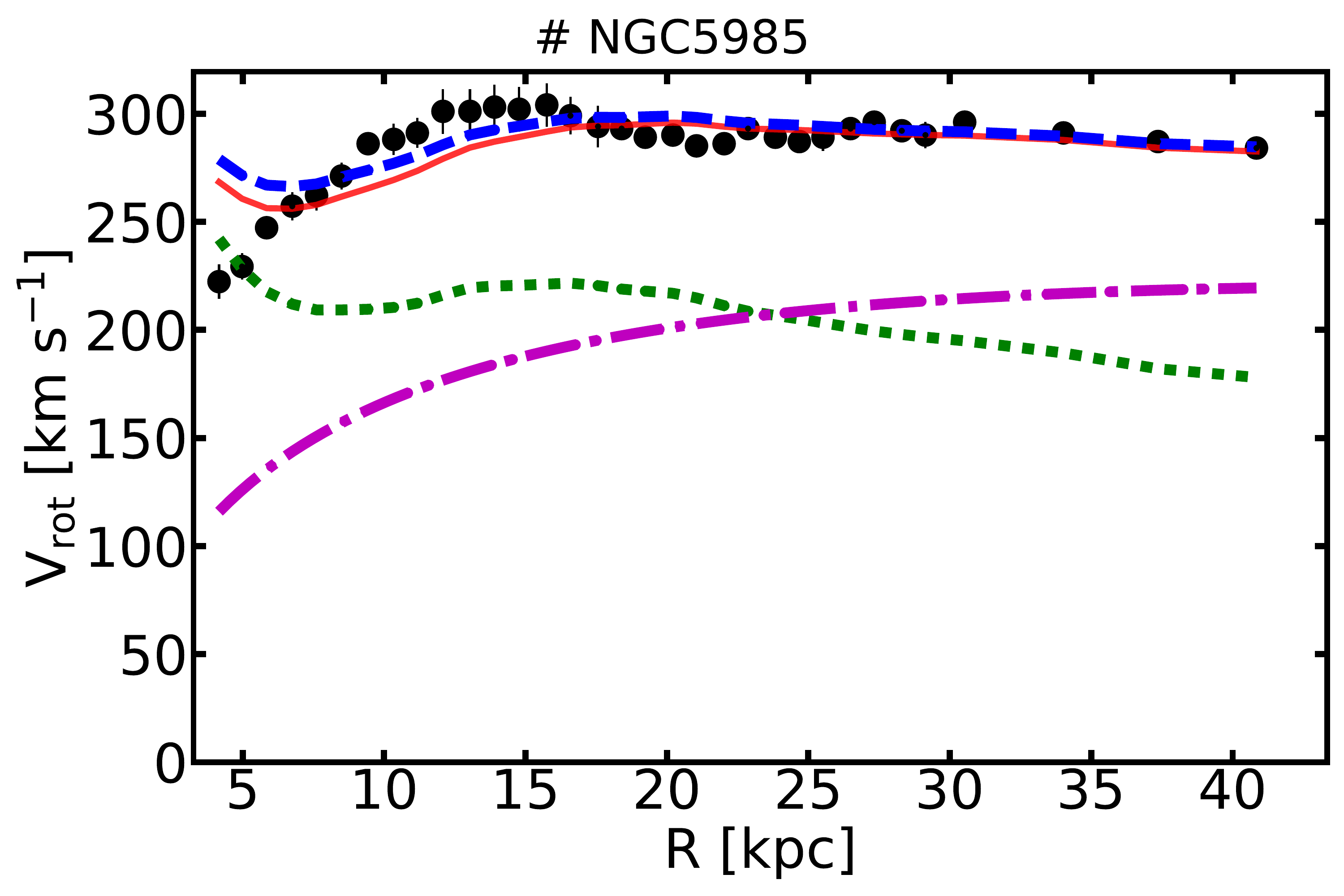}

\includegraphics[width=.33\textwidth]{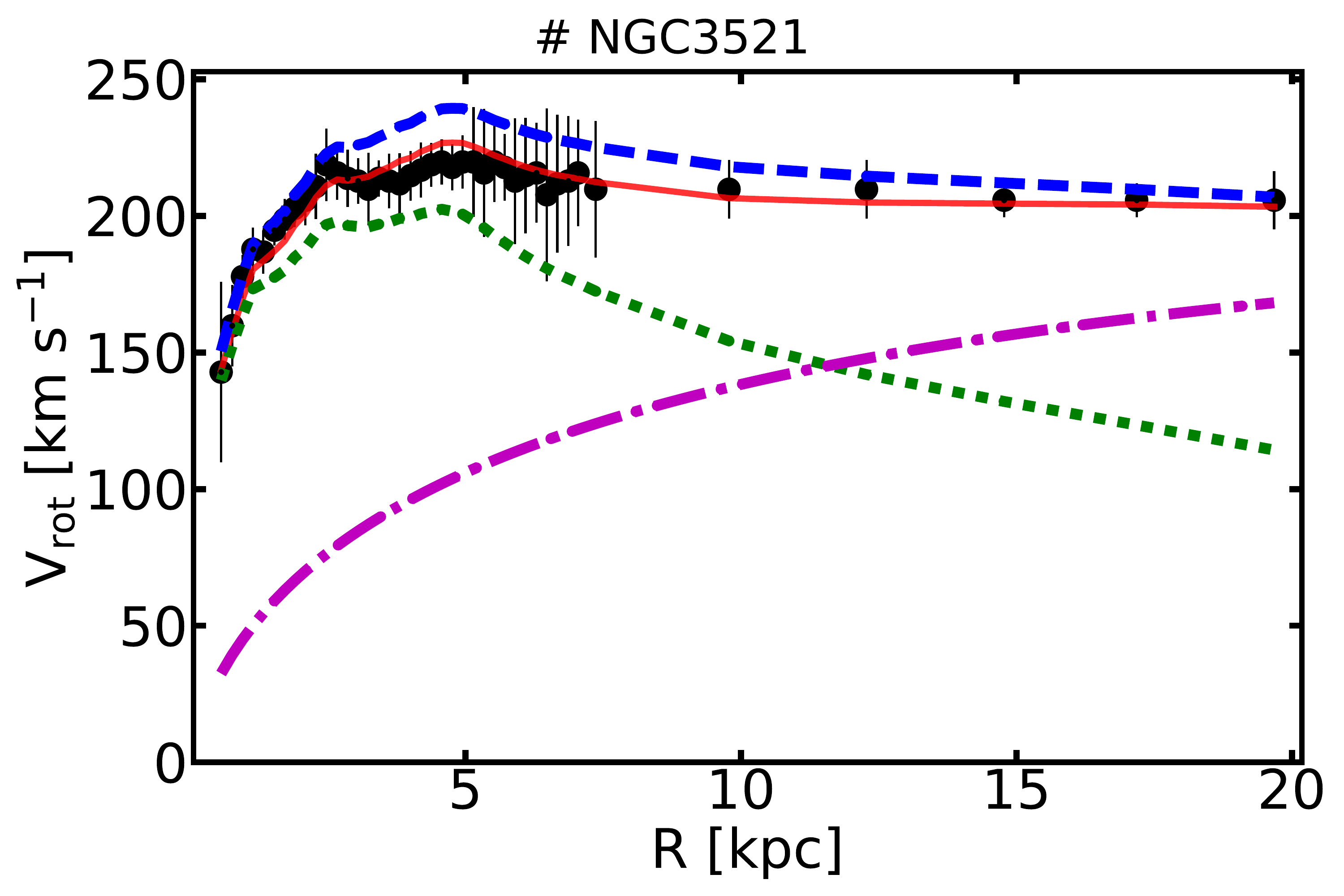}\hfill
\includegraphics[width=.33\textwidth]{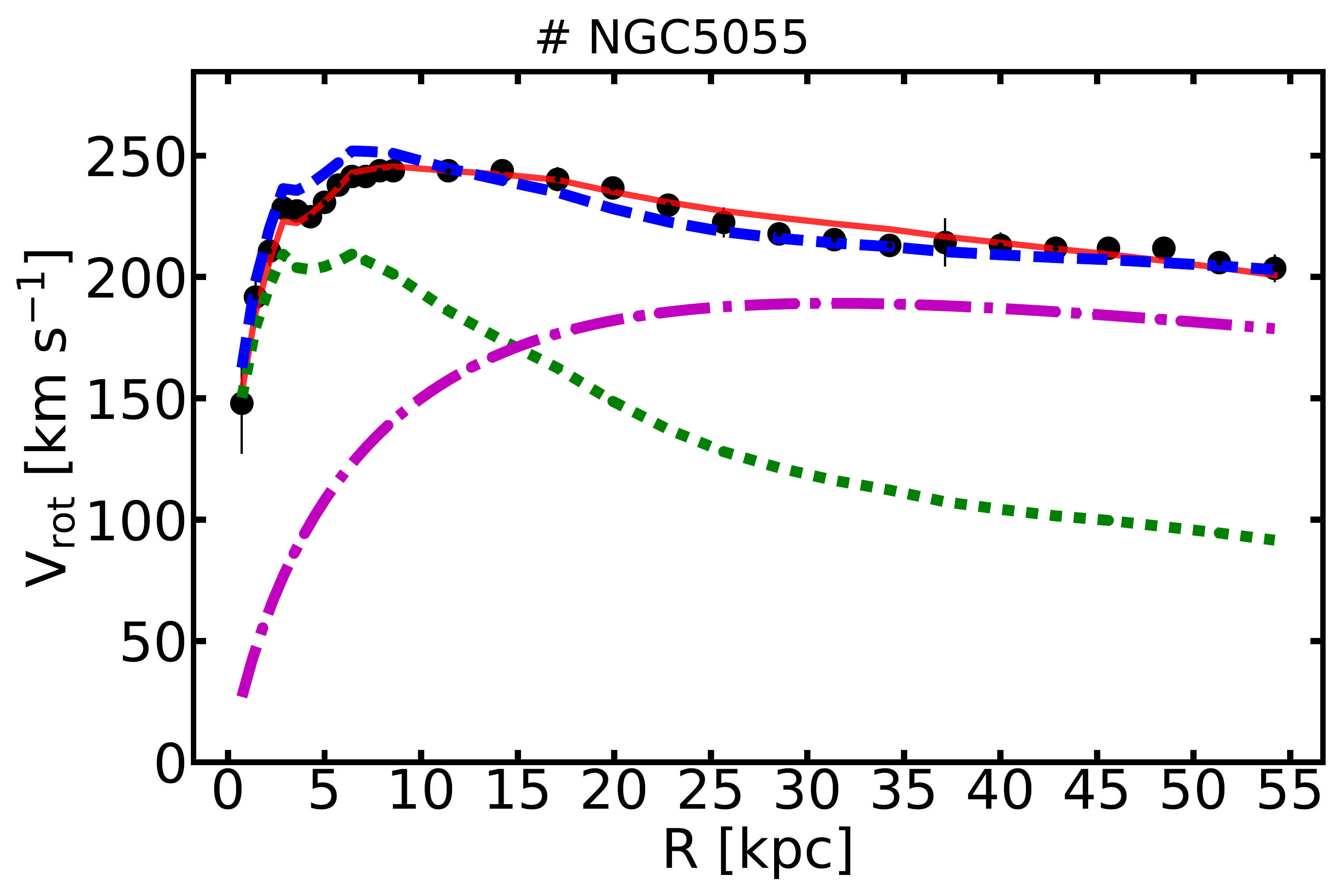}\hfill
\includegraphics[width=.33\textwidth]{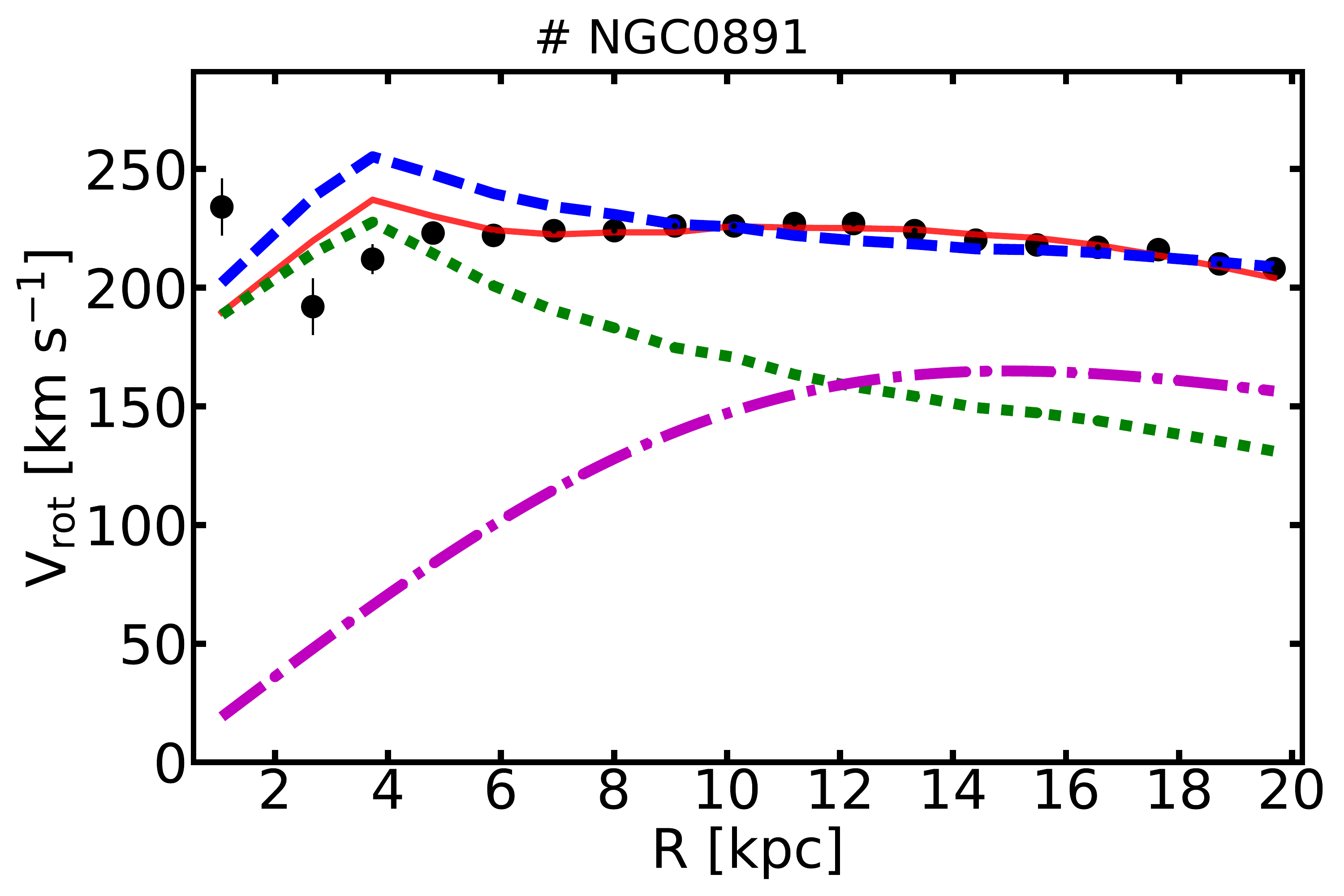}

\caption{Continued.}
\end{figure*}

\begin{figure*}

\centering
\ContinuedFloat

\includegraphics[width=.33\textwidth]{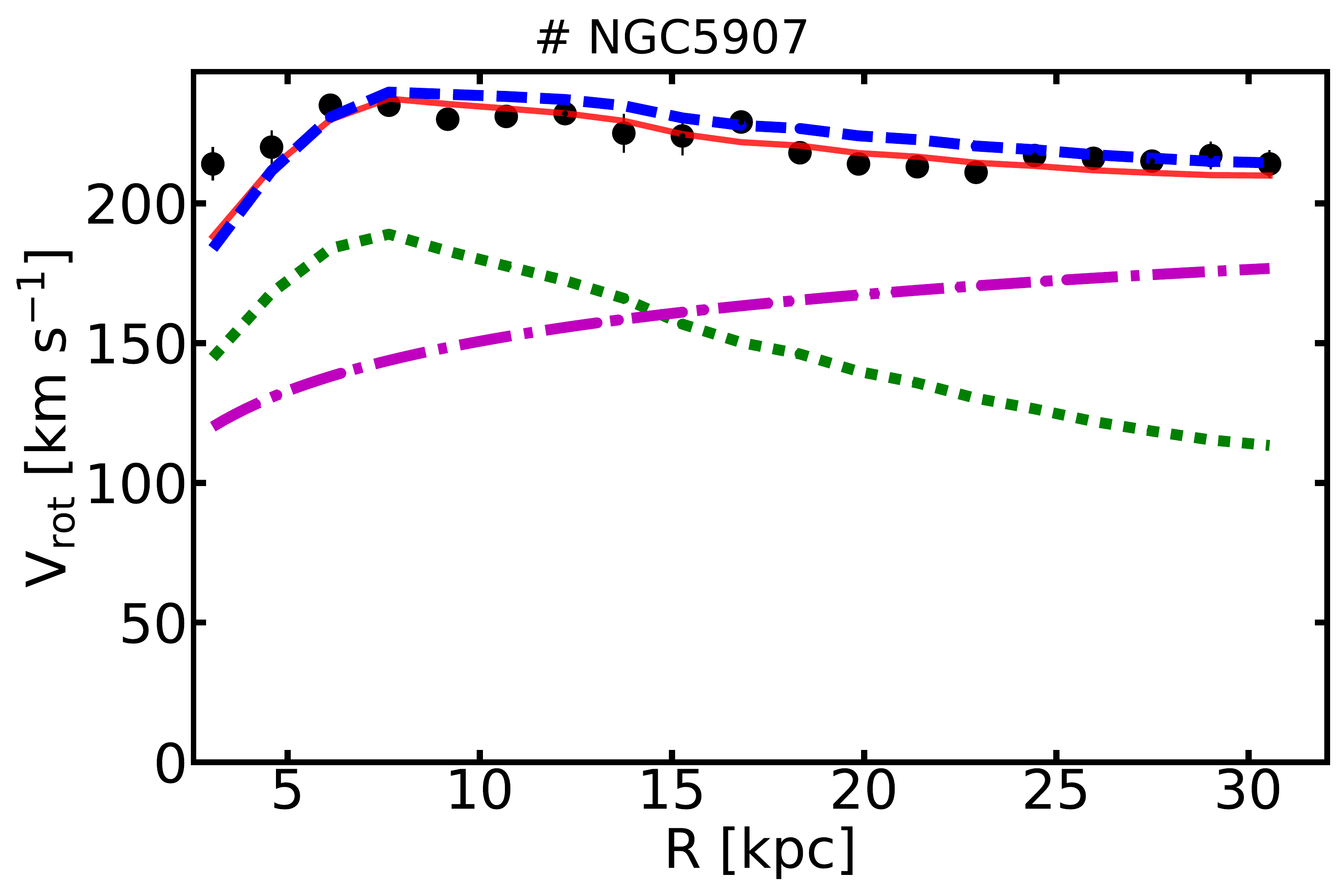}\hfill
\includegraphics[width=.33\textwidth]{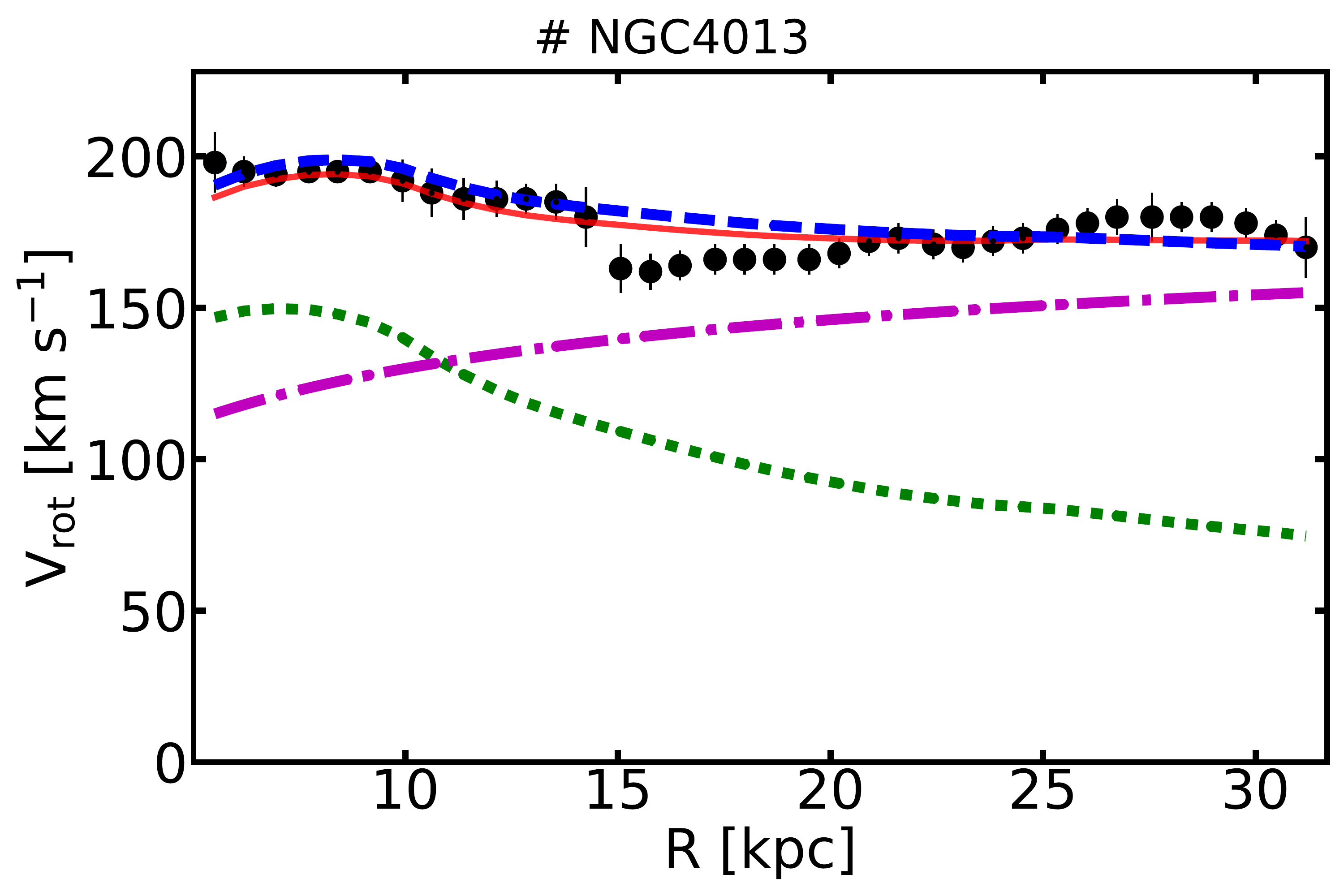}\hfill
\includegraphics[width=.33\textwidth]{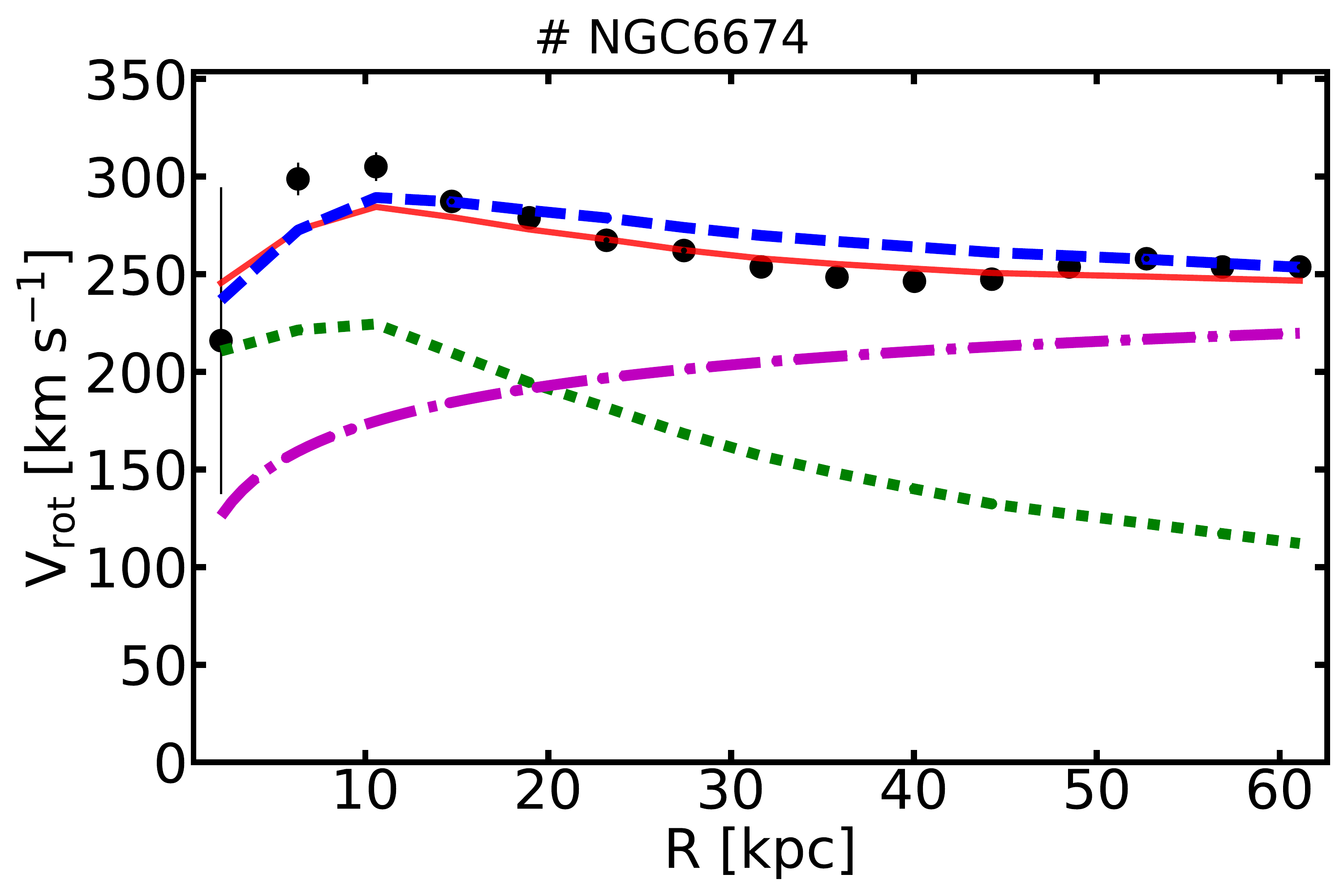}

\includegraphics[width=.33\textwidth]{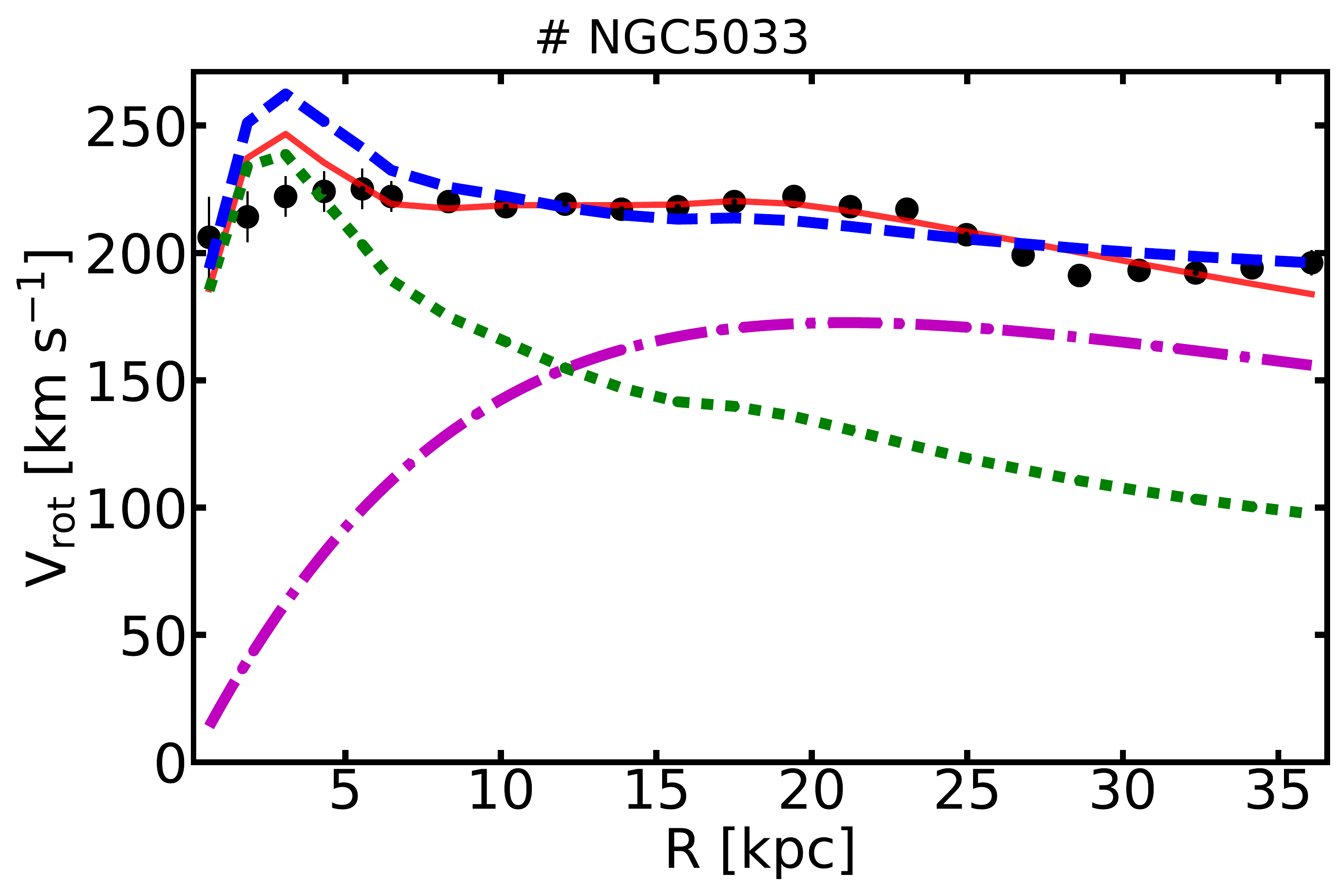}\hfill
\includegraphics[width=.33\textwidth]{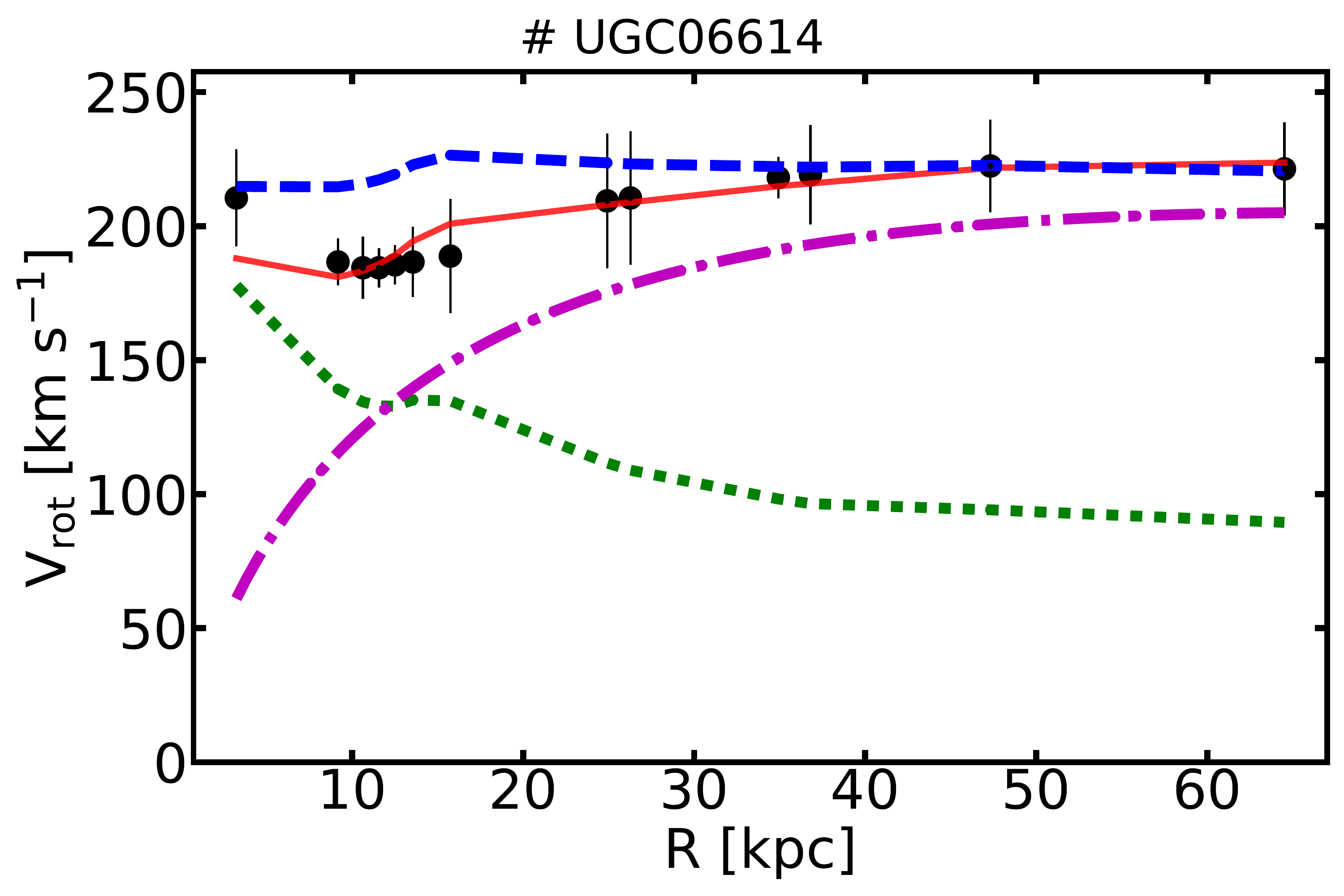}\hfill
\includegraphics[width=.33\textwidth]{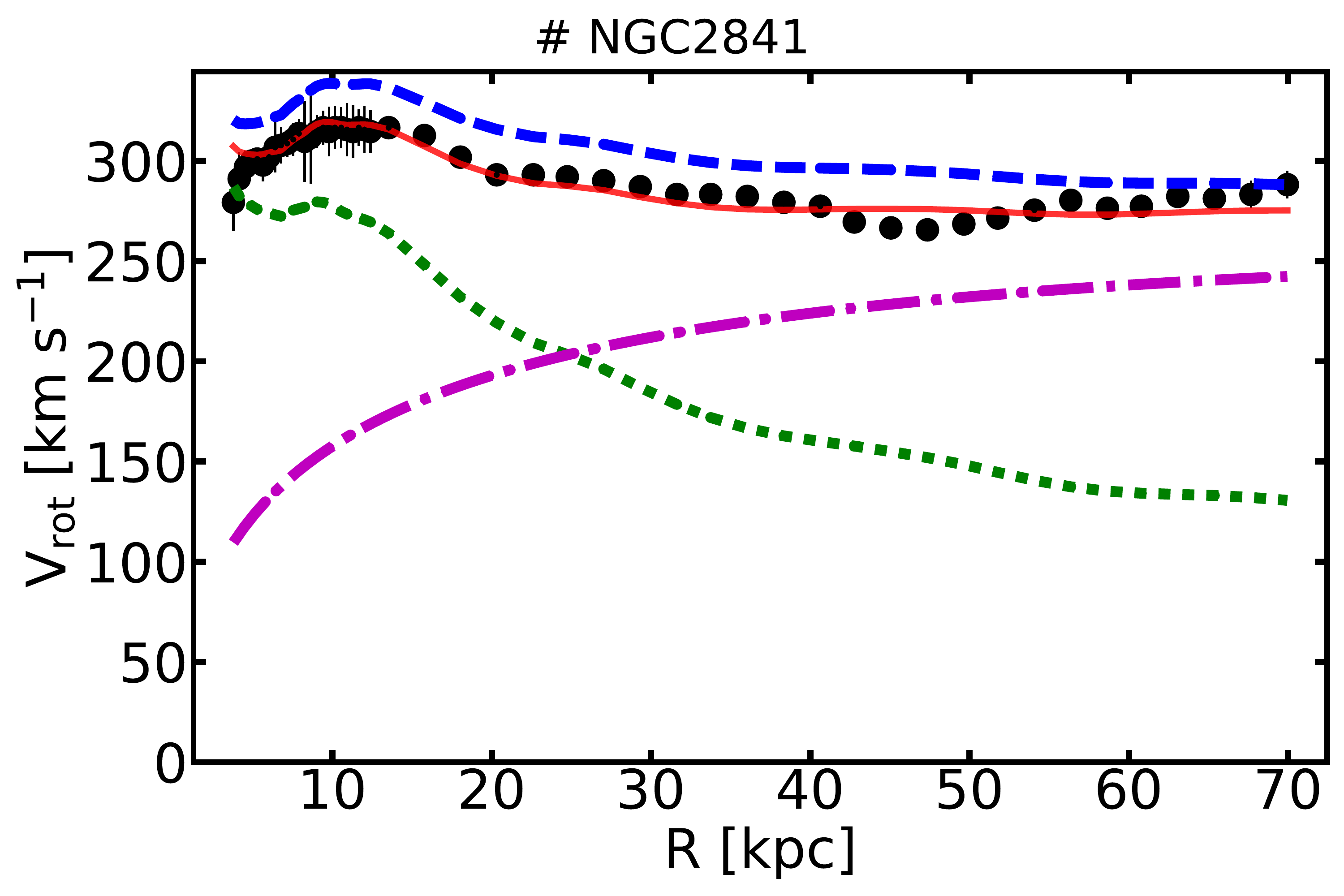}

\includegraphics[width=.33\textwidth]{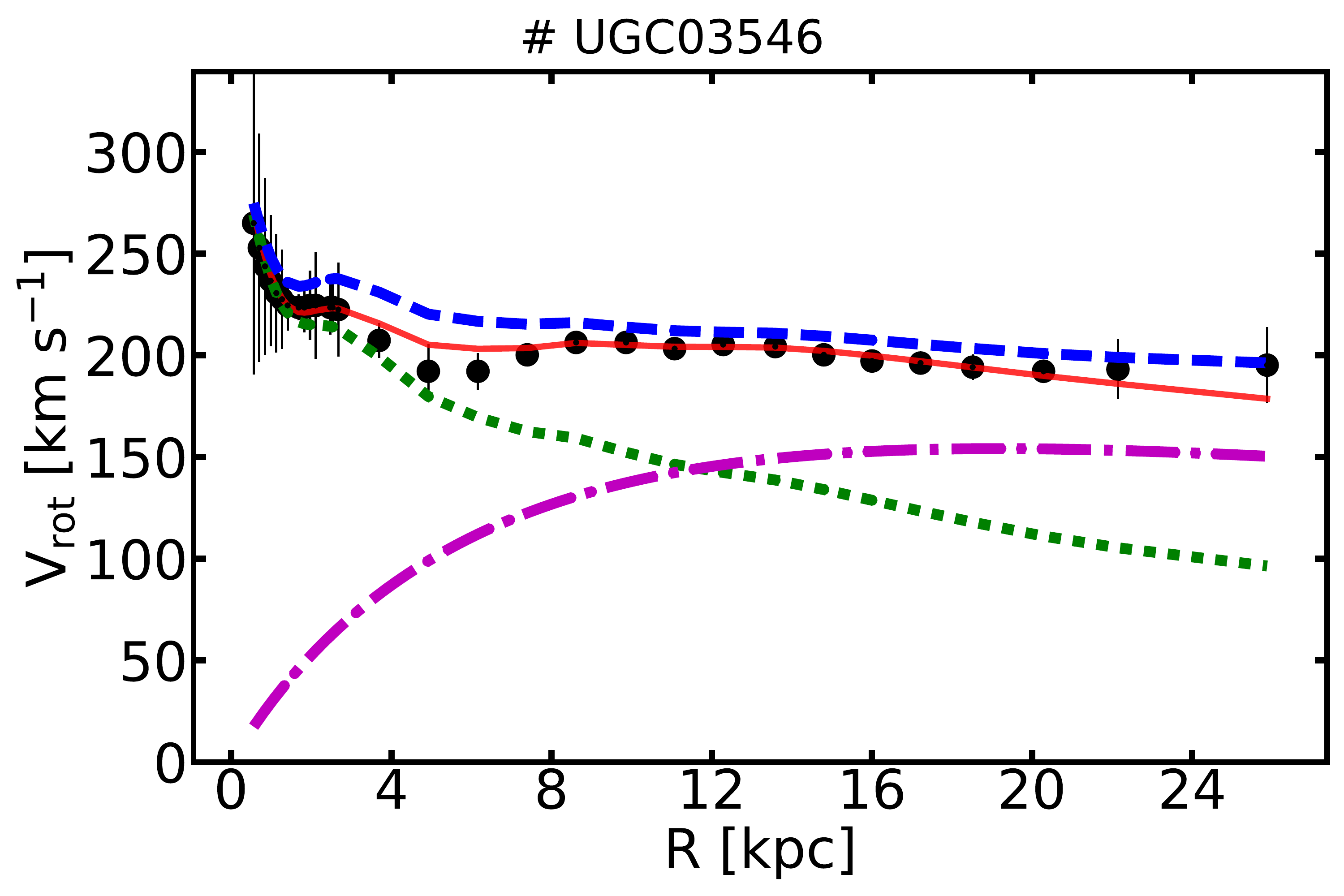}\hfill
\includegraphics[width=.33\textwidth]{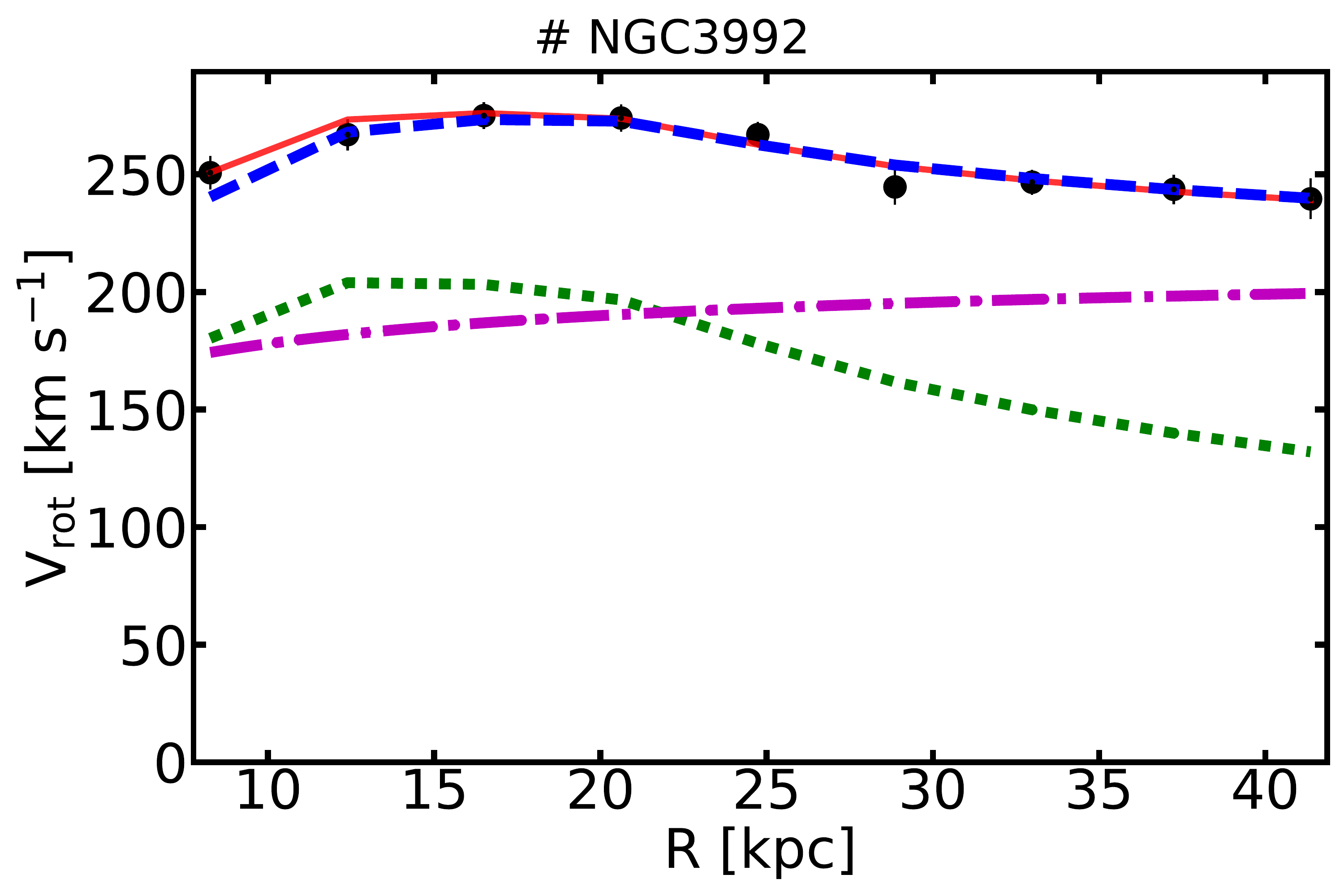}\hfill
\includegraphics[width=.33\textwidth]{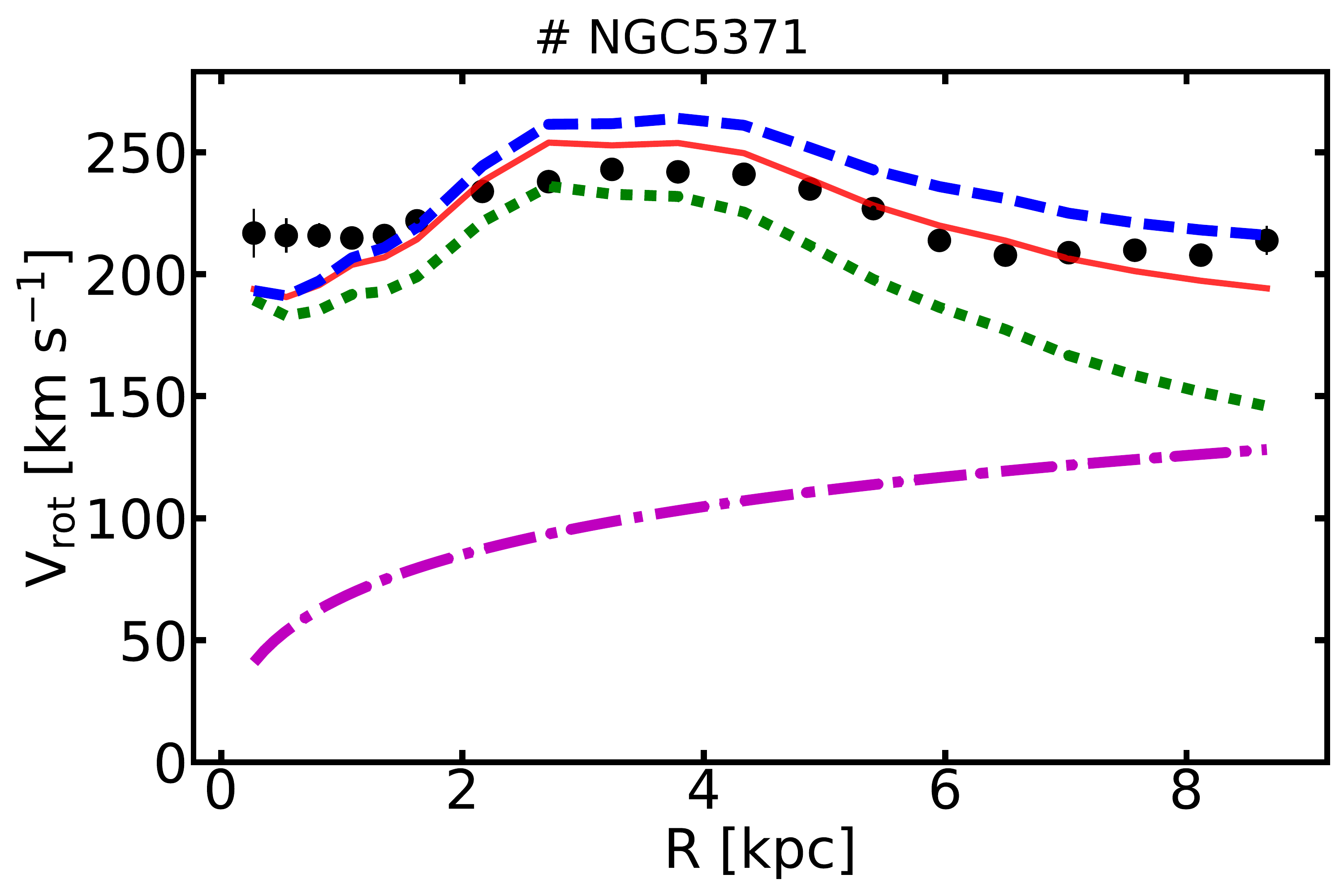}

\includegraphics[width=.33\textwidth]{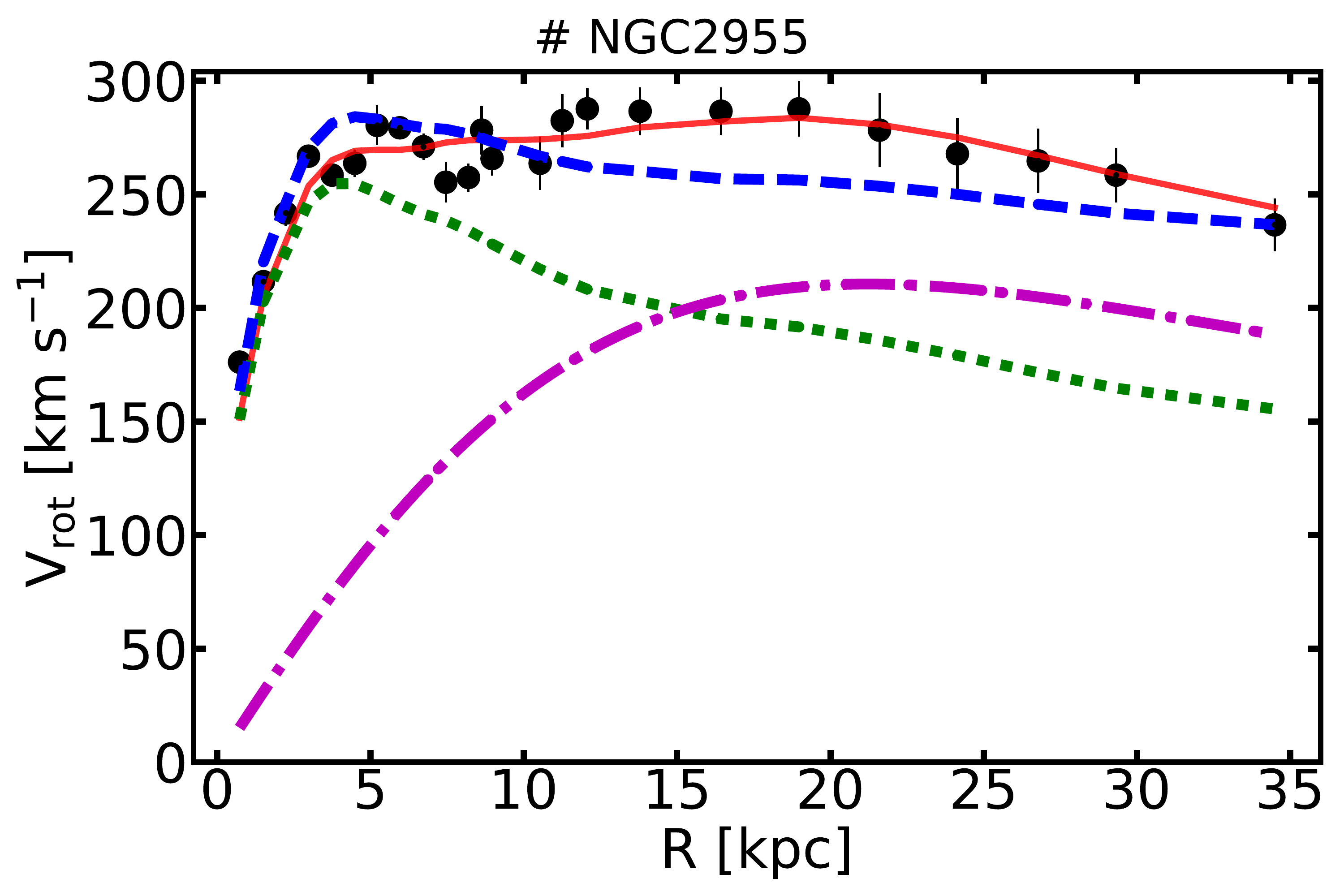}\hfill
\includegraphics[width=.33\textwidth]{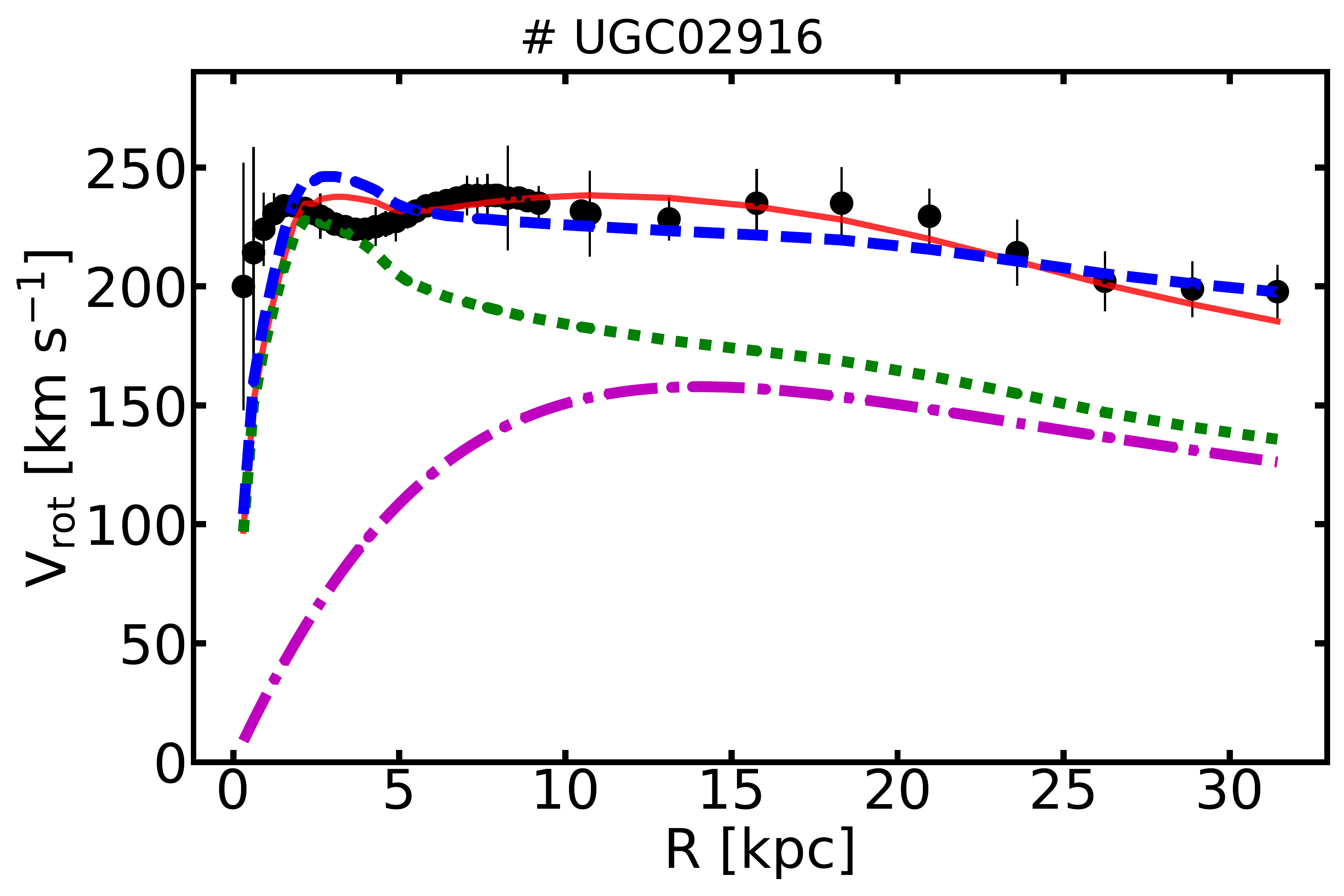}\hfill
\includegraphics[width=.33\textwidth]{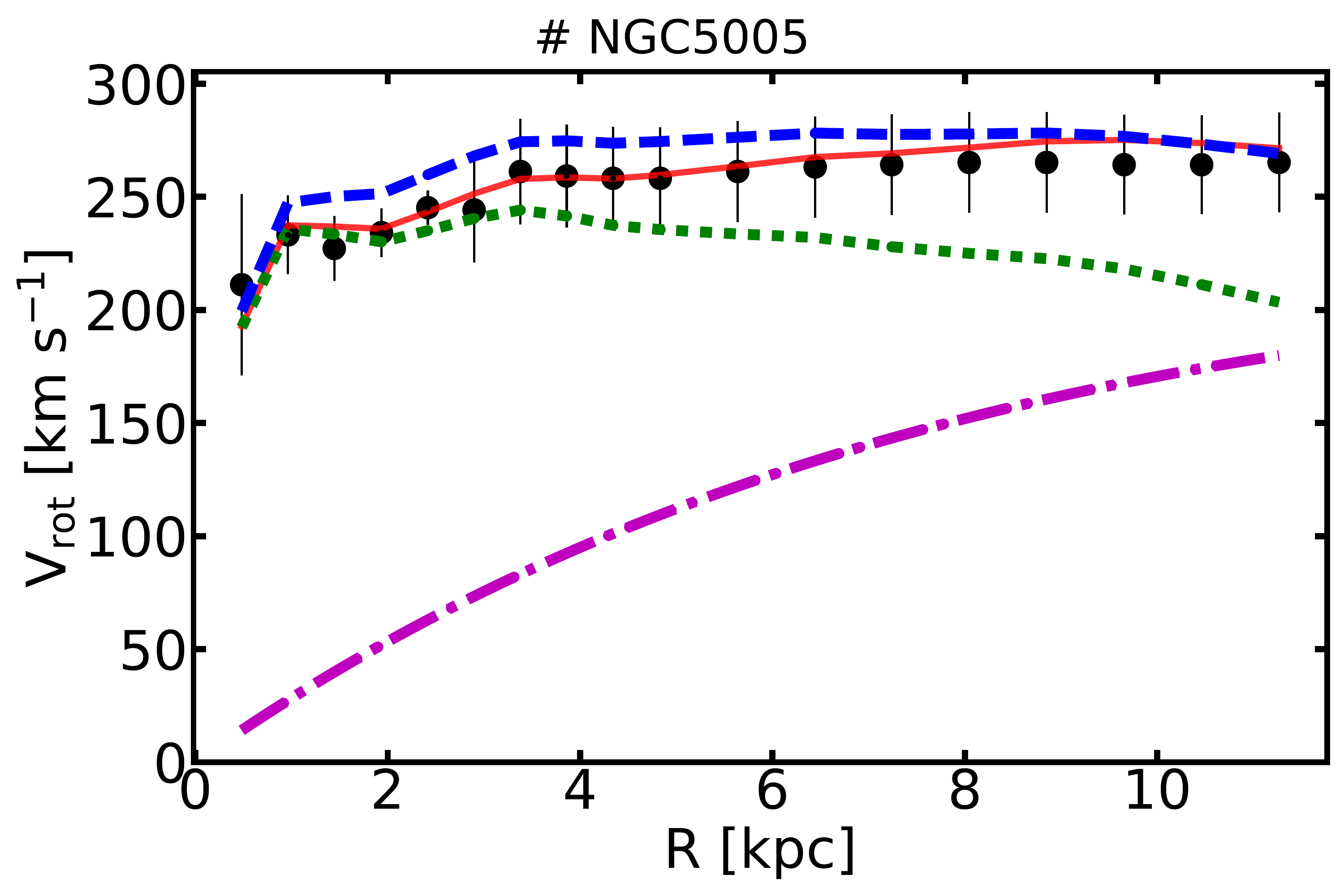}

\includegraphics[width=.33\textwidth]{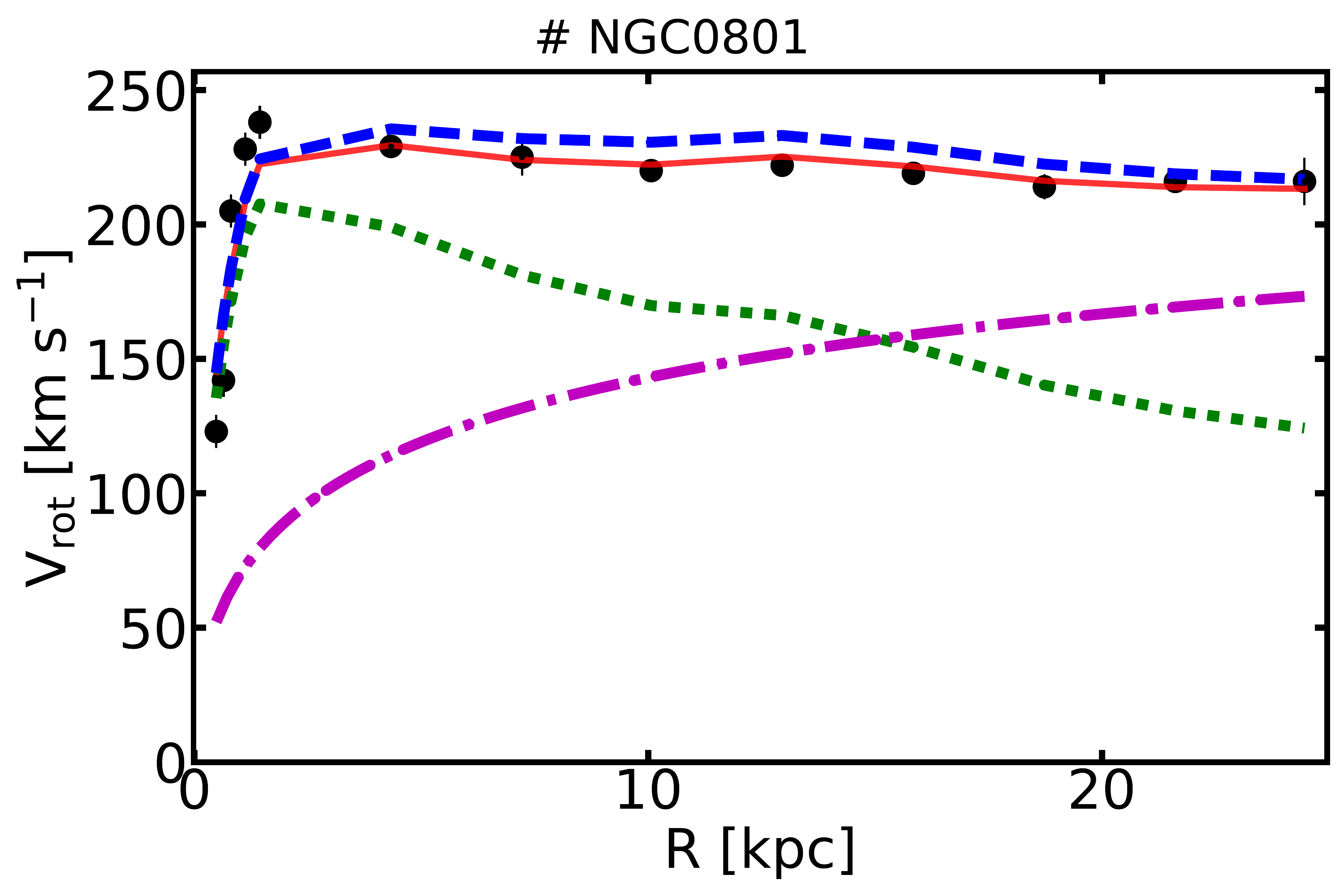}\hfill
\includegraphics[width=.33\textwidth]{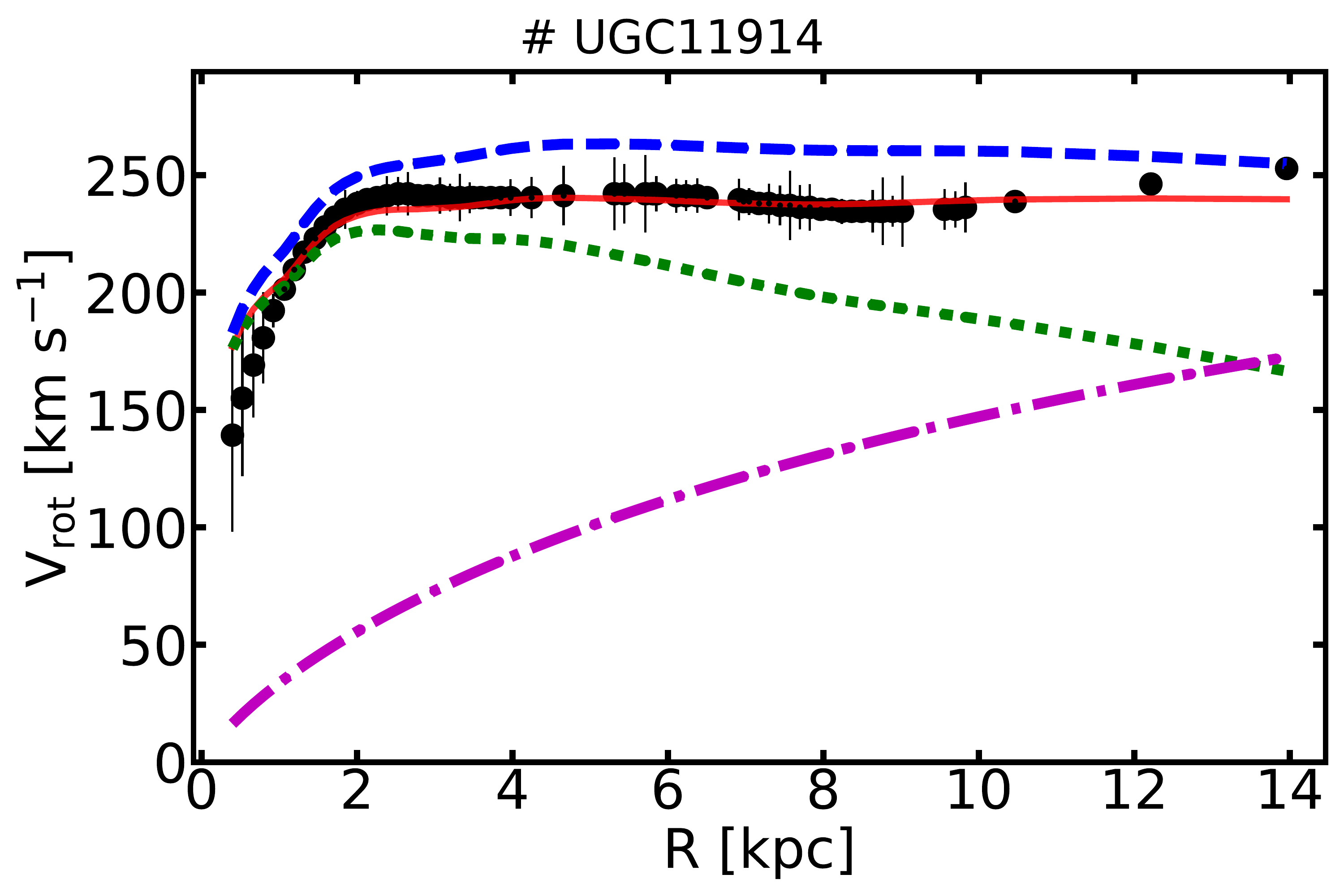}\hfill
\includegraphics[width=.33\textwidth]{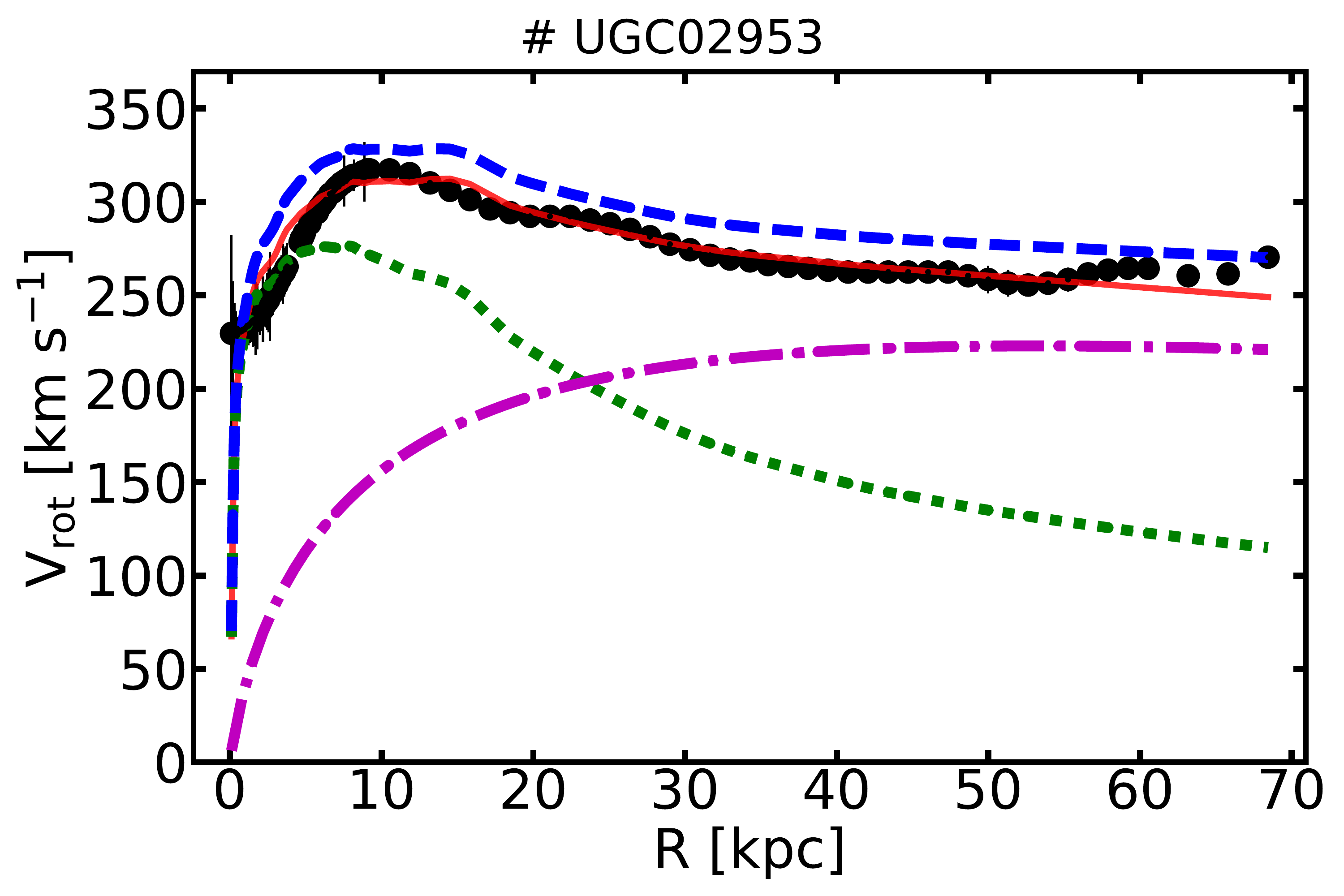}
\caption{Continued.}
\end{figure*}

\begin{figure*}

\centering
\ContinuedFloat
\includegraphics[width=.33\textwidth]{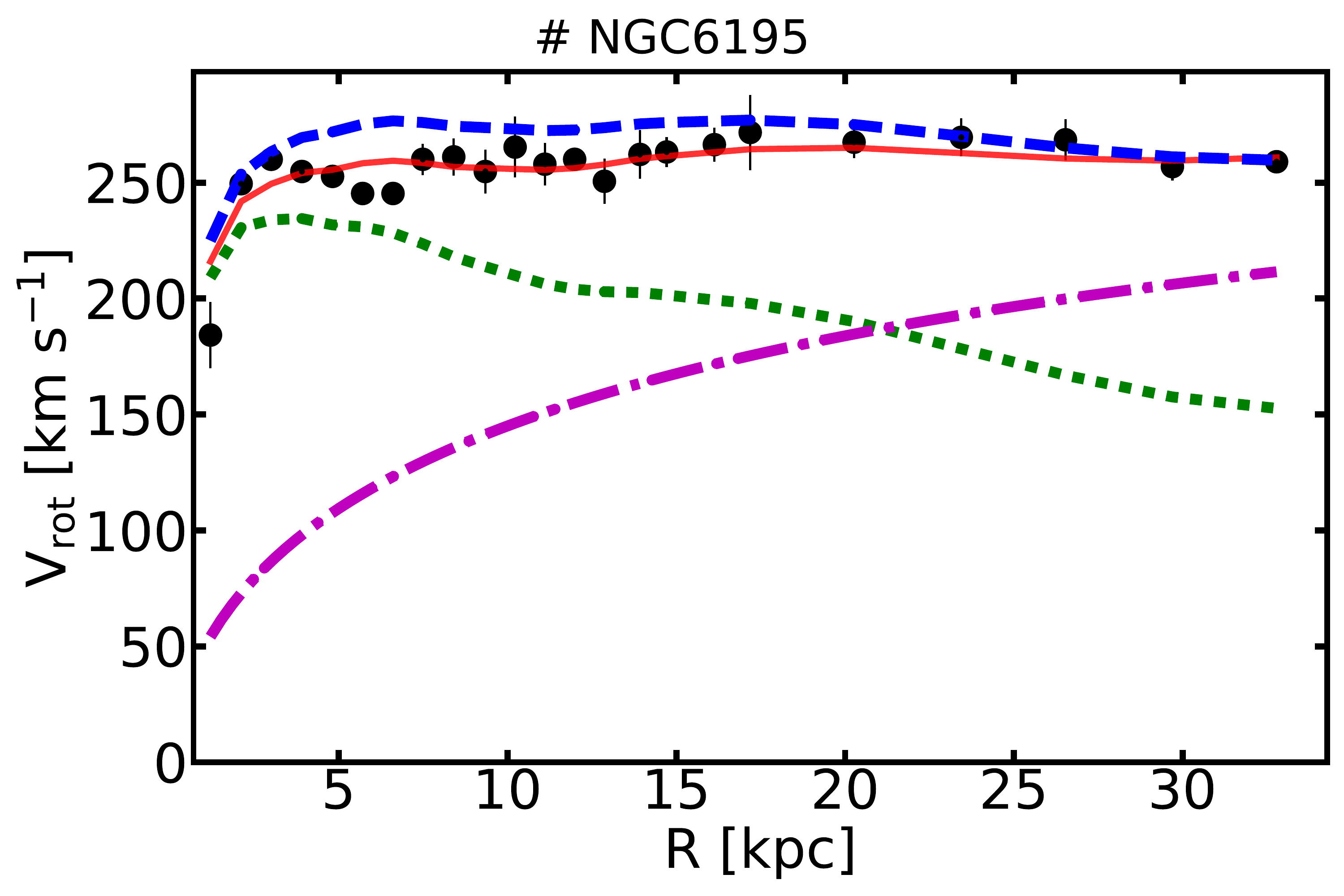}\hfill
\includegraphics[width=.33\textwidth]{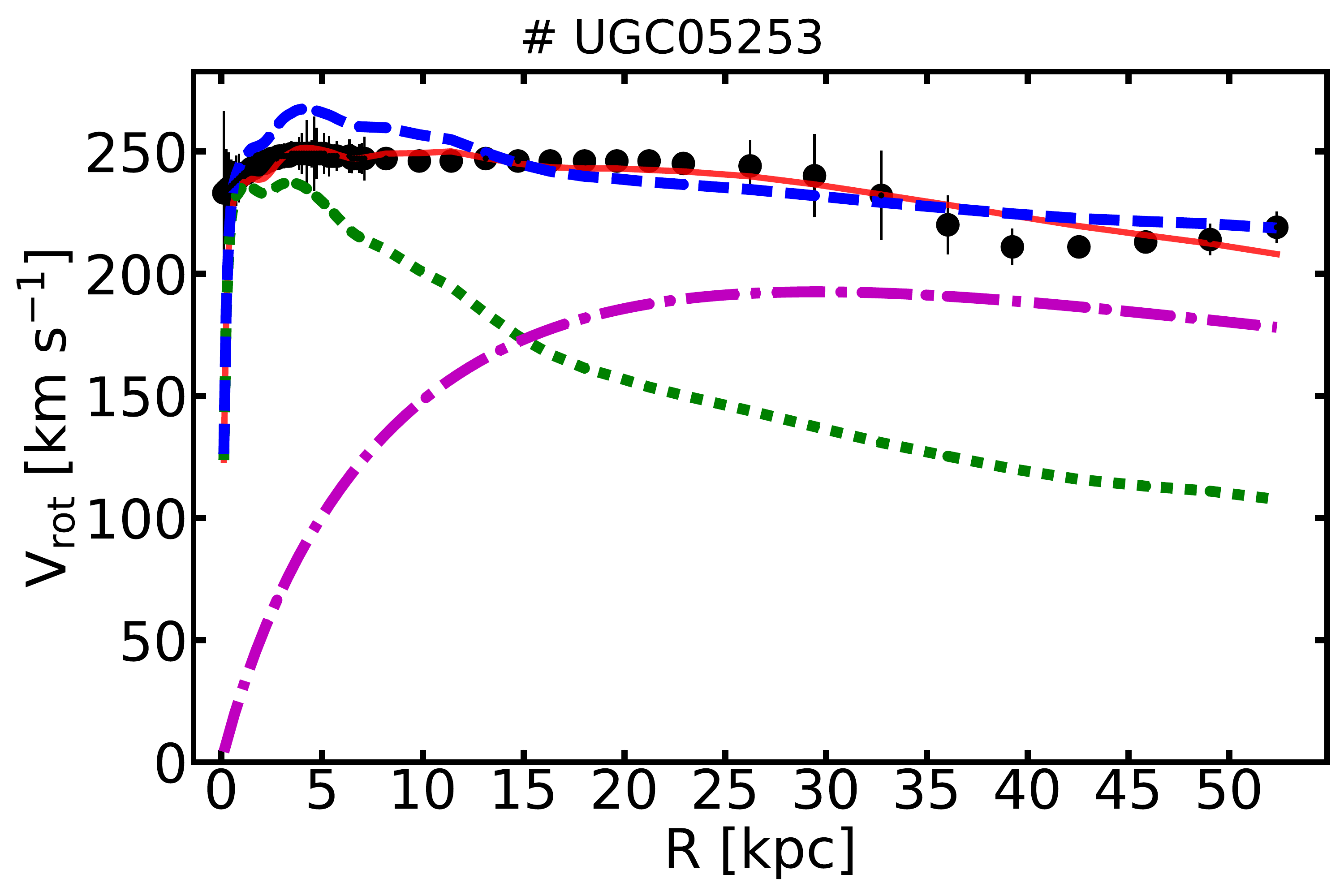}\hfill
\includegraphics[width=.33\textwidth]{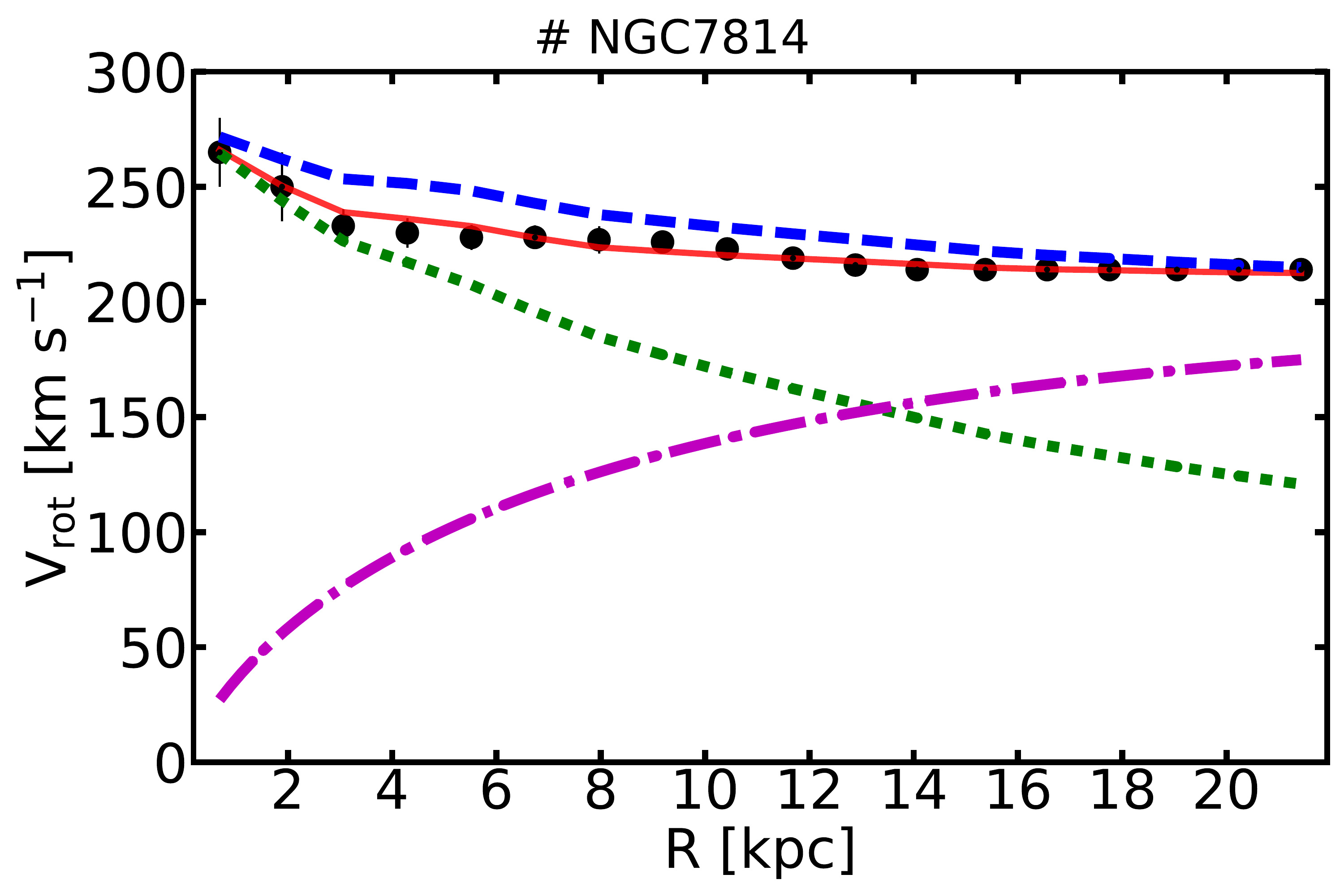}

\includegraphics[width=.33\textwidth]{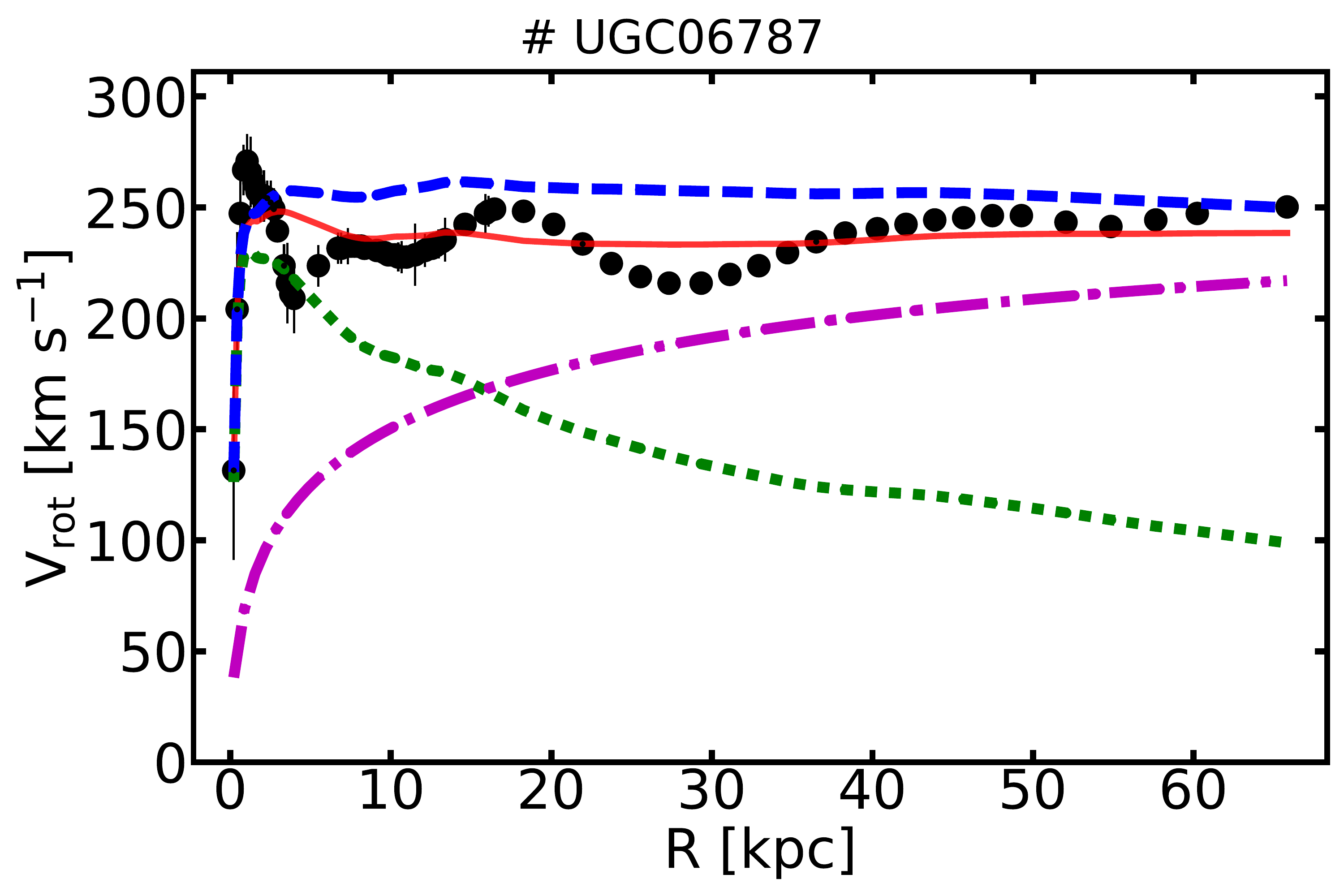}\hfill
\includegraphics[width=.33\textwidth]{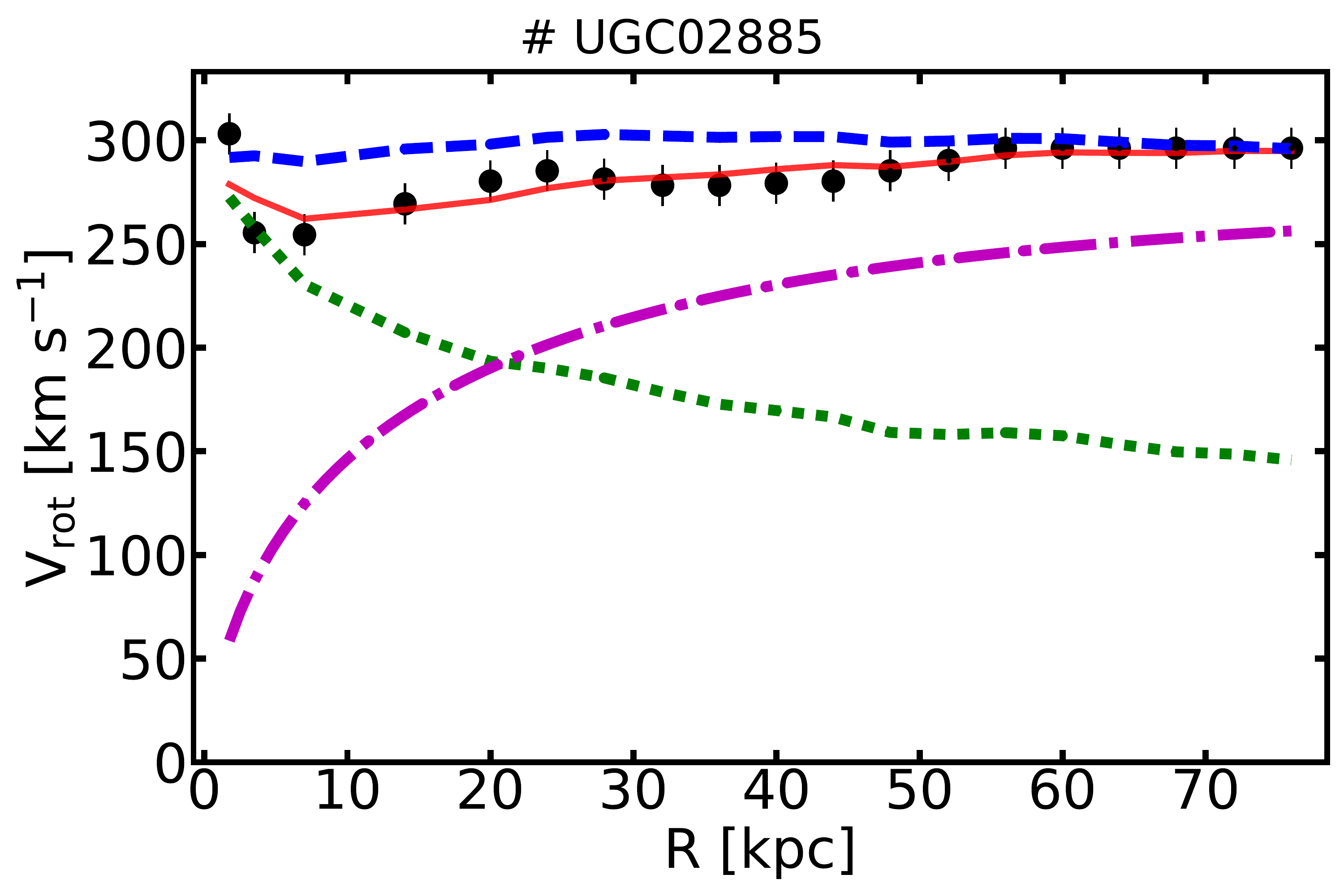}\hfill
\includegraphics[width=.33\textwidth]{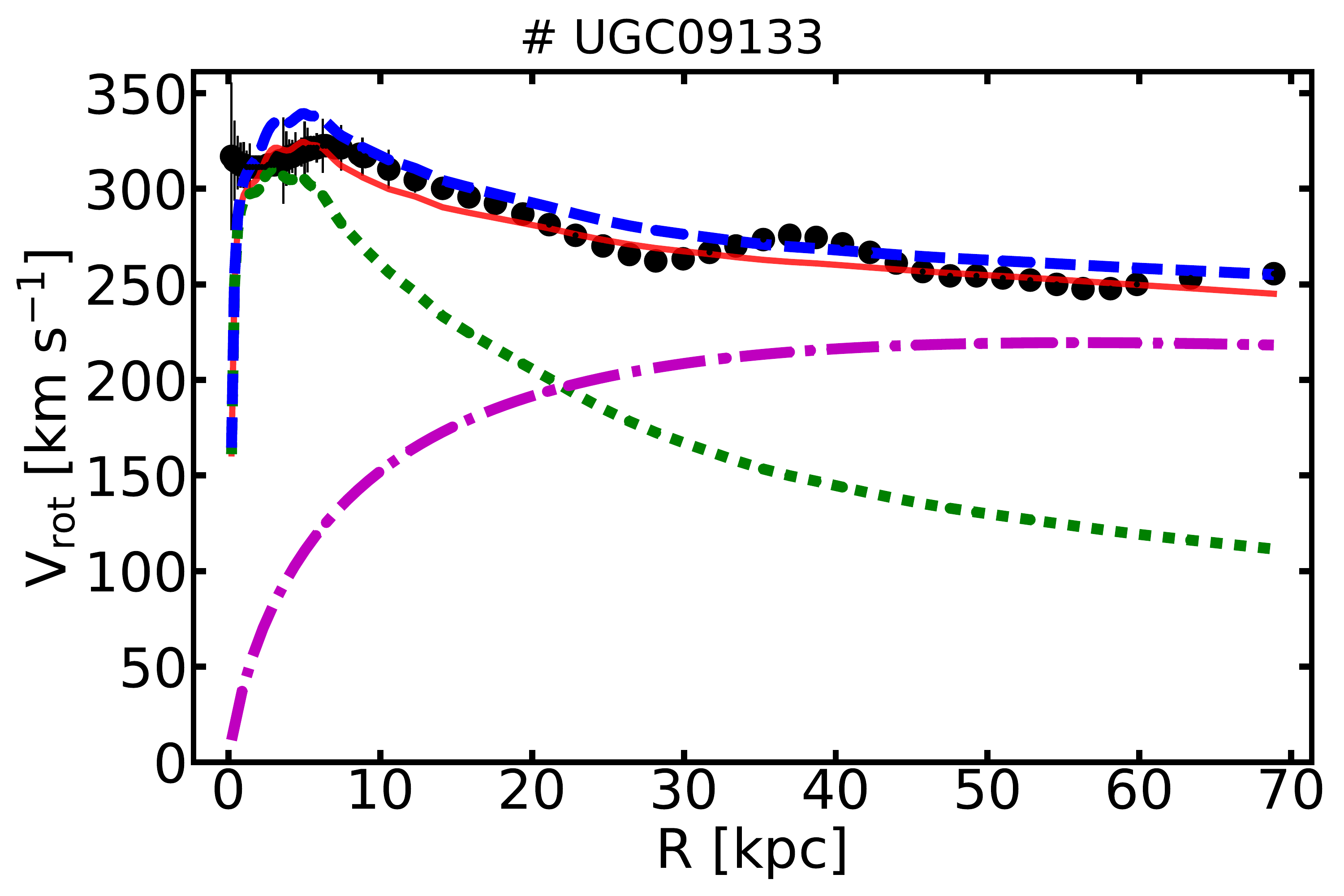}

\includegraphics[width=.33\textwidth]{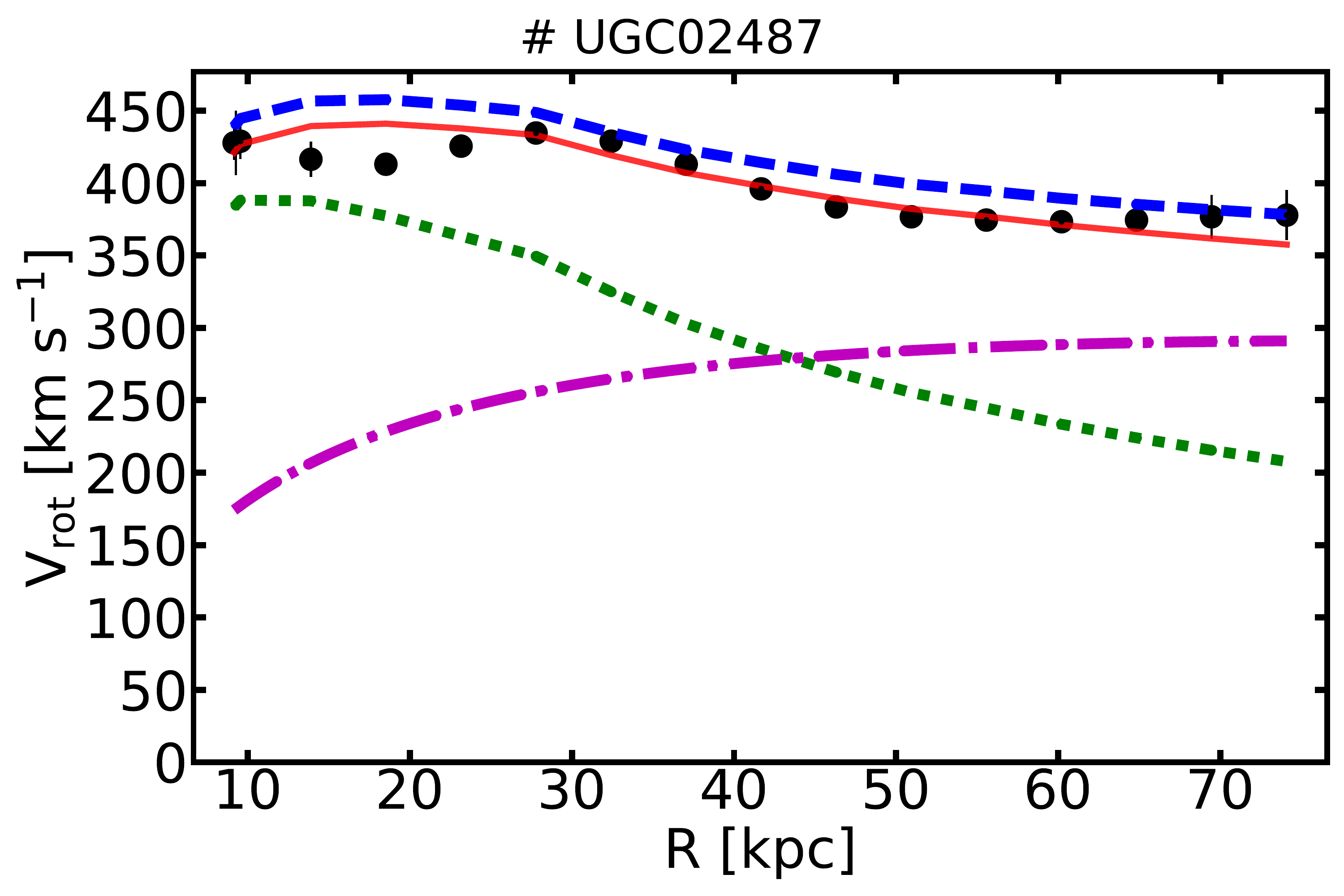}\hfill
\caption{Continued.}
\end{figure*}


\end{document}